\documentclass[12pt]{article}
\usepackage{amsfonts}
\usepackage{amsmath}
\usepackage{amssymb}
\usepackage{graphicx}
\usepackage{color}
\usepackage[all, knot]{xy}
\usepackage{tikz}

\usepackage{array}
\usepackage{hyperref}
\usepackage{booktabs}
\usepackage{needspace}

\usepackage{comment}

\usepackage[utf8]{inputenc}
\usepackage{epstopdf}
\usepackage{subcaption}
\usepackage{amsthm}
\usepackage{enumitem}
\usepackage{mathrsfs}
\usepackage{mathtools}

\usepackage[margin=3cm]{geometry}

\def \be {\begin{equation}}
\def \ee {\end{equation}}
\def \bea {\begin{eqnarray}}
\def \eea {\end{eqnarray}}

\def \rr {\raise.35ex\hbox{\small $\prime$}\kern-.17em{\mbox{\large $\imath$}}}

\def \dels {\partial\kern-.6em /\kern.1em}
\def \As {{A\kern-.5em / \kern.5em}}
\def \Ds {D\kern-.7em / \kern.5em}

\def \ks {k\kern-.5em /}
\def \ls {l\kern-.5em /}

\newcommand{\ci}[1]{}

\newcommand{\ba}{\begin{eqnarray}}
\newcommand{\ea}{\end{eqnarray}}
\newcommand{\bal}{\begin{align}}
\newcommand{\eal}{\end{align}}
\newcommand{\bay}[1]{\left(\begin{array}{#1}}
\newcommand{\eay}{\end{array}\right)}

\def \tr {{\rm Tr}}

\newcommand{\hide}[1]{}

\newlist{axioms}{enumerate}{2}
\setlist[axioms,1]{label=\textbf{A\arabic{axiomsi}.}, ref=A\arabic{axiomsi}}
\setlist[axioms,2]{label=\textbf{A\arabic{axiomsi}\rlap{\myEnumCounter{axiomsii}}.},%
                   ref=A\arabic{axiomsi}\myEnumCounter{axiomsii},%
                   align=parleft,%
                   leftmargin=0em,%
                   itemsep=1.4ex,%
                   before={\stepcounter{axiomsi}}}

  \usetikzlibrary{decorations.markings}

\usepackage{bm}
\usepackage{microtype}

\newcommand{\FF}{\mathcal F}
\newcommand{\GG}{\mathcal G}
\newcommand{\DD}{\mathcal D}
\newcommand{\XX}{\mathcal X}
\newcommand{\HH}{\mathcal H}
\newcommand{\RR}{\mathcal R}
\newcommand{\Id}{\mathsf I}

\hypersetup{pdftitle={Bulk Energy and Volume Response of Entanglement Wedge Polygons from Interval Entropies},pdfauthor={Chen-Te Ma},pdfborder={0 0 0}}
\graphicspath{{./}{figures/}}

\begin{document}

\hypersetup{pageanchor=false}
\begin{titlepage}
\begin{center}
\textbf{\LARGE
Bulk Energy and Volume Response of Entanglement Wedge Polygons from Interval Entropies
\vskip.3cm
}
\vskip .5in
{\large
Chen-Te Ma$^{a}$ \footnote{e-mail address: yefgst@gmail.com} 
\\
\vskip 1mm
}
{\sl
$^a$
Hetao Institute of Mathematics and Interdisciplinary Sciences, Shenzhen,\\ 
Guangdong 518017, China.
}
\\
\vskip 1mm
\vspace{40pt}
\end{center}
\begin{abstract}
\noindent
In three-dimensional anti-de Sitter gravity, six boundary entropies measure a quadrilateral's sides and diagonals.
Two excitations share the same six entropies and total energy up to quadratic order. Yet, one maximal volume grows while the other shrinks.
An entropy combination weights energy by position, whereas area counts enclosed energy uniformly.
Scanning intervals reconstruct energy density.
Two widths determine the volume response for the lowest dimension-$3/2$ scalar mode with a known center and definite time parity in a purely rotationally invariant family.
In vacuum, union entropies reveal paired endpoints and the area inside the union wedge left uncovered by individual wedges.
\end{abstract}
\end{titlepage}
\hypersetup{pageanchor=true}
\setcounter{tocdepth}{3}
{\hypersetup{linkcolor=black}\tableofcontents}
\newpage

\section{Introduction}
\label{sec:introduction}

\noindent
An observer measures the four sides and both diagonals of a convex quadrilateral.
On a flat sheet, the four sides and one diagonal fix the area because they determine the areas of the two triangles sharing that diagonal.
In gravity, matter can change the curvature inside those triangles.
We study this question in a setting where each length is measured by quantum entanglement on the boundary of space.
We construct two weak excitations that leave all six readings at their vacuum values through quadratic order.
The total energy increases by the same amount in both excitations at that order, yet the enclosed area grows in one and shrinks in the other.
\\

\noindent
The boundary system is a conformal field theory (CFT), a quantum theory whose spacetime symmetries preserve angles and light cones.
Its space is a circle, and its vacuum describes three-dimensional anti-de Sitter (AdS) spacetime with curvature radius $L$.
We call the gravitating interior the bulk.
Label four points on the circle in cyclic order at one time.
For $1\le i<j\le4$, tracing out the complement of arc $ij$ gives its reduced state $\rho_{ij}$ and its entropy $S_{ij}=-\operatorname{Tr}(\rho_{ij}\ln\rho_{ij})$.
In a pure state, complementary arcs have equal entropies, so four endpoints give six distinct pairs to measure.
On a slice invariant under time reflection, the Ryu--Takayanagi (RT) prescription identifies $4G_N S_{ij}$ with the length of the shortest bulk geodesic that bounds a region with the arc, where $G_N$ is Newton's constant \cite{Ryu:2006bv}.
A geodesic is locally straight in the spatial metric, and its length is regulated by stopping the curve just short of the boundary.
We use the same cutoff for all six readings.
The spacetime region determined by initial data between the arc and its geodesic is the entanglement wedge.
Four of the six geodesics join consecutive endpoints and bound the quadrilateral $Q$, and the other two form its diagonals \cite{Fujiki:2026ucr}. \\

\noindent
The vertices of $Q$ lie at the conformal boundary, where their interior angles vanish.
Such vertices are called \emph{ideal}.
For a geodesic polygon, the Gauss--Bonnet Theorem gives the integrated Gaussian curvature from the exterior turns at its corners.
At each ideal vertex, the exterior turn tends to $\pi$, so $\int_Q K\,\mathrm dA=2\pi-4\pi=-2\pi$ for the Gaussian curvature $K$.
In the vacuum, $K=-1/L^2$, giving $A_Q=2\pi L^2$ independently of the endpoint positions.
With matter present, Einstein's constraint on the reflection slice gives $K=-1/L^2+8\pi G_N\rho$, where $\rho$ is the energy density measured by an observer moving perpendicular to the slice.
Integrating the constraint gives $-2\pi=-A_Q/L^2+8\pi G_N\int_Q\rho\,\mathrm dA$, so every unit of enclosed proper energy changes the area by $8\pi G_N L^2$.
There are two spatial dimensions, so a spatial volume in this problem is an area.
Appendix~\ref{app:geometry} gives conditions under which every other admissible surface with the same edges has area at most $A_Q$.
Under those conditions, the curvature integral also determines maximal volume. \\

\noindent
We now combine the six readings so that their vacuum value is the same for any four endpoints.
Set $D_{ij}=\exp(3S_{ij}/c)$, where the central charge $c=3L/(2G_N)$ fixes the normalization of the left- and right-moving CFT stress tensors \cite{Brown:1986nw}.
In vacuum, the six $D_{ij}$ are straight chord lengths on a circle, up to a factor at each endpoint.
Ptolemy's identity for four points on a circle equates the product of the diagonal lengths to the sum of the two products of opposite side lengths \cite{Brown:2015waa,Levay:2020rzs}.
For a square of side $d$, this reads $(\sqrt{2}d)^2=d^2+d^2$.
Moving one side product to the left and dividing by the other gives the vacuum normalization for the ratio
\[
\Pi=\frac{D_{13}D_{24}-D_{14}D_{23}}{D_{12}D_{34}}
\]
which equals one in the vacuum.
To see why the ratio is independent of the regulator, multiply $D_{ij}$ by endpoint factors $q_iq_j$ with $q_i>0$ for $1\le i\le4$.
Every product acquires the same factor $q_1q_2q_3q_4$, which cancels.
Foertsch and Schroeder obtain a related inequality for boundary distances when triangles are no thicker than comparison triangles of curvature $-1$.
Other entropy combinations characterize CFT ground states through stationarity under small variations \cite{Lin:2023pvl,Kim:2024suq}. \\

\begin{figure}[t]
\centering
\includegraphics[width=.96\columnwidth]{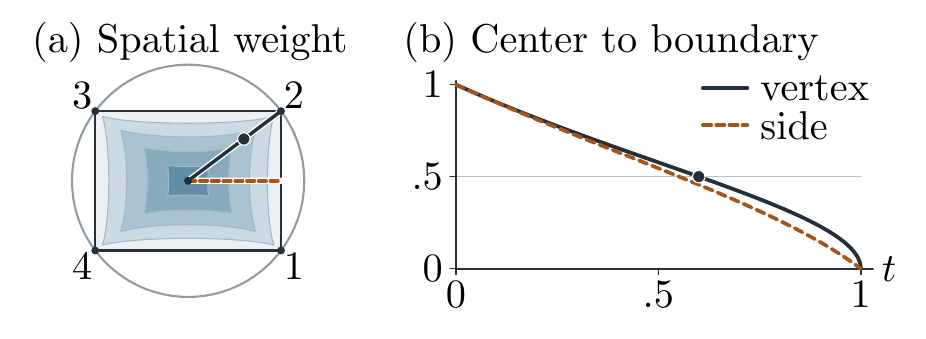}
\caption{Sensitivity of the six-interval entropy measurement to the location of energy.
In (a), geodesics appear straight in disk coordinates, and the contours inside the rectangle of half-widths $4/5$ and $3/5$ increase inward in steps of one fifth of the peak sensitivity.
In (b), the vertical axis gives the signal per unit energy relative to its value at the center, and $t$ gives the fraction of the plotted distance along each path.
The solid path goes to a vertex and the dashed path to a side midpoint.
The dots mark the same location in both panels, where the vertex path reaches $t=3/5$, and the signal is half its central value.}
\label{fig:quartetweight}
\end{figure}

\noindent
The area formula assigns the same weight to enclosed energy everywhere. However, a weak excitation changes $\Pi$ most strongly if it lies near the diagonal crossing.
At the marked point in Fig.~\ref{fig:quartetweight}, the weight falls to half its central value.
Thus, a small packet of energy $E$ near the crossing and one of energy $2E$ near the marked point give approximately the same change in $\Pi$.
Their area changes are in the ratio one to two.
The example concerns a single combination of six measurements.
Matching the measurements individually and also matching the global energy requires additional freedom in the state, constructed in Sec.~\ref{sec:energy} and Appendix~\ref{sec:matching-hessians}. \\

\noindent
We vary the state by exciting a free real scalar field while holding the boundary metric and its sources fixed.
In a coherent excitation, the mean field is $\lambda\phi_1+O(\lambda^2)$ and its fluctuations retain their vacuum values.
The amplitude $\lambda$ is dimensionless, and the scalar stress starts at order $\lambda^2$ because it is quadratic in the field.
For an observable $X$ with endpoints held fixed, the combination $[X(+\lambda)+X(-\lambda)-2X(0)]/\lambda^2$ cancels the linear change and tends to the second derivative $X''(0)$ as $\lambda\to0$.
We call this derivative the \emph{state Hessian}.
An oscillation observed at maximum displacement has vanishing momentum (the field phase). In contrast, one observed as it passes through zero has nonzero momentum (the momentum phase).
The leading field profile is even in time about the reflection slice in the field phase and odd in the momentum phase.
We call this distinction time parity.
Both preparations admit a time-reflection-symmetric geometry whose reflection slice has zero extrinsic curvature, the tensor measuring its bending in spacetime. \\

\noindent
For an explicit example, take the lowest rotationally invariant scalar mode with scaling dimension $3/2$, the exponent that describes how its boundary operator transforms under a rescaling of length.
In the field phase, the entropy of every proper interval increases quadratically.
The scalar has a negative mass squared in the stable AdS range, so its potential energy gives a negative central density. In contrast, the gradient energy dominates farther out.
A centered regular triangle encloses too little of that outer energy to balance the core, a square encloses just enough, and a pentagon encloses more.
Consequently, the triangle shrinks, the square has zero quadratic area response, and the pentagon grows (Fig.~\ref{fig:signs}).
As the state varies, we follow the same family of shortest geodesics, so no competing geodesic takes over.
This condition defines a fixed minimizing branch.
On this branch, we first take the state derivatives at $\lambda=0$ and then take the classical limit $c\to\infty$ with $\sqrt{G_N}\phi_1$ fixed.
For more general states, the quantum entropy of the bulk fields can also contribute to the boundary entropy \cite{Faulkner:2013ana,Belin:2018juv}. \\

\noindent
A change in an interval's entropy can also come from gravity in the absence of any scalar excitation.
A conformal reparametrization of the boundary circle produces such a state, called a vacuum descendant, and its bulk gravitational perturbation is called a boundary graviton.
We compare the excitation with a descendant having the same first-order boundary stress and reparametrization to isolate the matter contribution \cite{Calabrese:2004eu,Beach:2016ocq,deBoer:2016pqk,Hollands:2019czd,Panebianco:2019plp}.
For interval $I$, the relative entropy $\operatorname{Tr}[\rho_I(\ln\rho_I-\ln\rho_{I,0})]$ quantifies the distinguishability of the excited and vacuum states when only $I$ is accessible.
Define the vacuum modular Hamiltonian by $K_I=-\ln\rho_{I,0}$.
The two terms in the trace give $-S_I$ and $\langle K_I\rangle$, and in vacuum $\langle K_I\rangle_0=S_{I,0}$.
Thus, the relative entropy equals $\Delta\langle K_I\rangle-\Delta S_I$, with both changes taken from vacuum.
Boundary stress gives the first change and interval entropy gives the second.
In the bulk, the Hessian of this difference gives canonical energy, the quadratic energy associated with the evolution generated by $K_I$ \cite{Blanco:2013joa,Casini:2011kv,Cardy:2016fqc,Lashkari:2015hha,Jafferis:2015del}.
Earlier relations between quantum information and volume use different regions or perturbations \cite{Banerjee:2017qti,Bernamonti:2020bcf,Bhattacharya:2019zkb}.
Throughout the comparison, the states share the same boundary conditions, cutoff, and minimizing branch \cite{deHaro:2000vlm,Hartman:2013mia,Headrick:2010zt}. \\

\noindent
There are two ways to gain the information missing from a fixed quadrilateral.
The first is to measure more intervals.
When the four endpoints approach two distinct points in pairs, the quadrilateral becomes narrow, and its rescaled signal gives the energy integrated along the limiting geodesic.
Repeating the measurement for every position and orientation reconstructs the density from these line integrals, a procedure called tomography.
A related result in geometry, Lefeuvre's boundary rigidity theorem, determines a simply connected, negatively curved, asymptotically hyperbolic surface from all its renormalized geodesic lengths, up to a coordinate change fixing the boundary.
In holography, differential equations for interval entropies likewise relate their variations to curvature and stress \cite{Chae:2026wnp,Nozaki:2013vta,Lin:2014hva,Czech:2016tqr,Cao:2020uvb,Czech:2015qta,deBoer:2015kda}.
The second way is to supply independent information about the preparation.
For the nonzero lowest dimension-$3/2$ mode with a known center and definite time parity in a pure rotationally invariant family, the phase and squared amplitude are the only unknowns.
The semicircle response is smaller than the quarter-circle response in the field phase and larger in the momentum phase.
Once their ordering fixes the phase, either response fixes the squared amplitude (Fig.~\ref{fig:two-width-phase}).
A different inference is possible even in vacuum, where the scalar energy vanishes.
For separated intervals, the minimizing geodesics of their union can join endpoints across a gap.
If the minimizing pairing persists, shifting two endpoints separately and together tests whether they share a geodesic.
Repeating this test for every candidate pair gives the area inside the union wedge and outside the individual wedges (Sec.~\ref{sec:vacuum-pairing} and Fig.~\ref{fig:two-interval-pairing}).

\subsection{Summary of Results}
\label{sec:summary-results}

\noindent
We distinguish measurements that bound an energy from those that determine a polygon's volume response.
A double prime denotes the second derivative in the excitation amplitude $\lambda$ at fixed endpoints, so the quadratic change from vacuum is $\lambda^2X''(0)/2$ for an observable $X$.
The energy density seen perpendicular to the reflection slice is $\lambda^2e/L^3+o(\lambda^2)$, and $\mathrm d\mu=\mathrm dA_0/L^2$ is the vacuum area element in units of $L^2$.
\begin{itemize}
\item
At four endpoints, the six interval entropies define a ratio $\Pi$.
Its response and the area response of the quadrilateral $Q$ weight the same energy differently.
Secs.~\ref{sec:polygon-area-law} and \ref{sec:quartet-weights} express the responses as integrals over the same vacuum quadrilateral,
\begin{equation}
\frac c3\Pi''(0)=4\pi\int_Q K_Q e\,\mathrm d\mu,
\qquad
\frac{A_Q''(0)}{L^2}=\frac{24\pi}{c}\int_Q e\,\mathrm d\mu .
\label{eq:two-readings-intro}
\end{equation}
The entropy weight $K_Q$ peaks at the diagonal crossing, while the area weights every unit of energy equally (Fig.~\ref{fig:quartetweight}).
A bound on spatial variance therefore gives sharp bounds on the mean of a square-integrable density at fixed entropy response.
The area equals the maximal volume under the causal and curvature hypotheses of Appendices~\ref{app:causalclass} and \ref{app:globalmaximum}, with a scalar-field test in Appendix~\ref{app:potentialbound} \cite{Headrick:2014cta,Burnett:1995rf}.
Away from time reflection, a term measuring surface bending also enters (Appendix~\ref{app:detailA7}).
\item
A negative ratio response detects negative energy because $K_Q$ is positive inside $Q$.
Knowing the total inside $Q$ also bounds its integrated negative part sharply (Sec.~\ref{sec:signed-energy-bound}).
For unrestricted smooth densities and $N\ge4$ endpoints, $(N-2)(N-3)/2$ quartet readings are independent (Sec.~\ref{sec:quartet-rank}).
For positive measures, placing masses at chord crossings gives the least compatible energy (Sec.~\ref{sec:pentagon-program}).
Even six individual entropies and a global energy need not fix whether the volume grows.
For a scalar with negative mass squared above the AdS stability bound \cite{Breitenlohner:1982jf,Klebanov:1999tb}, Sec.~\ref{sec:energy} and Appendix~\ref{sec:matching-hessians} construct two families with opposite volume responses.
Both match all six vacuum entropies through quadratic order and have equal positive global-energy coefficients.
At small nonzero amplitude, exact classical interval data on any prescribed finite list and the initial boundary stress can still leave the area undetermined (Appendix~\ref{app:finiteexact}).
Tab.~\ref{tab:data-conditions} compares these finite measurements.

\begin{table}[tbp]
\centering
\caption{What can be inferred about a bulk polygon from finite boundary measurements.
In rows one through four, a response is the second derivative with respect to excitation amplitude at the vacuum, and energy means the coefficient of $\lambda^2$.
The last row compares entropy values at the same nonzero amplitude.}
\label{tab:data-conditions}
\small
\begin{tabular}{@{}p{.32\linewidth}p{.27\linewidth}p{.33\linewidth}@{}}
\toprule
Measurements & Assumptions & What follows \\
\midrule
Five entropy-ratio responses, each omitting one vertex of a regular pentagon
& Smooth nonnegative density and compatible responses, not all zero
& Enclosed energy has a sharp lower bound.
Every larger value occurs (Sec.~\ref{sec:pentagon-program}). \\[.6em]
Six separate entropy responses at four endpoints
& Scalar energy of either sign, with no bound on total energy
& Any finite volume response fits the readings (Appendix~\ref{sec:matching-hessians}). \\[.6em]
Six entropies agreeing with vacuum through order $\lambda^2$, plus total energy
& Initially stationary scalar with mass squared in $(-1/L^2,0)$
& At the same positive total-energy coefficient, either sign of volume change occurs (Sec.~\ref{sec:energy}). \\[.6em]
Semicircle and quarter-circle entropy responses
& Pure rotationally invariant family, nonzero lowest $\Delta=3/2$ mode, known center, definite time parity
& Phase, squared amplitude, and every polygon's volume response are fixed (Sec.~\ref{sec:square-calibration}). \\[.6em]
Finitely many interval entropies and the initial boundary stress
& Smooth nonnegative scalar density, fixed minimizing geodesic branch
& Identical data permit different polygon areas (Appendix~\ref{app:finiteexact}). \\
\bottomrule
\end{tabular}
\end{table}

\item
A continuous scan can recover the density.
Narrowing a quadrilateral onto one geodesic gives a line integral of energy after rescaling (Sec.~\ref{sec:chord-limit}).
In disk coordinates, these line integrals give the Radon transform of a rescaled density \cite{Facchi:2009nls}.
All offsets and directions determine the density under the conditions of Appendix~\ref{app:radoninversion}. In contrast, any finite set of directions can miss a change in polygon energy.
For a chosen polygon, the line integrals also give its area response directly, with error bounds for noisy grid data under a smoothness assumption (Appendices~\ref{app:detailC4} and \ref{app:detailC6a}).
The spreading of neighboring geodesics bounds the integrated curvature excess along a perturbed geodesic (Appendix~\ref{app:finitefocus}) \cite{Wald:1984rg}.
For a short interval centered where boundary stress vanishes but its slope is nonzero, the gravitational relative-entropy response scales as the sixth power of its length (Appendices~\ref{app:detailB2} and \ref{app:detailB3}).
Removing the stress contribution selects the normal energy (Appendix~\ref{app:localentropy}), while lightlike endpoint shifts probe other stress components \cite{Bhattacharya:2014vja}.
\item
Two interval widths can suffice when the excitation is a known scalar mode.
For the lowest rotational mode, the field and momentum phases correspond to maximum displacement and passage through zero.
Appendices~\ref{app:detailD2} and \ref{app:phasegap} derive the difference between their relative-entropy responses from the vacuum two-point function \cite{Lashkari:2015dia,Faulkner:2017tkh,Sarosi:2017rsq,Petz:1993er}.
In the $\Delta=3/2$ field phase, all proper arcs gain entropy to quadratic order, yet among centered regular polygons, the triangle shrinks, the square has zero response, and every polygon with more sides grows (Sec.~\ref{sec:interval-and-polygon-signs} and Appendix~\ref{app:onephase}).
The triangle calculation includes the motion of its sides (Appendix~\ref{app:fieldflux32}) \cite{Ghosh:2019jgd}.
For a pure rotationally invariant family in this nonzero mode with known center and definite time parity, the ratio of semicircle to quarter-circle responses cancels the amplitude and identifies the phase.
Either reading then fixes the squared amplitude and hence every polygon's volume response (Appendix~\ref{sec:two-width-proof} and Sec.~\ref{sec:square-calibration}).
Fig.~\ref{fig:two-width-phase} shows when the two phases remain distinguishable at finite precision.
\item
Sec.~\ref{sec:boosted-quartets} shows how the location of the lowest dimension-$3/2$ scalar mode affects what boundary intervals measure in the field phase.
For a centered excitation, scanning chords at different distances from the center reconstructs a negative-energy core surrounded by positive energy (Sec.~\ref{sec:angular-chords}).
Displacing the excitation relative to the endpoints also makes the readings depend on orientation.
Two perpendicular diameters can give energy integrals of opposite sign, and scanning all chord positions and orientations recovers the displaced density.
Sec.~\ref{sec:stationary-quartets} holds the excitation fixed and instead moves the four endpoints through a family with cross ratio $1/2$.
The quadratic volume response vanishes throughout this family because the enclosed positive and negative energies cancel. In contrast, the weighted combination of the six entropy responses changes sign exactly once.
At that sign change, both responses vanish even though the quadrilateral contains energy of both signs.
Fig.~\ref{fig:squarecertificate}(d) shows how replacing one endpoint by a fifth can give a negative reading and thereby certify negative energy somewhere inside the new quadrilateral.
\item
In vacuum, shifting two endpoints of $m\ge2$ separated intervals tests whether a geodesic joins them.
On a fixed minimizing branch, subtracting the joint change in union entropy from the sum of the separate changes isolates the shared edge, if present (Sec.~\ref{app:detailG1}).
Alternating geodesic partners with interval partners gives $C$ cycles and uncovered area $2\pi L^2(m-C)$ (Sec.~\ref{app:detailG5}) \cite{Headrick:2010zt,Abt:2017pmf,Abt:2018ywl}.
Sec.~\ref{app:detailG3} certifies $m-C$ if the error matrix of normalized readings has operator norm below $\sin(\pi/m)$.
The norm is the largest error on a unit vector.
A detection threshold tending to zero can count subleading entropy corrections as area (Appendix~\ref{app:detailG4}) \cite{Casini:2008wt}.
Two states can have the same two conditional mutual informations used here and still differ in how faithfully a missing subsystem can be recovered (Appendix~\ref{app:detailH2}) \cite{Lieb:1973zz,Fawzi:2015eym,Devetak:2008oqm}.
\end{itemize}

\noindent
Sec.~\ref{sec:energy-local} relates the area of an ideal polygon to its enclosed energy and specifies when this area measures maximal volume.
Sec.~\ref{sec:quartets-main} derives the weight with which four boundary endpoints sample that energy and uses overlapping quartets to bound a nonnegative density.
Sec.~\ref{sec:dual} takes the narrow-quartet limit and reconstructs the energy in a chosen polygon from a scan over geodesics.
Sec.~\ref{sec:scalar-phases} compares interval and polygon responses in the field and momentum phases of a scalar excitation.
Sec.~\ref{sec:square-calibration} determines the phase and squared amplitude of a known mode from two interval widths, then follows the signal as the center or four endpoints move (Sec.~\ref{sec:boosted-quartets}).
Sec.~\ref{sec:sharp} constructs states with matching finite boundary data and different enclosed energies.
Sec.~\ref{sec:vacuum-pairing} recovers the uncovered vacuum area from endpoint pairings, and Sec.~\ref{sec:discussion} discusses the results and open questions. 
Appendix~\ref{app:geometry} establishes the conditions under which a reflection-slice area gives the maximal volume.
Appendix~\ref{app:quartet-methods} derives the quartet weight from geodesic lengths, evaluates its moments, and proves the rank and sharp energy bounds for several quartets.
Appendix~\ref{app:geodesicfocus} relates interval entropy to geodesic focusing and quantifies the error at finite quartet width.
Appendix~\ref{app:gravity} derives the boundary-stress subtraction that isolates scalar energy.
Appendix~\ref{app:tomography} derives the inversion from line integrals, the weights for a chosen polygon, and error bounds for noisy or finitely sampled entropies.
Appendix~\ref{app:cft} obtains the two scalar phases' interval responses from the vacuum two-point function.
Appendix~\ref{app:scalar} calculates their polygon responses, accounts for the motion of the sides, and proves the angular reconstruction and the sign change for moving quartets.
Appendix~\ref{app:finite} constructs states with matching boundary data and derives the bounds from measurement errors and conserved energy.
Appendix~\ref{app:vacuum-support} derives the accuracy needed to distinguish vacuum pairings, treats changes of minimizing branch, and examines what the pairing data determine about matter and quantum recovery.

\section{Bulk Energy and Ideal Polygon Volume}
\label{sec:energy-local}

\noindent
The gravitational constraint relates the area of a polygon on the reflection slice to the proper energy it encloses.
Identifying this area with the maximal volume requires a second step, since other spacelike surfaces can span the same edges. \\

\noindent
How does matter change the area enclosed by geodesics whose endpoints are fixed at the boundary?
We vary a pure state on the boundary circle with a real amplitude $\lambda$, keeping the boundary metric and sources fixed.
At $\lambda=0$, the bulk is the vacuum of three-dimensional Einstein gravity with cosmological constant $-L^{-2}$.
For an observable $X$, we write $X_1=X'(0)$ and $X_2=X''(0)$, so its quadratic Taylor coefficient is $X_2/2$.
As $\lambda\to0$, the ratio of $O(|\lambda|^q)$ to $|\lambda|^q$ stays bounded, and that of $o(|\lambda|^q)$ tends to zero.
For a fixed endpoint pairing, the Ryu--Takayanagi (RT) prescription divides the length of the minimizing bulk geodesic by $4G_N$ to obtain the interval entropy, with $G_N$ Newton's constant.
We use the same boundary cutoff for all lengths because each diverges at its endpoints.
The matter is a free, minimally coupled real scalar with mass squared $M_\phi^2$ and potential $M_\phi^2\phi^2/2$.
In AdS, stability permits negative $M_\phi^2$ down to the Breitenlohner--Freedman bound $M_\phi^2L^2=-1$.
Here, we work strictly above the bound with vanishing boundary source.
On the time-reflection slice, the scalar can have either parity under time reversal.
Its stress tensor and the metric are even in both cases, so the extrinsic curvature, which measures the change of the induced metric along the slice normal, vanishes there.
We use coherent states, in which the mean field is displaced to $\phi_\lambda=\lambda\phi_1+O(\lambda^2)$. At the same time, the connected free-field covariance retains its vacuum value. \\

\noindent
Write $h_0=L^2q$ for the spatial vacuum metric, where $q$ has Gaussian curvature $-1$, covariant derivative $D$, and area measure $d\mu=dA_0/L^2$.
Let $n$ be the future unit normal in the physical metric, and let $\pi_\phi=L n^\mu\partial_\mu\phi_1$ be the rescaled normal derivative of the linear scalar.
Near the conformal boundary, choose a radial coordinate $z$ that is positive in the bulk, with $z=0$ and $dz\ne0$ at the boundary.
The rescaled metric $z^2h_0$ is regular there.
In standard quantization, the coefficient of the slower of the two scalar falloffs is the boundary source.
Setting it to zero leaves the faster falloff, so $\phi_1$, $|D\phi_1|_q$, and $\pi_\phi$ are $O(z^\Delta)$ as $z\downarrow0$, with $\Delta=1+\sqrt{1+M_\phi^2L^2}>1$.
Greek spacetime indices range over $0,1,2$, and spatial tensor indices $a,b$ range over $1,2$.
Repeated upper and lower tensor indices are contracted over these ranges.
An observer moving with the slice normal $n$ measures the normal energy density $\rho_\lambda=T_{\mu\nu}n^\mu n^\nu$.
The vacuum scalar vanishes, so the quadratic stress tensor first contributes at order $\lambda^2$.
The physical spatial gradient squared is $|D\phi_1|_q^2/L^2$, and the normal time derivative is $\pi_\phi/L$.
Substituting both into the scalar stress tensor gives
\begin{equation}
\rho_\lambda=T_{\mu\nu}n^\mu n^\nu=\frac{\lambda^2}{L^3}e+o(\lambda^2),
\qquad
\frac{2e}{L}=|D\phi_1|_q^2+\pi_\phi^2+M_\phi^2L^2\phi_1^2 .
\label{eq:energy-density}
\end{equation}
We write $\rho_2=e/L^3$ for the coefficient of $\lambda^2$ in $\rho_\lambda$ and $S_2=S''(0)$ for the entropy Hessian, so $\rho_\lambda''(0)=2\rho_2$.
The proper-energy coefficient $T_P=\int_{P_0}e\,d\mu$ sums the local density over the vacuum polygon $P_0$ with uniform weight per unit proper area.
To obtain the conserved energy of global AdS time, one weights each packet by $\cosh\chi$, where $L\chi$ is its proper distance from the origin.
This factor converts a time step on the clock at the origin into the proper time elapsed on a stationary clock at the packet.
Our hyperbolic-function conventions are $\cosh x=[\exp(x)+\exp(-x)]/2$, $\sinh x=[\exp(x)-\exp(-x)]/2$, and $\tanh x=\sinh x/\cosh x$.
In Sec.~\ref{sec:energy}, we use this distinction to bound $T_P$ from independent boundary energy data. \\

\noindent
The central charge $c=3L/(2G_N)$ of each chiral sector sets the strength of the leading classical entropy response \cite{Brown:1986nw}.
We take state derivatives at $\lambda=0$ before the large-$c$ limit, holding $\sqrt{G_N}\phi_1$ fixed in that limit.
A coherent displacement preserves the free field's entropy in a fixed vacuum region, leaving the classical change of geometry at leading order.
More general states require a separate calculation of bulk quantum entropy \cite{Faulkner:2013ana,Belin:2018juv}.
The scalar stress starts at order $\lambda^2$. However, the metric can also change at first order via a smooth reparametrization of the vacuum boundary.
We allow such a perturbation, called a boundary graviton, induced by a bulk diffeomorphism with the same boundary action in both chiral sectors.
The Hamiltonian constraint gives the enclosed area on the reflection slice.
To identify that area with volume, one must also establish that this slice maximizes the area among the allowed fillings.
Appendix~\ref{app:geometry} gives the causal and curvature conditions for this comparison and shows how the motion of the geodesic sides enters the area variation.

\subsection{From Hamiltonian Constraint to Polygon Area}
\label{sec:polygon-area-law}
\noindent
On the reflection slice, Einstein's constraint reads $K_{\rm G}=-L^{-2}+8\pi G_N\rho_\lambda$.
Its constant term integrates to $-A(\lambda)/L^2$, so solving for the area requires integrating the Gaussian curvature $K_{\rm G}$.
The Gauss--Bonnet Theorem gives that integral as $2\pi\chi(P_\lambda)$ minus the total boundary turn, where $\chi(P_\lambda)$ is the Euler characteristic.
A geodesic side contributes no turn along its length.
At each ideal vertex, the short cutoff arc has vanishing integrated curvature in the limit. However, the two joins with the sides each retain a right-angle turn.
Counting these vertices with multiplicity gives
\begin{align}
\int_{P_\lambda}K_{\rm G}\,dA_\lambda&=2\pi\chi(P_\lambda)-\pi N_v,\nonumber\\
\frac{A(\lambda)}{L^2}&=\pi[N_v-2\chi(P_\lambda)]
+8\pi G_N\int_{P_\lambda}\rho_\lambda\,dA_\lambda .
\label{eq:GBexact}
\end{align}
For a single quadrilateral $Q_\lambda$, the four vertices give $N_v=4$, and the disk topology gives $\chi=1$.
The vacuum area is therefore $2\pi L^2$ for every ideal quadrilateral, no matter how far apart its four endpoints lie.
In the presence of matter, the full relation becomes
\begin{equation}
A_Q(\lambda)/L^2=2\pi+8\pi G_N\int_{Q_\lambda}\rho_\lambda\,dA_\lambda .
\label{eq:gauss_bonnet_area}
\end{equation}
Write $Q=Q_0$ and $T_Q=\int_Q e\,d\mu$.
To see where to integrate the leading density, truncate the ideal vertices and use a smooth map $F_\lambda$ from the vacuum polygon to its perturbed shape, with $F_0$ the identity.
In the pulled-back integral, the density is $\rho_\lambda\circ F_\lambda=\lambda^2\rho_2+o(\lambda^2)$ and the area Jacobian tends to one.
Their product has coefficient $\rho_2$ on the vacuum polygon because the density has neither a constant nor a linear term.
Appendix~\ref{app:detailA6} justifies removal of the cutoff at each ideal tip.
The resulting area responses are
\begin{align}
\frac{A''(0)}{L^2}
&=16\pi G_N\int_{P_0}\rho_2\,dA_0
=\frac{24\pi}{c}\int_{P_0}e\,d\mu,\qquad \rho_2=e/L^3,
\label{eq:geometricresponse}\\
\frac{A_Q(\lambda)}{L^2}
&=2\pi+\frac{12\pi}{c}\lambda^2 T_Q+o(\lambda^2),
\qquad \frac{A_Q''(0)}{L^2}=\frac{24\pi}{c}T_Q .
\label{eq:area_response}
\end{align}
In Eq.~\eqref{eq:geometricresponse}, $\rho_2\,dA_0=e\,d\mu/L$ and $G_N/L=3/(2c)$ give the factor $24\pi/c$.
The coefficient of $\lambda^2$ in the area itself is half as large as in Eq.~\eqref{eq:area_response}. \\

\noindent
The area $A$ is measured on one time slice, whereas the volume $V_P$ allows a spacelike surface to bend in time while its boundary stays on the same RT edges.
We take the supremum of the areas of the Cauchy fillings on the prescribed side of the edges, called their homology class \cite{Fujiki:2026ucr,Headrick:2014cta}.
The reflection polygon is an admissible filling, so $V_P\ge A$.
For the opposite inequality, transport a small patch of the polygon along timelike geodesics leaving it orthogonally.
It starts with zero expansion, and nonnegative timelike Ricci curvature $\operatorname{Ric}(v,v)=R_{\mu\nu}v^\mu v^\nu$ for unit timelike $v$ keeps it from growing before a focus.
If the competing filling intersects the transported patch at a tilt, its area decreases further.
Taking the longest normal paths to that filling also accounts for crossings of distinct geodesics \cite{Burnett:1995rf}.
Appendix~\ref{app:geometry} gives the causal hypotheses for this argument and shows that the curvature test is needed only along those longest paths.
For the scalar potential above, $\operatorname{Ric}(v,v)=2/L^2+8\pi G_N[(v^\mu\partial_\mu\phi)^2-M_\phi^2\phi^2]$.
The curvature hypothesis is automatic for $M_\phi^2\le0$ and follows for positive $M_\phi^2$ from the uniform bound in Eq.~\eqref{eq:mainpotential}.
The area Hessian itself follows directly from the slice constraint, including in state families for which maximality has yet to be established.

\section{Quartet Weights and Bounds on Enclosed Energy}
\label{sec:quartets-main}

\noindent
We combine the six entropies at four endpoints to determine how strongly energy at each position contributes to the boundary signal.
We then use that weight to bound the total energy in a polygon from one or more quartet measurements. 
Four boundary endpoints determine six intervals, one for each pair.
In the vacuum, $\exp(3S_{ij}/c)$ is proportional to the Euclidean chord joining endpoints $i,j$, so Ptolemy's theorem fixes one relation among the six entropies.
Choose unwrapped angles $x_1<x_2<x_3<x_4<x_1+2\pi$ to follow arcs continuously across $2\pi$.
For $1\le i<j\le4$, $S_{ij}=-\operatorname{Tr}(\rho_{ij}\ln\rho_{ij})$ is the von Neumann entropy of the reduced state on the counterclockwise arc from $x_i$ to $x_j$.
Set $t_c=3/c$ and $D_{ij}=\exp(t_c S_{ij})$, and denote the diagonal product by $A=D_{13}D_{24}$ and the two opposite-side products by $B_s=D_{12}D_{34}$ and $C_s=D_{14}D_{23}$.
The vacuum relation $A=B_s+C_s$ motivates the ordered ratio
\begin{equation}
\Pi=\frac{A-C_s}{B_s}
=\frac{D_{13}D_{24}-D_{14}D_{23}}{D_{12}D_{34}} .
\label{eq:ptolemy_entropy_ratio}
\end{equation}
The ordered ratio uses $B_s$ as its reference.
For a comparison invariant under exchange of the two side products, divide $A$ by $B_s+C_s$ and take a logarithm.
The vacuum fraction assigned to $B_s$ is the cross ratio $\eta_Q$,
\begin{equation}
\exp(t_cR_Q)=\frac{A}{B_s+C_s},
\qquad \eta_Q=\frac{B_s^0}{A^0}\in(0,1).
\label{eq:bpi}
\end{equation}
A superscript $0$ denotes vacuum values, and we suppress $Q$ for a single quartet.
Every product contains each endpoint once, so its four local cutoff factors cancel in either ratio.
The vacuum chords $D^0_{ij}=2\sin[(x_j-x_i)/2]/\epsilon_{\rm UV}$ obey $A^0=B_s^0+C_s^0$, giving $\Pi(0)=1$ and $R_Q(0)=0$ \cite{Brown:2015waa,Levay:2020rzs}.
Away from the vacuum, both differences have the same numerator $A-B_s-C_s$.
The ordered one divides by $B_s$, and the symmetric one by $B_s+C_s$.
Their conversion factor is therefore
\begin{equation}
\exp(t_cR_Q)-1=p(\lambda)(\Pi-1),
\qquad p(\lambda)=\frac{B_s}{B_s+C_s},\qquad p(0)=\eta_Q.
\label{eq:quartet-conversion}
\end{equation}
Cyclic relabeling exchanges $B_s$ and $C_s$ while leaving $R_Q$ fixed.
The relabeled ordered ratio thus obeys $(1-p)(\Pi_{\rm cyc}-1)=p(\Pi-1)$. \\

\noindent 
Appendix~\ref{app:geometry} specifies when the resulting area bounds also constrain maximal volume. 
Appendix~\ref{app:quartet-methods} derives the quartet weight and proves the bounds for overlapping quartets. 
The matter contribution uses the boundary-stress subtraction in Appendix~\ref{app:gravity} and the normalization of the energy integral in Appendix~\ref{app:tomography}. 
Appendix~\ref{app:finite} constructs smooth scalar data used to test sharpness.

\subsection{State Hessian and Boundary-Graviton Subtraction}
\label{sec:quartet-state-hessian}
\noindent
In the logarithm of $B_s+C_s$, the logarithmic derivatives of the side products enter with their respective shares of the sum.
For six interval quantities $U_{ij}$, write $C_{\eta_Q}$ for the resulting combination of diagonal and side pairs, and divide it by $\eta_Q$ for the ordered combination $\mathcal L_Q$,
\begin{align}
C_{\eta_Q} U=\eta_Q\mathcal L_Q U={}&U_{13}+U_{24}
-(1-\eta_Q)(U_{14}+U_{23})\nonumber\\
&-\eta_Q(U_{12}+U_{34}).
\label{eq:quartet_linear_main}
\end{align}
In a pure state, differences of the same six entropies also give two conditional mutual informations.
Each measures correlations between two separated arcs after conditioning on an intervening arc,
\begin{align}
I_{1,3}&=S_{13}+S_{24}-S_{14}-S_{23},\nonumber\\
I_{2,4}&=S_{13}+S_{24}-S_{12}-S_{34}.
\label{eq:measured_cmis}
\end{align}
Here, $I(A:C\mid B)=S(AB)+S(BC)-S(B)-S(ABC)$, with juxtaposition denoting union.
If $A_j$ denotes the arc $(x_j,x_{j+1})$ for $j=1,2,3,4$, with $x_5=x_1+2\pi$, purity identifies the two quantities in Eq.~\eqref{eq:measured_cmis} with $I(A_1:A_3\mid A_2)$ and $I(A_2:A_4\mid A_3)$, respectively.
Strong subadditivity makes each nonnegative \cite{Czech:2015qta,Lieb:1973zz}.
Write $I_1=I_{1,3}$ and $I_2=I_{2,4}$.
From their definitions, $\exp(-t_cI_1)=C_s/A$ and $\exp(-t_cI_2)=B_s/A$.
Their sum is $\exp(-t_cR_Q)$, and the relative weight of the second summand is $p$ in Eq.~\eqref{eq:quartet-conversion}.
Differentiating at a general $\lambda$ gives $R_Q'=(1-p)I_1'+pI_2'$ and $p'=t_cp(1-p)(I_1'-I_2')$.
At the vacuum, these identities yield
\begin{align}
R_Q''&=(1-\eta_Q)I_1''+\eta_Q I_2''
-t_c\eta_Q(1-\eta_Q)(I_1'-I_2')^2 ,
\label{eq:measured_quartet_hessian}\\
&= C_{\eta_Q}S_2-t_c\eta_Q(1-\eta_Q)(I_1'-I_2')^2 .
\label{eq:ptolemy_hessian}
\end{align}
Here, the state Hessian is the second derivative along the specified family at $\lambda=0$.
Differentiating the weights in $R_Q'$ contributes $p'(I_2'-I_1')$, which is the negative square above and vanishes for equal initial slopes. \\

\noindent
For the increasing boundary reparametrization $f_\lambda$ that matches the linear metric and stress tangents, the exact descendant interval entropy is \cite{Calabrese:2004eu,Beach:2016ocq,deBoer:2016pqk}
\begin{equation}
S_{ij,f}=\frac c3\ln\!\left[
\frac{2\sin[(f_\lambda(x_j)-f_\lambda(x_i))/2]}
{\epsilon_{\rm UV}\sqrt{f_\lambda'(x_i) f_\lambda'(x_j)}}\right],
\label{eq:descendant_entropy}
\end{equation}
where the primes on $f_\lambda$ in this equation denote spatial derivatives.
Its normalized quartet has $\Pi_f=1$ for all $\lambda$ by vacuum Ptolemy and cancellation of the four endpoint derivatives.
For a reparametrized vacuum, the interval relative entropy is nonzero whenever the reparametrization changes the reduced state \cite{Hollands:2019czd,Panebianco:2019plp}.
By the entanglement first law, matching the first-order boundary stress matches each of the six interval entropy slopes.
The descendant obeys the chord identity at all amplitudes, so agreement of the slopes gives $R_Q'(0)=\Pi'(0)=0$ for the state as well.
Subtracting the descendant Hessian in Eq.~\eqref{eq:ptolemy_hessian} cancels the squared slope difference even when individual slopes are nonzero.
To convert the remaining Hessian, differentiate Eq.~\eqref{eq:quartet-conversion} twice at the vacuum, obtaining $t_cR_Q''+t_c^2(R_Q')^2=\eta_Q\Pi''+2p'(0)\Pi'$ because $\Pi(0)=1$.
Both first derivatives vanish, leaving
\begin{equation}
R_Q''(0)=C_{\eta_Q}(S_2-S_{2,f})=\frac c3\eta_Q\Pi''(0).
\label{eq:bC}
\end{equation}
In particular, under a cyclic relabeling $\Pi''(0)$ rescales by $\eta_Q/(1-\eta_Q)$.
For a rotationally invariant pure state on an equally spaced quartet, $R_{\square}=2[S(\pi/2)-S(\pi/4)]-(c/3)\ln2$, with $S(\alpha)$ indexed by the half-opening.
Its Hessian yields the two-width measurement in Eq.~\eqref{eq:square_readout_main}. \\

\noindent
For a square, the two terms in Eq.~\eqref{eq:measured_quartet_hessian} can be nonzero even for a reparametrized vacuum.
Take $x_j=(j-1)\pi/2$ for $j=1,2,3,4$ and $f_\lambda(x)=x+\lambda\cos(2x)$ with $|\lambda|<1/2$.
Both diagonals remain diameters, while the openings of arcs $12,34$ become $\pi/2-2\lambda$, giving $p=\sin^2(\pi/4-\lambda)$.
The conditional mutual informations $-(c/3)\ln(1-p)$ and $-(c/3)\ln p$ have opposite first derivatives $-2c/3,2c/3$ and equal second derivatives $4c/3$ at zero.
The squared difference of the slopes in Eq.~\eqref{eq:measured_quartet_hessian} subtracts precisely $4c/3$, giving $R_Q''(0)=0$ as required for a reparametrized vacuum.

\subsection{Bulk and Boundary Weights on Klein Disk}
\label{sec:quartet-weights}
\noindent
To find which bulk energy each quartet detects, choose coordinates that turn its geodesics into straight chords.
Central projection of the unit future hyperboloid onto a plane gives the Klein disk.
A vacuum isometry brings the diagonal crossing to the disk center, where both diagonals become diameters.
After a rotation, their four endpoints are the corners $(a,-b),(a,b),(-a,b),(-a,-b)$ of a Euclidean rectangle.
Each vertical side has chord length $2b$ and each diagonal length $2$, so the vacuum cross ratio is $\eta_Q=(2b)^2/2^2=b^2$.
Thus, $b=b_Q=\sqrt{\eta_Q}$ and $a=a_Q=\sqrt{1-\eta_Q}$.
Setting $d=\sqrt{1-|y|^2}$ at $y=(X,Y)$ gives the unit-radius metric, area measure, and hyperboloid embedding
\begin{equation}
q_{ab}=\delta_{ab}/d^2+y_a y_b/d^4,\qquad
d\mu=dX\,dY/d^3,\qquad \mathsf X=(1,y)/d .
\label{eq:klein_metric}
\end{equation}
The embedding product has signature $(-,+,+)$, and we denote the positive part of a real number by $[s]_+=\max(s,0)$.
For the counterclockwise arc of midpoint $\theta$ and half-opening $\alpha\in(0,\pi)$, write $n_\theta=(\cos\theta,\sin\theta)$ and $p_0=\cos\alpha$.
The chord $n_\theta\cdot y=p_0$ separates the interval's wedge on the slice from the rest of the disk.
The vacuum modular flow of the arc acts in the bulk as a boost that fixes the chord.
Its rate of advance in local proper time, divided by $2\pi L$, is the signed normal component $w$, or boost lapse, on the slice.
Each interval probes only its own wedge, which is why the energy integral will use the positive part $[w]_+$.
The unit spacelike vector $N=(p_0,n_\theta)/\sqrt{1-p_0^2}$ is normal to the plane of the chord in the embedding of Eq.~\eqref{eq:klein_metric}.
Writing $w=\mathsf X\cdot N$, we obtain
\begin{equation}
w=\frac{n_\theta\cdot y-p_0}{d\sqrt{1-p_0^2}},\qquad
\mathsf C=\frac{\mathsf X-wN}{\sqrt{1+w^2}},\qquad
\mathsf X=\cosh\ell\,\mathsf C+\sinh\ell\,N,\qquad
\ell=\operatorname{arsinh}w .
\label{eq:chord-normal-distance}
\end{equation}
Indeed, $\mathsf X-wN$ is orthogonal to $N$ and has squared norm $-1-w^2$, so $\mathsf C$ lies on the unit hyperboloid and on the chord.
The curve $s\mapsto\cosh s\,\mathsf C+\sinh s\,N$ is a unit-speed geodesic because its acceleration is normal to the hyperboloid.
At $s=0$, its tangent is $N$, perpendicular to the chord, so $L\ell$ is the signed proper distance from the chord, positive on the side $n_\theta\cdot y>p_0$. \\

\noindent
For the arc $(i,j)$, the vacuum modular energy is the expectation of $-\ln\rho^0_{ij}$, where $\rho^0_{ij}$ is its reduced vacuum state.
Relative entropy is the difference between the change in this expectation and the change in entropy.
For each of the six arcs, $M_{2,ij}$ denotes the second state derivative of the modular energy, and $M_{2,f,ij}$ denotes the corresponding derivative for the matched descendant.
The difference between the relative-entropy identities for the two families gives
\[
S_{2,ij}-S_{2,f,ij}
=M_{2,ij}-M_{2,f,ij}-(\FF_{ij}-\GG_{ij}).
\]
Here, $\FF_{ij}$ and $\GG_{ij}$ are the relative-entropy Hessians of the state and its matched descendant.
The gravitational canonical energy is quadratic in their common linear metric perturbation and cancels in the difference.
The remaining scalar boost energy is $\FF_{ij}-\GG_{ij}=4\pi\int[w_{ij}]_+ e\,d\mu$, supported on the interval's wedge \cite{Lashkari:2015hha,Czech:2016tqr}.
Eq.~\eqref{eq:weightedenergy} derives the factor $4\pi$. \\

\noindent
To find the overlap of the six measurements, first remove their common denominator $d$.
For the sides $(12,34,23,14)$ in that order, $d w_{ij}$ takes the values $\big((X-a_Q)/b_Q,(-X-a_Q)/b_Q,(Y-b_Q)/a_Q,(Y+b_Q)/a_Q\big)$.
For the diagonals $13$ and $24$, the values are $a_QY+b_QX$ and $a_QY-b_QX$.
For example, to the right of side $12$, one has $X>a_Q$ and therefore $|Y|<b_Q$ inside the disk.
Only $w_{12},w_{13},w_{14}$ are positive, and $w_{13}=\eta_Q w_{12}+(1-\eta_Q)w_{14}$ cancels the entire contribution of a packet there, irrespective of its energy.
To cover all four exterior caps and the interior at once, use $[s+t]_++[s-t]_+=s+\max(|s|,|t|)$.
With $u_{ij}=d[w_{ij}]_+$, $\xi_x=b_Q|X|$, $\xi_y=a_Q|Y|$, and $\xi_0=a_Qb_Q$, the three opposite-pair sums are
\[
\begin{aligned}
u_{13}+u_{24}&=a_QY+\max(\xi_x,\xi_y),\\
\eta_Q(u_{12}+u_{34})&=[\xi_x-\xi_0]_+,\\
(1-\eta_Q)(u_{14}+u_{23})&=a_QY+\max(\xi_y,\xi_0).
\end{aligned}
\]
The terms linear in $Y$ cancel in Eq.~\eqref{eq:quartet_linear_main}, leaving
\[
-\eta_Q d\,\mathcal L_Q[w]_+
=\max(\xi_y,\xi_0)+[\xi_x-\xi_0]_+-\max(\xi_x,\xi_y).
\]
Inside $Q$, $\xi_x,\xi_y\le \xi_0$ and the expression is $\xi_0-\max(\xi_x,\xi_y)$.
Outside $Q$, the disk condition $X^2+Y^2<1=a_Q^2+b_Q^2$ prevents both $\xi_x$ and $\xi_y$ from exceeding $\xi_0$.
For $\xi_x>\xi_0$, the expression is $\xi_0+(\xi_x-\xi_0)-\xi_x=0$, and for $\xi_y>\xi_0$, it is $\xi_y+0-\xi_y=0$.
Together, these cases give $-\mathcal L_Q[w]_+=K_Q$. \\

\noindent
To connect the bulk cancellation to the boundary stress, let $y_x=(\cos x,\sin x)$ be the boundary point with angle $0\leq x<2\pi$.
For the arc $ij$ with center $\theta_{ij}$ and half-opening $0<\alpha_{ij}<\pi$, its vacuum modular-energy weight on the unit cylinder is $2\pi k_{ij}(x)$, with
\[
k_{ij}(x)=\left[\frac{\cos(x-\theta_{ij})-\cos\alpha_{ij}}
{\sin\alpha_{ij}}\right]_+
=\lim_{r\uparrow1}d(r y_x)[w_{ij}(r y_x)]_+,
\qquad d(y)=\sqrt{1-|y|^2}.
\]
The positive part restricts the boundary weight to the arc, and $n_{\theta_{ij}}\cdot y_x=\cos(x-\theta_{ij})$ gives the radial limit.
Along every ray except those ending at the four vertices, the bulk weights cancel in an exterior cap before the ray reaches the circle.
Thus, $\mathcal L_Q k(x)=0$ there and, by continuity, at the vertices as well.
Let $\delta t_2(x)$ be the difference in the second state derivatives of $\langle T_{\tau\tau}(x)\rangle$ between the state and its matched descendant.
Since the vacuum modular Hamiltonian is linear in this boundary energy density,
\[
\mathcal L_Q(M_2-M_{2,f})
=2\pi\int_0^{2\pi}\delta t_2(x)\,\mathcal L_Q k(x)\,dx=0 .
\]
Even different second-order boundary stress profiles therefore give zero after the quartet combination, angle by angle.
The entropy difference is left with the minus sign from the relative-entropy identity,
\[
\frac c3\Pi''(0)
=-\mathcal L_Q(\FF-\GG)
=-4\pi\int_{\mathbb H^2} \bigl(\mathcal L_Q[w]_+\bigr)e\,d\mu
=4\pi\int_Q K_Q e\,d\mu .
\]
This equality holds for signed densities as soon as the weighted integral converges absolutely. \\

\begin{figure}[t]
\centering
\includegraphics[width=.96\columnwidth]{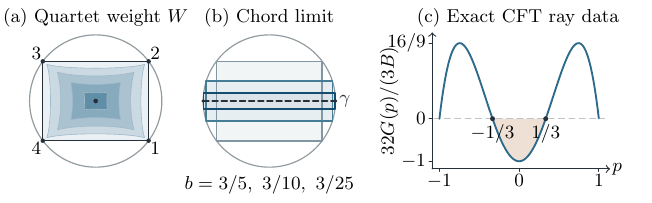}
\caption{As the quadrilateral narrows, its weighted integral of energy approaches a line integral after rescaling.
Panel (a) shows the weight $W$ as a fraction of its maximum $M$, with contours $W/M=1/5,2/5,3/5,4/5$ for half-widths $(a,b)=(4/5,3/5)$.
In (b), the rectangles with $a=\sqrt{1-b^2}$ and $b=3/5,3/10,3/25$ narrow toward the diameter $\gamma$.
The rescaled reading $R''/b^2$ approaches $4\pi I_\gamma$, where $I_\gamma$ is the energy integrated along that diameter.
In (c), orange marks the chords along which the integrated energy is negative in the dimension-$3/2$ scalar example, where the CFT predicts $32G(p)/(3B)=(1-p^2)(9p^2-1)$ as a function of chord offset $p$ in Eq.~\eqref{eq:mode_abel}.}
\label{fig:bweight}
\end{figure}

\noindent
In the coordinates of the rectangle, $m=\max(|X|/a,|Y|/b)$ is the fraction of the distance traveled along a straight ray from the crossing to the boundary of $Q$.
Thus, $m=0$ at the crossing and $m=1$ on a side, and the ordered measurement has the weight
\begin{equation}
K_Q=\frac{a}{b}\frac{[1-m]_+}{d},
\qquad \frac c3\Pi''(0)=4\pi\int_Q K_Q e\,d\mu .
\label{eq:localized_main}
\end{equation}
In terms of the symmetric weight $W=W_Q=\eta_QK_Q$ and its peak $M=M_Q=ab$, the same identity reads
\begin{equation}
W=\frac M d[1-m]_+,\qquad R_Q''(0)=4\pi\int_Q We\,d\mu .
\label{eq:bkernel}
\end{equation}
The dimensionless normalized weight is $k=W/M$.
At unit-radius distance $\rho$ from the intersection, $|y|=\tanh\rho\le m$ because $|y|^2\le m^2(a^2+b^2)$.
Thus, $k\le(1-|y|)/\sqrt{1-|y|^2}=\exp(-\rho)$.
The weight peaks at the diagonal crossing and falls to zero at the sides.
Since it is positive inside and zero outside, a negative reading requires negative energy on a set of positive area inside the quartet.
Figs.~\ref{fig:quartetweight}(a) and \ref{fig:bweight}(a) show contours of equal sensitivity for the same rectangle.
On a ray $y=t v$ to $v\in\partial Q$, $m=t$, hence $k=(1-t)/\sqrt{1-t^2|v|^2}$.
For $0<\kappa<1$, solving $k=\kappa$ for its root in $[0,1)$ gives
\[
t_\kappa=\frac{1-\kappa^2}{1+\kappa\sqrt{1-(1-\kappa^2)|v|^2}} .
\]
Fig.~\ref{fig:quartetweight}(b) compares points at the same fraction $t$ along rays to a vertex and to a side midpoint.
On the vertex ray, $|v|=1$ gives $k=\sqrt{(1-t)/(1+t)}$.
On the ray to the positive $X$-axis midpoint, $|v|=a_Q<1$ and $k/(1-t)\to(1-a_Q^2)^{-1/2}$ as $t\to1^-$.
Both weights vanish at the edge, but the vertex ray remains more sensitive for every $0<t<1$ because its denominator $\sqrt{1-t^2|v|^2}$ is smaller. \\

\noindent
There is also a direct geometric way to get the weight.
The perturbed geodesics move identically to first order in the state and the matched vacuum descendant.
Subtracting their length Hessians cancels the effect of that shared motion and leaves the length change linear in the difference of their second metric derivatives.
Integration by parts in the scalar-curvature constraint then transfers the metric derivatives to the weight, producing terms supported on the six geodesics.
Appendix~\ref{app:quartetlocalization} derives Eq.~\eqref{eq:bkernel} by this method and controls the boundary terms at the ideal tips. \\

\noindent
The quartet weight also has a simple normalization.
For $0<x<1$, let $H(x)=-x\ln x-(1-x)\ln(1-x)$ denote the binary entropy in natural units.
Integrating the symmetric weight over the four triangles formed by the diagonals gives $\int_Q W_Q\,d\mu=H(\eta_Q)$, or $\int_Q K_Q\,d\mu=H(\eta_Q)/\eta_Q$.
Appendix~\ref{app:quartet-moments} evaluates these integrals and the second moment needed below. \\

\noindent
The measured integral divided by this normalization gives the mean density
\begin{equation}
\frac{\int_Q K_Q e\,d\mu}{\int_Q K_Q\,d\mu}
=\frac{R_Q''(0)}{4\pi H(\eta_Q)} .
\label{eq:measured_mean_density}
\end{equation}
For a density constant throughout $Q$, the right-hand side gives that constant exactly.
For a varying density, it gives an average in which energy at the diagonal crossing receives more weight than the same energy near a side.
Both $R_Q''$ and $H(\eta_Q)$ are invariant under exchange of the two opposite-side products. \\

\noindent
To compare the weighted reading with the area response, we also need the second moment of the weight, $\mathcal J_Q=\int_Q W_Q^2\,d\mu$.
It depends only on the vacuum cross ratio. \\

\noindent
An ideal quadrilateral has dimensionless area $2\pi$, so write $\langle f\rangle_Q=(2\pi)^{-1}\int_Q f\,d\mu$ for an area average.
For any square-integrable density $e$ on $Q$, let $\sigma_e^2=\langle (e-\langle e\rangle_Q)^2\rangle_Q$ denote its variance across the quadrilateral.
Since $\langle W_Q\rangle_Q=H(\eta_Q)/(2\pi)>0$ is the area average of $W_Q$, expansion around the two means gives
\[
\frac{R_Q''(0)}{4\pi H(\eta_Q)}-\langle e\rangle_Q
=\frac{\langle (W_Q-\langle W_Q\rangle_Q)(e-\langle e\rangle_Q)\rangle_Q}{\langle W_Q\rangle_Q}.
\]
The numerator is the spatial covariance of the density and the quartet weight.
When high densities occur preferentially near the diagonal crossing, the covariance is positive, and the entropy estimate exceeds the area mean.
For a constant density, the covariance vanishes, and either measurement gives that same density. \\

\noindent
By the Cauchy--Schwarz inequality, the absolute covariance is bounded by $\sigma_e$ times the standard deviation of $W_Q$, namely $[\mathcal J_Q/(2\pi)-\langle W_Q\rangle_Q^2]^{1/2}$.
Since $T_Q=2\pi\langle e\rangle_Q$, the difference between the two density averages obeys
\[
\left|\frac{T_Q}{2\pi}-\frac{R_Q''(0)}{4\pi H(\eta_Q)}\right|
\leq \sigma_e\sqrt{\frac{2\pi\mathcal J_Q}{H(\eta_Q)^2}-1}.
\]
The geometric factor is $C_Q=\sigma_W/\overline W$, where $\overline W=\langle W_Q\rangle_Q$ and $\sigma_W^2=\mathcal J_Q/(2\pi)-\overline W^2$.
Both $\overline W$ and $\sigma_W$ are strictly positive because $W_Q$ is positive and nonconstant in the interior.
For a measured weighted mean $m_Q=R_Q''(0)/(4\pi H(\eta_Q))$ and a variance bound $\sigma_e\leq s_*$ with $s_*\geq0$, the inequality places the unweighted mean in $[m_Q-s_* C_Q,m_Q+s_* C_Q]$. \\

\noindent
The interval is sharp for unrestricted real square-integrable densities.
Adding a multiple of the centered weight increases or decreases the area mean, and subtracting a constant restores the measured mean.
Appendix~\ref{app:quartet-moments} constructs the saturating densities and shows why the bound weakens for very narrow quartets.
For any finite quartet satisfying $s_*C_Q<|m_Q|$, the interval of possible means excludes zero and fixes the sign of $T_Q$.
It therefore fixes the sign of $A_Q''(0)/L^2=24\pi T_Q/c$ and, under the maximality hypotheses of Appendix~\ref{app:geometry}, that of $V_Q''(0)$.

\subsection{Positive Energy and Several Quartets}
\label{sec:positive-quartets}

\noindent
The bounds below use the sign of the density and the spatial weight derived in Sec.~\ref{sec:quartet-weights}.
For the resulting lower bound on the volume Hessian, maximality of the vacuum polygon suffices when the Hessian exists. \\

\noindent
A fixed amount of positive energy gives the largest quartet signal at the diagonal crossing, where the weight peaks.
With overlapping quartets, the same energy can produce several signals.
We determine how much energy one reading requires, which additional readings are independent, and how to combine them into a sharp bound.

\subsubsection{Sharp Bound for Single Quartet}
\label{sec:single-energy-bound}
\noindent
For $e\ge0$ on $Q$, the peak weight bounds the reading by $0\le\int_Q W e\,d\mu\le M_Q T_Q$.
A packet near the crossing approaches this upper limit for fixed energy. In contrast, one near a side produces a smaller signal.
For a lower bound on the volume Hessian, it suffices that the reflection slice maximize the vacuum area.
Indeed, $V_Q-A_Q$ then has a minimum of zero at $\lambda=0$, so $V_Q''(0)\ge A_Q''(0)$ whenever both derivatives exist.
Combining this inequality with the reading gives
\begin{align}
\frac{V_Q''(0)}{L^2}
&\ge \frac{A_Q''(0)}{L^2}\ge 2\frac{b_Q}{a_Q}\Pi''(0)\ge0 ,
\label{eq:sharp_volume_main}\\
\frac{V_Q''(0)}{L^2}
&\ge\frac{6R_Q''(0)}{c\sqrt{\eta_Q(1-\eta_Q)}}.
\label{eq:finite_quartet_lower_bound}
\end{align}
Even if differentiability of $V_Q$ is unknown, vacuum maximality and $A_Q'(0)=0$ give
\[
\liminf_{\substack{\lambda\to0\\\lambda\ne0}}\frac{2[V_Q(\lambda)-V_Q(0)]}{\lambda^2L^2}
\ge \frac{A_Q''(0)}{L^2}.
\]
Subtracting the common vacuum value $V_Q(0)=A_Q(0)$ from $V_Q(\lambda)\ge A_Q(\lambda)$ before division by $\lambda^2$ gives this limit.
For signed densities, $0\le W\le M_Q$ yields the negative-energy bound $\int_Q[-e]_+\,d\mu\ge[-R_Q''(0)]_+/(4\pi M_Q)$.
Sec.~\ref{sec:sharp} derives the sharp bound when the unweighted energy is also known. \\

\noindent
Since $W>0$ throughout the interior, a zero reading from a nonnegative density forces $e=0$ almost everywhere in $Q$.
For signed energy and a negative reading, $\int_Q W[-e]_+\,d\mu$ must be at least the magnitude of the net weighted signal.
Using $W\le M_Q$ and converting to the ordered response gives
\[
\int_Q[-e]_+\,d\mu\ \ge\ \frac{c}{12\pi}
\sqrt{\frac{\eta_Q}{1-\eta_Q}}\,[-\Pi''(0)]_+ .
\]
These inequalities concern the second state derivative at the vacuum for fixed endpoints. \\

\noindent
To obtain the coefficient in Eq.~\eqref{eq:finite_quartet_lower_bound}, insert $T_Q\ge R_Q''(0)/(4\pi M_Q)$ into $A_Q''(0)/L^2=24\pi T_Q/c$.
Here, $M_Q=\sqrt{\eta_Q(1-\eta_Q)}$ is the peak of the symmetric weight $W_Q=\eta_Q K_Q$.
The symmetric response is $Z_Q''(0)=(3/c)R_Q''(0)=\eta_Q\Pi''(0)$, so its ratio to the area depends only on the mean weight $\langle W_Q\rangle_e=\int_QW_Qe\,d\mu/E_Q$ of the energy, with $E_Q=T_Q>0$.
For a nonnegative density, the mean is positive. It lies strictly below the peak because a single point carries zero area.
Consequently,
\[
\frac{A_Q''(0)}{L^2Z_Q''(0)}
=\frac{2}{\langle W_Q\rangle_e},
\qquad 0<\langle W_Q\rangle_e<M_Q .
\]
The difference from equality also bounds the energy away from the crossing.
For a dimensionless radius $s>0$, let $E_{\rm out}(s)=\int_{Q\cap\{\rho\geq s\}}e\,d\mu$, where $L\rho$ is the proper distance from the crossing.
The pointwise estimate $W_Q/M_Q\leq\exp(-\rho)$ after Eq.~\eqref{eq:bkernel} gives $1-W_Q/M_Q\geq1-\exp(-s)$ on this set.
Integrating there and using $1-W_Q/M_Q\geq0$ on the rest of $Q$ yields
\[
\frac{E_{\rm out}(s)}{E_Q}
\leq\frac{1-R_Q''(0)/(4\pi M_Q E_Q)}{1-\exp(-s)} .
\]
At proper distance $L\ln 2$, the weight is at most half its peak.
A reading of at least $95\%$ of $4\pi M_Q E_Q$ therefore places at least $90\%$ of the energy inside that radius.
As $R_Q''(0)/(4\pi M_Q E_Q)\to1$, the energy outside any fixed disk about the crossing becomes a vanishing fraction of the total.
Only a positive point mass at the crossing can attain equality at nonzero energy. \\

\noindent
For a fixed positive quartet signal, the dimensionless ratio $A_Q''(0)/(L^2 Z_Q''(0))$ can take any value in $(2/M_Q,\infty)$ with a smooth density.
One packet near the crossing and another near a side sample different weights, so redistributing energy between them changes the mean weight.
Appendix~\ref{app:quartet-sharpness} gives the amplitudes that hold the reading fixed and shows how the same signal averages the energies of nested quadrilaterals.

\subsubsection{Rank of Quartet Data}
\label{sec:quartet-rank}
\noindent
Quartets sharing endpoints reuse some of the same interval entropies, so counting quartets overcounts the independent measurements.
For example, any three consecutive readings from five equally spaced endpoints determine the other two.
Fix $N\ge4$ distinct angles $x_1<\cdots<x_N<x_1+2\pi$ as the state varies.
For $1\le i<j<k<l\le N$, measure the failure of the vacuum chord identity by
\[
T_{ijkl}=D_{ik}D_{jl}-D_{ij}D_{kl}-D_{il}D_{jk}
=D_{ij}D_{kl}(\Pi_{ijkl}-1).
\]
The vacuum chord identity and first-order matching to a common vacuum descendant give $T_{ijkl}(0)=T'_{ijkl}(0)=0$.
Therefore, $T''_{ijkl}(0)=D^0_{ij}D^0_{kl}\Pi''_{ijkl}(0)$.
A common short-distance regulator can be kept in this algebra and removed in the normalized formulas below. \\

\noindent
Choose adjacent endpoints $1,2$ as anchors.
For each remaining pair $3\le i<j\le N$, applying the vacuum chord identity to $(1,2,i,j)$ predicts $D_{ij}$ from the chords touching the anchors.
The predicted entry minus the measured one is $T_{12ij}/D_{12}$.
Since the residuals and their first derivatives vanish in vacuum, their second derivatives enter every quartet response linearly.
Thus, the $\binom{N-2}{2}$ readings that contain the two anchors determine all quartet responses.
They are independent when smooth energy profiles can vary freely because only the weight for $(1,2,i,j)$ has a normal-derivative jump along $\gamma_{ij}$ away from other chord crossings.
Appendix~\ref{app:quartet-rank-proof} gives the explicit linear reconstruction and the independence proof. \\

\noindent
For five equally spaced points, let $r_j$ be the symmetric reading $R''(0)$ from the four points left after omitting $j$, with $1\le j\le5$ and indices continued cyclically.
All five vacuum products of the two diagonals are equal, and the ratio of a pentagon diagonal to a pentagon side is $\varphi=2\cos(\pi/5)=(1+\sqrt5)/2$.
Let $\ell_{ij}=2\sin[(x_j-x_i)/2]$ denote the vacuum chords for $1\le i<j\le5$.
The reconstruction in Appendix~\ref{app:quartet-rank-proof} gives $\ell_{15}r_1-\ell_{25}r_2+\ell_{35}r_3-\ell_{45}r_4=0$.
Here, $\ell_{15}=\ell_{45}$ is a side and $\ell_{25}=\ell_{35}=\varphi\ell_{15}$ is a diagonal.
Cyclic rotation of the five endpoints gives
\begin{equation}
r_{j+3}=r_j+\varphi(r_{j+2}-r_{j+1}),\qquad j=1,\ldots,5 .
\label{eq:pentagon-cyclic-readings}
\end{equation}
Any three consecutive readings therefore predict the other two.

\subsubsection{Linear Program and Regular Pentagon}
\label{sec:pentagon-program}
\noindent
For $m\ge1$ measured quartets $Q_i$ in a convex ideal vacuum polygon $P_0$, how small can $E_P=\int_{P_0}e\,d\mu$ be at fixed readings?
Assume $e\ge0$ on $P_0$ and write each reading as $b_i=R_{Q_i}''(0)/(4\pi)=\int_{P_0}W_i e\,d\mu$, where $W_i=\eta_i K_{Q_i}$ and $\eta_i$ is its vacuum cross ratio for $1\le i\le m$.
Each weight vanishes outside its own quartet, so the energy outside $P_0$ does not affect the bound.
We can minimize over finitely many chord crossings because masses placed there reproduce the readings from any other packet using at most its energy. \\

\noindent
Draw the measured sides and diagonals inside $P_0$, and call the resulting finite vertices $\mathcal V$, excluding vertices at the ideal boundary.
On each cell, the product $d(y)W_i(y)$ is affine in the Klein coordinates $y$, where $d(y)=\sqrt{1-|y|^2}$.
Triangulating a cell expresses every weight at $y$ as a nonnegative combination of its values at the vertices, with the same coefficients $t_v(y)$ for all measurements and $\sum_{v\in\mathcal V}t_v(y)\leq1$ (Appendix~\ref{app:quartet-optimization}).
A unit of energy at $y$ can thus be replaced by masses $t_v(y)$ at the vertices without changing any reading or increasing the total mass.
For a finite nonnegative measure $d\nu$, this replacement gives $m_v=\int t_v\,d\nu$.
Conversely, feasible vertex masses already give a measure with the prescribed readings.
Minimizing over these masses gives the finite linear program
\begin{equation}
E_{\min}(b)=\min\left\{\sum_{v\in\mathcal V}m_v:
m_v\ge0,\quad
\sum_{v\in\mathcal V}m_vW_i(v)=b_i\quad(1\le i\le m)\right\}.
\label{eq:quartet-vertex-primal}
\end{equation}
For feasible data, restricting the total mass to that of any one feasible assignment leaves a closed bounded set, so the minimum is attained.
A vertex where every weight vanishes carries zero mass at the minimum.
The minimizing measure can put its mass at isolated points. However, its mass bounds the energy of every smooth nonnegative density.
Concentrating smooth squared momentum profiles near those points approximates their readings by continuity of the weights.
For the regular pentagon, one can approach the minimum with the five readings held exactly fixed. \\

\noindent
In a regular pentagon, the finite vertices are the five diagonal crossings, and each quartet weight has the same peak $M=\varphi^{-3/2}$ [Fig.~\ref{fig:quad}(c)].
At a crossing the five weights are $M$, $M/\varphi$, $M/\varphi$, $0$, and $0$, so their sum is $\sqrt5\,M$.
Thus, for a feasible measure supported on these crossings, summing its five readings counts the total energy with this common factor.
With $r_j=R''_{Q_j}(0)$, the result is
\[
E_{\min}=\frac{\sum_{j=1}^5r_j}{4\pi\sqrt5\,M},\qquad
\frac{V_P''(0)}{L^2}\ge
\frac{6}{c\sqrt5\,M}\sum_{j=1}^5r_j .
\]
For example, when all five readings equal $r>0$, one quartet alone gives the lower bound $r/(4\pi M)$ on the pentagon energy.
The five readings together raise it to $\sqrt5\,r/(4\pi M)$, a factor $\sqrt5$ because the five regions overlap.
Equal masses at the five crossings realize these data and attain the joint bound in the space of measures.
Converting the energy bound to the displayed volume bound uses $V_P''(0)\ge A_P''(0)$ under the admissibility, vacuum maximality, and differentiability hypotheses of Eq.~\eqref{eq:sharp_volume_main}.
For a smooth nonzero density in the polygon, the first term in Eq.~\eqref{eq:quartet-mass-excess} is strictly positive, since the finite set of vertices has zero area.
The range of attainable energies also gives the range of volume responses whenever the comparison hypotheses ensure $V_P=A_P$. \\

\noindent
The five readings are feasible for nonnegative energy precisely when all are nonnegative and obey the cyclic relations in Eq.~\eqref{eq:pentagon-cyclic-readings}.
For any nonzero feasible set, smooth nonnegative densities realize every $E_P>E_{\min}$, while the minimum requires point masses at crossings.
If all readings vanish, a smooth nonnegative density vanishes almost everywhere inside the pentagon.
Appendix~\ref{app:quartet-optimization} derives the feasibility test and constructs the smooth densities with all five readings held fixed.
These statements leave the individual interval Hessians and the conserved boundary energy free.

\section{Polygon Volume from Narrow Quartets}
\label{sec:dual}

\noindent
Move the four endpoints into two close pairs, one at either end of a chosen interval, as in Fig.~\ref{fig:quad}(a).
At quadratic order about the vacuum, the stress-subtracted signal gives the integral of normal energy along the geodesic joining the pairs.
Repeating the measurement across the disk determines the energy in a specified polygon and hence its quadratic area response. \\

\noindent
Appendix~\ref{app:geodesicfocus} derives the geodesic limit and estimates the error at finite width.
Appendix~\ref{app:gravity} derives the boundary-stress subtraction, and Appendix~\ref{app:tomography} gives the inversion and finite-sampling error estimates.
The area response gives the maximal-volume response under the comparison conditions in Appendix~\ref{app:geometry}.

\subsection{Coalescing Endpoints and Interval Observable}
\label{sec:endpoint-limit}
\begin{figure}[t]
\centering
\includegraphics[width=.86\linewidth]{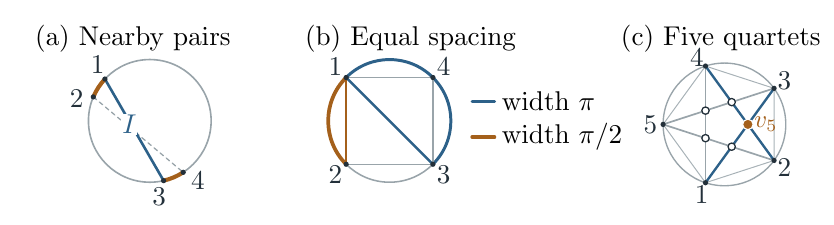}
\caption{Geodesics in the Klein disk are straight chords, and the boundary labels increase counterclockwise.
(a) Two adjacent pairs of endpoints shrink onto the ends of the limiting chord $I$.
In (b), a side and a diagonal of the square select arcs with full openings $\pi/2$ and $\pi$, respectively.
In (c), omitting each vertex of a regular pentagon in turn gives five quadrilaterals, each with its own diagonal crossing.
The filled marker $v_5$ is the crossing of chords $13$ and $24$ for the quartet omitting vertex $5$.}
\label{fig:quad}
\end{figure}

\noindent
Choose real angles $u<v<u+2\pi$ so that the arc $I=(u,v)$ stays continuous across the angular origin.
Its center is $\theta=(u+v)/2$ modulo $2\pi$, and its half-opening is $\alpha=(v-u)/2\in(0,\pi)$.
The four endpoints $(u,u+h,v,v+k)$ remain in cyclic order for $0<h<v-u$ and $0<k<2\pi-v+u$.
Shrinking $h$ and $k$ independently leaves the geodesic from $u$ to $v$ in Fig.~\ref{fig:quad}(a).
Measure all four entropies $S(a,b)$ in one state $\lambda$ on one smooth RT branch, and write $S_{uv}$ for the mixed endpoint derivative at $(u,v)$.
Extending the right end by $k$ changes the entropy by $S(u,v+k)-S(u,v)$.
Moving the left end by $h$ changes that gain by the conditional mutual information of the two short arcs given the intervening arc,
\begin{align}
I_c&=S(u,v)+S(u+h,v+k)-S(u,v+k)-S(u+h,v)\nonumber\\
&=\int_0^h\!da\int_0^k\!db\,
(\partial_u\partial_vS)(u+a,v+b) .
\label{eq:rectangular_entropy_average}
\end{align}
Any term depending on just one endpoint cancels in this difference.
Integrating once in each endpoint gives the second line, so $I_c/(hk)$ is the average of $S_{uv}$ on the rectangle.
Strong subadditivity gives $I_c\ge0$ \cite{Lieb:1973zz}.
Continuity of $S_{uv}$ at $(u,v)$ ensures that the average tends to its value there, even when $h$ and $k$ shrink at different rates. \\

\noindent
Remove the common short-distance cutoff while the four endpoints are distinct and then take $h,k\downarrow0$.
In the quartet ratio $\Pi$, the two short sides supply a factor proportional to $hk$, while the difference of long-side products supplies the mixed derivative measured by $I_c/(hk)$.
Using the vacuum values $S_{0,uv}=c/(12\sin^2\alpha)$ and $\Pi=1$, the resulting observable is
\begin{equation}
\mathcal A_I(\lambda)\equiv\lim_{h,k\downarrow0}\Pi
=\exp\!\left[6(S-S_0)/c\right]\frac{S_{uv}}{S_{0,uv}}.
\label{eq:ratio}
\end{equation}
For a vacuum reparametrization common to both chiral sectors, the change in the mixed derivative exactly cancels the exponential, so $\mathcal A_I=1$ on every proper interval.
Appendix~\ref{app:quartet-resolution} gives the endpoint calculation, including cancellation of local cutoff factors.

\subsection{State Hessian and Chord Integral}
\label{sec:chord-limit}
\noindent
To extract the scalar response, first differentiate in state amplitude $\lambda$ at four fixed endpoints and then bring the pairs together.
Let $S_j(u,v)=\left.\partial_\lambda^j S(u,v;\lambda)\right|_0$ for $j=1,2$, with $S_1,S_2$ and the mixed endpoint derivative $S_{2,uv}$ continuous near the target and $S_1,S_2$ tending uniformly to zero on short arcs near its endpoints.
For a continuous scalar energy coefficient $e$ with the near-boundary falloff of Appendix~\ref{app:quartet-resolution}, the normalized quartet weight becomes proper arclength along the vacuum geodesic $\gamma_I$, in units of $L$.
The narrowing support is shown in Fig.~\ref{fig:bweight}(b).
Writing $R$ for the defect in Eq.~\eqref{eq:bpi} and $\eta_Q$ for the vacuum cross ratio gives
\begin{equation}
\lim_{\eta_Q\downarrow0}\frac{R''(0)}{\eta_Q}
=4\pi I_\gamma,\qquad
I_\gamma=(\XX e)(I)=\int_{\gamma_I}e\,d\sigma .
\label{eq:chord_limit}
\end{equation}
Although the vacuum quadrilateral always has area $2\pi L^2$, its normalized response singles out one chord.
For the state families considered here, $\Pi'(0)=0$ because their first-order interval entropies agree with a vacuum descendant. \\

\noindent
The corresponding boundary reading follows from Eq.~\eqref{eq:ratio}, or by applying the state derivative before the endpoint limit in Appendix~\ref{app:quartet-resolution}.
\begin{align}
\mathcal Q_I
&\equiv\frac c3\,\partial_\lambda^2\ln\mathcal A_I\big|_0
=4\sin^2\alpha\,\partial_u\partial_v S_2+2S_2-\frac{12}{c}S_1^2
=4\pi (\XX e)(I), \label{eq:entropysource}\\
&=2S_2+\sin^2\alpha(\partial_\theta^2-\partial_\alpha^2)S_2
-\frac{12}{c}S_1^2 . \label{eq:raycurvature}
\end{align}
In the matched descendant, differentiating Eq.~\eqref{eq:ratio} with $\mathcal A_I=1$ gives $S_{1,uv}/S_{0,uv}=-6S_1/c$.
The actual state shares this slope, so the second derivative of $\ln S_{uv}$ contributes $-(c/3)(S_{1,uv}/S_{0,uv})^2=-12S_1^2/c$ in Eq.~\eqref{eq:entropysource}.
The second line compares sliding an interval at fixed width with changing its width at fixed center; the two derivatives are in $\DD=\sin^2\alpha(\partial_\alpha^2-\partial_\theta^2)-2$.
Write $\FF_I$ and $\GG_I$ for the relative-entropy Hessians in the perturbed state and its matched vacuum descendant, respectively.
The modular energy $M$ is the change in the expectation of the fixed vacuum operator $-\ln\varrho_{I,0}$, where $\varrho_{I,0}$ is the reduced vacuum state of $I$.
It solves $\DD M=0$, the first law fixes $S_1$, and $\DD\GG_I=12 S_1^2/c$.
Thus, $\mathcal Q_I=\DD(\FF_I-\GG_I)$ can be evaluated either from ordinary entropy derivatives or by making the gravitational subtraction in relative entropy \cite{Hollands:2019czd,Panebianco:2019plp,Blanco:2013joa,Lashkari:2015hha,Czech:2016tqr,deBoer:2015kda}.
Appendix~\ref{app:gravity} gives the explicit stress kernels, their null directions, and the treatment of complementary arcs.
Appendix~\ref{app:tomography} obtains the normalization independently by applying $\DD$ to the vacuum modular boost weight, whose second derivative on the wedge boundary picks out the chord.
Relative-entropy tomography and its applications to volume provide related constructions \cite{Banerjee:2017qti,Lin:2014hva}. \\

\noindent
There is also an interpretation at fixed, finite amplitude, when classical entropy is given by a unique minimizing geodesic and the cutoff curvature has the integrability and convergence properties of Appendix~\ref{app:geodesicfocus}.
Along the actual geodesic, let $\kappa=1+L^2\mathcal K=8\pi G_N L^2\rho$ be its excess ambient curvature and let $\tau_I=\tfrac12\int_{\gamma_I(\lambda)}\kappa\,d\sigma$.
Its transverse Jacobi equation gives $\tau_I\le\ln\mathcal A_I$ for either sign of $\kappa$, so $\mathcal A_I<1$ implies negative integrated energy on that chord.
If $\kappa\ge0$ along the regulated geodesics, the additional lower bound is $1-\mathcal A_I^{-1}\le\tau_I$.
The exact remainder and the leading nonlinear correction are given in Eqs.~\eqref{eq:app_focus_deficit} and~\eqref{eq:app_chordquadraticcorrection}.
The shifts used here move both endpoints along one spatial slice and probe the energy measured normal to that slice.
Moving an endpoint along a light ray instead changes both its time and position and probes different stress components \cite{Bhattacharya:2014vja}.

\subsection{Reconstruction from all Chords}
\label{sec:all-chord-reconstruction}
\noindent
One chord reading in Eq.~\eqref{eq:entropysource} sums the energy along a path and leaves its location on that path undetermined.
The complete set gives $\mathcal Q=4\pi\XX e$, so applying the inverse line transform $\XX^{-1}$ and dividing by $4\pi$ recovers the density.
Integrating that density over the vacuum polygon gives its area response by Eq.~\eqref{eq:geometricresponse}.
When the volume comparison of Appendix~\ref{app:globalmaximum} holds in a neighborhood of $\lambda=0$, this area response also gives $V''(0)=A''(0)$ and
\begin{equation}
\frac{V''(0)}{L^2}=\frac6c\int_{P_0}\XX^{-1}[\mathcal Q](y)\,d\mu(y).
\label{eq:mainresult}
\end{equation}
In the Klein disk, the interval center sets the chord normal $n_\theta=(\cos\theta,\sin\theta)$ and its half-opening sets the signed offset $p=\cos\alpha$ from the origin.
For example, $\alpha=\pi/2$ selects a diameter at every center.
Write $s=\sin\alpha$ and $n_\theta^\perp=(-\sin\theta,\cos\theta)$, so $y=pn_\theta+t n_\theta^\perp$ runs along the chord.
The arclength factor $d\sigma=s\,dt/(1-|y|^2)$ separates into a factor fixed by the endpoints and a factor absorbed into $g=e/(1-|y|^2)$ inside the disk, extended by zero outside.
Writing $\RR_E$ for the Euclidean line integral then gives
\begin{equation}
G(p,\theta)=I_\gamma/s=\RR_E g(p,\theta),
\qquad J(p,\theta)=\mathcal Q_I/s=4\pi G(p,\theta),\qquad |p|<1 .
\label{eq:chord_data_conventions}
\end{equation}
Set both functions to zero at $|p|\ge1$, where the line misses the open disk.
At fixed normal $n_\theta$, the Radon transform $\RR_E g$ integrates $g$ along parallel lines at signed distances $p$ from the origin.
A Fourier transform in $p$ gives the density's Fourier components on the single line of wave vectors $\sigma n_\theta$, where $\sigma$ is conjugate to $p$.
Combining directions fills the Fourier plane, with polar area factor $|\sigma|$ accounting for the filter $\widetilde{\Lambda H}(\sigma,\theta)=|\sigma|\widetilde H(\sigma,\theta)$.
We take the forward transform without a factor of $2\pi$, with one factor $1/(2\pi)$ per dimension in the inverse transform.
The angular range $0\le\theta<2\pi$ and signed $\sigma\in\mathbb R$ cover every nonzero wave vector twice, yielding $g=(16\pi^2)^{-1}\int_0^{2\pi}(\Lambda J)(y\cdot n_\theta,\theta)\,d\theta$ \cite{Facchi:2009nls}. \\

\noindent
For a specified polygon, the target energy is $\int_{P_0}e\,d\mu=\int_{\mathbb R^2}g h_P\,d^2y$, where $h_P=1/\sqrt{1-|y|^2}$ inside $P_0$ and zero outside.
The weight of each line, $W_P=\RR_E h_P$, integrates the known metric factor over the portion inside the polygon.
The same vertices give the same weights in every state compared with the vacuum.
Multiplying the inverse transform by $h_P$ and integrating over the polygon therefore gives a linear functional of the filtered chord readings, with a weight fixed by the vertices alone,
\begin{equation}
\frac{V''(0)}{L^2}=\frac3{2\pi c}
\int_0^{2\pi}\!d\theta\int_{-1}^{1}\!dp\,
W_P(p,\theta)(\Lambda J)(p,\theta).
\label{eq:boundarywindow}
\end{equation}
For an ideal $m$-gon, the corresponding expansion is $V(\lambda)=(m-2)\pi L^2+\lambda^2V''(0)/2+o(\lambda^2)$.
For smooth compactly supported densities, the integral is ordinary.
For square-integrable $g$, Appendix~\ref{app:continuum_stability} controls the pairing with the polygon window, and Appendix~\ref{app:ordinary_entropy_weights} gives a regularized formula. \\

\noindent
For any finite list of directions, profiles with the same line data can have different polygon energies even if every offset in those directions is measured (Appendix~\ref{app:radoninversion}).
With sufficiently dense angular and width sampling and known smoothness bounds, a finite grid instead gives a controlled estimate.
Measurements at three nearby amplitudes per interval estimate $S_1$ and $S_2$, and smoothing places the endpoint derivatives on the known polygon weights.
Eq.~\eqref{eq:grid-error-budget} accounts for measurement errors, unsampled intervals, smoothing and numerical errors in the weights.
If the resulting interval excludes zero, it fixes the sign of the area response and, when the reflection slice remains maximal, the sign of the volume response.

\section{CFT Interval Response and Bulk Scalar Phases}
\label{sec:scalar-phases}

\noindent
An interval measures the response of its reduced quantum state, while a polygon's area measures the energy enclosed on the reflection slice.
We compare these readings for two phases of a single scalar oscillation. \\

\noindent
Consider the same scalar oscillator at maximum displacement and a quarter period later as it passes through zero.
At equal amplitude, the two preparations have equal total energy but distribute it differently in space.
We use boundary interval entropies to find the energy along each bulk geodesic and to decide which phase deposits more energy inside a chosen polygon. \\

\noindent
Let $\mathcal O$ be a Hermitian scalar primary on the unit circle, a spin-zero operator of scaling dimension $\Delta>0$ under a local rescaling.
In radial quantization, its insertion at the origin of the plane creates the unit-normalized state $|\mathcal O\rangle$.
We choose state derivatives at the vacuum of $\beta|\mathcal O\rangle$ and $\mathrm{i}\beta|\mathcal O\rangle$, labeled $+$ and $-$, with real $\beta\ne0$ and $\mathrm{i}^2=-1$.
The relative phase changes interference with the vacuum at the same excitation probability.
We set $B=\beta^2$ and take both state families to be pure and rotationally invariant through second order. \\

\noindent
The vacuum two-point function determines the phase dependence in the CFT.
The interpretation in terms of a bulk field requires a primary dual to the lowest mode of a canonical scalar, with $M_\phi^2L^2=\Delta(\Delta-2)$ and $\Delta=1+\sqrt{1+M_\phi^2L^2}>1$ in standard quantization.
With the boundary source fixed at zero, standard quantization keeps the more rapidly decaying radial solution \cite{Klebanov:1999tb}.
In the bulk description, $A$ is the field amplitude and $a=A\sqrt L$ its dimensionless counterpart.
Both waves have the same radial profile $\operatorname{sech}^\Delta\chi$, where $\operatorname{sech}\chi=1/\cosh\chi$ and $L\chi$ is the proper radial distance, $0\leq\chi<\infty$.
Their time dependence is $\cos(\Delta\tau)$ and $\sin(\Delta\tau)$, respectively, with $\tau\in\mathbb R$ the dimensionless global time.
A shift of one quarter period, $\pi/(2\Delta)$, interchanges the two profiles up to sign.
On the common slice $\tau=0$, the cosine wave has zero velocity throughout space and the sine wave has zero field throughout space.
We express both profiles in the Klein radius $r=\tanh\chi\in[0,1)$, for which $D=1-r^2$.
State derivatives are taken at fixed $c$, before the semiclassical limit with $B/c$ fixed. \\

\noindent
Appendix~\ref{app:cft} derives the two interval responses from the vacuum two-point function and matches the boundary and bulk amplitudes.
Appendix~\ref{app:scalar} evaluates the energy enclosed by each polygon and checks the area response by following its moving sides.

\subsection{Modular Energy and Phase Difference}
\noindent
The cylinder Hamiltonian above vacuum is $\widehat H_0=\widehat H_{\rm cyl}+c/12$.
The identity $\widehat H_0|0\rangle=0$ removes the terms containing the second derivative of the state, and $\widehat H_0|\mathcal O\rangle=\Delta|\mathcal O\rangle$ gives
\begin{equation}
\partial_\lambda^2\langle \widehat H_0\rangle\big|_0
=2\langle\psi_0'|\widehat H_0|\psi_0'\rangle=2\Delta B .
\label{eq:primary_energy}
\end{equation}
Rotational symmetry spreads the energy Hessian $2\Delta B$ uniformly around the circle of circumference $2\pi$, giving density $\Delta B/\pi$ in both phases.
For an interval $I=(-\alpha,\alpha)$ of half-opening $0<\alpha<\pi$, the vacuum modular Hamiltonian $-\ln\sigma_I$ weights the local energy by $2\pi(\cos x-\cos\alpha)/\sin\alpha$ \cite{Casini:2011kv,Cardy:2016fqc}.
This weight peaks at the midpoint and vanishes at the endpoints, so integrating the common density gives
\begin{equation}
\frac{M_I''}{B}=\frac{2\Delta}{\sin\alpha}
\int_{-\alpha}^{\alpha}(\cos x-\cos\alpha)\,\mathrm dx
=4\Delta(1-\alpha\cot\alpha).
\label{eq:primary_modular_energy}
\end{equation}
The linear change in boundary stress is proportional to the vacuum two-point function of the stress tensor with $\mathcal O$, which vanishes.
The first law of entanglement therefore gives $S_{1,\pm}=0$, since the entropy and the modular energy have the same first derivative.
At second order, the identity between relative entropy, modular energy, and ordinary entropy gives $S_{2,\pm}=M_I''-\FF_{I,\pm}$.
Here $\FF_{I,\pm}$ measures how sharply the perturbed interval state can be distinguished from the vacuum using measurements confined to the interval.
Rotational symmetry through second order makes the energy density uniform around the boundary circle, so the two preparations give the same $M_I''$ even though the modular weight varies across the interval. \\

\noindent
On the thermal cylinder for the interval, the perturbed ket and bra lie at imaginary modular times $\pi+\alpha$ and $\pi-\alpha$.
Their contributions at frequency $\omega$ combine to $2\cosh(\alpha\omega)$ for the real tangent and $2\mathrm{i}\sinh(\alpha\omega)$ for the imaginary tangent \cite{Lashkari:2015dia,Faulkner:2017tkh,Sarosi:2017rsq}.
The Bogoliubov--Kubo--Mori form weights the squared magnitudes with the same vacuum spectrum, giving each relative-entropy Hessian \cite{Petz:1993er} (Appendix~\ref{app:detailD2}).
At zero frequency, the two insertions add for the field phase and cancel for the momentum phase.
At any frequency, their squared magnitudes differ by the same amount, four, since $\cosh^2 x-\sinh^2 x=1$.
Subtracting the ordinary entropy Hessians cancels the common modular energy and reverses the order of the relative-entropy Hessians, $S_{2,-}-S_{2,+}=\FF_{I,+}-\FF_{I,-}$.
What remains is a single vacuum integral $\beta^2s^{2\Delta}\int_{\mathbb R}\operatorname{sech}^{2\Delta+2}(t/2)\,dt$ over real modular time $t$, with $s=\sin\alpha$.
The substitution $z=\tanh(t/2)$ gives $dt=2\,dz/(1-z^2)$, and the endpoints $t=\pm\infty$ become $z=\pm1$.
Consequently,
\begin{align}
S_{2,-}-S_{2,+}
&=2\beta^2 I_\Delta s^{2\Delta}>0 .
\label{eq:phasegap_main}
\end{align}
The moment $I_b=\int_{-1}^{1}(1-z^2)^b\,dz$ exists for $b>-1$.
At $b=3/2$, the substitution $z=\sin u$ gives $I_{3/2}=2\int_0^{\pi/2}\cos^4u\,du=3\pi/8$.
The boundary result holds for every $\Delta>0$ (Appendix~\ref{app:phasegap}).
Combining it with the phase average (Appendix~\ref{app:elementaryentropy}) gives each entropy, in elementary form for $\Delta=3/2$ (Appendix~\ref{app:cft_three_halves}).
Normalizing one bulk quantum fixes $\beta^2=\pi a^2/2$ (Appendix~\ref{app:kg}).
Varying the state amplitude at fixed endpoints gives the interval response $S_{2,\pm}$, while comparing openings at the same time gives $\mathcal Q_\pm=-(s^2\partial_\alpha^2-2)S_{2,\pm}$.
For rotationally invariant states, the latter measures the energy along a geodesic (Appendix~\ref{app:chordphasegeneral}), with independent stress perturbations treated in Appendix~\ref{app:detailD8}.
On a chord with signed offset $p=\cos\alpha$, write its longitudinal Klein coordinate as $t_{\rm ch}=sz$, $-1<z<1$.
Then, $r^2=p^2+t_{\rm ch}^2$, $D=s^2(1-z^2)$, and the unit hyperbolic line element is $d\sigma=dz/(1-z^2)$.
Integration along the entire chord gives
\begin{equation}
(\XX D^\nu)(p)=I_{\nu-1}s^{2\nu}\qquad(\nu>0).
\label{eq:powerchords}
\end{equation}
For rotational data, $\DD=s^2\partial_\alpha^2-2$ sends $s^k$ to $(k-2)(k+1)s^k-k^2s^{k+2}$.
Applying $-\DD/(4\pi)$ to the measured phase difference in Eq.~\eqref{eq:phasegap_main} gives
\[
\XX(e_--e_+)
=a^2\Delta(1-\Delta) I_{\Delta-1}s^{2\Delta}
+a^2\Delta^2 I_\Delta s^{2\Delta+2}.
\]
To simplify the coefficients, integrate the derivative of $z(1-z^2)^\Delta$ over $-1<z<1$.
The vanishing endpoint terms give $2\Delta I_{\Delta-1}=(2\Delta+1)I_\Delta$.
Eq.~\eqref{eq:powerchords} maps a radial power $D^b$ into a known coefficient times a power of $s$.
The two observed powers $s^{2\Delta}$ and $s^{2\Delta+2}$ change differently as the interval width varies, so their coefficients can be read off separately.
Matching both coefficients gives the density difference $a^2\Delta(1-\Delta)D^\Delta+a^2\Delta^2D^{\Delta+1}$, or, in terms of the radius,
\begin{equation}
e_--e_+=a^2\Delta(1-\Delta r^2)D^\Delta .
\label{eq:genericcross}
\end{equation}
For $\Delta>1$ and nonzero amplitude, the momentum phase has the larger density at $r<1/\sqrt{\Delta}$ and the field phase at $r>1/\sqrt{\Delta}$.
When a polygon straddles this circle, its energy difference depends on how much positive and negative density it encloses.
For a polygon containing the center, each ray stays inside from the origin to a single boundary radius $R$, so we first integrate along the ray and set $u=R^2$.
With the positive factor $a^2\Delta/2$ removed, the accumulated difference in Eq.~\eqref{eq:raygap} is
\[
\int_0^u (1-\Delta w)(1-w)^{\Delta-3/2}\,dw
= u(1-u)^{\Delta-1/2}
+\tfrac12\int_0^u w(1-w)^{\Delta-3/2}\,dw .
\]
Differentiating with respect to $u$ gives the integrand on the left, and both sides vanish at $u=0$.
The right-hand side is strictly positive for $0<u\leq1$ and $\Delta>1$, including the limit $u\uparrow1$.
Each ray through the interior of a convex polygon containing the origin accumulates a positive difference, even if its outer segment crosses the sign-change circle.
The angular integral then gives $V_-''>V_+''$ under the volume comparison hypotheses.
For a polygon lying wholly at $r>1/\sqrt\Delta$, every contribution is negative, and the ordering reverses.
To compare with the conserved energy, one must also account for the rate of the boundary clock $\tau$ relative to a local observer.
An observer normal to the slice accumulates proper time proportional to $\cosh\chi$ per unit $\tau$, so this factor, the lapse, weights the local density in the conserved charge.
Using $w=r^2$ and $d\mu=(1-r^2)^{-3/2}r\,dr\,d\varphi$, the difference of the two charges is proportional to
\[
\int_0^1 (1-\Delta w)(1-w)^{\Delta-2}\,dw
=\big[w(1-w)^{\Delta-1}\big]_{w=0}^{w=1}=0 ,
\qquad \Delta>1 .
\]
Equal total energy uses the full disk and the radial weight of the global clock. In contrast, the polygon integrates with unit weight per proper area only over its interior.

\subsection{Triangle Signs and Positive Interval Entropies}
\label{sec:interval-and-polygon-signs}
\noindent
In the known field phase at $\Delta=3/2$, one interval reading fixes $\beta^2$ whenever the entropy response per unit amplitude squared is nonzero.
To determine whether it stays positive for every opening, divide by $s^2$ and start from the relative-entropy Hessians, writing $t_\alpha=\tan(\alpha/2)$,
\begin{equation}
\frac{\FF_{I,\pm}}{\beta^2}=3\pi\left(t_\alpha-\tfrac12s\ \pm\tfrac18s^3\right).
\label{eq:closed_fisher_main}
\end{equation}
For a short arc, we set $b=s^2(1-z^2)$, $-1<z<1$.
The common modular-energy integral has integrand $2\Delta\beta^2s^2z^2/(1-b)$.
The phase-averaged relative entropy inserts a factor $b^{\Delta-1}$ in that integrand (Appendix~\ref{app:elementaryentropy}).
At $\Delta=3/2$, their subtraction gives $(1-\sqrt b)/(1-b)=1/(1+\sqrt b)$, with the value at $b=1$ fixed by continuity.
The field term subtracts $\beta^2 I_{3/2}s^3$, giving
\begin{equation}
\frac{S_{2,+}}{\beta^2s^2}
=3\!\int_{-1}^1\!\frac{z^2\,dz}{1+s\sqrt{1-z^2}}-\frac{3\pi}{8}s
\ge 6-\frac{15\pi}{8}>0 .
\label{eq:positive32}
\end{equation}
Purity gives complementary arcs the same entropy Hessian, so it suffices to vary $s$ from zero to one.
As $s$ increases, the integral decreases and $-3\pi s/8$ becomes more negative.
The minimum occurs at the semicircle, where $\int_{-1}^{1}(1-\sqrt{1-z^2})dz=2-\pi/2$.
Thus, $S_{2,+}(\pi/2)=\beta^2(6-15\pi/8)$, and the response per unit squared amplitude is strictly positive for every proper arc. \\

\noindent
Since $\DD(t_\alpha-s/2)=s^3/2$ and $\DD s^3=4s^3-9s^5$, Eq.~\eqref{eq:closed_fisher_main} gives $(\XX e_+)(p)=(3\beta^2/32)s^3(9p^2-1)$.
Although the ordinary entropy Hessian is positive on every proper arc, the energy integrated along its geodesic changes sign at $|p|=1/3$.
Using $I_{1/2}=\pi/2$ and $I_{3/2}=3\pi/8$ in Eq.~\eqref{eq:powerchords} gives the local density, $e_+d\mu=(3a^2/8)(3r^2-1)d^2y$.
Inside $r=1/\sqrt3$, the negative mass-squared term outweighs the gradient energy.
A chord tangent to this circle samples positive density on either side of its point of contact, so cancellation along a whole chord occurs only for a smaller offset.
The mode still satisfies the AdS$_3$ stability bound $M_\phi^2L^2\geq-1$, since $M_\phi^2L^2=-3/4$ \cite{Breitenlohner:1982jf}.
The mass term in the stress tensor is proportional to the metric, so it drops out when contracted twice with a null vector $k$ satisfying $g_{\mu\nu}k^\mu k^\nu=0$.
Thus, $T_{\mu\nu}k^\mu k^\nu=(k^\mu\partial_\mu\phi)^2\ge0$ even where a timelike observer measures negative density.
Here and below, Greek spacetime indices range from $0$ to $2$, with repeated upper and lower indices summed. \\

\begin{figure}[t]
\centering
\includegraphics[width=.96\columnwidth]{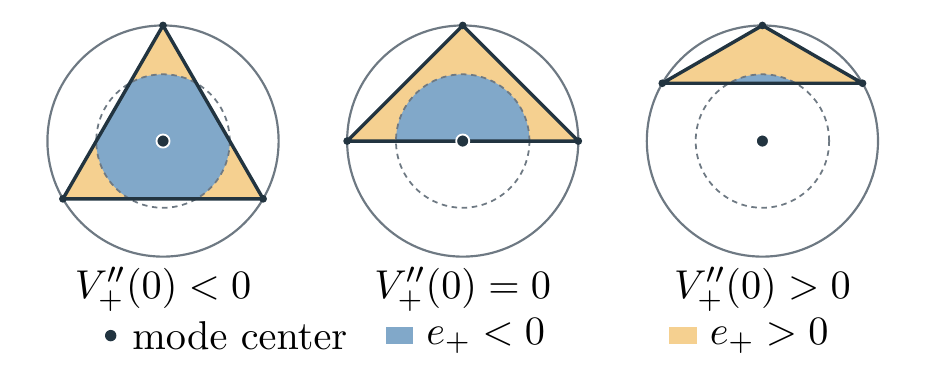}
\caption{Three ideal triangles in the same $\Delta=3/2$ field excitation have identical vacuum area $\pi L^2$.
From left to right, the dot marking the mode center lies inside the triangle, on a side, and outside the triangle.
The corresponding volume responses at quadratic order are negative, zero, and positive.
The dashed circle marks $r=1/\sqrt3$, where the normal energy density changes sign.}
\label{fig:triangle-signs}
\end{figure}

\noindent
The Euclidean field $(3r^2-2)y/4$ has divergence $3r^2-1$, reducing the integral to side fluxes (Appendix~\ref{app:onephase}).
For a convex polygon $P=P_0$ with $N_v\ge3$ vertices, the intervening arcs have openings $2\alpha_j$, $0<\alpha_j<\pi$, $j=1,\ldots,N_v$, and $\sum_{j=1}^{N_v}\alpha_j=\pi$.
On side $j$, the flux $\int_{-\sin\alpha_j}^{\sin\alpha_j}(p_j/4)[3(p_j^2+t^2)-2]dt$ equals $\sin(4\alpha_j)/8$, with $p_j=\cos\alpha_j$.
Thus, $\int_Pe_+d\mu=(3a^2/64)\sum_{j=1}^{N_v}\sin(4\alpha_j)$.
Eq.~\eqref{eq:geometricresponse} and the semicircle Hessian eliminate $a^2$, giving
\begin{equation}
\frac{V_+''(0)}{L^2}=
\frac{6S_{2,+}(\pi/2)}{c(16-5\pi)}
\sum_{j=1}^{N_v}\sin(4\alpha_j).
\label{eq:semicircle_calibration}
\end{equation}
For a triangle, $\sum_{j=1}^3x_j=2\pi$ with $x_j=2\alpha_j$, so $\sum_{j=1}^3\sin(2x_j)=-4\prod_{j=1}^3\sin x_j$ (Appendix~\ref{app:polygonextremes}).
The half-plane inequalities $\boldsymbol n_j\cdot y\leq\cos(x_j/2)$, with outward Euclidean unit normals $\boldsymbol n_j$ for $j=1,2,3$, put the origin inside precisely when every gap is smaller than $\pi$.
The three panels of Fig.~\ref{fig:triangle-signs} give the possible signs.
In the left panel, three positive sines give a negative volume Hessian.
In the middle panel, a diameter gives zero, and in the right panel, one gap above $\pi$ gives a positive Hessian. \\

\begin{figure}[t]
\centering
\includegraphics[width=.59\linewidth]{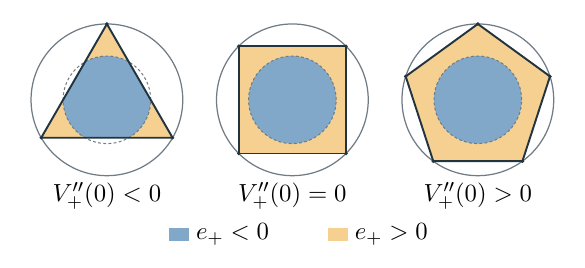}
\caption{Every proper interval has a positive entropy Hessian in this field-phase excitation at $\Delta=3/2$.
Regular polygons with three, four, and five sides have negative, zero, and positive quadratic volume response, respectively.
Inside the dashed circle $r=1/\sqrt3$, the normal energy density $e_+$ is negative; outside it, the density is positive.
The square encloses exactly enough positive energy to cancel the negative contribution.}
\label{fig:signs}
\end{figure}

\noindent
The regular triangle minimizes the shape sum $\sum_{j=1}^{N_v}\sin(4\alpha_j)$ over all convex ideal polygons, with value $-3\sqrt3/2$ (Appendix~\ref{app:polygonextremes}).
For a centered regular polygon, the sign can also be read from the mean squared radius in one of the $N_v$ congruent Euclidean triangles with apex at the origin.
Its boundary vertex vectors $v_1,v_2$ have unit length and $v_1\cdot v_2=\cos\delta$, $\delta=2\pi/N_v$.
Writing $y=u v_1+v v_2$ gives the constant-Jacobian average $2\int_0^1du\int_0^{1-u}f(u,v)\,dv$ for any integrable $f$.
The averages of $u^2,v^2,uv$ are $1/6,1/6,1/12$, respectively.
Using $r^2=u^2+v^2+2uv\cos\delta$ and writing $A_{\rm E}(P)$ for Euclidean area gives
\[
\frac{1}{A_{\rm E}(P)}\int_P r^2\,d^2y=\frac{2+\cos\delta}{6},
\qquad
\frac{1}{A_{\rm E}(P)}\int_P e_+\,d\mu=\frac{3a^2}{16}\cos\delta .
\]
For the nonzero $\Delta=3/2$ field mode, the negative central density outweighs the outer contribution in a regular triangle, cancels it in a square, and is outweighed by it for $N_v\ge5$ (Fig.~\ref{fig:signs}).
More generally, the lowest field mode gives positive entropy Hessians on every proper arc and a shrinking regular triangle precisely for $\Delta_*<\Delta<\Delta_T$ (Appendix~\ref{app:onephase}).
The endpoints mark distinct zeros, the semicircle Hessian at $\Delta_*\simeq1.462674$ and the triangle energy at $\Delta_T\simeq1.609144$.

\subsection{Motion of Geodesic Sides}
\noindent
A change in the metric affects both a geodesic's length and its position between fixed endpoints \cite{Ghosh:2019jgd}.
For radial matter with no linear metric response, boundary-anchored conformal coordinates give $h_\lambda=\exp(2\lambda^2\psi+o(\lambda^2))h_0$.
A small deformation of a vacuum geodesic with fixed endpoints has vanishing first-order effect on its length, by the defining stationarity of the geodesic.
Here the displacement is of order $\lambda^2$, so its contribution to length in the vacuum metric starts at order $\lambda^4$.
At order $\lambda^2$, it suffices to integrate the metric perturbation on the original geodesic, giving $S_2=(c/3)\int_{\gamma_I(0)}\psi\,d\sigma$.
For the enclosed area, the same displacement adds or removes a strip whose width is already of order $\lambda^2$.
On side $j$, write this signed outward width as $\lambda^2\zeta_j(\sigma)+o(\lambda^2)$, with vacuum arclength $\sigma\in\mathbb R$ and outward normal distance $n_\perp$ both in units of $L$.
Neighboring geodesics perpendicular to the side start at unit separation with zero rate of separation.
In unit negative curvature, their separation grows as $\cosh n_\perp$, giving vacuum metric $dn_\perp^2+\cosh^2 n_\perp\,d\sigma^2$.
Write $h_\lambda/L^2=\exp(2\omega_\lambda)h_0/L^2$, with $\omega_\lambda=\lambda^2\psi+o(\lambda^2)$ locally together with its first derivatives.
For a normal graph $n_\perp=q_j(\sigma)$, the dimensionless length integrand is
\[
\mathscr L(\sigma,q_j,q_j')
=\exp[\omega_\lambda(\sigma,q_j)]
\sqrt{(q_j')^2+\cosh^2 q_j},
\qquad q_j=\lambda^2\zeta_j+o(\lambda^2).
\]
Primes denote derivatives with respect to arclength $\sigma$.
The slope dependence of the length integrand gives $\partial_{q_j'}\mathscr L=\lambda^2\zeta_j'+o(\lambda^2)$.
Its position dependence gives $\partial_{q_j}\mathscr L=\lambda^2(\partial_{\boldsymbol n}\psi+\zeta_j)+o(\lambda^2)$, with the first term from the outward gradient of the conformal factor and the second from differentiating $\cosh^2q_j$ under the square root.
Substitution into $d(\partial_{q_j'}\mathscr L)/d\sigma=\partial_{q_j}\mathscr L$ and division by $\lambda^2$ give the linear displacement equation $\zeta_j''-\zeta_j=\partial_{\boldsymbol n}\psi$.
The homogeneous solutions $\exp(\pm\sigma)$ each grow at one end, so fixed endpoints and normalizable falloff select a unique solution decaying at both ends.
A unit point source produces the decaying displacement $-\exp(-|\sigma|)/2$, whose first derivative jumps by one at $\sigma=0$.
Its integral is $-1$, so the signed area swept per unit $\lambda^2$ is $\int_{\mathbb R}\zeta_j\,d\sigma=-\int_{\gamma_j(0)}\partial_{\boldsymbol n}\psi\,d\sigma$.
In two spatial dimensions, the area element changes by $2\lambda^2\psi\,d\mu$, and side $j$ sweeps the signed area $\lambda^2\int_{\mathbb R}\zeta_j\,d\sigma$.
Summing over $j=1,\ldots,N_v$ and taking two derivatives in $\lambda$ gives
\begin{equation}
\frac{V''(0)}{L^2}=4\int_{P_0}\psi\,d\mu
-2\int_{\partial P_0}\partial_{\boldsymbol n}\psi\,d\sigma .
\label{eq:edgefluxmain}
\end{equation}
For the regular triangle at $\Delta=3/2$, the two terms are approximately $+0.218a^2/c$ and $-9.400a^2/c$ in this gauge (Appendix~\ref{app:fieldflux32}). \\

\begin{figure}[tbp]
\centering
\includegraphics[width=.80\linewidth]{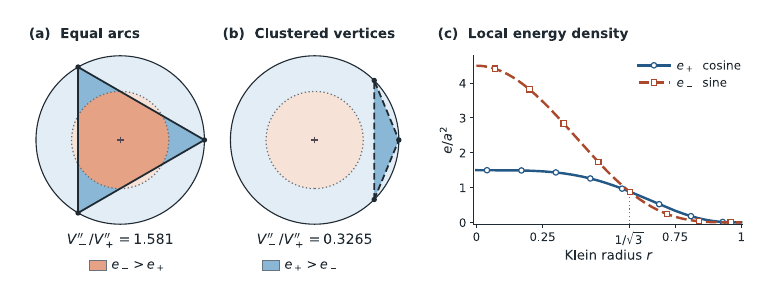}
\caption{The same two scalar states reverse their volume-response ordering when the boundary partition changes.
The darker region is the vacuum triangle in Klein coordinates for (a) three equal arcs and (b) endpoints at $-\pi/4,0,\pi/4$.
The dotted circle $r=1/\sqrt3$ separates the inner region with $e_->e_+$ from the outer region with $e_+>e_-$.
(c) At scaling dimension $\Delta=3$, the lines are the bulk cosine ($+$) and sine ($-$) densities.
The open circles and squares show the densities reconstructed from the boundary interval entropies by inverting their radial line integrals.}
\label{fig:matter}
\end{figure}

\noindent
At $\Delta=3$, the scalar initial data give two explicit radial energy densities (Appendix~\ref{app:detailE1}).
Both inversion of the chord data (Appendix~\ref{app:detailE2}) and filtering the interval entropies (Appendix~\ref{app:detailE3}) recover those profiles (Fig.~\ref{fig:matter}).
Integrating either directly over a triangle (Appendix~\ref{app:detailE4}) or over the disk with its exterior caps removed (Appendix~\ref{app:partition_ordering}) agrees with the area obtained by following the moving sides (Appendix~\ref{app:directarea}).
For three equal arcs, the momentum phase gives about $58\%$ more volume response than the field phase at the same amplitude, with $V_-''/V_+''\simeq1.58093$.
For these positive-mass data, Appendix~\ref{app:scalarmax} gives conditions under which this slice area is the maximal volume.

\section{Scalar Volume from Two Widths and Displaced Quartets}
\label{sec:square-calibration}

\noindent
For a pure family in the known, nonzero lowest scalar mode of dimension $3/2$, two interval entropy responses determine the phase and squared amplitude when the center is known and the perturbation has definite time parity.
We first use two interval widths to determine the phase and squared amplitude, then examine how a reading changes when the scalar center or the boundary endpoints move. \\

\noindent
Suppose the position and radial shape of a scalar excitation are known. However, its amplitude and its time-reflection parity are unknown.
For the lowest rotational mode of dimension $\Delta=3/2$, measuring the entropy response at two interval widths suffices to predict the volume response of any polygon with specified vertices.
In this pure family, rotational symmetry through second order makes equal widths give equal entropy Hessians, the second derivatives with respect to the state amplitude.
It also makes the boundary energy uniform, so its integral fixes the modular-energy response at every width.
An independent second-order boundary reparametrization would shift these readings according to Eq.~\eqref{eq:endpointjets} and is excluded from this mode family.
Write $B=\beta^2$ for the squared amplitude and $\varsigma\in\{+,-\}$ for the parity.
At the reflection time, the field phase $(+)$ has only a field displacement, and the momentum phase $(-)$ has only normal momentum.
For an interval of full opening $2\alpha$, the entropy Hessian is $S_{2,\varsigma}(\alpha)=B s^2H_\varsigma(s)$, where $s=\sin\alpha$ and $H_-(s)>H_+(s)>0$.
Entropy Hessians depend on $\beta$ through its square.
Moreover, either phase fits any positive reading $S_2$ at one width by choosing $B_\varsigma=S_2/[s^2H_\varsigma(s)]$.
For example, the choice $B_-/B_+=(16-5\pi)/(16-3\pi)$ gives identical semicircle Hessians and opposite signs for the regular-triangle volume responses.
If the total cylinder-energy Hessian is also measured, Eq.~\eqref{eq:primary_energy} fixes $B$, and one proper interval distinguishes the phases.
A second interval width can supply the missing information because changing the width affects the two phases differently. \\

\noindent
The two shapes $H_+(s)$ and $H_-(s)$ can be read directly from the field-phase response and the phase difference.
At scaling dimension $3/2$, Eq.~\eqref{eq:positive32} gives
\[
H_+(s)=3\int_{-1}^{1}\frac{z^2\,dz}{1+s\sqrt{1-z^2}}-\frac{3\pi}{8}s,
\qquad
H_-(s)=H_+(s)+\frac{3\pi}{4}s,\qquad 0<s\leq1 .
\]
The second expression follows by dividing the entropy difference in Eq.~\eqref{eq:phasegap_main} by $B s^2$.
Increasing $s$ decreases the integral and the term $-3\pi s/8$ in $H_+$, while $H_--H_+$ increases linearly.
Consequently, the two phase profiles cannot remain proportional as the interval width varies. \\

\noindent
Choose two half-openings $0<\alpha_j<\pi$ with distinct sines, ordered so that $s_1<s_2$, where $s_j=\sin\alpha_j$ for $j=1,2$.
At each width, the predicted reading per unit squared amplitude is $h_{j,\varsigma}=s_j^2H_\varsigma(s_j)>0$.
As $B$ varies, the pair of measurements $(Bh_{1,\varsigma},Bh_{2,\varsigma})$ traces a ray.
Its slope $r_\varsigma=h_{2,\varsigma}/h_{1,\varsigma}$ can identify the phase because it is independent of $B$.
The single-width ratio $H_-(s)/H_+(s)=1+(3\pi/4)s/H_+(s)$ increases with $s$ because $H_+(s)$ is positive and decreasing by Eq.~\eqref{eq:positive32}.
Consequently,
\[
\frac{r_-}{r_+}=
\frac{H_-(s_2)/H_+(s_2)}{H_-(s_1)/H_+(s_1)}>1.
\]
For exact readings and $B>0$, the observed ratio $S_2(\alpha_2)/S_2(\alpha_1)$ selects one of these two values and $B=S_2(\alpha_1)/h_{1,\varsigma}$ fixes the normalization.
Complementary half-openings $\alpha_2=\pi-\alpha_1$ have equal sines and produce the same reading by rotational invariance and purity.
If $B=0$, both Hessians vanish and the scalar contributes zero polygon response at this order, with no distinguishable parity. \\

\noindent
Suppose the reported readings are $\widetilde y_j$ with absolute errors at most $\epsilon_j\geq0$ for $j=1,2$.
The two error bars form a rectangle in the plane of readings.
A phase fits a nonzero excitation wherever its ray meets that rectangle away from the origin.
Equivalently, dividing each error interval by $h_{j,\varsigma}>0$ gives the squared amplitudes allowed by that width.
The intersection required by a common amplitude is
\[
\mathcal I_\varsigma=(0,\infty)\cap
\left[\max_{1\leq j\leq2}\frac{\widetilde y_j-\epsilon_j}{h_{j,\varsigma}},
\ \min_{1\leq j\leq2}\frac{\widetilde y_j+\epsilon_j}{h_{j,\varsigma}}\right],
\]
where the maximum and minimum enforce both error bars.
Each nonempty $\mathcal I_\varsigma$ gives all amplitudes for that phase.
If both intervals are nonempty, the parity is ambiguous.
If both are empty, no nonzero excitation of the assumed mode fits the readings. \\

\noindent
For a state of parity $\varsigma$ and squared amplitude $B>0$, the error tolerances can also be set before taking data.
A state of the other parity $\tau$ and squared amplitude $B'>0$ could produce the same pair of readings precisely when
\[
|B h_{j,\varsigma}-B' h_{j,\tau}|\leq 2\epsilon_j,
\qquad j=1,2 .
\]
The factor of two accounts for errors in either direction, since predictions $2\epsilon_j$ apart can give the same reported midpoint.
Dividing the two inequalities by $h_{j,\tau}>0$ gives intervals for $B'$ with centers $B h_{j,\varsigma}/h_{j,\tau}$ and half-widths $2\epsilon_j/h_{j,\tau}$.
Their upper endpoints are positive, so a common positive $B'$ exists if and only if the distance between their centers is no greater than the sum of their half-widths.
Consequently, every allowed measurement error still permits a unique parity assignment if and only if
\begin{equation}
B\left|h_{1,\varsigma}h_{2,\tau}-h_{2,\varsigma}h_{1,\tau}\right|
>2\left(\epsilon_1 h_{2,\tau}+\epsilon_2 h_{1,\tau}\right).
\label{eq:two-width-error-separation}
\end{equation}
The determinant in Eq.~\eqref{eq:two-width-error-separation} has magnitude $h_{1,\varsigma}h_{1,\tau}|r_\tau-r_\varsigma|$.
When the sines approach the same nonzero value, both slopes approach one, and the guaranteed error tolerance tends to zero.
At equality, the intervals for $B'$ touch, so some readings fit both phases.
Actual readings can still identify a unique phase through $\mathcal I_\pm$ when the advance guarantee fails. \\

\noindent
The phase test uses the interval profiles of Appendix~\ref{app:cft} and excludes the independent boundary reparametrizations treated in Appendix~\ref{sec:matching-hessians}.
Appendix~\ref{app:scalar} converts the inferred phase and amplitude into polygon responses and derives the sign change when the endpoints move.
For a displaced mode, the reconstruction uses the line-integral inversion in Appendix~\ref{app:tomography}.
The volume interpretation uses the geometric conditions in Appendix~\ref{app:geometry}.

\subsection{Square Measurement and Polygon Prediction}
\begin{figure}[tbp]
\centering
\includegraphics{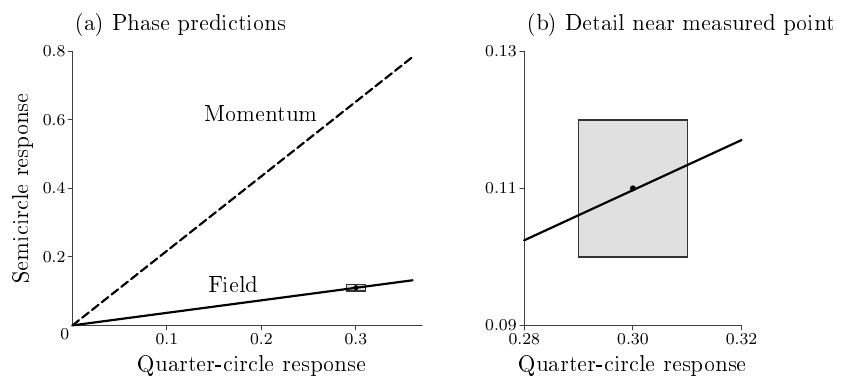}
\caption{Identifying a scalar phase from two interval-entropy responses.
In (a), each ray gives the pairs of readings predicted for one phase as the squared amplitude increases from zero.
In (b), the same data are shown on a finer scale.
The dot marks the reported pair of readings, and the shaded rectangle covers their allowed errors of $0.01$ in each direction.}
\label{fig:two-width-phase}
\end{figure}

\noindent
A square provides both widths in a single endpoint configuration, as in Fig.~\ref{fig:quad}(b).
Its four sides subtend quarter circles with full opening $\pi/2$, and its two diagonals subtend semicircles with full opening $\pi$.
For an exactly rotationally invariant completion, denote the two entropies by $S_{\rm q}=S(\pi/4,\lambda)$ and $S_{\rm h}=S(\pi/2,\lambda)$, using the half-opening as the first argument.
Purity and rotation symmetry reduce each side product in Eq.~\eqref{eq:ptolemy_entropy_ratio} to $\exp(6S_{\rm q}/c)$ and the diagonal product to $\exp(6S_{\rm h}/c)$, giving
\begin{equation}
1+\Pi_\square=\exp[6(S_{\rm h}-S_{\rm q})/c].
\label{eq:square_entropy_ratio}
\end{equation}
If rotation invariance is imposed only through second order, this reduction holds to that order, which is sufficient here.
In the vacuum, $S_{\rm h}-S_{\rm q}=(c/3)\ln[\sin(\pi/2)/\sin(\pi/4)]=(c/6)\ln 2$.
The exponential in Eq.~\eqref{eq:square_entropy_ratio} therefore equals two at $\lambda=0$.
For $D=S_{\rm h}-S_{\rm q}$, the second derivative of the exponential contains a term proportional to $D''$ and a term proportional to $(D')^2$.
In the vacuum, $D'=0$, so only the difference of the two entropy Hessians remains, multiplied by the vacuum factor of two.
Thus, the square supplies the observable
\begin{equation}
d_\square\equiv S_2(\pi/2)-S_2(\pi/4)
=\frac{c}{12}\Pi_\square''(0)=B k_\varsigma .
\label{eq:square_readout_main}
\end{equation}
The two constants $k_\varsigma$ can be obtained by evaluating the modular energy and the CFT Hessians in Eqs.~\eqref{eq:primary_modular_energy} and~\eqref{eq:closed_fisher_main} at the two half-openings.
Explicitly,
\[
\begin{array}{c|cc}
\varsigma & U_\varsigma=S_{2,\varsigma}(\pi/2)/B
& S_{2,\varsigma}(\pi/4)/B \\ \hline
+&6-15\pi/8& 6+3\pi/2-75\sqrt2\,\pi/32\\
-&6-9\pi/8&6+3\pi/2-69\sqrt2\,\pi/32 .
\end{array}
\]
Subtracting the last column from the middle one gives $k_+=(3\pi/32)(25\sqrt2-36)<0$ and $k_-=(3\pi/32)(23\sqrt2-28)>0$.
Thus, for any nonzero excitation in either phase, a semicircle response below the quarter-circle response selects the field phase, and the reverse ordering selects the momentum phase.
The magnitude of their difference then gives
\begin{equation}
B=\frac{d_\square}{k_\varsigma}=\frac{c\Pi_\square''(0)}{12k_\varsigma}.
\label{eq:square_phase_amplitude}
\end{equation}
Measuring $d_\square$ alone selects the phase by its sign and fixes the squared amplitude by its magnitude.
The separate semicircle reading provides a consistency test, since the selected phase must also predict $d_\square/S_2(\pi/2)=k_\varsigma/U_\varsigma$.
For measured square readings $\widetilde y_{\rm h},\widetilde y_{\rm q}$ with respective error bounds $\epsilon_{\rm h},\epsilon_{\rm q}$, set $\widetilde d_\square=\widetilde y_{\rm h}-\widetilde y_{\rm q}$.
The triangle inequality gives $|\widetilde d_\square-d_\square|\leq\epsilon_{\rm h}+\epsilon_{\rm q}$.
A difference larger in magnitude than the sum of the error bounds fixes its true sign.
For example, suppose the semicircle and quarter-circle Hessians are reported as $0.11$ and $0.30$, respectively, each with absolute error at most $0.01$.
Their true difference lies in $[-0.21,-0.17]$, which selects the field phase if the measurements fit this mode.
In that phase, the two Hessians per unit $B$ are approximately $0.10951$ and $0.29938$, so dividing each error interval by its predicted coefficient gives overlapping ranges.
Their intersection has endpoints approximately $0.969$ and $1.035$ for the squared amplitude $B$.
In Fig.~\ref{fig:two-width-phase}, the field-phase prediction passes through the measured range, while the momentum-phase prediction lies above it. \\

\noindent
Once the two square readings fix the phase and squared amplitude, the only remaining input for a polygon $P$ is its geometry relative to the scalar center.
Let $2\alpha_j$ be its successive angular gaps for $1\le j\le N_v$ and form the sums
\[
F(P)=\sum_{j=1}^{N_v}\sin(4\alpha_j),\qquad A_{\rm E}(P)=\tfrac12\sum_{j=1}^{N_v}\sin(2\alpha_j).
\]
The oriented Euclidean triangle formed by the origin and the endpoints of side $j$ has area $\sin(2\alpha_j)/2$.
The signed sum gives the area $A_{\rm E}(P)$ of the Klein polygon, including when the origin lies outside, and the exterior triangles cancel.
This Euclidean area enters because the two phases at $\Delta=3/2$ have a simple sum at equal amplitude.
Their respective densities satisfy $e_+\,d\mu=3a^2(3r^2-1)d^2y/8$ and $e_-\,d\mu=9a^2(1-r^2)d^2y/8$.
The terms proportional to $r^2$ cancel pointwise, so the summed density is a constant $3a^2/4$ times the Euclidean area element.
The field-phase flux in Eq.~\eqref{eq:polygon32} gives $E_+(P)=3a^2F(P)/64$, while the sum fixes $E_-(P)=3a^2[16A_{\rm E}(P)-F(P)]/64$, with $E_\varsigma(P)=\int_P e_\varsigma\,\mathrm d\mu$.
Multiplication by $24\pi/c$, followed by $a^2=2B/\pi$, gives the volume calibrated by the boundary amplitude,
\begin{equation}
\frac{4c}{9B}\frac{V_{\varsigma}''(P)}{L^2}
=\begin{cases}
F(P),&\varsigma=+,\\
16A_{\rm E}(P)-F(P),&\varsigma=-.
\end{cases}
\label{eq:square_to_polygon}
\end{equation}
Using Eq.~\eqref{eq:square_phase_amplitude} eliminates $B$ and gives the direct square-to-polygon prediction
\begin{equation}
\frac{V''(P)}{L^2}=\frac{3\Pi_\square''(0)}{16}
\begin{cases}
F(P)/k_+, &\text{field},\\
(16 A_{\rm E}(P)-F(P))/k_-,&\text{momentum}.
\end{cases}
\label{eq:direct_square_volume}
\end{equation}
At finite precision, Eq.~\eqref{eq:square_to_polygon} maps each allowed $B\in\mathcal I_\varsigma$ to a volume Hessian by multiplication by the known polygon factor.
If both parities fit, take the union of the two predicted intervals.
If the zero-amplitude tangent is admitted, include zero whenever both measurement intervals contain zero. \\

\noindent
For the centered and exterior triangles in Fig.~\ref{fig:triangle-signs}, $(F,A_{\rm E})=(-3\sqrt3/2,3\sqrt3/4)$ and $(3\sqrt3/2,\sqrt3/4)$, respectively.
The same two square readings predict different responses for these two triangles.
In the field phase, the centered triangle shrinks, and the exterior triangle grows by the same amount at quadratic order.
In the momentum phase, both triangles grow, with the exterior triangle gaining $5/27$ as much area as the centered one to this order.
The calculation in Appendix~\ref{app:onephase} also derives the two square coefficients by integrating the bulk energy against the quartet kernel.
For a tangent containing both field and momentum quadratures, two widths still determine their squared amplitudes within the single-primary model, as shown in Appendix~\ref{sec:two-width-proof}.
The entropy Hessians contain only squares of amplitudes, so they cannot recover their relative sign.
With both quadratures present, the bulk momentum flux can be nonzero, and the slice can lose time-reflection symmetry.
Identifying the polygon result with a maximal volume therefore uses definite parity and the volume-comparison hypotheses of Appendix~\ref{app:geometry}.

\subsection{Displaced Quartets and Angular Reconstruction}
\label{sec:boosted-quartets}

\noindent
We now distinguish two ways of changing the relative geometry of measurement and state.
First, displace the center of the excitation while keeping the boundary endpoints fixed.
Second, hold the excitation fixed and change which four endpoints enter the entropy contrast.
In both experiments, each state derivative is taken at fixed endpoint positions and fixed displacement.

\subsubsection{Energy from Chord Positions and Orientations}
\label{sec:angular-chords}

\begin{figure}[t]
\centering
\includegraphics[width=.96\columnwidth]{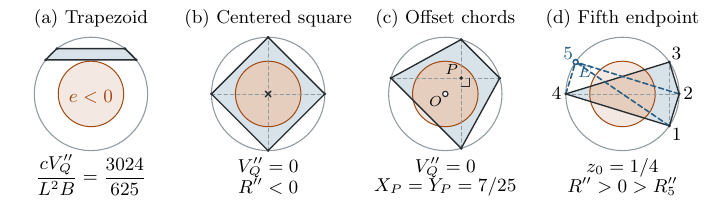}
\caption{For the dimension-$3/2$ mode, blue marks $Q$ and orange the disk $e<0$.
The volume response is positive in (a) and zero in (b,c,d).
In (c), the diagonals cross at a right angle in the Euclidean coordinates of the mode-centered Klein disk.
In (d), the two quartets use five endpoints, $(\pm5\sqrt7/16,9/16)$, $(5\sqrt7/16,-9/16)$ and $(\pm1,0)$, and give readings of opposite sign.
Here $E=Q_5\setminus Q$ marks the portion of the new polygon beyond the old one.
The reading with the fifth endpoint certifies that negative energy lies somewhere in the full quadrilateral $Q_5$.}
\label{fig:squarecertificate}
\end{figure}

\noindent
For the nonzero field phase at $\Delta=3/2$, let the origin of the Klein disk be the mode center, $r^2=X^2+Y^2$, and $B=\pi a^2/2$.
Multiplying the normal energy density by the hyperbolic area element gives
\begin{equation}
e_+\,\mathrm d\mu=\frac{3B}{4\pi}(3r^2-1)\,\mathrm dX\,\mathrm dY .
\label{eq:modedensity}
\end{equation}
The disk $r<1/\sqrt3$ has negative density, while the outer region has positive density.
Fig.~\ref{fig:squarecertificate}(a,b) shows a trapezoid that samples only the positive region and a square in which the two contributions cancel. \\

\noindent
An interval of half-opening $\alpha$ measures a chord of signed offset $p=\cos\alpha$, with $-1<p<1$ and $s=\sqrt{1-p^2}$.
Let $I_\gamma$ be the integral of normal energy density along that chord in hyperbolic arclength.
In the centered state, rotational symmetry makes $G(p)=I_\gamma/s$ a function only of the offset.
Let $t$ be signed Euclidean distance along the chord from its midpoint, so $r^2=p^2+t^2$.
Along each half, increasing $r$ from $|p|$ to $1$ gives $|\mathrm dt|=r\,\mathrm dr/\sqrt{r^2-p^2}$.
With $d=\sqrt{1-r^2}$, hyperbolic arclength is $\mathrm d\sigma=s\,|\mathrm dt|/d^2$, so $G$ is the Euclidean line integral of $e_+/d^2$.
The resulting radial line transform and its inverse are
\begin{equation}
G(p)=\frac{I_\gamma}{s}=\frac{3B}{32}(1-p^2)(9p^2-1),
\qquad
\frac{e_+}{d^2}=-\frac1\pi\int_r^1\frac{\partial_pG}{\sqrt{p^2-r^2}}dp
=\frac{3B}{4\pi}(3r^2-1)d.
\label{eq:mode_abel}
\end{equation}
The line signal changes sign at $|p|=1/3$, although the local density changes sign at $r=1/\sqrt3$.
A chord between these offsets samples some negative density and still gives a positive integral. \\

\noindent
Move the mode center to $(v,0)$ in the observer's disk, where $|v|<1$.
The disk isometry that returns the center to the origin takes each observed chord to an image chord and preserves both the normal energy and hyperbolic arclength.
Thus, the original line integral equals the centered integral at the offset of the image chord.
The diameter through the displaced center maps to a diameter and has negative integral, whereas the perpendicular observer diameter maps to an offset of magnitude $|v|$ and has positive integral when $|v|>1/3$.
With every offset and orientation, the line data determine the full displaced profile by the inversion in Appendix~\ref{app:radoninversion}.
Appendix~\ref{app:angular-chord-derivation} calculates the image chord, derives the angular response from interval entropies, and checks it against integration of the displaced bulk density.

\subsubsection{Stationary Volume and Opposite Quartet Readings}
\label{sec:stationary-quartets}

\noindent
Keep the field excitation centered and specify a polygon $P$ by the successive angular gaps $\delta_j$ between its boundary vertices, $1\leq j\leq N_v$.
Eq.~\eqref{eq:square_to_polygon} gives its quadratic volume response directly from these gaps,
\begin{equation}
\frac{cV_{+}''(P)}{L^2B}
=18\int_P(3r^2-1)\,\mathrm dX\,\mathrm dY
=\frac94\sum_{j=1}^{N_v}\sin(2\delta_j).
\label{eq:modepolygon}
\end{equation}
For the four cyclically ordered endpoints $(-g,0,g,\pi)$ with $0<g\leq\pi/2$, the gaps are $g,g,\pi-g,\pi-g$.
The two pairs contribute $2\sin(2g)$ and $-2\sin(2g)$, so the quadratic volume response vanishes for every $g$.
The same cancellation occurs if either diagonal passes through the mode center or the diagonals are perpendicular in the mode-centered Klein chart, as in Fig.~\ref{fig:squarecertificate}(c). \\

\noindent
For the family parametrized by $g$, symmetry keeps the cross ratio at $1/2$.
The six-entropy contrast weights the two diagonal entropy Hessians with coefficient $+1$ and the four side Hessians with coefficient $-1/2$.
As $g$ increases, this weighted response changes from positive to negative exactly once, at $g\simeq0.85096$ radians (Appendix~\ref{app:quartet-sign-derivation}).
At that value, both the contrast and the volume Hessian vanish, although positive and negative energy are present inside the quadrilateral.
Fig.~\ref{fig:squarecertificate}(d) gives two distinct quartets with opposite readings for the same preparation.
Replacing one endpoint by a fifth produces a negative contrast, certifying negative energy somewhere inside the new quadrilateral.

\section{Finite Interval Data and Polygon Energy}
\label{sec:sharp}

\noindent
Two constructions test how much polygon energy can vary at fixed boundary data.
Sec.~\ref{sec:square-calibration} used two widths to identify the phase and amplitude of a known scalar mode.
Now allow the spatial energy density itself to be unknown.
Each quartet supplies one spatially weighted integral, while the polygon volume responds to the unweighted integral over its interior.
A negative reading establishes that the polygon contains negative energy because the weight is positive there.
The magnitude of the reading also depends on where the energy lies. \\

\noindent
To decide whether the measured integrals determine the desired one, first allow arbitrary smooth real density variations $f$ of compact support in the open vacuum disk $\mathbb D$.
For $q\geq1$ measured quartet weights $W_j$, collect the readings into $\mathcal M(f)$ and denote the unweighted polygon energy by $\mathcal T(f)$,
\[
\mathcal M(f)_j=\int_{\mathbb D}W_j f\,d\mu\quad(1\leq j\leq q),
\qquad
\mathcal T(f)=\int_{P_0} f\,d\mu .
\]
Two profiles yield the same readings precisely when their difference $f$ belongs to $\ker\mathcal M$, the set on which all $q$ measured integrals vanish.
If such a difference has $\mathcal T(f)\ne0$, adding it to a profile changes the polygon energy without affecting any measurement.
Otherwise, each attainable vector of readings has a unique target value.
This assignment is linear because both $\mathcal M$ and $\mathcal T$ are linear.
Extending it from the attainable vectors to $\mathbb R^q$ gives coefficients $b_j$ such that $\mathcal T(f)=\sum_{j=1}^{q}b_j\mathcal M(f)_j$ for every test profile.
Thus, a finite set determines the energy on this whole space precisely when its weights combine to give one inside $P_0$ and zero outside, in the sense of integration against every smooth test density. \\

\noindent
To realize this ambiguity in gravity, the density and metric must also satisfy the scalar and gravitational constraints.
Appendix~\ref{app:finiteexact} gives a pair of physical constraint solutions with identical sampled entropies and different areas.
Tab.~\ref{tab:data-conditions} distinguishes this finite-amplitude result from the example with four endpoints at quadratic order.
For four endpoints, an exterior packet gives two families the same global-energy coefficient. In contrast, their polygon energies have opposite signs (Sec.~\ref{sec:energy}).
A boundary reparametrization then matches all six individual entropies to their vacuum values through second order (Appendix~\ref{sec:matching-hessians}).
Agreement of their first and second derivatives at the vacuum leaves their entropies at nonzero amplitude undetermined. \\

\noindent
The scalar calibration for the fifth-endpoint test is derived in Appendix~\ref{app:angular-chord-derivation}.
Appendices~\ref{app:finite_measurement_errors} and~\ref{app:finite_charge_estimates} give the bounds from measurement errors and conserved energy, respectively.
Appendix~\ref{app:geometry} states when the distinct reflection-slice areas are also distinct maximal volumes.

\subsection{Sharp Bounds on Integrated Negative Energy}
\label{sec:signed-energy-bound}
\noindent
Suppose a quartet provides its entropy contrast and an independent measurement supplies the signed energy inside the same ideal quadrilateral $Q$.
How much negative energy must be present, allowing an arbitrary spatial distribution of both signs?
Use $d\mu$ for unit hyperbolic area and $e$ for the coefficient of $\lambda^2$ in the normal energy density in units of $L^{-3}$.
The contrast measures an integral with spatial weight $W$, whose maximum is $M=a_Qb_Q=\sqrt{\eta_Q(1-\eta_Q)}$.
Rescaling to $k=W/M$ puts the weighted reading $u$, the unweighted energy $T_Q$, and the target negative-energy magnitude $E_Q^-$ in the same units,
\begin{equation}
u=\frac{R''}{4\pi M}=\int_Q k e\,d\mu,\qquad
T_Q=\int_Q e\,d\mu,\qquad E_Q^-=\int_Q[-e]_+\,d\mu .
\label{eq:bbound}
\end{equation}
Here, $[x]_+=\max(x,0)$, $0\le k\le1$, and $V_Q''/L^2=24\pi T_Q/c$ when the reflection polygon maximizes throughout the family.
Under uniform integrable falloff, the physical negative energy inside the evolving polygon is $\lambda^2E_Q^-/L+o(\lambda^2)$.
For any $w\ge0$, write $e=[e]_+-[-e]_+$.
If $\int_Q we\,d\mu<0$, the negative contribution must supply at least the magnitude of that reading, giving $\int_Q w[-e]_+\,d\mu\ge[-\int_Q we\,d\mu]_+$ for either sign.
With $w=k\le1$, this gives $E_Q^-\ge[-u]_+$.
If $T_Q$ is also known, the complementary weight $1-k$ has reading $T_Q-u$.
Since $k+(1-k)=1$ pointwise, adding the two inequalities counts each unit of negative energy once,
\begin{equation}
E_Q^-\ge N_*=[-u]_++[u-T_Q]_+
=\max(0,-u,u-T_Q,-T_Q).
\label{eq:joint}
\end{equation}
The gap between the bound and the actual negative energy has a direct interpretation.
A measurement sees only the excess of one sign over the other, so an equal positive and negative contribution to that measurement can go undetected.
For $w\in\{k,1-k\}$, write $P_w=\int_Qw[e]_+\,d\mu$ and $N_w=\int_Qw[-e]_+\,d\mu$.
Since $\int_Q we\,d\mu=P_w-N_w$, the elementary identity $N_w-[N_w-P_w]_+=\min(P_w,N_w)$ gives
\[
E_Q^--N_*=\min(P_k,N_k)+\min(P_{1-k},N_{1-k}).
\]
Each minimum counts the amount of opposite-sign energy that cancels in one reading.
The weight $k$ is positive in the interior, while $1-k$ is positive there except at the diagonal crossing.
Thus, both weights give positive integrals for each sign of a density whose positive and negative parts each have nonzero mass.
Consequently, Eq.~\eqref{eq:joint} is attained precisely by densities of one sign almost everywhere, including the zero density.
In particular, a zero weighted reading for an integrable density of one known sign forces that density to vanish almost everywhere inside $Q$.
If both signs are allowed, a zero reading admits cancellation, and any such nonzero density has strictly larger $E_Q^-$ than the bound. \\

\noindent
When $u$ and $T_Q-u$ have opposite signs, approaching the bound forces the two signs of energy toward different parts of $Q$.
For $u>0$, positive energy concentrates near the maximum of $k$ and negative energy near its zero set, with the locations reversed for $u<0$.
Appendix~\ref{app:finite_measurement_errors} gives a quantitative estimate. \\

\noindent
Although Eq.~\eqref{eq:joint} is strict for smooth sign-changing densities, its lower endpoint can be approached by a canonical scalar with $-1<M_\phi^2L^2<0$, standard source-free falloff, and time-even field data.
The classical limit is taken before the localization limit.
The construction in Appendix~\ref{app:finite} concentrates negative energy near a point with $k=s_-$ and positive energy near another with $k=s_+$.
A pair of adjustable amplitudes fixes both $T_Q$ and $u$ exactly while the negative-energy content approaches $N_*$.
If $T_Q$ is unspecified, choosing $T_Q\ge u$ eliminates the second positive part in Eq.~\eqref{eq:joint}.
The infimum at fixed $u$ is $[-u]_+$. \\

\noindent
With uncertainties in the measured derivatives, Eq.~\eqref{eq:joint} gives a guaranteed lower bound after minimizing its right-hand side over the allowed data.
For example, when the first metric and stress variations vanish, an absolute error bound $\sigma$ on each of the six entropy Hessians gives $|\widetilde u-u|\leq\sigma/(\pi M)$ because the absolute coefficients in their contrast sum to four.
Appendix~\ref{app:finite_measurement_errors} treats an independently bounded $T_Q$, the extraction of derivatives from entropy samples, and a nonzero first-order boundary stress. \\

\noindent
Now prescribe the six individual entropy curvatures $d_{ij}$ and hence their contrast $u=C_{\eta_Q}d/(4\pi M)$.
Appendix~\ref{app:finite} realizes this $u$ at any prescribed finite $T_Q$.
The discrepancies between the resulting six Hessians and the six targets have zero contrast.
In both chiral sectors, the boundary map $f_\lambda(x)=x+\lambda^2v(x)/2$ corrects precisely these five independent discrepancies.
Eq.~\eqref{eq:endpointjets} sets them through the values and first derivatives of $v$ at the endpoints.
The map preserves the scalar tangent and hence $u$ and $T_Q$.
Thus, without an independent global-energy bound, the six fixed expansions $S_{ij}(0)+\lambda^2d_{ij}/2+o(\lambda^2)$ allow any finite-volume Hessian under the volume comparison.
The examples share their first-order metric and boundary stress, while their scalar tangents and global energies can differ.

\subsection{Fifth Endpoint and Conserved Boundary Energy}
\label{sec:energy}

\noindent
There are two ways to supplement the original quartet data.
Adding a fifth endpoint gives a second localized quartet weight. In contrast, the conserved boundary charge weights energy over the whole disk.
For the first measurement, take the four angles $(-g_0,0,g_0,\pi)$ in the chart centered on the $\Delta=3/2$ field mode, with $g_0=\arcsin(9/16)$.
The original quartet has zero signed energy $T_Q$ and a positive entropy reading $u/B=3007/196608$, where $B$ is the squared amplitude.
With $x_1=-g_0$, $x_2=0$, $x_3=g_0$, and $x_4=\pi$, add $x_5=\pi-g_0$.
The new entropies for endpoint pairs $(1,5),(2,5),(4,5)$ complete the quartet $Q_5=(1,2,5,4)$ in cyclic order, as in Fig.~\ref{fig:squarecertificate}(d).
Its two diagonals are diameters.
The four successive gaps are $g_0,\pi-g_0,g_0,\pi-g_0$, so the terms $\sin(2g_0)$ and $\sin(2\pi-2g_0)$ cancel pairwise.
Eq.~\eqref{eq:modepolygon} therefore gives $T_{Q_5}=0$.
The squared sine of half the short angular gap gives its cross ratio,
\[
\eta_5=\sin^2(g_0/2)=\frac{16-5\sqrt7}{32},\qquad
M_5=\sqrt{\eta_5(1-\eta_5)}=\frac9{32}.
\]
Writing $s_2(\alpha)=S''(\alpha)/B$ for the field-phase entropy Hessian per unit squared amplitude of an arc of half-width $\alpha$, purity gives
\[
\frac{R_5''}{B}
=2s_2(\pi/2)-2(1-\eta_5)s_2((\pi-g_0)/2)-2\eta_5s_2(g_0/2).
\]
Substituting Eq.~\eqref{eq:entropy_calibration}, with $z_0=\sin(g_0/2)+\cos(g_0/2)-1=1/4$, gives
\begin{align}
\frac{R_5''}{B}&=-\frac{3\pi}{16}z_0^2
(z_0^3+5z_0^2+2z_0+2)=-\frac{543\pi}{16384},\label{eq:fifth_contrast}\\
E^-_{Q_5}&\ge -\frac{R_5''}{4\pi M_5}
=\frac{181}{6144}\,B.\label{eq:exteriorenergy}
\end{align}
The negative contrast certifies at least $181B/6144$ of integrated negative energy somewhere in $Q_5$.
For an unknown density, subtracting the old reading would isolate $Q_5\setminus Q$ only if the weights agreed on the overlap and contributions from $Q\setminus Q_5$ also canceled.
Neither reading isolates the added region, so negative energy in the overlap can contribute to the observed sign change.
A pure vacuum descendant has $R_f''=0$, with $C_{\eta_5}S_f''\ge0$ from the nonnegative first-slope square in the logarithmic contrast.
A measured $C_{\eta_5}\widetilde S''<-4\sigma$ excludes such a descendant even after allowing an error $\sigma$ in each entropy Hessian. \\

\noindent
A fifth endpoint measures a different part of the disk. At the same time, the conserved boundary energy constrains the whole disk with a single integral.
To identify this charge with a matter integral at quadratic order, we hold the boundary sources fixed and assume the first metric variation vanishes.
Write $\widehat H_0=\widehat H_{\mathrm{cyl}}+c/12$ for the dimensionless cylinder Hamiltonian with the vacuum energy subtracted.
The measured coefficient is $\mathcal E=\partial_\lambda^2\langle\widehat H_0\rangle|_0/2$.
On the unit hyperbolic disk, let $D$ denote the covariant derivative and $\chi$ the proper distance from its center in units of $L$.
The lapse $N=\cosh\chi$ converts a boundary time interval into proper time at that radius, so a unit of local energy farther from the center contributes more to the boundary charge.
The charge identity gives $\mathcal E=\int_{\mathbb H^2}Ne\,d\mu$ \cite{Brown:1986nw,Lashkari:2015hha,deHaro:2000vlm}.
For standard source-free falloff with $\Delta>1$, the extra boundary term is $O(z^{2\Delta-2})$ and vanishes as the boundary-defining function $z$ approaches zero.
Define $\mu_*^2=-M_\phi^2L^2\in(0,1)$, and denote the dimensionless scalar momentum by $\pi_\phi=Ln^\mu\partial_\mu\phi_1$, with $n$ the unit future normal and the spacetime index $\mu$ summed over $0,1,2$.
The negative mass squared permits a local negative energy density through the potential term.
Because the disk lapse obeys $N\geq1$, upper bounds on the full positive and negative energy control both $T_Q$ and $T_Q-u$, which use weights $1$ and $1-k$ in $Q$.
Completing a square in the scalar gradient bounds these positive and negative parts in terms of the measured global charge, as shown in Appendix~\ref{app:finite_charge_estimates}.
Intersecting the resulting bounds for the two readings gives
\begin{equation}
[u]_+-\frac{\mu_*^2\mathcal E}{1-\mu_*^2}
\le T_Q\le
\frac{\mathcal E}{1-\mu_*^2}-[-u]_+ .
\label{eq:cappedvolume}
\end{equation}
Retaining the radius dependence in the bound on the negative contribution gives the stronger one-sided estimate
\begin{equation}
T_Q\geq [u]_+-\frac{\mu_*^2\mathcal E}{2\sqrt{1-\mu_*^2}} .
\label{eq:global-negative-improved}
\end{equation}
For example, at $M_\phi^2L^2=-3/4$, the bound reads $T_Q\geq[u]_+-3\mathcal E/4$.
Thus, even a positive global charge and a vanishing quartet reading still permit a negative value of $T_Q$ within this bound.
For general $0<\mu_*^2<1$, a reading $u>\mu_*^2\mathcal E/(2\sqrt{1-\mu_*^2})$ guarantees positive reflection-area response and, under the global comparison hypotheses, positive volume response.
An upper bound on $\mathcal E$ suffices for this sign test. \\

\noindent
Suppose the measured boundary charge is strictly positive and fixed.
We can still change the sign of the proper energy inside the polygon while keeping all six interval entropy Hessians at their vacuum values.
For each $\sigma\in\{-,+\}$, choose a compact time-even scalar profile $\Phi^{(\sigma)}$ from Appendix~\ref{sec:energy-packets} with $u=0$, proper energy $T^{(\sigma)}$, and global charge $\mathcal E_\sigma>0$, taking $T^{(-)}<0<T^{(+)}$.
Choose a nonzero smooth compact profile $\Phi_{\rm out}$ outside $\overline Q$ and disjoint from both supports, and denote its positive charge by $\mathcal E_{\rm out}$.
Prescribe the same measured global-energy coefficient $\mathcal E_0>0$ for both branches, with $\mathcal E_0/c$ fixed in the classical limit.
The quartet weight and the polygon indicator both vanish outside $\overline Q$, whereas the global charge still measures the exterior packet.
We can therefore vary the amplitude of the first profile and compensate the change in charge with the exterior one.
For $0\leq t\leq\mathcal E_0/\mathcal E_\sigma$, choose the time-even scalar tangent
\[
\Phi^{(\sigma)}_t=\sqrt t\,\Phi^{(\sigma)}
+\sqrt{\frac{\mathcal E_0-t\mathcal E_\sigma}{\mathcal E_{\rm out}}}\,\Phi_{\rm out}.
\]
The square roots appear because both the local density and the global charge are quadratic in the field amplitude.
Disjoint supports remove their cross terms, so the first profile supplies $t\mathcal E_\sigma$ and the exterior packet supplies exactly the balance $\mathcal E_0-t\mathcal E_\sigma$.
Inside $Q$, only the first term contributes, giving $u=0$ and $T_Q=tT^{(\sigma)}$.
Varying $t$ on the two branches therefore realizes every proper energy in the interval
\[
\left[\frac{\mathcal E_0T^{(-)}}{\mathcal E_-},
\frac{\mathcal E_0T^{(+)}}{\mathcal E_+}\right].
\]
This interval is realized by the two chosen profiles, while Eq.~\eqref{eq:cappedvolume} gives bounds for all profiles at that charge.
Fixing the charge and setting the quartet contrast to zero still leaves the six individual entropy Hessians to match.
Their departures from the vacuum satisfy the one linear relation required by Eq.~\eqref{eq:endpointjets}.
A second-order boundary reparametrization supplies the five remaining independent adjustments and sets all six Hessians to zero.
The identity $\widehat H_0|0\rangle=0$ for the reference vacuum gives
\begin{equation}
\mathcal E=\langle\psi_0'|\widehat H_0|\psi_0'\rangle ,
\label{eq:energy_tangent}
\end{equation}
where $|\psi_0'\rangle=\partial_\lambda|\psi_\lambda\rangle|_{\lambda=0}$ is the Hilbert-space tangent.
Differentiating the energy expectation twice gives one term containing two first-state derivatives and two terms containing a second-state derivative.
In each of the latter, $\widehat H_0$ acts directly on the vacuum ket or bra and gives zero.
The conformal map starts at order $\lambda^2$, so it preserves the first state derivative and hence the measured energy coefficient.
Both branches now have the vacuum values, slopes and Hessians for the six entropies, together with the same prescribed positive global-energy coefficient.
For nonzero $t$ on both branches, their polygon energies have opposite signs and give opposite volume Hessians under the comparison conditions.
This construction fixes Taylor coefficients at $\lambda=0$.
Higher-order terms could distinguish these preparations in measurements performed at nonzero $\lambda$.

\subsection{Exact Classical Ambiguity at Finite Amplitude}
\noindent
Suppose an observer is given the classical entropies of a finite list of intervals at the same small nonzero amplitude and the time-zero boundary stress.
Finitely many sampled geodesics and polygon edges leave an open disk $U$ inside a positive-area polygon.
Choose a weak pure-momentum scalar background with positive density throughout $\overline U$.
At quadratic order, each sampled interval weight satisfies $(2-\bar\Delta)[w_{ij}]_+=0$ on $U$, with $\bar\Delta$ the unit hyperbolic Laplacian, whereas the polygon weight satisfies $(2-\bar\Delta)1=2$.
For smooth nonnegative $g$ supported inside $U$, integration by parts transfers this operator from $g$ to either weight.
The density variation $(2-\bar\Delta)g$ contributes zero to every sampled moment and $2\int_U g\,d\mu>0$ to the polygon if $g$ is nonzero. \\

\noindent
Equality of the entropies at finite amplitude follows from a length comparison in Appendix~\ref{app:finite}.
Multiply the baseline metric by $\exp(2\varepsilon\lambda^2g)$, taking $\varepsilon>0$ small enough that both scalar densities remain nonnegative.
The original shortest curves miss the support of $g$ and keep their lengths, while every competing curve can only lengthen.
Consequently, the minimum lengths and their classical entropies agree throughout open neighborhoods of the queried endpoints.
The unchanged boundary collar fixes the same stress, so the vacuum modular expectation and hence the leading relative entropy to vacuum also agree for each queried single interval.
Inside $U$, the area element increases, and the reflection-area Hessians differ by $4\varepsilon\int_Ug\,dA_0>0$.
This difference also applies to maximal-volume Hessians under the assumed coherent-state realization, smooth RT branch, and regular causal comparison.

\section{Vacuum Pairings from Finite Union Entropies}
\label{sec:vacuum-pairing}
\label{app:topology}

\noindent
In the vacuum disk, the geodesic partners of the interval endpoints determine both the positions and the areas of the uncovered polygons.
Which ends of separated intervals are joined by the geodesics that determine the entropy of their union?
In the vacuum disk, these connections enclose an area, measured outside the individual interval wedges, in units of $2\pi L^2$ \cite{Fujiki:2026ucr,Abt:2017pmf}. \\

\noindent
One can test a pair of endpoints by measuring the entropy before either endpoint moves, after moving each endpoint separately, and after moving both together.
If different geodesics carry the two endpoints, their length changes add.
If one geodesic carries both, its joint change differs from that sum by a known single-interval signal.
This distinction works when the same assignment of endpoint partners gives the shortest total length for all four measurements, and the signal is large enough to resolve.
Appendix~\ref{app:vacuum-support} derives bounds on the endpoint shifts that preserve the minimizing pairing and on the entropy errors for which the measurements still determine the uncovered area.

\subsection{Endpoint Shifts and Pairwise Cancellation}
\label{app:detailG1}

\noindent
On a spatial boundary line, take $m\geq2$ intervals
\begin{equation}
I_i=[u_i,v_i],\qquad
u_1<v_1<\cdots<u_m<v_m,\qquad U=\bigcup_{i=1}^m I_i .
\end{equation}
Extend the left end of $I_i$ by the exterior strip $X_i=[u_i-a_i,u_i]$ and the right end of $I_j$ by $Y_j=[v_j,v_j+b_j]$.
Here, $i,j\in\{1,\ldots,m\}$, the widths $a_i,b_j$ are positive, and $b_i+a_{i+1}<u_{i+1}-v_i$ for $1\leq i<m$.
These restrictions keep the strips disjoint and preserve the order of the shifted endpoints.
All endpoints and widths, as well as $m$, are fixed as the chiral central charge $c$ increases. \\

\noindent
All entropies use a common short-distance cutoff.
Write $S_c(W)$ for the entropy of a region $W$, and $s_c(u,v)=s_c(v,u)$ for that of a single interval with endpoints $u,v$.
Juxtaposed region labels denote their union.
Adding $X$ to $B$ changes its entropy by $S(BX)-S(B)$.
If $Y$ is already included, the same addition changes the entropy by $S(BXY)-S(BY)$.
Subtracting the second change from the first gives the conditional mutual information $I(X:Y\mid B)$.
For each $(i,j)$, the numerator below measures this difference for the union, and the denominator gives the signal expected for a geodesic joining the tested ends,
\begin{align}
N_{ij}^{(c)}&=I(X_i:Y_j\mid U),\nonumber\\
D_{ij}^{(c)}&=s_c(u_i-a_i,v_j)+s_c(u_i,v_j+b_j)
-s_c(u_i,v_j)-s_c(u_i-a_i,v_j+b_j),\nonumber\\
R_{ij}^{(c)}&=N_{ij}^{(c)}/D_{ij}^{(c)} .
\label{eq:top_def}
\end{align}
Testing all $m^2$ pairs uses the same unshifted entropy and the same $2m$ single-shift entropies throughout.
With one additional doubly shifted union per pair, at most $(m+1)^2$ distinct union measurements suffice.
For $u_i<v_j$, the denominator is $I(X_i:Y_j\mid[u_i,v_j])$.
When $v_j<u_i$, the region between the strips is $[v_j+b_j,u_i-a_i]$, which gives the same expression for $D_{ij}^{(c)}$ using the symmetry of $s_c$.
Strong subadditivity, the entropy inequality $S(BX)+S(BY)\geq S(B)+S(BXY)$, implies the nonnegativity of both conditional mutual information \cite{Lieb:1973zz}. \\

\noindent
In the vacuum on the line, $s_c(u,v)=(c/3)\ln(|v-u|/\epsilon)+\mathrm{const.}$, with ultraviolet cutoff $\epsilon$.
For the signed separation $d=v_j-u_i$, substitution gives
\begin{equation}
\frac{3D_{ij}^{(c)}}c
=\ln\frac{(d+a_i)(d+b_j)}{d(d+a_i+b_j)}
=\ln\!\left(1+\frac{a_i b_j}{d(d+a_i+b_j)}\right)>0 .
\label{eq:pair-strip-denominator}
\end{equation}
Because the strips are disjoint, $d$ and $d+a_i+b_j$ have the same sign, even if the tested right endpoint lies to the left of the tested left endpoint.
Their product is positive, so the logarithm has an argument greater than one and defines a positive reference signal proportional to $c$ for fixed widths.
A cutoff term associated with either endpoint alone appears with both signs and cancels. \\

\noindent
For now, choose a minimizing RT pairing on $U$ and suppose its assignment of left and right partners still minimizes on every sampled $UX_i$, $UY_j$, and $UX_iY_j$.
The chord lengths change as the endpoints move, but the partner labels must stay the same.
It is enough that one such pairing minimizes in every configuration, even if another pairing ties with it in some of them.
To see why partners carry opposite labels, consider any chord in a minimizing pairing.
Every endpoint on either side must pair with another endpoint on that side, since minimizing chords do not cross.
The chord thus skips an even number of endpoints in the alternating left--right list and joins a left endpoint to a right endpoint.
Write $\pi(k)$ for the index of the right endpoint joined to the left endpoint $u_k^W$ in the shifted or unshifted region $W$.
Each right endpoint occurs once, so the assignment is a permutation with matrix $P_{ij}=\delta_{j,\pi(i)}$, where the Kronecker delta equals one for equal indices and zero otherwise.
For each sampled region $W$, define the remainder after subtracting the single-interval entropies for this pairing,
\begin{equation}
e_c(W)=S_c(W)-\sum_{k=1}^{m}s_c(u_k^W,v_{\pi(k)}^W).
\end{equation}
For any function $f$ of the sampled regions, write
\begin{equation}
\Delta_{ij}f=f(UX_i)+f(UY_j)-f(U)-f(UX_iY_j).
\end{equation}
To see what $\Delta_{ij}$ measures, group the four lengths of each edge before summing over edges.
If the edge misses $v_j$, adding $Y_j$ leaves both its original length and its length after adding $X_i$ unchanged, so both lengths cancel.
The same argument with $i$ and $j$ interchanged removes every edge that misses $u_i$.
Only an edge joining $u_i$ to $v_j$ contributes its single-interval response $D_{ij}^{(c)}$ to the measured difference $N_{ij}^{(c)}$, giving
\begin{equation}
N_{ij}^{(c)}=P_{ij}D_{ij}^{(c)}+\Delta_{ij}e_c,
\qquad
R_{ij}^{(c)}-P_{ij}=\frac{\Delta_{ij}e_c}{D_{ij}^{(c)}} .
\label{eq:top_exact}
\end{equation}
If $|\Delta_{ij}e_c|/D_{ij}^{(c)}$ tends to zero for each pair as $c$ increases, Eq.~\eqref{eq:top_exact} gives one entry equal to one in each limiting row, in the column of its partner.
All other entries in that row tend to zero. \\

\begin{figure}[tbp]
\centering
\includegraphics{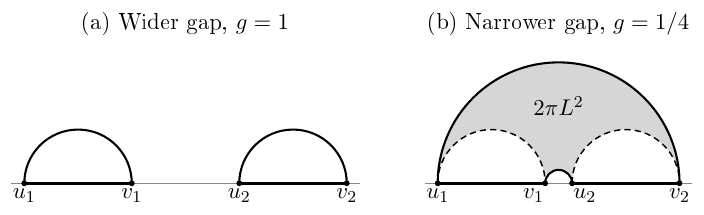}
\caption{Vacuum geodesics for two boundary intervals of unit length.
The heavy horizontal segments are the intervals, and solid semicircles are the geodesics selected by their union entropy.
In (a), each interval keeps its own geodesic, so the two individual wedges fill the union wedge.
In (b), the outer endpoints pair with each other, as do the endpoints across the gap.
The shaded region lies inside the union wedge and outside both individual wedges, whose curved boundaries are dashed.}
\label{fig:two-interval-pairing}
\end{figure}

\noindent
For a two-interval example, use the dimensionless boundary coordinate in Fig.~\ref{fig:two-interval-pairing} and set $I_1=[0,1]$ and $I_2=[1+g,2+g]$, with gap $g>0$.
Each leading entropy is a sum of logarithms, so comparing the two candidate entropies amounts to comparing their products of endpoint separations.
Joining the ends of each interval gives product $1$, while joining the two outer ends and the two inner ends gives $g(2+g)$.
At $g=1$, the first product is smaller, each interval retains its own geodesic, and the uncovered area is zero.
At $g=1/4$, the second product is $9/16$, and the two union geodesics together with the two individual-interval geodesics bound an ideal quadrilateral of area $2\pi L^2$.
To see the difference in the actual entropy readings, shift each outer endpoint outward by $1/4$, separately and together.
At $g=1$, the two outer ends remain on separate geodesics in all four configurations, so their leading mixed response vanishes.
At $g=1/4$, the outer edge spans $9/4$, then $10/4$ under either single shift, then $11/4$ under both. At the same time, the coordinate separation across the inner gap stays at $1/4$.
The inner-edge contribution cancels, leaving $N_{12}^{(c)}=(c/3)\ln(100/99)+o(c)$, the single-edge signal that gives $R_{12}^{(c)}\to1$ as $c\to+\infty$.

\subsection{Cycle Count with Finite Entropy Errors}
\label{app:detailG3}

\noindent
Each identified partner tells us which interval to visit next.
Start at $u_i$, follow its union geodesic to $v_{\pi(i)}$, and return along that interval's individual geodesic to $u_{\pi(i)}$.
Repeating the walk partitions the $m$ intervals into $C(P)$ closed cycles, including intervals visited alone.
In the two-interval example, separate partners give two one-step cycles. In contrast, the shared outer and inner edges give one two-step cycle.
The null equation $(\Id-P)x=0$ sets $x_i=x_{\pi(i)}$, leaving exactly one freely chosen value on each cycle.
Thus, the rank $\nu=m-C(P)$ counts units of vacuum area, provided we can distinguish a genuinely positive singular value from a zero displaced by entropy errors.
Within a cycle of length $k$, number its members $r=0,\ldots,k-1$ in the order visited.
Applying $P$ advances $r$ by one modulo $k$.
For $a=0,\ldots,k-1$, assign the value $\exp(2\pi\mathrm{i}ar/k)/\sqrt{k}$ to position $r$, where $\mathrm{i}^2=-1$.
Advancing one place multiplies this pattern by $\exp(2\pi\mathrm{i}a/k)$, so the difference operator $\Id-P$ multiplies it by $1-\exp(2\pi\mathrm{i}a/k)$.
These $k$ patterns form an orthonormal basis, and the magnitudes of the multipliers are the singular values $2|\sin(\pi a/k)|$.
On a cycle of length $k\ge2$, only the constant mode gives zero, and the smallest positive singular value is $2\sin(\pi/k)$.
Since $2\le k\le m$, this value is at least $\gamma_m=2\sin(\pi/m)$ regardless of the endpoint positions.
For a measured real $m\times m$ matrix $R$, count the singular values above half this gap,
\begin{equation}
H_m(R)=\#\{a\in\{1,\ldots,m\}:\sigma_a(\Id-R)>\gamma_m/2\}.
\label{eq:pair-threshold}
\end{equation}
Here, $\sigma_a(M)$ is the $a$th largest square root of the eigenvalues of $M^{\mathsf T}M$, with multiplicity and $1\le a\le m$.
Write $\|R-P\|_2$ for the operator norm of the error, the largest value of $\|(R-P)x\|_2$ over vectors $x$ of unit Euclidean length.
For any unit vector $x$, replacing $P$ by $R$ changes the length of $(\Id-P)x$ by at most $\|R-P\|_2$.
Taking the extrema of these lengths on subspaces of each dimension gives the same bound for each ordered singular value.
For the matrix $R$ obtained from the measured entropies, a bound $\|R-P\|_2<\sin(\pi/m)=\gamma_m/2$ therefore keeps all displaced zeros below the threshold and all originally positive values above it.
Consequently, $H_m(R)=m-C(P)$, even if the measured matrix has no exact zero modes.
Appendix~\ref{app:pairing-error-proofs} derives a sufficient bound on this norm from the entropy uncertainties and an independent estimate of the finite-central-charge remainder.
For three intervals, exchanging partners $1$ and $2$ gives one quadrilateral, just as exchanging $2$ and $3$ does.
The count assigns the same area to these two configurations because it keeps the number of cycles and discards their endpoint labels.

\subsection{Vacuum Polygon Area}
\label{app:detailG5}

\noindent
In the pure vacuum, the complement $O$ and the union $U=I_1\cdots I_m$ share an RT surface.
Draw the union chords together with the individual-interval chords on the disk.
Following a union chord from $u_i$ to $v_{\pi(i)}$ and returning along the individual chord to $u_{\pi(i)}$ advances one step around a cycle.
A cycle of length $k>1$ therefore visits $2k$ different ideal vertices before closing.
The geodesics of the union do not cross one another, and the individual-interval geodesics join consecutive boundary endpoints.
In the Klein disk, these chords are straight segments.
A segment that skipped a vertex of the cycle in circular order would leave vertices on both sides, forcing the return path to cross that segment.
Each cycle thus bounds a convex ideal polygon.
Distinct cycles cannot nest because the only points of a closed polygon on the ideal circle are its own vertices.
Removing from the union wedge the part between each boundary arc and its chord leaves precisely the interiors bounded by these cycles. \\

\noindent
For an ideal polygon of area $A_k$, the interior curvature $-1/L^2$ integrates to $-A_k/L^2$, and the geodesic sides have zero geodesic curvature.
Each of its $2k$ corners has zero interior angle and contributes an exterior turn of $\pi$, so the Gauss--Bonnet Theorem gives $-A_k/L^2+2k\pi=2\pi$.
A cycle of length $k>1$ therefore contributes $2\pi L^2(k-1)$, and the same expression gives zero for the coincident chords of a one-step cycle.
The sum of the cycle lengths counts all $m$ intervals once, while the subtraction of one per cycle gives the total uncovered volume
\begin{equation}
V_{I_1:\cdots:I_m:O}=2\pi L^2\,[m-C(P)].
\label{eq:pairing-vacuum-area}
\end{equation}
For two intervals, separate one-step cycles leave no uncovered area. In contrast, a single two-step cycle encloses a quadrilateral of area $2\pi L^2$.
The cycle count gives the total, and the partners specify its location. \\

\noindent
At a specified central charge, certifying the inferred area requires an independent bound on the combined entropy corrections and measurement errors.
For fixed widths and a common minimizing branch, convergence to the leading RT entropies gives eventual agreement with the cycle count, but supplies no value of $c$ at which that agreement must begin.
When weak matter is added on the same branch, the labeled cycles specify which region to integrate over, and the enclosed energy sets its quadratic area change as in Eq.~\eqref{eq:cycle-matter-expansion}.
For quantum recovery, Appendix~\ref{app:detailH2} constructs two states with the same two conditional mutual informations and different optimal fidelities for recreating a missing register.

\section{Discussion and Conclusion}
\label{sec:discussion}

\noindent
Can boundary interval entropies determine how the volume of a bulk polygon changes when matter is added?
In general, a fixed finite set of intervals leaves the answer undetermined.
The reason is visible at quadratic order on a reflection slice.
By the Gauss--Bonnet Theorem and the gravitational constraint, the polygon area counts every unit of enclosed proper energy equally.
The six-entropy combination for a quadrilateral gives greater weight to energy near its diagonal crossing than to energy near its sides.
We recover the missing information either by scanning intervals through the bulk or by using independent knowledge of the preparation to restrict the allowed energy profiles.
Both routes give the maximal-volume response if the reflection slice remains maximal throughout the family.
If maximality is known only at the vacuum, the area Hessian bounds the volume Hessian from below whenever both exist~\eqref{eq:sharp_volume_main}. \\

\noindent
Information is lost even if an observer keeps all six entropy readings separately and measures the global energy as well.
We construct two scalar families with those six entropies equal to vacuum through quadratic order and the same global-energy increase, yet opposite signs of the maximal-volume change.
An exterior packet matches the global energies without changing the energy inside the polygon, and a boundary reparametrization matches the entropy responses.
At small finite amplitude, a second construction applies to any prescribed finite list of intervals.
Their geodesics miss an interior patch where the metric can change the polygon area while preserving nonnegative energy density, the initial boundary stress, and every queried classical entropy even in neighborhoods of the sampled endpoints.
Any common estimate of the polygon area misses at least one of the two values by at least half their separation. \\

\noindent
A fixed collection of readings can still certify negative energy or bound a positive distribution.
Because the quartet weight is positive in the interior, a negative reading requires negative normal energy on a region of positive area, even for a state with positive global energy.
For nonnegative energy in a regular pentagon, five overlapping quartets give a sharp lower bound, larger by a factor $\sqrt5$ than the single-quartet bound when all five readings are equal and positive.
Smooth profiles realize every energy above this bound for feasible nonzero data (Sec.~\ref{sec:positive-quartets}).
For signed densities, a quartet reading and the unweighted polygon energy together bound the amount of negative energy, and Eq.~\eqref{eq:global-negative-improved} gives the extra constraint from a global charge. \\

\noindent
For an unknown density, reconstruction uses the freedom to move both endpoints.
As a quartet narrows to one geodesic, its rescaled response, after subtraction of the boundary-stress term, measures energy integrated along that geodesic.
Even measuring every offset at finitely many orientations can leave two different polygon energies compatible with the data (Appendix~\ref{app:radoninversion}).
Scanning all orientations and offsets supplies the line integrals needed to recover the density, or just its integral over a chosen polygon.
On a finite grid, smoothing the readings reduces sensitivity to noise. However, it gives some weight to energy just outside the polygon (Appendix~\ref{app:ordinary_entropy_weights}).
The estimate in Appendix~\ref{app:detailC6a} includes that boundary error, the entropy uncertainty, finite sampling, and errors in the numerical weights~\eqref{eq:grid-entropy-volume}.
The reconstruction determines energy, which can be identical for scalar fields with different relative signs on disjoint momentum packets because their interference term vanishes (Appendix~\ref{app:detailH1}). \\

\noindent
With a restricted preparation, two interval sizes can suffice.
For a centered, rotationally invariant excitation of the lowest scalar mode of dimension $3/2$, assume definite time parity and nonzero amplitude.
The semicircle and quarter-circle entropy Hessians identify field or momentum initial data and determine the squared amplitude, which fixes the quadratic volume response for every polygon in the family.
In the field phase, every proper interval has a positive entropy Hessian. Yet, a centered regular triangle shrinks, a square has zero quadratic response, and a pentagon grows.
The positive energy enclosed outside the negative core is too small in the triangle, exactly balances the core in the square, and outweighs it in the pentagon. \\

\noindent
A finite set of entropy shifts also determines a discrete property of the vacuum disk when the same endpoint pairing minimizes throughout the measurements.
Following the partners of $m\ge2$ intervals gives $C$ closed cycles, which enclose a total uncovered area of $2\pi L^2(m-C)$.
The threshold in Eq.~\eqref{eq:pair-threshold} certifies that integer if the combined measurement and finite-$c$ error in the ratio matrix has operator norm below $\sin(\pi/m)$.
The partner labels locate the polygon, and the cycle count fixes its vacuum area.
Neither number specifies how faithfully a missing quantum subsystem can be recovered, since even both absolute conditional mutual informations entering one ratio can agree for states with different optimal recovery fidelities. \\

\noindent
Beyond time-reflection symmetry, the extremal curves enclose a surface that can bend in time.
Can interval entropies measured at neighboring boundary times detect this bending well enough to determine the maximal volume?
At finite central charge, which additional boundary measurement could separate geodesic length from the bulk quantum entropy that contributes to the same interval reading?
For finite samples, one can ask whether matching examples saturate our error estimate and thus fix the smallest resolvable volume change for a given grid spacing, entropy precision, and density smoothness.
A quantum interpretation would need a concrete transfer or recovery task whose accessible registers and optimal rate or fidelity can be inferred from a finite-amplitude polygon.

\section*{Acknowledgments}
\noindent
The author thanks Ma-Ke Yuan for her helpful discussion and Nan-Peng Ma for his encouragement.

\appendix
\counterwithin{equation}{section}
\renewcommand*{\theHequation}{\Alph{section}.\arabic{equation}}

\section{Geometric Hypotheses and Polygon Response}
\label{app:geometry}

\noindent
Time reflection makes the polygon stationary under infinitesimal displacements in time.
To compare its area with the largest area of any filling, we identify the spacetime region the fillings share and follow freely falling observers from the polygon across that region.
A bound on the scalar field there controls focusing. In contrast, decay at each ideal vertex controls the change in area as the state varies.

\subsection{Classical Sector and Normalization}
\label{app:detailA1}

\noindent
The relative normalization of curvature and matter in the area formula follows from the action for Einstein gravity with cosmological constant $-L^{-2}$ and a canonical minimally coupled scalar with vacuum value zero,
\begin{equation}
I=\frac{1}{16\pi G_N}\int \sqrt{-g}\,(R+2/L^2)\,d^3x
-\frac12\int\sqrt{-g}\,\left[\sum_{\mu,\nu=0}^2g^{\mu\nu}\partial_\mu\phi\partial_\nu\phi+M_\phi^2\phi^2\right]d^3x.
\end{equation}
An even scalar interaction beginning at fourth order would leave the quadratic response unchanged.
For the global curvature test, we use the displayed quadratic potential, whose $M_\phi^2\phi^2$ term in the trace-reversed Einstein equation becomes $2U(\phi)$ for a general potential $U$.
As the state varies, we hold the CFT, its sources, and the boundary metric fixed.
Here, $g$ is the determinant of the spacetime metric, $R$ its scalar curvature, $G_N$ Newton's constant, and $M_\phi^2$ the scalar mass squared.
With signature $(-,+,+)$, a unit future normal $n$ measures $\rho=\sum_{\mu,\nu=0}^2T_{\mu\nu}n^\mu n^\nu$.
Decomposing the field gradient into its normal and spatial parts gives
\begin{equation}
\rho=\tfrac12\big[(n\phi)^2+|\nabla_h\phi|_h^2+M_\phi^2\phi^2\big].
\end{equation}
The derivative $n\phi$ is taken along the observer's clock, and the gradient norm uses the positive spatial metric $h$ in the two remaining directions.
All three terms are quadratic around the vacuum $\phi=0$. \\

\noindent
A smooth one-parameter family through the vacuum has the expansions
\begin{equation}
g_\lambda=g_{\rm AdS}+\lambda h_1+O(\lambda^2),\qquad
\phi_\lambda=\lambda\phi_1+O(\lambda^2),\qquad
\rho_\lambda=\lambda^2\rho_2+o(\lambda^2).
\label{eq:app_family}
\end{equation}
The spacetime metric variation $h_1$ solves the homogeneous linearized Einstein equation because the scalar stress begins at order $\lambda^2$.
Its restriction to the reflection slice is the corresponding first variation of the spatial metric.
We restrict it to the regular vacuum boundary-graviton orbit, excluding perturbations that independently change the global parallel transport around a noncontractible loop (its holonomy), introduce a conical singularity, or change a black-hole mass.
A boundary graviton is obtained from AdS by a diffeomorphism inducing a conformal reparametrization at infinity.
Time reflection exchanges the left- and right-moving boundary stress profiles and reverses the slice normal $n$.
It fixes the induced metric $h$, so $K_{ab}=\tfrac12(\mathcal L_n h)_{ab}=0$ for $a,b=1,2$, where $\mathcal L_n$ is the Lie derivative along $n$.
An even scalar is invariant under time reversal, and an odd scalar is invariant when reversal is combined with $\phi\mapsto-\phi$.
Both give reflection-symmetric stress and metric. \\

\noindent
The bulk energy associated with one interval also has a boundary definition.
Let $\varrho_I(\lambda)$ be the reduced state on interval $I$ and $\sigma_I=\varrho_I(0)$ its vacuum value.
Their relative entropy measures how distinguishable the perturbed state is from the vacuum using measurements in $I$.
The leading quadratic measure of this distinguishability is the interval quantum Fisher information used here,
\[
\left.\frac{d^2}{d\lambda^2}
S\!\left(\varrho_I(\lambda)\Vert\sigma_I\right)\right|_{\lambda=0},
\]
with $\sigma_I$ held fixed.
For a diagonal example, regulate the vacuum support to dimension $d$ and let $p_j(\lambda)>0$, $1\le j\le d$, be the normalized probabilities in its fixed eigenbasis.
The relative entropy is $\sum_{j=1}^d p_j(\lambda)\ln[p_j(\lambda)/p_j(0)]$.
Its derivative is $\sum_{j=1}^d p'_j(\lambda)\ln[p_j(\lambda)/p_j(0)]$ because normalization gives $\sum_{j=1}^d p'_j(\lambda)=0$.
At zero, each logarithm vanishes, so a second derivative leaves $\sum_{j=1}^d[p'_j(0)]^2/p_j(0)$.
Thus, equal changes in a probability carry greater weight when that outcome was initially rarer.
For a general tangent with off-diagonal matrix elements, Appendix~\ref{app:detailD2} derives the corresponding quadratic form in Eq.~\eqref{eq:cftbkm}.
The vacuum interval state generates an evolution called modular flow, with generator $-\ln\sigma_I$.
The energy of a linear bulk perturbation with respect to this flow is quadratic in its amplitude and is called canonical energy.
At leading classical order, it equals the interval quantum Fisher information \cite{Lashkari:2015hha}.
We keep $A^2L/c$ fixed for a mode of amplitude $A$ as $c$ increases, while $\lambda$ controls the change of the geometry.
With this scaling, the classical contribution to the interval entropy response is of order $c$ and the bulk quantum corrections are of order one.

\subsection{Causal Domain and Admissible Homology Class}
\label{app:causalclass}

\noindent
At fixed $\lambda$, let $\Gamma_\lambda=\partial P_\lambda$ consist of the RT sides enclosing the polygon and their limiting endpoints at conformal infinity.
A shared edge cancels when two homology regions meet with no intervening polygon interior.
If every edge cancels, the polygon is empty, and its volume is zero.
The EWP maximization compares surfaces with this same $\Gamma_\lambda$ in a single state \cite{Fujiki:2026ucr}.
Between states, only the boundary endpoints stay fixed, so we determine the bulk RT sides anew.
An admissible filling $Q$ is smooth and spacelike in the physical interior, with a compact acausal conformal closure, boundary $\Gamma_\lambda$, and the same oriented side as $P_\lambda$.
Acausality means that no causal curve joins distinct points of the closure, and the prescribed class excludes extra closed components and asymptotic ends.
We assume a regular globally hyperbolic conformal initial development and include its timelike conformal boundary when defining causal curves. \\

\noindent
We require each filling, including its ideal vertices, to extend to an ambient Cauchy slice of the conformal development.
Such a slice meets every inextendible causal curve exactly once.
The closed edge divides the slice into the open polygon side $P$ and its open complement $P^c$.
Their domains of dependence obey
\begin{equation}
\widetilde M\setminus
\left[J^+(\Gamma_\lambda)\cup J^-(\Gamma_\lambda)\right]
=D(P)\,\dot\cup\,D(P^c).
\label{eq:app_causaldecomposition}
\end{equation}
In the conformal development $\widetilde M$, $J^\pm(S)$ are the causal future and past of $S$, and $D(S)$ consists of points through which every inextendible causal curve meets $S$.
For a point $x$ in the future of the slice, consider all possible starting points of causal curves from the slice to $x$.
They form a connected set on the slice.
Projecting any such curve back along a timelike flow tangent to the conformal boundary gives a path to the common projection of $x$.
Each projected point stays in $J^-(x)$ because it can follow the flow forward to the original curve and then continue to $x$.
A path between starting points on opposite sides would therefore meet $\Gamma_\lambda$ and put $x$ in $J^+(\Gamma_\lambda)$.
Time reversal gives the past case, so outside $J^+(\Gamma_\lambda)\cup J^-(\Gamma_\lambda)$ every inextendible causal curve through $x$ meets the same side of the slice \cite{Headrick:2014cta}. \\

\noindent
Removing the causal future and past of the edge leaves connected components fixed by $\Gamma_\lambda$ alone, as in Eq.~\eqref{eq:app_causaldecomposition}.
A timelike flow connects every point in each component of $D(P)$ to that component's spatial section.
The choice of oriented side at the edges therefore identifies the same components for both fillings, giving
\begin{equation}
D(Q)=D(P_\lambda),\qquad
\mathcal N_\lambda=\operatorname{int}D(P_\lambda),
\label{eq:app_samedomain}
\end{equation}
so the two interiors are Cauchy hypersurfaces of the same globally hyperbolic physical region $\mathcal N_\lambda$.
This open region excludes the RT edges and has no timelike conformal boundary.
The polygon meets conformal infinity only at isolated ideal vertices, and causal curves along the boundary can avoid its interior.
The same argument identifies the common region for each component of a finite union.

\subsection{Global Volume Comparison Including Caustics}
\label{app:globalmaximum}

\noindent
Follow normal geodesics from a finite-area Cauchy surface $P$ to a competing one, $Q$.
Curvature bounds the area carried by each small patch up to its first focus, but different patches can reach the same point of $Q$.
Counting their preimages extends the compact-surface comparison of Ref.~\cite{Burnett:1995rf} to an open polygon.
Assume a smooth, oriented, globally hyperbolic spacetime $\mathcal N$ of dimension $2+1$ and a smooth spacelike Cauchy surface $P$ of finite area and zero mean curvature.
If
\begin{equation}
\operatorname{Ric}(v,v)\geq0
\quad\hbox{for every unit timelike }v\hbox{ in }\mathcal N,
\label{eq:app_TCC}
\end{equation}
then every smooth spacelike Cauchy hypersurface $Q$ satisfies
\begin{equation}
\operatorname{Area}(Q)\leq\operatorname{Area}(P).
\label{eq:app_globalcomparison}
\end{equation}
Strict positivity in Eq.~\eqref{eq:app_TCC} makes $P$ the unique maximizer. \\

\noindent
For $p\in P$, the curve $\gamma_p(s)$ is the inextendible future-oriented unit timelike geodesic normal to $P$ with $\gamma_p(0)=p$.
Its unique intersection with the Cauchy hypersurface $Q$ defines
\begin{equation}
\Phi(p)=\gamma_p(T(p))\in Q.
\end{equation}
The intersection is transverse because $Q$ is spacelike, so its signed proper time $T$ and the map $\Phi$ are smooth.
The time $T$ vanishes on $P\cap Q$ and can have either sign and unbounded magnitude elsewhere.
Distinct normals can reach the same point of $Q$. \\

\noindent
Follow neighboring normal geodesics for equal proper times $s$ and denote the ratio of their transverse patch area to its initial area by $\nu(s,p)$.
At a focal point, a transverse direction collapses, whereas distinct normals can cross even when both transverse patches remain regular.
The normal geodesics form a bundle with zero vorticity (local rotation) because they start orthogonally to a surface.
Along a normal with tangent $n$, parallel transport an orthonormal basis $e_1,e_2$ of $T_pP$.
For $b=1,2$, the Jacobi field $Y_b$ is the change in position at fixed $s$ per unit displacement of the starting point along $e_b(0)$.
It stays transverse because $g(Y_b,n)$ starts at zero and $\partial_s g(Y_b,n)=g(\nabla_{Y_b}n,n)=0$ for unit-speed geodesics.
Let $\mathsf J_{ab}=g(e_a,Y_b)$ and $\mathsf R_{ab}=g(e_a,R(e_b,n)n)$ for $a,b=1,2$, where $R$ is the spacetime Riemann curvature operator with the convention $\mathsf J''=-\mathsf R\mathsf J$ for geodesic deviation.
Primes here denote proper-time derivatives, and $\operatorname{tr}\mathsf R=\operatorname{Ric}(n,n)$.
Before the first focal point, $\nu=\det\mathsf J>0$ since $\mathsf J(0)$ is the identity.
The matrix $\mathsf S=\mathsf J'\mathsf J^{-1}$ gives the transverse rate of deformation, so differentiating the determinant and inverse gives
\[
\vartheta=\partial_s\ln\nu=\operatorname{tr}\mathsf S,
\qquad \mathsf S'=-\mathsf R-\mathsf S^2 .
\]
If the principal lengths of a patch grow at fractional rates $s_1,s_2$, its area grows at fractional rate $\vartheta=s_1+s_2$.
Subtracting the mean rate from each direction leaves the shear $\sigma_{ab}=\mathsf S_{ab}-\tfrac12\vartheta\delta_{ab}$, which changes the shape at fixed area to first order, with $\delta_{ab}$ the Kronecker delta.
In the two spatial directions, $\operatorname{tr}(\mathsf S^2)=\vartheta^2/2+\sum_{a,b=1}^2\sigma_{ab}\sigma^{ab}$, so the trace of the evolution equation gives the Raychaudhuri equation
\begin{equation}
\partial_s\vartheta
=-\frac{\vartheta^2}{2}
-\sum_{a,b=1}^2\sigma_{ab}\sigma^{ab}
-\operatorname{Ric}(n,n).
\label{eq:maximality_expansion}
\end{equation}
Here, $n=\dot\gamma_p$ is future-directed, and the shear norm is a sum of squares because both directions transverse to $n$ are spatial.
For $u=\sqrt{\nu}$, the relations $u'/u=\vartheta/2$ and $u''/u=\partial_s\vartheta/2+\vartheta^2/4$ cancel the square of the expansion, giving
\begin{equation}
u''=-\frac12\left[
\operatorname{Ric}(n,n)+
\sum_{a,b=1}^2\sigma_{ab}\sigma^{ab}\right]u\leq0,
\qquad u(0)=1,\quad u'(0)=0.
\label{eq:app_Jacobiconcavity}
\end{equation}
Zero mean curvature gives $u'(0)=0$, so concavity bounds $u$ by $u(0)=1$ up to the first focus.
Let $u_p(r)=\sqrt{\nu}$ at proper time $r\ge0$ from $P$ in either direction.
Reversing the tangent preserves $\mathcal F_p=\operatorname{Ric}(n,n)+\sum_{a,b=1}^2\sigma_{ab}\sigma^{ab}$, so integrating Eq.~\eqref{eq:app_Jacobiconcavity} twice gives
\[
1-u_p(r)=\frac12\int_0^r(r-x)\,\mathcal F_p(x)\,u_p(x)\,dx .
\]
Before focusing, $u_p>0$ and timelike Ricci convergence make the integrand nonnegative.
Hence, $0<\nu\le1$, and continuity permits equality at $r>0$ only if $\mathcal F_p$ vanishes along the entire segment. \\

\noindent
For the area comparison, retain the normal segments that realize the greatest proper time from $P$ to their endpoint.
Let $d_{\rm L}(x,y)$ denote the Lorentzian time separation, the supremum of future-directed causal lengths from $x$ to $y$, set to zero if no such curve exists.
For $q\in Q$, the greatest proper time on its side of $P$ is
\[
\ell_P(q)=\sup_{p'\in P}\big[d_{\rm L}(p',q)+d_{\rm L}(q,p')\big],
\qquad
E=\{p\in P:\ |T(p)|=\ell_P(\Phi(p))\}.
\]
Only one time separation contributes for $q\notin P$, and both vanish on $P\cap Q$.
Global hyperbolicity makes time separation continuous, so a countable dense subset of $P$ gives the same supremum and makes $\ell_P$, hence $E$, measurable.
A longest segment can first focus at its endpoint, never in its interior.
At a regular $p\in E$, the fixed-time displacement $Y_X$ from varying $p$ along $X\in T_pP$ is transverse to $n$.
The arrival time also varies, so the chain rule gives $d\Phi_p(X)=Y_X(T(p))+dT_p(X)n$.
Taking its inner product with a second such displacement gives
\begin{equation}
\Phi^*h_Q=h_T-dT\otimes dT,
\end{equation}
where $h_T$ is the metric of the fixed-time displacements and the minus sign comes from $g(n,n)=-1$.
In common coordinates at $p$, the transverse area ratio is $\nu=\sqrt{\det h_T/\det h_P}$, with $h_P$ the induced metric on $P$.
The area Jacobian for arrival on $Q$ is
\begin{equation}
J_\Phi(p)
=\nu(T(p),p)\sqrt{1-|dT|_{h_T}^2}
=\frac{\nu(T(p),p)}{-g(n,N_Q)}
\leq\nu(T(p),p)\leq1.
\label{eq:app_normalmapJacobian}
\end{equation}
For the first equality, the two-dimensional metric matrix $H$ of $h_T$ is reduced to $H-vv^{\mathsf T}$, where $v$ is the column of derivatives of $T$.
Its determinant is $\det(H)(1-v^{\mathsf T}H^{-1}v)$.
For the second equality, the future unit normal $N_Q$ is orthogonal to both tangents of the graph, which gives $-g(n,N_Q)=(1-|dT|_{h_T}^2)^{-1/2}$ after normalization.
For example, in a locally orthonormal transverse frame with $dT=(v,0)$ and $|v|<1$, the two metric coefficients on $Q$ are $1-v^2$ and $1$.
The factor $\sqrt{1-v^2}$ is the area loss due to tilt alone. \\

\noindent
To see how tilt and focusing enter separately, hold the spacetime fixed and consider the normal graph $Q_t$ given by $s=t f(p)$, where $f$ is a smooth function with compact support in the interior of $P$ and $t$ is a real deformation parameter.
For sufficiently small $|t|$, this graph is spacelike, and its area density relative to $P$ is $\nu(tf,p)\sqrt{1-t^2|df|_{h_{tf}}^2}$ by Eq.~\eqref{eq:app_normalmapJacobian}.
At $s=0$, the vanishing expansion gives $\partial_s\nu(0,p)=0$, and the shear equals the second fundamental form $K_{ab}$ because its trace vanishes.
Let $h_{\rm ind}$ denote the induced metric on $P$ at this fixed state.
Consequently, Eq.~\eqref{eq:maximality_expansion} gives $\partial_s^2\nu(0,p)=-\operatorname{Ric}(n,n)-|K|_{h_{\rm ind}}^2$, where $|K|_{h_{\rm ind}}^2=\sum_{a,b=1}^2 K_{ab}K^{ab}$ with both indices raised using $h_{\rm ind}$.
The square root contributes $-|df|_{h_{\rm ind}}^2$ to the second derivative of the density, so
\[
\left.\frac{d^2}{dt^2}\operatorname{Area}(Q_t)\right|_{t=0}
=-\int_P\left\{
|df|_{h_{\rm ind}}^2+
\big[\operatorname{Ric}(n,n)+|K|_{h_{\rm ind}}^2\big] f^2
\right\}\,dA_P .
\]
Only the interior is displaced because $f$ vanishes near the edges.
The derivative term penalizes neighboring patches displaced by different times, while the term proportional to $f^2$ measures focusing along their trajectories.
Under Eq.~\eqref{eq:app_TCC}, this derivative is nonpositive.
The state response $A_Q''(0)$ in Sec.~\ref{sec:energy-local} instead differentiates the geometry and its geodesic edges with respect to the state amplitude $\lambda$. \\

\noindent
For $q$ in the chronological future $I^+(P)$, a causal curve from $P$ to $q$ must start in $P\cap J^-(q)$.
This set is closed in the physical spacetime $\mathcal N$ because $P$ is Cauchy and $J^-(q)$ is closed by global hyperbolicity.
To see compactness, suppose a sequence of starting points escapes every compact set.
In a complete auxiliary Riemannian metric, causal curves from $q$ back to these points then have unbounded length.
Parametrization by this length and passage to a subsequence on each compact parameter interval give a past-inextendible causal curve in $P\cup I^+(P)$.
The Cauchy property forces that curve to cross the spacelike $P$ into $I^-(P)$, a contradiction.
The continuous time separation from $q$ therefore attains its positive maximum on $P\cap J^-(q)$, and global hyperbolicity supplies a realizing timelike geodesic.
At its starting point $p$, first variation of proper time along any $X\in T_pP$ gives $g(v,X)=0$ for the unit tangent $v$ toward $q$, so the geodesic leaves $P$ normally.
Time reversal treats $q\in I^-(P)$, and $T=0$ on $P\cap Q$, proving $\Phi(E)=Q$.
At a focal endpoint, some nonzero starting displacement has zero fixed-time Jacobi field.
For that displacement, $d\Phi$ is proportional to the timelike normal but tangent to the spacelike $Q$, so it vanishes.
Such a starting point belongs to the critical set $C$ where $d\Phi$ loses rank.
Sard's theorem gives its image zero area.
Every other point of $Q$ is reached at least once by a regular longest segment, so counting its preimages gives the first inequality and $J_\Phi\le1$ gives the second,
\begin{equation}
\operatorname{Area}(Q)
\leq\int_{E\setminus C}J_\Phi\,dA_P
\leq\int_{E\setminus C}dA_P
\leq\operatorname{Area}(P).
\label{eq:app_areabound}
\end{equation}
On the open polygon, apply the area formula on increasing compact patches and use monotone convergence for the nonnegative integrands.
For $B=E\setminus C$, let $m_B(q)$ count its preimages of $q$.
Surjectivity away from the zero-area critical values gives $m_B\ge1$ almost everywhere, while the area formula gives $\int_B J_\Phi\,dA_P=\int_Q m_B\,dA_Q$.
The difference between the two areas is therefore
\[
\begin{aligned}
\operatorname{Area}(P)-\operatorname{Area}(Q)
={}&\int_{P\setminus B}dA_P+\int_B(1-J_\Phi)\,dA_P\\
&+\int_Q\big(m_B(q)-1\big)\,dA_Q .
\end{aligned}
\]
The three nonnegative contributions count unused starting patches, area lost to focusing or tilt, and excess coverage on $Q$, respectively.
For example, two regular maximizing normals reaching one patch of $Q$ count its area twice, so the last term removes the extra copy when recovering $\operatorname{Area}(Q)$.
Only the middle term uses the curvature bound, which can therefore be imposed only along the maximizing segments.
Under strict timelike convergence, $J_\Phi<1$ on $B$ wherever $T\ne0$.
If the fillings differ, the open set $Q\setminus P$ has positive area and its preimage satisfies $\int_{B\cap\Phi^{-1}(Q\setminus P)}J_\Phi\,dA_P\geq\operatorname{Area}(Q\setminus P)>0$.
The middle term is then strictly positive on a set of positive area, proving uniqueness. \\

\noindent
For the reflection polygon, $K_{ab}=0$ implies zero mean curvature, so its own area attains the supremum in $\mathcal N_\lambda$ under the comparison hypotheses.
Any admitted weak limit whose area equals the limit of smooth filling areas obeys the same bound.

\subsection{Scalar Bound on Causal Development}
\label{app:potentialbound}

\noindent
To apply the comparison to a scalar field, we need the Ricci curvature along the observer's velocity.
Einstein's equation fixes it from the observer's energy density and the trace $T=g^{\mu\nu}T_{\mu\nu}$, with spacetime indices summed over $0,1,2$.
Writing $X=g^{\mu\nu}\partial_\mu\phi\,\partial_\nu\phi$, the scalar stress tensor is $T_{\mu\nu}=\partial_\mu\phi\,\partial_\nu\phi-g_{\mu\nu}(X+M_\phi^2\phi^2)/2$ and its trace is $T=-(X+3M_\phi^2\phi^2)/2$.
Tracing $R_{\mu\nu}-R g_{\mu\nu}/2-L^{-2}g_{\mu\nu}=8\pi G_N T_{\mu\nu}$ in three dimensions gives $R=-6/L^2-16\pi G_N T$.
On substituting this result, the $Xg_{\mu\nu}$ terms cancel in $T_{\mu\nu}-g_{\mu\nu}T$, leaving
\begin{equation}
R_{\mu\nu}
=-\frac2{L^2}g_{\mu\nu}
+8\pi G_N\left[
\partial_\mu\phi\,\partial_\nu\phi
+M_\phi^2\phi^2g_{\mu\nu}\right],
\qquad \mu,\nu=0,1,2.
\end{equation}
For a unit timelike vector $v$, its directional derivative is $v\phi=\sum_{\mu=0}^2v^\mu\partial_\mu\phi$.
Contraction gives
\begin{equation}
\operatorname{Ric}(v,v)=\frac2{L^2}+8\pi G_N\big[(v\phi_\lambda)^2-M_\phi^2\phi_\lambda^2\big] .
\end{equation}
To determine exactly when timelike Ricci convergence holds at a point of this solution, write $w=\nabla\phi_\lambda$, so $X=g(w,w)$, and put $[-X]_+=\max(0,-X)$.
Since $v\phi_\lambda=g(v,w)$, the smallest possible contribution of the derivative term is
\[
\inf_{\substack{g(v,v)=-1\\v\ \text{future directed}}}
[g(v,w)]^2=[-X]_+ .
\]
For timelike $w$, the rest frame of $w$ gives $[g(v,w)]^2=(-X)(v^0)^2\ge -X$, with equality for $v$ parallel or antiparallel to $w$.
For spacelike $w$, a timelike unit vector orthogonal to $w$ gives zero, and for $w=0$, every direction gives zero.
For nonzero null $w$, choose an orthonormal frame with $w=\kappa(1,1,0)$, where $\kappa\ne0$.
The unit timelike vector $v=(\cosh\zeta,\sinh\zeta,0)$ gives $[g(v,w)]^2=\kappa^2\exp(-2\zeta)\to0$ as the real rapidity $\zeta\to+\infty$, so the infimum in this case is approached without being attained.
At each point, the condition $\operatorname{Ric}(v,v)\ge0$ in every timelike direction is therefore equivalent to
\[
M_\phi^2\phi_\lambda^2\le \frac{1}{4\pi G_N L^2}+[-X]_+ .
\]
At positive $M_\phi^2$, a timelike field gradient can offset part of the potential because every observer measures a nonzero time derivative in that case.
Dropping the nonnegative $[-X]_+$ gives a sufficient test involving the field amplitude alone,
\begin{equation}
q_\lambda=4\pi G_N L^2\max(M_\phi^2,0)\sup_{\mathcal N_\lambda}|\phi_\lambda|^2<1 .
\label{eq:mainpotential}
\end{equation}
Since $(v\phi)^2\geq0$ for every timelike direction, this bound implies
\begin{equation}
\operatorname{Ric}(v,v)\geq\frac{2(1-q_\lambda)}{L^2}>0.
\label{eq:app_Riccibound}
\end{equation}
The supremum concerns the entire causal development at the chosen amplitude.
For a regular solution with $M_\phi^2\le0$, the mass term strengthens focusing and $q_\lambda=0$.
For instance, a field with negative mass squared and zero first derivatives at a point contributes negative local energy there when its value is nonzero, yet increases $\operatorname{Ric}(v,v)$ for every observer.
For $M_\phi^2>0$, Eq.~\eqref{eq:app_Riccibound} compares the loss of focusing from the potential with the positive vacuum curvature contribution.
The full spacetime can therefore focus even when the matter alone violates the strong energy condition, which here requires a nonnegative matter contribution for every timelike direction. \\

\noindent
The same calculation quantifies the contraction when $q_\lambda<1$.
Here, $\omega_\lambda=\sqrt{1-q_\lambda}/L$, and $s\ge0$ denotes proper time measured away from the reflection slice along either orientation of a normal geodesic.
Eqs.~\eqref{eq:app_Jacobiconcavity} and \eqref{eq:app_Riccibound} give $u''+\omega_\lambda^2u\le0$ before the first focal point.
While $\cos(\omega_\lambda s)>0$, the derivative of $u'\cos(\omega_\lambda s)+\omega_\lambda u\sin(\omega_\lambda s)$ equals $(u''+\omega_\lambda^2u)\cos(\omega_\lambda s)\le0$.
This expression vanishes at the slice, so $u/\cos(\omega_\lambda s)$ is nonincreasing from its initial value of one.
Thus, a regular maximizing segment has $\nu(s)\le\cos^2(\omega_\lambda s)$ before focality and within $0\le s<\pi/(2\omega_\lambda)$.
In pure AdS, the totally geodesic slice has zero shear along its normals and equality holds up to the focal time. \\

\noindent
The supremum in Eq.~\eqref{eq:mainpotential} involves the exact field throughout $\mathcal N_\lambda$.
A sufficient perturbative formulation uses the dimensionless rescaled field
\begin{equation}
\Psi_\lambda=\sqrt{8\pi G_N}\,\phi_\lambda,\qquad
q_\lambda=\frac{L^2(M_\phi^2)_+}{2}
\|\Psi_\lambda\|_{L^\infty(\mathcal N_\lambda)}^2.
\label{eq:app_rescaledbound}
\end{equation}
Suppose all developments lie in a common finite interval of conformal time, and $\Psi_\lambda$ approaches $\Psi_0=0$ uniformly there as $\lambda\to0$.
Then $q_\lambda<1$ for every sufficiently small $|\lambda|$ by the norm bound above.
This condition concerns the exact solutions only over that time interval.
Appendix~\ref{app:scalarmax} gives a criterion on such an interval for the ideal triangle.
The slice constraint still gives the area response if this control or the causal hypotheses are unavailable, but identifying it with the largest filling requires a separate comparison.

\subsection{Gauss--Bonnet Theorem with Ideal Vertices}
\label{app:detailA5}

\noindent
An ideal vertex lies at infinite distance along its two sides.
To find its contribution to the Gauss--Bonnet Theorem, close the sides with a short cutoff arc and track both the arc curvature and the turns at its ends.
For a finite union of polygons $P_\lambda$, the vertex count $N_v$ includes multiplicities, and the Euler characteristic $\chi$ adds one for each disk component.
The minimizing RT pairing, $N_v$, and $\chi$ stay fixed as the geometry varies.
The Hamiltonian constraint on the slice reads
\begin{equation}
R^{(2)}+K^2-\sum_{a,b=1}^2K_{ab}K^{ab}
=-\frac2{L^2}+16\pi G_N\rho.
\label{eq:app_constraint}
\end{equation}
Here, $K=\sum_{a,b=1}^2h^{ab}K_{ab}$ and the Gaussian curvature is $\mathcal K=R^{(2)}/2$.
On the reflection slice, $K_{ab}=0$, so $\mathcal K=-L^{-2}+8\pi G_N\rho$.
A short cutoff arc at each ideal vertex gives a compact polygon to which the Gauss--Bonnet Theorem applies.
The geodesic edges contribute no geodesic curvature.
We assume an asymptotically hyperbolic end, smooth in the interior, whose conformally compactified metric is twice continuously differentiable at infinity.
In the asymptotically AdS collar used here, choose boundary arclength $s$ and a defining function $z$ such that $h=L^2z^{-2}(dz^2+b(s,z)^2ds^2)$, with $b=1+O(z^2)$ and $\partial_z b=O(z)$.
Each side approaches its ideal endpoint $s_*$ orthogonally in the compactified metric, so near that endpoint it is a graph $s=s(z)$ with $s'(z)\to0$ as $z\downarrow0$.
To find the rate of approach, put $s'=ds/dz$ and vary the dimensionless length integrand $F=\sqrt{1+b^2(s')^2}/z$ for this graph.
The geodesic equation $d(\partial F/\partial s')/dz=\partial F/\partial s$ gives the tangential momentum $p_s$ and its derivative,
\[
p_s=\frac{b^2s'}{z\sqrt{1+b^2(s')^2}},\qquad
\frac{dp_s}{dz}=\frac{b(\partial_s b)(s')^2}{z\sqrt{1+b^2(s')^2}} .
\]
Since $b(s,0)=1$ and $b$ has continuous second derivatives, $\partial_s b=O(z)$ and the right-hand side is bounded.
Thus, $p_s=p_*+O(z)$ for a finite endpoint value $p_*$ for each side.
Substitution into the first equality gives $s'=p_*z+O(z^2)$ and integration gives $s(z)=s_*+p_*z^2/2+O(z^3)$.
At $z=\epsilon$, the two sides differ in $s$ by $O(\epsilon^2)$, so their connecting arc has physical length $O(L\epsilon)$ after multiplying by $Lb/\epsilon$.
The magnitude of its geodesic curvature is $|1-z\partial_z\ln b|/L$, obtained by projecting the acceleration of its unit tangent $(z/Lb)\partial_s$ onto $(z/L)\partial_z$.
Its integrated geodesic curvature thus tends to zero.
Conformal rescaling preserves angles, so the normal approach of the two sides to $z=0$ makes both interior corner angles tend to $\pi/2$.
Their exterior turns are the supplements of these angles and sum to $\pi$.
At height $z$, the removed tip has coordinate width $O(z^2)$, and its area density is $L^2 b/z^2$, giving a removed area of order $L^2\epsilon$. \\

\noindent
To pass from the compact cutoff polygon to the curvature integral, one must also control the curvature on these tips.
Write $\widehat h=dz^2+b^2 ds^2$ so that $h=(L/z)^2\widehat h$.
The compact metric has Gaussian curvature $-\partial_z^2 b/b$, and its scalar Laplacian on a function $f(z)$ is $b^{-1}\partial_z(b\partial_z f)$.
The two-dimensional conformal curvature formula then gives
\[
L^2\mathcal K
=z^2\left[-\frac{\partial_z^2 b}{b}
-\frac1b\partial_z\!\left(b\,\partial_z\ln(L/z)\right)\right]
=-1+z\frac{\partial_z b}{b}-z^2\frac{\partial_z^2 b}{b} .
\]
Since $\partial_z b=O(z)$ and $\partial_z^2 b$ is bounded, $L^2\mathcal K=-1+O(z^2)$, so each removed tip contributes $O(\epsilon)$ to the integral of $|\mathcal K|$.
Let $P_{\lambda,\epsilon}$ denote the retained compact region, let $C_{v,\epsilon}$ be its cutoff arc at vertex $v=1,\ldots,N_v$, and let $\theta_{v,j}$ be the interior angle at end $j=1,2$ of that arc.
Truncating each tip leaves the Euler characteristic equal to $\chi$.
With $k_g$ the signed geodesic curvature for the positively oriented boundary and $d\ell$ its physical line element, the Gauss--Bonnet Theorem for the truncated polygon gives
\[
\int_{P_{\lambda,\epsilon}}\mathcal K_\lambda\,dA_\lambda
+\sum_{v=1}^{N_v}\int_{C_{v,\epsilon}}k_g\,d\ell
+\sum_{v=1}^{N_v}\sum_{j=1}^{2}(\pi-\theta_{v,j})
=2\pi\chi .
\]
As $\epsilon\downarrow0$, the arc integrals vanish, the two turns at each vertex contribute $\pi$ together, and the omitted curvature integral tends to zero.
The remaining terms give
\begin{equation}
\int_{P_\lambda}\mathcal K_\lambda\,dA_\lambda=2\pi\chi-\pi N_v.
\end{equation}
The curvature from the constraint then gives the exact reflection-slice identity
\begin{equation}
\frac{A(\lambda)}{L^2}=\pi(N_v-2\chi)+8\pi G_N\int_{P_\lambda}\rho_\lambda\,dA_\lambda .
\label{eq:appendixGB}
\end{equation}
Under the causal and timelike-curvature hypotheses of Appendix~\ref{app:globalmaximum}, $A(\lambda)=\operatorname{Area}(P_\lambda)$ also equals $V(\lambda)$.
In constant curvature, the formula recovers the usual ideal-polygon area \cite{Fujiki:2026ucr,Abt:2017pmf}.
For example, an ideal triangle and an ideal quadrilateral have vacuum areas $\pi L^2$ and $2\pi L^2$, respectively.
The theorem therefore includes odd vertex counts as well as the even counts arising for separated intervals and their purifier.

\subsection{Regularity and Shape Derivative}
\label{app:detailA6}

\noindent
Two different regularity questions arise at an ideal vertex.
For the area response, the integral over the omitted tip must tend to zero uniformly in the varying state.
For reconstruction from line integrals, one also needs regularity of the density extended beyond the Klein disk.
Write $h_0=L^2\bar h$ with $\bar h=q$ of curvature $-1$ and $r=|y|$ in Klein coordinates.
Two useful classes for the leading density $e=L^3\rho_2$ are
\begin{equation}
e\in C_c^\infty(\mathbb H^2),\qquad\hbox{or}\qquad
e(y)=(1-r^2)^3 a_0(y),\quad a_0\in C^\infty(\overline{\mathbb D}).
\label{eq:app_sourceclasses}
\end{equation}
Here, $C_c^\infty$ denotes smooth functions vanishing outside a compact set, and $a_0$ is smooth on the closed disk $\overline{\mathbb D}=\{y\in\mathbb R^2:|y|\le1\}$.
The function $a_0$ is distinct from the scalar amplitude $a$.
Reconstruction uses $g=e/(1-r^2)$ inside the disk and zero outside.
In the first class, $g$ already vanishes in a collar of the circle.
In the second, $g=(1-r^2)^2a_0$ meets zero with vanishing first derivatives, so its second weak derivatives have no singular part on the circle and are square integrable.
In particular, the extension lies in $C^1\cap H^2(\mathbb R^2)$, where $H^2$ requires square integrability through two weak derivatives.
The two $\Delta=3$ profiles of Appendix~\ref{app:scalar} obey these conditions.
For fractional $\Delta>1$, the source-free falloff gives a weaker decay estimate, but the zero extension $g=e/(1-r^2)=O((1-r^2)^{\Delta-1})$ still belongs to $L^1\cap L^2$.
By the Fourier slice identity, transforming the measured parallel line integrals in their offset recovers the Fourier transform of $g$ along one radial line.
Knowledge of all directions therefore determines the density, and $L^2$ convergence gives the regularized polygon integral (Appendix~\ref{app:ordinary_entropy_weights}).
The linear-in-smoothing-parameter bias estimate additionally requires $\nabla g\in L^2$. \\

\noindent
Identify nearby slices in coordinates anchored at the boundary, write $J_\lambda=dA_\lambda/dA_0$, and let $\mathbf1_P$ equal one in $P$ and zero outside.
Assume the areas are twice differentiable at zero and that, as $\lambda\to0$ from either side, $\mathbf1_{P_\lambda}\to\mathbf1_{P_0}$ almost everywhere, $J_\lambda\to1$, and $\rho_\lambda/\lambda^2\to\rho_2$ in the interior pointwise.
A single $dA_0$-integrable bound on $|\mathbf1_{P_\lambda}\lambda^{-2}\rho_\lambda J_\lambda|$ for all small nonzero $\lambda$ then suffices by dominated convergence.
Alternatively, one can use integrable bounds on each retained region $z\ge\epsilon$ and estimate the omitted tips directly.
Assume the collar metric and tip width of Appendix~\ref{app:detailA5} are bounded uniformly for small $|\lambda|$ and that $|\rho_\lambda|/\lambda^2\le C z^\beta$ there, with fixed $C>0$ and $\beta>-1$.
The $O(z^2)$ coordinate width cancels the $O(L^2z^{-2})$ area factor, so summing the finitely many tips gives a uniform $C'>0$ such that
\[
\frac{1}{\lambda^2}\int_{P_\lambda\cap\{0<z<\epsilon\}}
|\rho_\lambda|\,dA_\lambda
\le C'\int_0^\epsilon z^\beta\,dz
=\frac{C'}{\beta+1}\epsilon^{\beta+1}.
\]
Since $\beta+1>0$, the omitted integral tends to zero uniformly in $\lambda$.
For a family supported in a fixed compact interior set, it is already zero for small enough $\epsilon$.
With the tip estimate, one can take $\lambda\to0$ on each retained region by dominated convergence and then let $\epsilon\downarrow0$, obtaining
\begin{equation}
\lambda^{-2}\int_{P_\lambda}\rho_\lambda\,dA_\lambda
\longrightarrow\int_{P_0}\rho_2\,dA_0\qquad(\lambda\to0,\ \lambda\ne0).
\end{equation}
Substitution in Eq.~\eqref{eq:appendixGB} yields
\begin{equation}
A''(0)=16\pi G_NL^2\int_{P_0}\rho_2\,dA_0
=16\pi G_NL\int_{P_0}e\,d\mu.
\label{eq:app_shape}
\end{equation}
With $G_N/L=3/(2c)$, this gives $A''(0)/L^2=(24\pi/c)\int_{P_0}e\,d\mu$.
If the slice is maximal throughout the family, the volume has the same Hessian.
At fixed cutoff, smooth side motion sweeps out strips of area $O(|\lambda|)$, and the fractional change in the area element is $O(|\lambda|)$.
Both multiply a density of order $\lambda^2$ and hence enter at order $O(|\lambda|^3)$ there.
After removing the cutoff, the uniform tip estimate gives the $o(\lambda^2)$ remainder needed in Eq.~\eqref{eq:app_shape}. \\

\noindent
A direct differentiation of the geometric area must account for its moving boundary because the vacuum area element is already nonzero.
For example, take a conformal metric variation $h_\epsilon=\exp(2\epsilon f)h_0$ with compatible asymptotics and follow the geodesic sides of its polygon.
With the convention $\Delta_h=\operatorname{div}_h\nabla_h$, the two-dimensional conformal identity $\mathcal K[\exp(2u)h]=\exp(-2u)(\mathcal K[h]-\Delta_hu)$ gives $\delta\mathcal K=2f/L^2-\Delta_{h_0}f$.
The quantity $\int_{P_\epsilon}(\mathcal K_\epsilon+L^{-2})\,dA_\epsilon$ has zero vacuum integrand.
Its first derivative is $\int_{P_0}\delta\mathcal K\,dA_0$, and the Gauss--Bonnet Theorem gives
\begin{equation}
\left.\frac{d\operatorname{Area}(P_\epsilon)}{d\epsilon}\right|_0
=\int_{P_0}\left(2f-L^2\Delta_{h_0}f\right)dA_0.
\end{equation}
At fixed coordinates, the two-dimensional area element scales by $\exp(2\epsilon f)$, accounting for the term $2f$.
The Laplacian term measures the area swept out when the geodesic edges move.
To derive the edge contribution directly, take $f$ smooth and compactly supported in the interior of the hyperbolic disk, allowing its support to meet the sides.
Along one vacuum side $\gamma$, let $s\in\mathbb R$ denote physical arclength, $T$ its unit tangent, $\nu$ the outward unit normal within the slice, and $u(s)$ the outward displacement per unit $\epsilon$ at $\epsilon=0$.
The normal part of the variation of geodesic acceleration in the vacuum metric is $u''+\mathcal K_0u=u''-u/L^2$, where primes denote $s$ derivatives.
The conformal change of connection contributes $2(Tf)T-\nabla_{h_0}f$ per unit $\epsilon$, whose normal projection is $-\partial_\nu f$ with $\partial_\nu f=\langle\nabla_{h_0}f,\nu\rangle_{h_0}$.
The perturbed edge is geodesic, so the total normal acceleration vanishes and
\begin{equation}
\frac{d^2u}{ds^2}-\frac{u}{L^2}=\partial_\nu f,
\qquad
u(s)=-\frac L2\int_{-\infty}^{\infty}
\exp\!\left(-\frac{|s-t|}{L}\right)\partial_\nu f(t)\,dt .
\label{eq:conformal-side-displacement}
\end{equation}
Outside the support of $f$, the perturbed and vacuum sides are vacuum geodesics with a common ideal endpoint.
Their $O(z^2)$ coordinate separation in Appendix~\ref{app:detailA5} gives an $O(z)$ physical displacement, imposing decay at both infinite ends of the edge.
These conditions select the decaying kernel in Eq.~\eqref{eq:conformal-side-displacement}.
Its first derivative jumps by one at $s=t$, giving the required unit point source. \\

\noindent
A side displaced outward by $\epsilon u(s)$ sweeps out signed area $\epsilon\int_\gamma u\,ds$ to first order.
Integrating the exponential kernel over $s$ gives $2L$, so Eq.~\eqref{eq:conformal-side-displacement} yields $\int_\gamma u\,ds=-L^2\int_\gamma\partial_\nu f\,ds$.
Summing the sides gives $-L^2\int_{P_0}\Delta_{h_0}f\,dA_0$ by the divergence theorem, with zero cutoff flux because $f$ vanishes in the collar.
The change in area is thus the sum of this boundary motion and the change $2\int_{P_0}f\,dA_0$ of the area element.

\subsection{Extrinsic Curvature Away from Time Reflection}
\label{app:detailA7}

\noindent
Time reflection sets all components of the extrinsic curvature to zero. However, stationarity under compactly supported displacements of a smooth spacelike filling sets only its trace $K$ to zero.
The filling can still bend in opposite senses in its two principal directions, with equal and opposite principal curvatures.
Assume the chosen spacelike extremal geodesics lie on this filling with the prescribed pairing and ideal endpoints.
For their unit tangent $t$, the zero spacetime acceleration splits into a tangent part and a normal part.
The first makes the side an intrinsic geodesic, and the second sets $K_{ab}t^at^b=0$ along that side. \\

\noindent
Let $A_P$ be the intrinsic area of the filling.
Using $R^{(2)}=2\mathcal K$, the Hamiltonian constraint gives
\[
\mathcal K+L^{-2}=8\pi G_N\rho
+\tfrac12\big(K_{ab}K^{ab}-K^2\big).
\]
Here, repeated spatial indices are summed over $1,2$ with the metric induced on the filling.
For example, in an orthonormal basis of principal directions with principal curvatures $\kappa_1,\kappa_2$, the last term is $-\kappa_1\kappa_2$.
When $K=0$, it equals $\kappa_1^2=K_{ab}K^{ab}/2$ and is nonnegative.
With the same ideal-corner limits as in Appendix~\ref{app:detailA5}, integration of $\mathcal K$ gives $2\pi\chi-\pi N_v$.
Solving for $A_P$ yields
\begin{equation}
\frac{A_P}{L^2}=\pi(N_v-2\chi)+8\pi G_N\int_P\rho\,dA
+\frac12\int_P\sum_{a,b=1}^2K_{ab}K^{ab}\,dA.
\label{eq:app_extrinsic}
\end{equation}
On a branch proven to attain the volume supremum for these edges, $V=A_P$ and Eq.~\eqref{eq:app_extrinsic} gives the maximal volume as well. \\

\noindent
Along a smooth branch through the totally geodesic vacuum slice, write $K_{ab}=\lambda k_{1,ab}+O(\lambda^2)$ and $\rho_\lambda=\lambda^2\rho_2+o(\lambda^2)$.
The Hessian of the squared bending depends only on $k_{1,ab}$, which receives contributions from the ambient metric and the displacement of the filling.
To separate them locally, use Gaussian normal coordinates $(s,x^1,x^2)$ based on a smoothly chosen reference slice, with future proper time $s$ and metric $-ds^2+h_\lambda(s,x)$.
Let $h_0=h_{\lambda=0}(0,x)$ be the vacuum slice metric and $D$ its covariant derivative.
In vacuum AdS, each transverse separation of normal geodesics satisfies the harmonic-oscillator equation with frequency $1/L$ and starts with zero derivative on the totally geodesic slice.
The separations scale as $\cos(s/L)$, so their metric is $h_{\lambda=0}(s,x)=\cos^2(s/L)h_0$.
With $K_{ab}=\tfrac12(\mathcal L_n h)_{ab}$, the reference extrinsic curvature is $\tfrac12\partial_s h_{\lambda,ab}$.
Its derivative with respect to position at the vacuum slice is $-h_{0,ab}/L^2$, and its derivative with respect to state at fixed position is $B_{ab}=\tfrac12\partial_\lambda\partial_s h_{\lambda,ab}|_{\lambda=s=0}$. \\

\noindent
Write the moving filling locally as $s=\lambda f(x)+O(\lambda^2)$, where $f$ is its proper-time displacement per unit $\lambda$.
Its tangent vectors are $\partial_a+\lambda(\partial_a f)\partial_s+O(\lambda^2)$ for $a=1,2$, so orthogonality and unit normalization give the future normal $n=\partial_s+\lambda\nabla_{h_0}f+O(\lambda^2)$ on the graph.
Differentiating this tilted normal tangentially contributes $D_aD_b f$, while evaluating the reference extrinsic curvature at $s=\lambda f$ contributes $-f h_{0,ab}/L^2$.
Combining these terms and taking the trace gives
\begin{equation}
\begin{aligned}
k_{1,ab}&=B_{ab}+D_aD_b f-\frac{f}{L^2}h_{0,ab},\\
\left(\Delta_{h_0}-\frac{2}{L^2}\right)f
&=-\operatorname{tr}_{h_0}B .
\end{aligned}
\label{eq:stationary-filling-displacement}
\end{equation}
Here, $\Delta_{h_0}f=\sum_{a,b=1}^2 h_0^{ab}D_aD_b f$ and $\operatorname{tr}_{h_0}B=\sum_{a,b=1}^2 h_0^{ab}B_{ab}$.
For a stationary filling, the first derivative of $K$ is the vacuum trace of $k_1$ because $K_{ab}=0$ at $\lambda=0$, and this trace vanishes.
The spacetime positions of the chosen extremal edges supply the boundary displacement for Eq.~\eqref{eq:stationary-filling-displacement}.
Consequently, evaluating the squared-curvature contribution requires the first-order filling and the normal matter density. \\

\noindent
Since $K_{ab}$ starts at order $\lambda$, raising its indices with $h_\lambda^{-1}=h_0^{-1}+O(|\lambda|)$ gives $K_{ab}K^{ab}=\lambda^2k_{1,ab}k_1^{ab}+o(\lambda^2)$ with indices on $k_1$ raised by $h_0$.
If $\lambda^{-2}\rho_\lambda$ and $\lambda^{-2}K_{ab}K^{ab}$, including their area Jacobian, satisfy the integrability conditions in Appendix~\ref{app:detailA6}, we can integrate both leading terms on $P_0$.
The resulting Hessian is
\[
\frac{A_P''(0)}{L^2}
=16\pi G_N\int_{P_0}\rho_2\,dA_0
+\int_{P_0}k_{1,ab}k_1^{ab}\,dA_0 .
\]
The factor of two from differentiating $\lambda^2$ cancels the factor $1/2$ in the extrinsic term.
When $k_{1,ab}=0$, the squared extrinsic curvature is pointwise $O(\lambda^4)$ and gives no Hessian contribution under the stated integral control.
The embedding of the filling determines this term, and the boundary-graviton Fisher subtraction in Appendix~\ref{app:gravity} addresses a separate contribution to interval relative entropy.
For a nonreflecting family, this argument identifies the area Hessian with $V''(0)$ when the extremal edges lie on the chosen filling and the filling maximizes area for all sufficiently small amplitudes.

\section{Quartet Weights and Bounds from Finite Measurements}
\label{app:quartet-methods}

\noindent
The same spatial weights underlie the entropy calculation in Sec.~\ref{sec:quartets-main} and the positive-energy bounds in Sec.~\ref{sec:positive-quartets}.
We derive the weight from geodesic lengths, evaluate the moments that bound the mean energy, and then prove the sharpness, rank, and feasibility results for finite sets of measurements.

\subsection{Intrinsic Metric Derivation of Quartet Weight}
\label{app:quartetlocalization}

\noindent
A second derivation relates two effects of the same spatial metric change.
Curvature measures the interior energy, while length measures an interval entropy.
Use coordinates anchored to the same boundary points and subtract a vacuum descendant matching both the metric and its first state derivative at zero.
The remaining metric change is $h_{ab}=L^{-2}\partial_\lambda^2(h^\Sigma_{\lambda,ab}-h^\Sigma_{f,\lambda,ab})|_0$.
Use the dimensionless vacuum metric $q$ to raise indices, with trace $H=q^{ab}h_{ab}$ and Laplacian $\bar\Delta=D^aD_a$.
The linear scalar-curvature variation is $D^aD^bh_{ab}-\bar\Delta H-\operatorname{Ric}^{ab}(q)h_{ab}$, whose last term equals $+H$ because $\operatorname{Ric}_{ab}(q)=-q_{ab}$.
Subtracting the constraints and inserting $\rho''(0)=2e/L^3$ gives
\[
\mathcal L h=D^aD^bh_{ab}-\bar\Delta H+H=32\pi G_N e/L .
\]
Terms quadratic in the first metric derivative cancel because they are shared by the two states.
At fixed regulated endpoints, let $\mathscr D_{ij}(g)$ denote the chord length in units of $L$ for the spatial metric $g=h^\Sigma/L^2$ on the chosen branch.
In the vacuum metric of curvature $-1$, a transverse displacement $j$ between neighboring geodesics satisfies $j''-j=0$ in unit arclength.
Zero displacement at both endpoints forces $j=0$, so there is no transverse zero mode obstructing smooth variation of the geodesic with the metric.
For a metric perturbation $u_{ab}$, stationarity removes the path variation, leaving the derivative of the line element itself,
\[
D\mathscr D_{ij}(q)[u]
=\frac12\int_{\gamma_{ij}}u_{ab}\tau^a\tau^b\,d\sigma ,
\]
where $\tau$ and $d\sigma$ are the vacuum unit tangent and arclength in $q$.
In this calculation, $D\mathscr D_{ij}$ and $D^2\mathscr D_{ij}$ denote the first and second metric derivatives of this length functional.
For the actual metric family $g_\lambda$, the chain rule gives
\[
\left.\frac{d^2}{d\lambda^2}\mathscr D_{ij}(g_\lambda)\right|_0
=D\mathscr D_{ij}(q)[g''_0]
+D^2\mathscr D_{ij}(q)[g'_0,g'_0].
\]
The second term includes the motion of the geodesic under $g'_0$ and is identical for the two families.
Subtracting their length Hessians leaves the first metric derivative $D\mathscr D_{ij}(q)[h]$, which samples $h$ along the vacuum geodesic.
Restoring the factor $L/(4G_N)$ and removing the common regulator gives $S_{2,ij}-S_{2,f,ij}=(L/8G_N)\int_{\gamma_{ij}}h_{ab}\tau^a\tau^b\,d\sigma$.
Under $h_{ab}\mapsto h_{ab}+D_a\xi_b+D_b\xi_a$, the dimensionless length variation acquires the endpoint term $[\xi_a\tau^a]_{i}^{j}$ because $D_\tau\tau=0$.
At finite cutoff, this term is zero for a vector field $\xi$ that vanishes at the two regulated endpoints.
At conformal infinity, an endpoint can keep the same angular coordinate even when the radial cutoff scale changes.
The corresponding displacement is $\xi^z=z\omega(s)+o(z)$, where $s$ labels boundary arclength and $\omega$ is the infinitesimal rescaling.
In the collar of Appendix~\ref{app:detailA5}, orient the chord from $i$ to $j$ and normalize its tangent with respect to the unit metric $q$.
The endpoint expansion $ds/dz=O(z)$ gives $\tau^z/z\to+1$ at $i$ and $\tau^z/z\to-1$ at $j$.
Since $q_{zz}=z^{-2}$ and the tangential component $\xi^s$ tends to zero, the resulting endpoint term is $-\omega(x_i)-\omega(x_j)$.
For each interval separately, the gauge that also preserves the cutoff scale has $\omega=0$.
In the quartet, even a common change of cutoff scale at each endpoint cancels because the three incident chords in $C_{\eta_Q}$ have coefficients $1$, $-\eta_Q$ and $-(1-\eta_Q)$.
We assume a twice differentiable conformal compactification and convergence of the regulated, vacuum-subtracted length variations in the common cutoff convention. \\

\noindent
To turn the curvature constraint into a statement about lengths, integrate by parts twice against a smooth compactly supported function $f$.
This transfers the derivatives from the unknown metric $h$ to the chosen weight $f$, giving its adjoint tensor $P^{ab}f=D^aD^bf+q^{ab}(f-\bar\Delta f)$.
The identity $D_aD_b\mathsf X=q_{ab}\mathsf X$ gives $P^{ab}w=0$ for a signed lapse.
Its positive part $[w]_+$ equals $w$ on one side of the chord and zero on the other, so only the change in slope on the chord survives.
We can compute that contribution by smoothing the corner with $F_\delta(w)=(w+\sqrt{w^2+\delta^2})/2$, where $\delta>0$.
Using $D^aD^b w=q^{ab}w$ and $\bar\Delta w=2w$, the chain rule gives
\[
P^{ab}F_\delta(w)=q^{ab}(F_\delta-w F_\delta')
+F_\delta''(D^awD^bw-q^{ab}|Dw|^2).
\]
Here, primes denote ordinary derivatives of $F_\delta$ with respect to its scalar argument.
The first coefficient is bounded by $\delta/2$, and $F_\delta''(w)=\delta^2/[2(w^2+\delta^2)^{3/2}]$ concentrates at zero with unit integral as $\delta\downarrow0$.
In normal distance $\ell$ and arclength $\sigma$ along the chord, $w=\sinh\ell$ gives $d\mu=\cosh\ell\,d\ell\,d\sigma=dw\,d\sigma$.
Thus, the limiting line distribution has normalization $\int f\delta_\gamma\,d\mu=\int_\gamma f\,d\sigma$, and on its support $D^aw=\nu^a$ is the unit normal toward $w>0$.
Linearity adds the separate chord contributions even where they cross.
The metric splits into its tangent and normal projections, so $\nu^a\nu^b-q^{ab}=-\tau^a\tau^b$ selects precisely the tangent component of $h$ that changes a length.
For a single chord and for the quartet weight, respectively, this gives
\begin{equation}
P^{ab}[w]_+=-\tau^a\tau^b\delta_\gamma,qquad
P^{ab}W=C_{\eta_Q}(\tau^a\tau^b\delta_\gamma),\qquad
W=-C_{\eta_Q}[w]_+ .
\label{eq:adjoint}
\end{equation}
Contracting with $h_{ab}$, integrating over the disk, and using Eq.~\eqref{eq:bC} gives $R_Q''=(L/8G_N)\int_Q W\mathcal Lh\,d\mu=4\pi\int_Q We\,d\mu$.
At an ideal vertex, cut off the polygon at $z=\epsilon$ in the chart of Appendix~\ref{app:detailA5}.
The cutoff has physical length $O(\epsilon)$ and $W=O(\epsilon)$ there.
To see which derivatives enter the boundary term, the product rule gives the distributional Green identity
\[
W\mathcal L h-h_{ab}P^{ab}W=D_aB^a,\qquad
B^a=W(D_bh^{ab}-D^aH)-h^{ab}D_bW+H D^aW .
\]
On each smooth piece of the cutoff, bounded $DW,h,Dh$ in an orthonormal frame give a bounded normal flux density.
Only an arc of length $O(\epsilon)$ meets the support of $W$ or $DW$, so the integrated flux vanishes.
Equivalently, placing an ideal vertex at infinity in the upper half-plane makes the polygon a strip there, with $W=p(x)/y$ and $p$ bounded and piecewise linear.
For an individual canonical-energy integral, $[w_{ij}]_+=O(z^{-1})$, $e=O(z^{2\Delta})$, and $d\mu=O(z^{-2})\,dz\,dx$ give $O(z^{2\Delta-3})\,dz$ after integration across a finite boundary arc.
This is integrable for every $\Delta>1$. \\

\noindent
Eq.~\eqref{eq:localized_main} also expresses the quartet weight in terms of the signed distances to the four sides.
In each sector, one of four rescaled lapse functions sets the weight.
Restricting to the interior and expanding the two positive parts of diagonal lapses gives
\begin{equation}
K_Q=\min\left\{-w_{12},-w_{34},-\frac{1-\eta_Q}{\eta_Q}w_{23},
\frac{1-\eta_Q}{\eta_Q}w_{14}\right\}.
\label{eq:finite_quartet_window}
\end{equation}
The three identities used here are $w_{13}=(1-\eta_Q)w_{14}+\eta_Q w_{12}$, $w_{24}=(1-\eta_Q)w_{14}+\eta_Q w_{34}$, and $w_{13}+w_{24}=(1-\eta_Q)(w_{14}+w_{23})$.
They follow directly from the six affine numerators in the Klein chart. \\

\noindent
For a boundary-anchored conformal perturbation $h_\lambda=L^2\exp(2\lambda^2\psi+o(\lambda^2))\bar h$, the same weight can be derived from the lengths of the six fixed-endpoint chords.
Here, $\bar h$ is the vacuum metric of unit curvature radius, the linear metric variation vanishes, and $S_1=0$.
At fixed cutoff, stationarity makes the first variation of length with respect to the geodesic path vanish, leaving only the change in its line element.
That element scales as $1+\lambda^2\psi+o(\lambda^2)$, so the entropy Hessian is $S_{2,ij}=(c/3)\int_{\gamma_{ij}}\psi\,d\sigma$ using $L/(2G_N)=c/3$.
The curvature variation in the Hamiltonian constraint gives $(-\bar\nabla^2+2)\psi=(8\pi G_N/L)e$, and we assume the regulated length variations have a convergent renormalized limit. \\

\noindent
Let $\mathsf C(\sigma)$ run along the chord with unit speed in its arclength $\sigma$.
In Eq.~\eqref{eq:chord-normal-distance}, $\mathsf C'(\sigma)$ is orthogonal to both $\mathsf C(\sigma)$ and $N$, so varying $\sigma$ at fixed $\ell$ gives a tangent of norm $\cosh\ell$ orthogonal to the unit $\ell$ tangent.
Thus, the unit metric is $d\ell^2+\cosh^2\ell\,d\sigma^2$, and the area element is $\cosh\ell\,d\ell\,d\sigma$.
For $w=\sinh\ell$, the Laplacian reduces to $(\cosh\ell)^{-1}\partial_\ell(\cosh\ell\,\partial_\ell w)=2w$, and the normal derivative $\partial_\ell w$ equals one at the chord.
Its positive part therefore satisfies $(\bar\nabla^2-2)[w_{ij}]_+=\delta_{\gamma_{ij}}$ in distributions, where $\int f\delta_{\gamma_{ij}}d\mu=\int_{\gamma_{ij}} f\,d\sigma$.
The signed combination $K_Q=-\mathcal L_Q[w]_+$ consequently satisfies $(-\bar\nabla^2+2)K_Q=\mathcal L_Q\delta_\gamma$.
Green's identity gives
\[
\frac c3\Pi''(0)=\mathcal L_Q S_2
=\frac c3\int\psi(-\bar\nabla^2+2)K_Q\,d\mu
=\frac c3\int_Q K_Q(-\bar\nabla^2+2)\psi\,d\mu
=4\pi\int_Q K_Q e\,d\mu .
\]
The line sources include the jumps across the sides and diagonals, leaving only the ideal tips to check in Green's identity.
Assume absolute convergence of the chord integrals and bounded $\psi$ and unit normal derivative near each vertex.
Send one vertex to infinity in an upper-half-plane chart, so its two incident sides are vertical lines of finite separation.
Crossing any of the three vertical chords ending there changes the expression for $K_Q$ by a multiple of $(x-x_j)/y$.
Starting from $K_Q=0$ outside thus gives $K_Q=p(x)/y$ high in the strip, with $p$ bounded and piecewise linear.
At the horizontal cutoff $y=R$, the arc length, $K_Q$, and its unit normal derivative are all $O(R^{-1})$, so the remaining boundary flux vanishes as $R\to\infty$. \\

\noindent
Integrating the same Poisson equation also fixes the normalization of the weight directly from the boundary chords.
With $\Delta_{\mathbb H}=\operatorname{div}\nabla$ and $W_{ij}=(w_{ij})_+$, the derivative jump of one at the chord gives $\Delta_{\mathbb H}W_{ij}=2W_{ij}+\delta_{\gamma_{ij}}$.
Consequently, $K=K_Q=-\mathcal L_Q W$ obeys $(2-\Delta_{\mathbb H})K=\mathcal L_Q\delta_\gamma$.
Since $K$ is bounded on a quadrilateral of finite area and vanishes outside, $K\in L^1(d\mu)$.
A smooth radial cutoff $\chi_R$, equal to one at distance $\rho\le R$ and zero at $\rho\ge R+1$, can be chosen so that its Laplacian is uniformly bounded in the outer annulus.
Thus, $\int K\Delta_{\mathbb H}\chi_R\,d\mu\to0$ as $R\to+\infty$.
The length of each chord inside $\rho\le R$ is $2\operatorname{arcosh}(s_{ij}\cosh R)$ for large $R$, where $s_{ij}=\sin[(x_j-x_i)/2]$.
Here, $\operatorname{arcosh} z=\ln(z+\sqrt{z^2-1})$ for $z\ge1$ is the inverse of the hyperbolic cosine on $[0,\infty)$.
The length follows from $\cosh\rho=\cosh\sigma/s_{ij}$, with $\sigma$ the signed arclength from the closest point of the chord to the disk origin.
Its expansion is $2R+2\ln s_{ij}+o(1)$, and smoothing the radial cutoff adds the same constant to every chord up to $o(1)$.
The six coefficients of $\mathcal L_Q$ sum to zero.
The vacuum identities $s_{12}s_{34}=\eta_Q s_{13}s_{24}$ and $s_{14}s_{23}=(1-\eta_Q)s_{13}s_{24}$ then give
\[
\int_Q K\,d\mu=\mathcal L_Q\ln s
=\frac{-\eta_Q\ln\eta_Q-(1-\eta_Q)\ln(1-\eta_Q)}{\eta_Q}.
\]
The last expression is positive and depends only on the four vacuum endpoints.
Dividing $c\Pi''(0)/(12\pi)$ by it converts the measured energy integral into a weighted mean of the density.
For a square, $\eta_Q=1/2$, so the conversion amounts to dividing by $2\ln 2$. \\

\noindent
In the collapsing-rectangle limit $b_Q\downarrow0$, write $X=a_Qs$, $Y=b_Qt$, $-1<s,t<1$.
Writing $\alpha_s=|s|$, symmetry in $t$ gives
\[
\int_{-1}^{1}[1-\max(|s|,|t|)]\,dt
=2\alpha_s(1-\alpha_s)+2\int_{\alpha_s}^{1}(1-t)\,dt=1-s^2 .
\]
Under this change of variables, $W_Q\,d\mu$ contributes a prefactor $(a_Qb_Q)^2$ and a denominator $d^4$.
Dividing by $b_Q^2$ leaves $a_Q^2\to1$, and for fixed $|s|<1$ the denominator tends to $(1-s^2)^2$.
The transverse integral above cancels one power of $1-s^2$, leaving the hyperbolic arclength $ds/(1-s^2)$ along the limiting diameter.
For a continuous profile with $|e|\le C(1-X^2-Y^2)^\nu$, $\nu>0$, use $d^2=a_Q^2(1-s^2)+b_Q^2(1-t^2)$ and $1-\max(|s|,|t|)\le1-|s|$.
These bound the rescaled integrand by a constant times $(1-|s|)^{\min(\nu,2)-1}$ near both ideal endpoints, uniformly for small $b_Q$.
Its exponent exceeds $-1$, so integration commutes with the limit, and Eq.~\eqref{eq:chord_limit} applies even when the decaying source extends to the circle.

\subsection{Weight Moments and Sharp Variance Bound}
\label{app:quartet-moments}

\noindent
Sec.~\ref{sec:quartet-weights} bounds the ordinary mean density in a quadrilateral from its weighted mean and spatial variance.
We compute the two required moments by integrating over one triangle at a time, and then construct densities attaining the endpoints of the bound. \\

\noindent
The diagonals divide $Q$ into four triangles, each with one polygon side as its base.
At inward distance $\ell$ from that base, the ordered weight is $K_Q=\sinh\ell$ for a vertical base and $K_Q=(a_Q^2/b_Q^2)\sinh\ell$ for a horizontal base. \\

\noindent
An upper-half-plane isometry puts the side at $x=0$ and the crossing at $(u,v)$, with $u,v>0$ and $u^2+v^2=1$.
The other edges join $(u,v)$ to the ideal endpoints $\infty$ and $0$, respectively.
The first is $x=u$.
The second is the circle through $(0,0)$ and $(u,v)$ with center on $z=0$, hence $z^2=x/u-x^2$.
Since $\sinh\ell=x/z$ and $d\mu=dx\,dz/z^2$, the triangle integral is
\[
\int_0^u x\,dx\int_{\sqrt{x/u-x^2}}^\infty \frac{dz}{z^3}
=\frac12\int_0^u\frac{dx}{1/u-x}=-\ln v .
\]
At the crossing, the signed distance formula gives $\sinh\ell=a_Q/b_Q$ for a vertical side and $b_Q/a_Q$ for a horizontal one.
Since $\cosh^2\ell=1+\sinh^2\ell$ and $a_Q^2+b_Q^2=1$, the height $v=1/\cosh\ell$ in the upper-half-plane calculation equals $b_Q$ or $a_Q$, respectively.
Adding the two triangles of each type gives $\eta_Q\int_Q K_Q\,d\mu=-2\eta_Q\ln b_Q-2(1-\eta_Q)\ln a_Q=H(\eta_Q)$.
Here, $H(x)=-x\ln x-(1-x)\ln(1-x)$ for $0<x<1$ is the binary entropy in natural units.
For a square, all four triangles contribute equally and $\int_Q K_Q\,d\mu=2\ln 2$.
The intrinsic metric calculation gives the same normalization by a chord integral. \\

\noindent
The area response depends on the ordinary spatial mean of $e$, whereas the quartet reports the weighted mean in Eq.~\eqref{eq:measured_mean_density}.
Their difference depends on how strongly the density varies and on where its larger values lie relative to the crossing.
The first moment of the weight is $H(\eta_Q)$, and its second moment is $\mathcal J_Q=\int_Q W_Q^2\,d\mu$.
Integrating the squared hyperbolic sine over $z$ first in the same upper-half-plane triangle gives
\[
\int_{\triangle}\sinh^2\ell\,d\mu
=\frac13\int_0^u\frac{x^2\,dx}{(x/u-x^2)^{3/2}}
=\frac13\int_0^{u^2}\frac{\sqrt t\,dt}{(1-t)^{3/2}} .
\]
The first equality uses $\int_{z_0}^{\infty}z^{-4}dz=1/(3z_0^3)$ for $z_0>0$, and the second uses $t=ux$.
With $t=\sin^2\theta$ and $0\le\theta\le\arctan(u/v)$, the last integral becomes $(2/3)\int_0^{\arctan(u/v)}\tan^2\theta\,d\theta=(2/3)[u/v-\arctan(u/v)]$.
Here, $\arctan$ is the inverse tangent with values in $(-\pi/2,\pi/2)$. \\

\noindent
Define $j_2(\xi)=\xi-\arctan\xi$ for $\xi\ge0$.
On each vertical-side triangle, $W_Q=\eta_Q\sinh\ell$ and $u/v=a_Q/b_Q$.
On each horizontal-side triangle, $W_Q=(1-\eta_Q)\sinh\ell$ and $u/v=b_Q/a_Q$.
There are two triangles of each kind, hence
\[
\mathcal J_Q=\frac43\left[
\eta_Q^2 j_2(a_Q/b_Q)+(1-\eta_Q)^2 j_2(b_Q/a_Q)
\right].
\]
For the square, $a_Q=b_Q=1/\sqrt{2}$ and $j_2(1)=1-\pi/4$, giving $\mathcal J_Q=(4-\pi)/6$. \\

\noindent
Let $m_Q$ be the measured weighted mean and $s_*$ the upper bound on the spatial standard deviation of the density.
If $\overline W$ and $\sigma_W$ are the spatial mean and standard deviation of $W_Q$, then $C_Q=\sigma_W/\overline W$.
To attain each ordinary mean in $[m_Q-s_* C_Q,m_Q+s_* C_Q]$ with a real square-integrable density, define the standardized weight $U_Q=(W_Q-\overline W)/\sigma_W$ and consider
\[
e_t(y)=m_Q+t[U_Q(y)-C_Q],
\qquad -s_*\leq t\leq s_* .
\]
By construction, $U_Q$ has area mean zero and variance one.
Its weighted mean is $\langle W_Q U_Q\rangle_Q/\overline W=\sigma_W/\overline W=C_Q$, so $U_Q-C_Q$ changes the area mean without changing the measurement.
Thus, $e_t$ has measured mean $m_Q$, area mean $m_Q-tC_Q$, and standard deviation $|t|$.
Varying $t$ fills the entire interval for real $L^2$ densities. However, a fixed mode shape or a positivity condition can restrict the attainable values. \\

\noindent
The same expression explains what happens to the variance bound as a quartet narrows.
Set $x=\min(\eta_Q,1-\eta_Q)$ and use the symmetry of $\mathcal J_Q$ under exchange of the side pairs.
As $x\downarrow0$, the term $x^2j_2(\sqrt{(1-x)/x})$ is asymptotic to $x^{3/2}$, since $j_2(z)/z\to1$ as $z\to+\infty$.
The other term is asymptotic to $x^{3/2}/3$, since $j_2(z)=z^3/3+O(z^5)$ as $z\downarrow0$.
Using $H(x)=x[\ln(1/x)+1]+O(x^2)$ gives
\[
\mathcal J_Q\sim\frac{16}{9}x^{3/2},\qquad
C_Q=\sqrt{\frac{2\pi\mathcal J_Q}{H(x)^2}-1}
\sim\frac{4\sqrt{2\pi}}{3\,x^{1/4}\ln(1/x)}
\qquad (x\downarrow0).
\]
Here, $\sim$ denotes a ratio tending to one.
As the quartet narrows, the growing $C_Q$ weakens what its reading and a fixed variance bound say about the area mean.
For comparison, the geodesic measurement in Sec.~\ref{sec:dual} rescales the same signal to extract a line integral.

\subsection{Sharpness and Nested Regions for Single Quartet}
\label{app:quartet-sharpness}
\noindent
At fixed positive $Z_Q''(0)$, the attainable values of the ratio $A_Q''(0)/(L^2 Z_Q''(0))$ in Sec.~\ref{sec:positive-quartets} fill the entire open interval $(2/M_Q,\infty)$.
For any chosen $w\in(0,M_Q)$, continuity of $W_Q$ supplies disjoint open disks inside $Q$ on which $W_Q>w$ and $W_Q<w$, one near the crossing and the other near a side.
The disks support real smooth functions $f_+$ and $f_-$ of compact support, normalized by $\int_Q f_\pm^2\,d\mu=1$.
The numbers $w_\pm=\int_QW_Q f_\pm^2\,d\mu$ obey $w_-<w<w_+$.
For $t=(w-w_-)/(w_+-w_-)\in(0,1)$ and $A>0$, the scalar profile
\[
e=A\big[t f_+^2+(1-t)f_-^2\big]
\]
has $E_Q=A$ and $\langle W_Q\rangle_e=w$.
The choice $A=cZ_Q''(0)/(12\pi w)$ gives the prescribed readout.
These densities arise from initial data $\phi_\lambda=0$, $K_{ab}=0$, and $n^\mu\partial_\mu\phi_\lambda=\lambda L^{-3/2}\sqrt{2A}\,(\sqrt t\, f_++\sqrt{1-t}\,f_-)$, since the two supports are disjoint.
The momentum constraint is then automatic.
The Hamiltonian constraint has the small-amplitude solution of Appendix~\ref{app:detailF1} for each fixed profile.
The derivatives are evaluated at $\lambda=0$ before the profiles approach the crossing or a side.
Appendix~\ref{app:finiteexact} gives the coherent-state and causal assumptions needed to identify the area with maximal volume.
Here, fixing $Z_Q''(0)$ allows both the individual entropy Hessians and the boundary energy to vary. \\

\noindent
The spatial weight has an integral interpretation in terms of nested quadrilaterals.
For $0<s<1$, the finite-vertex rectangle $Q_s$ has $|X|<s a_Q$ and $|Y|<s b_Q$ in the Klein disk.
A point of $Q$ belongs to $Q_s$ precisely for $m<s<1$, so $[1-m]_+=\int_0^1\mathbf1_{Q_s}\,ds$.
The vacuum time translation centered at the diagonal crossing has dimensionless lapse $\cosh\rho=1/d$, where $\rho$ is the unit-radius hyperbolic distance from the crossing.
Eq.~\eqref{eq:localized_main} therefore gives
\[
\frac{c}{12\pi}\frac{b_Q}{a_Q}\Pi''(0)
=\int_0^1 ds\,\mathcal E_Q(s),\qquad
\mathcal E_Q(s)=\int_{Q_s} e\cosh\rho\,d\mu .
\]
A point with threshold $m$ contributes for the fraction $1-m$ of the integration range, with the same lapse $\cosh\rho$ in each rectangle.
Finiteness of $\int_Q K_Q|e|\,d\mu$ justifies exchanging the integrals.
A negative average therefore requires $\mathcal E_Q(s)<0$ for a set of $s$ of positive measure, even without a separately conserved energy in each $Q_s$.

\subsection{Reconstruction and Independence of Quartet Responses}
\label{app:quartet-rank-proof}
\noindent
For the ordered residuals $T_{ijkl}$ of Sec.~\ref{sec:quartet-rank}, both $T_{ijkl}(0)$ and $T'_{ijkl}(0)$ vanish.
The following determinant calculation expresses their Hessians in terms of quartets incident on two anchors and establishes independence for unrestricted smooth profiles. \\

\noindent
Choose neighboring endpoints $1,2$ as anchors and retain every measured chord incident on either.
The quartet $(1,2,i,j)$, with $3\le i<j\le N$, then has just one unretained entry, $D_{ij}$.
The value it would have if the vacuum chord identity held is $\bar D_{ij}=(D_{1i}D_{2j}-D_{1j}D_{2i})/D_{12}$.
Its error therefore measures how far that quartet departs from the vacuum identity.
To see why the predictions satisfy all quartet identities simultaneously, write their numerators as determinants of two-component vectors.
For this algebra alone, extend the measured positive entries by $D_{ji}=-D_{ij}$ for $i<j$ and $D_{ii}=0$ to record the order of determinant columns.
With $v_i=(D_{1i},D_{2i})$, the prediction becomes $\bar D_{ij}=\det(v_i,v_j)/D_{12}$ for all $1\le i<j\le N$.
In particular, $v_1=(0,-D_{12})$ and $v_2=(D_{12},0)$ reproduce both retained rows.
Writing $[ij]=\det(v_i,v_j)$, the two-dimensional vector identity $[ij]v_k-[ik]v_j+[jk]v_i=0$ follows by expanding its two components.
Taking the determinant of this vector with $v_l$ gives $[ik][jl]=[ij][kl]+[il][jk]$, which makes every quartet residual of $\bar D$ vanish.
The exact residual on any other pair is
\[
D_{ij}-\bar D_{ij}=-T_{12ij}/D_{12}\qquad(3\le i<j\le N).
\]
For $3\le i<j\le N$, let $r_{ij}(\lambda)=D_{ij}(\lambda)-\bar D_{ij}(\lambda)$.
Both $r_{ij}(0)$ and $r'_{ij}(0)$ vanish, so Taylor expansion gives $r_{ij}(\lambda)=\tfrac12\lambda^2r''_{ij}(0)+o(\lambda^2)$.
In a product of two chords, a term containing both residuals is $O(\lambda^4)$.
To get the Hessian of a quartet residual, it therefore suffices to replace one chord at a time by its residual Hessian and evaluate the partner at its vacuum value. \\

\noindent
We use the regulator-free vacuum chord $\ell_{ij}=2\sin[(x_j-x_i)/2]$ for $1\le i<j\le N$ and set $z_{ijkl}=\ell_{ik}\ell_{jl}R''_{ijkl}(0)$ for the cyclically invariant response in Eq.~\eqref{eq:measured_quartet_hessian}.
The equality $R''_{ijkl}(0)=(c/3)\eta_{ijkl}\Pi''_{ijkl}(0)$ shows that $z_{ijkl}=(c/3)\epsilon^2T''_{ijkl}(0)$ in the common vacuum cutoff convention.
With the shorthand $z_{12ij}=0$ for repeated indices, the six terms of the polynomial derivative give
\[
\begin{aligned}
\ell_{12}z_{ijkl}={}&\ell_{kl}z_{12ij}+\ell_{ij}z_{12kl}
+\ell_{jk}z_{12il}+\ell_{il}z_{12jk}\\
&-\ell_{jl}z_{12ik}-\ell_{ik}z_{12jl}.
\end{aligned}
\]
Once the endpoint positions are known, the six coefficients on the right are fixed vacuum lengths.
Thus, measuring the $\binom{N-2}{2}$ quartets containing the chosen pair suffices to predict all quartet responses.
For six endpoints, for instance, there are fifteen quartets, but choosing two other points from $3,4,5,6$ leaves only six readings to take.
For example, take a quartet $(1,i,j,k)$ with $3\leq i<j<k\leq N$ and use the repeated-index convention.
Only three terms survive, giving
\[
z_{1ijk}=
\frac{\ell_{1i}z_{12jk}+\ell_{1k}z_{12ij}
-\ell_{1j}z_{12ik}}{\ell_{12}} .
\]
Thus, even a quartet that omits endpoint $2$ has a response fixed by the retained quartets.
The division uses the nonzero vacuum separation $\ell_{12}$, so the rank statement is for distinct endpoints. \\

\noindent
The reconstruction holds even before integration over an energy profile.
To test it at one bulk point $y$, introduce an auxiliary change of the vacuum entries with slope $\partial_tD_{ij}|_0=D^0_{ij}[w_{ij}(y)]_+$ in a parameter $t$.
The first derivative of the determinant identity then contains $\partial_tT_{ijkl}|_0=D^0_{ij}D^0_{kl}\mathcal L_{ijkl}[w(y)]_+=-D^0_{ij}D^0_{kl}K_{ijkl}(y)$.
All residuals vanish at $t=0$, so their first derivatives obey the same linear reconstruction as the physical second derivatives above.
Hence, the upper bound holds for integrals against any smooth test density. \\

\noindent
To show the count is exact, test a hypothetical vanishing linear combination of the retained quartet weights on a line where just one weight changes slope.
For $(1,2,i,j)$, the term $[w_{ij}]_+$ in $K_{12ij}=-\mathcal L_{12ij}[w]_+$ has coefficient one.
Its normal derivative jumps across $\gamma_{ij}$ because $w_{ij}$ is the hyperbolic sine of signed distance and the positive part sets it to zero on one side.
No other retained quartet contains both $i$ and $j$, so its weight is smooth across $\gamma_{ij}$ away from the finitely many crossings.
Only the coefficient of $K_{12ij}$ can contribute to that jump, and a vanishing combination must have zero coefficient.
Repeating this argument for each $3\le i<j\le N$ proves independence, also after multiplication by the positive factors that normalize $R''$.
The distinction is measurable with smooth compactly supported scalar momentum data.
A nonzero continuous combination has a fixed strict sign on some open disk, and a squared smooth bump supported there integrates it to a nonzero value.
The resulting rank counts the independent energy integrals when profiles can vary freely. At the same time, a specified state family can yield fewer. \\

\noindent
For the regular pentagon, three consecutive readings determine the remaining two by Eq.~\eqref{eq:pentagon-cyclic-readings}.
To see which two real constraints this imposes, take a cyclic mode $r_j=z^{j-1}$, where $z^5=1$.
Eq.~\eqref{eq:pentagon-cyclic-readings} gives $(z-1)[z^2+(1-\varphi)z+1]=0$.
Since $\varphi-1=2\cos(2\pi/5)$, the three allowed roots are $1$ and $\exp(\pm2\pi\mathrm{i}/5)$, where $\mathrm{i}^2=-1$ is the imaginary unit.
The five cyclic modes form a basis, and real readings have conjugate Fourier coefficients.
Thus, the two real constraints are the real and imaginary parts of the single complex equation
\begin{equation}
\sum_{j=1}^5\exp[4\pi\mathrm{i}(j-1)/5]\,r_j=0 .
\label{eq:pentagon-fourier-readings}
\end{equation}
Adding or subtracting a common reading leaves these constraints intact because the five complex coefficients sum to zero.
The three independent measurements give the common mean and the cosine and sine amplitudes of one oscillation around the pentagon.

\subsection{Optimization and Feasibility for Positive Energy}
\label{app:quartet-optimization}
\noindent
The optimization problem in Sec.~\ref{sec:pentagon-program} minimizes the mass of a nonnegative measure at fixed quartet readings $b_i=\int_{P_0}W_i\,d\nu$ for $1\le i\le m$.
We first derive the interpolation formula used there, then obtain equality conditions and show how smooth densities can approach the sharp pentagon bound with unchanged readings. \\

\noindent
We draw all measured sides and diagonals in one Klein disk and intersect the resulting cells with $P_0$.
The set $\mathcal V$ consists of their distinct vertices at radius $|v|<1$.
A repeated crossing is included once.
With $d(y)=\sqrt{1-|y|^2}$, the function $d(y)W_i(y)$ is affine on each cell because the sign of every chord lapse is fixed there and $d w_{ab}$ is affine.
On the closure of a cell, this affine function vanishes at every ideal vertex because $d$ tends to zero and $W_i$ is bounded.
Every finite vertex on the boundary of $P_0$ also has zero weight, because a contained quartet has no interior point on that boundary. \\

\noindent
Triangulate each cell with Euclidean diagonals.
In a triangle with vertices $\mathcal T_y$, the barycentric coordinates of $y$ are numbers $\lambda_v\ge0$ satisfying $\sum_{v\in\mathcal T_y}\lambda_v=1$ and $y=\sum_{v\in\mathcal T_y}\lambda_v v$.
The affine function $dW_i$ equals the corresponding weighted average of its vertex values.
Dividing by $d(y)>0$ recovers $W_i(y)$, and concavity gives $\sum_{v\in\mathcal T_y}\lambda_v d(v)\le d(y)$.
Thus, the same coefficients reproduce every reading with total weight at most one,
\begin{equation}
W_i(y)=\sum_{v\in\mathcal V}t_v(y)W_i(v),\qquad
t_v(y)=\frac{\lambda_v d(v)}{d(y)},\qquad
\sum_{v\in\mathcal V}t_v(y)\le1 .
\label{eq:quartet-cell-interpolation}
\end{equation}
Here, a vertex outside the chosen triangle has $\lambda_v=0$.
Strict concavity makes the last inequality strict at every point that is not a vertex.
Boundaries shared by cells can be assigned consistently with the triangulation. \\

\noindent
For the regular pentagon, let $v_j$ be the diagonal crossing of the quartet omitting point $j$, and let $W_j$ be its symmetric weight, with $1\le j\le5$ and cyclic indices.
The five $v_j$ exhaust the finite vertices of the arrangement [Fig.~\ref{fig:quad}(c)].
The ratio of a diagonal to a side is $\varphi=(1+\sqrt5)/2$, so $\eta_j$ is $\varphi^{-2}$ or its complement according to the starting vertex.
All five weights have the same peak $M=\sqrt{\eta_j(1-\eta_j)}=\varphi^{-3/2}$.
At $v_j$, $W_j=M$ and $W_{j+2}=W_{j-2}=0$ because two other quartets have sides through that point.
Eq.~\eqref{eq:pentagon-cyclic-readings} also holds pointwise for the five weights.
Indeed, $r_j=4\pi\int W_j\,d\nu$ for the energy measure $\nu$, and a smooth nonnegative density of unit mass supported in a shrinking disk about an interior point tests the continuous weights at that point.
At $v_j$, reflection symmetry gives $W_{j+1}=W_{j-1}$.
Applying Eq.~\eqref{eq:pentagon-cyclic-readings} with starting index $j-1$ gives $0=W_{j-1}+\varphi(W_{j+1}-M)$, so each neighboring weight is $\varphi M/(1+\varphi)=M/\varphi$.
Thus, each vertex has the same sum of weights, $M(1+2/\varphi)=\sqrt5\,M$. \\

\noindent
A lower bound on the mass can be read directly from the measurements.
Choose real coefficients $u_i$, $1\leq i\leq m$, with $\sum_{i=1}^m u_i W_i(v)\leq1$ at every finite vertex $v\in\mathcal V$.
For any $y$ in a cell, the interpolation coefficients $t_v(y)$ are nonnegative, so
\[
\sum_{i=1}^m u_i W_i(y)
=\sum_{v\in\mathcal V}t_v(y)\sum_{i=1}^m u_i W_i(v)
\leq\sum_{v\in\mathcal V}t_v(y)\leq1 .
\]
Integrating against $e\,d\mu$ for $e\geq0$ yields the observable bound $\sum_{i=1}^m u_i b_i\leq E_P$. \\

\noindent
To find coefficients that attain the optimal bound, start with a minimizing mass assignment $m_v^*$ and let $\mathbf a_v=(W_1(v),\ldots,W_m(v))$ be the readings from one unit of mass at $v$.
The measured vector is their mass-weighted sum $\mathbf b=(b_1,\ldots,b_m)$.
An infinitesimal mass change $\delta_v$ can have either sign at an occupied vertex but must be nonnegative at an empty one.
The corresponding changes in the readings and in total mass form the cone
\[
C=\left\{\left(\sum_{v\in\mathcal V}\delta_v\mathbf a_v,\
\sum_{v\in\mathcal V}\delta_v\right):
\delta_v\ge0\text{ if }m_v^*=0\right\}.
\]
Each mass increase gives a direction $(\mathbf a_v,1)$, and each occupied vertex also allows its negative.
Their finitely generated cone is closed and convex.
It excludes $(\mathbf0,-1)$ because that direction would decrease total mass at fixed readings.
A separating linear functional is nonpositive on the cone and strictly positive at $(\mathbf0,-1)$.
The latter inequality makes the mass coefficient negative, so normalize it to $-1$ and write the functional as $(\mathbf u,-1)$.
Testing the increase gives $\mathbf u\cdot\mathbf a_v\le1$ at every vertex, and testing the decrease gives equality wherever $m_v^*>0$.
Multiplying by $m_v^*$ and summing yields $\mathbf u\cdot\mathbf b=\sum_{v\in\mathcal V}m_v^*=E_{\min}(b)$, even for data on the boundary of the feasible set.
For any other finite positive measure $\nu$ realizing $\mathbf b$, let $m_v=\int_{P_0}t_v(y)\,d\nu(y)$ be its equivalent vertex masses.
Subtracting the attained bound from the mass of $\nu$ and applying Eq.~\eqref{eq:quartet-cell-interpolation} gives
\begin{equation}
\nu(P_0)-E_{\min}(b)
=\int_{P_0}\!\left(1-\sum_{v\in\mathcal V}t_v(y)\right)d\nu(y)
+\sum_{v\in\mathcal V}\bigl(1-\mathbf u\cdot\mathbf a_v\bigr)m_v .
\label{eq:quartet-mass-excess}
\end{equation}
The first term counts the mass saved by moving energy to the finite vertices and keeping each reading fixed.
By strict concavity of $d$, it vanishes only for a measure supported on those vertices.
The second counts mass at vertices where each unit contributes less than one unit to the lower bound.
It vanishes precisely when $\mathbf u\cdot\mathbf a_v=1$ at every occupied vertex.
Together these characterize equality for the original measure and the finite program in Eq.~\eqref{eq:quartet-vertex-primal}. \\

\noindent
Among the minimizing measures, choose one occupying the fewest vertices.
If its occupied columns $\mathbf a_v$ were linearly dependent, there would be a nonzero change supported there with $\sum_{v\in\mathcal V}\delta_v\mathbf a_v=\mathbf0$.
Taking the dot product with the optimal $\mathbf u$ gives $\sum_{v\in\mathcal V}\delta_v=0$, so the change has both signs and also preserves the mass.
Increasing its size until the first mass reaches zero would yield a minimizer on fewer vertices.
Thus, the number of occupied crossings needed by a minimizer is at most the rank of the matrix with columns $\mathbf a_v$, even when the original density has extended support. \\

\noindent
For the regular pentagon, choosing $u_i=1/(\sqrt5\,M)$ for each of the five readings gives equality at every finite crossing because their column sums are $\sqrt5\,M$.
Consequently, every feasible measure supported there has the minimum mass stated in Sec.~\ref{sec:pentagon-program}.
We now characterize the data for which such a measure exists. \\

\noindent
For the regular pentagon, feasibility itself can also be decided from the readings.
It is equivalent to $r_j\geq0$ for every $1\leq j\leq5$ together with the real and imaginary parts of Eq.~\eqref{eq:pentagon-fourier-readings}.
To prove sufficiency for nonzero readings, set $s_r=\sum_{j=1}^{5}r_j$, $p_j=r_j/s_r$, $\zeta=\exp(2\pi\mathrm{i}/5)$, and $z=\sum_{j=1}^{5}p_j\zeta^{j-1}$.
The linear relations remove the second Fourier harmonic, and reality removes its conjugate.
Inverting the five-point Fourier transform gives
\[
p_j=\frac{1+2\operatorname{Re}(z\zeta^{-(j-1)})}{5},
\qquad 1\leq j\leq5 .
\]
The complex number $z$ describes the pattern of readings after their common scale has been removed.
Each condition $p_j\ge0$ restricts $z$ to a half-plane whose boundary is at distance $1/2$ from the origin.
These five half-planes meet in a regular pentagon with circumradius $(1/2)/\cos(\pi/5)=1/\varphi$.
At the vertex $\zeta^{j-1}/\varphi$, the two readings with indices $j\pm2$ vanish, with indices taken modulo five.
The normalized readings from a unit mass at the physical crossing $v_j$ give precisely this point in the plane of readings,
\[
\zeta^{j-1}
\frac{1+2\cos(2\pi/5)/\varphi}{1+2/\varphi}
=\frac{\zeta^{j-1}}{\varphi}.
\]
The inverse Fourier formula gives an explicit construction of masses for every allowed $z$.
Write $Z_j=\zeta^{j-1}/\varphi$ for $1\leq j\leq5$, and split the pentagon in the $z$ plane into the three triangles $(Z_1,Z_k,Z_{k+1})$ for $k=2,3,4$.
In the triangle containing $z$, solve $z-Z_1=\omega_k(Z_k-Z_1)+\omega_{k+1}(Z_{k+1}-Z_1)$ for its two real coefficients.
For complex numbers $u,v$, put $[u,v]=\operatorname{Im}(\overline u v)$ and let $D_k=[Z_k-Z_1,Z_{k+1}-Z_1]>0$.
Taking the oriented area with each side gives
\[
\omega_k=\frac{[z-Z_1,Z_{k+1}-Z_1]}{D_k},\qquad
\omega_{k+1}=\frac{[Z_k-Z_1,z-Z_1]}{D_k}.
\]
The containing triangles are exactly those for which $\omega_k,\omega_{k+1}\geq0$ and $\omega_k+\omega_{k+1}\leq1$.
With $E_{\min}=s_r/(4\pi\sqrt5\,M)$, assign masses $E_{\min}(1-\omega_k-\omega_{k+1})$, $E_{\min}\omega_k$, and $E_{\min}\omega_{k+1}$ to the physical crossings $v_1,v_k,v_{k+1}$, respectively.
Their normalized readings have first harmonic $z$ and sum to one, so the inverse Fourier formula recovers every $p_j$, including zeros on the boundary of the feasible pentagon. \\

\noindent
If all five readings vanish, the zero measure realizes them.
For a smooth nonnegative density, these data imply $E_P=0$.
Indeed, triangulate the physical pentagon using its ideal vertices.
The interior of each triangle lies inside a measured quadrilateral obtained by adjoining a fourth vertex, so that at least one $W_j$ is strictly positive there.
Only finitely many edges are omitted, a set of zero hyperbolic area.
It follows from $\int_{P_0}(\sum_{j=1}^5 W_j)e\,d\mu=0$ that $e=0$ almost everywhere in the polygon. \\

\noindent
It remains to replace the optimal point masses with smooth profiles while keeping each of the five readings fixed.
Each crossing $v_j$ borders an outer triangle with two ideal vertices and no other crossing in its interior.
Only $v_j$ has a nonzero column of weights there, so every point $y$ in the triangle gives the same five relative responses, multiplied by the common factor $t_j(y)$.
This factor tends to one at $v_j$ and to zero on the open outer side, allowing us to adjust the total energy at fixed readings.
Given any $0<\tau<1$, choose two real smooth bumps $f_{j,-}$ and $f_{j,+}$ with disjoint compact supports inside this triangle, one where $t_j<\tau$ and the other where $t_j>\tau$.
Normalize both by $\int f_{j,\pm}^2\,d\mu=1$ and set $\tau_{j,\pm}=\int t_j f_{j,\pm}^2\,d\mu$, with both integrals over the triangle.
Then $\theta_j=(\tau-\tau_{j,-})/(\tau_{j,+}-\tau_{j,-})$ lies strictly between zero and one.
For crossing mass $m_j>0$, take the density
\[
e_j=\frac{m_j}{\tau}\big[(1-\theta_j)f_{j,-}^2+\theta_j f_{j,+}^2\big].
\]
The density has energy $m_j/\tau$ and $\int t_j e_j\,d\mu=m_j$, so its contribution to each reading is exactly $m_jW_i(v_j)$ for $1\le i\le5$.
This also preserves zero readings, including the two that vanish when the relative readings coincide with those of a single crossing (an extreme ray).
Omit zero masses and sum the profiles in the five disjoint triangle interiors.
The readings remain fixed, and the total energy becomes $E_{\min}/\tau$.
Varying $0<\tau<1$ realizes the entire range $(E_{\min},\infty)$ with smooth nonnegative densities for any feasible nonzero readings.
The lower endpoint requires concentration at the crossings, taking the state derivatives at $\lambda=0$ for each profile before narrowing it.
Thus, fixing all five quartet Hessians still leaves room for different individual interval Hessians and conserved global energies.

\section{Geodesic Focusing and Interval Entropy}
\label{app:geodesicfocus}

\noindent
Geodesic length determines interval entropy, but how does that length respond when the endpoints move?
Two geodesics launched in slightly different directions have a separation set by spatial curvature.
To quadratic order, the entropy response measures the integrated curvature excess along their limiting vacuum path.
At finite amplitude, we obtain bounds on its integral along the geodesic of the changed geometry.
Appendix~\ref{app:quartet-resolution} gives the derivation and precision of the four-endpoint measurement before the chord is reached.

\subsection{Geodesic Deviation and Entropy Source}
\label{app:detailC8}

\noindent
Launch two geodesics from a common point at a small relative angle $d\varphi$, and write their transverse separation at distance $t$ in units of $L$ as $j(t)d\varphi$ to first order in the angle.
In vacuum, $j(t)=\sinh t$.
For a cutoff geodesic of dimensionless length $\ell$, a small curvature excess $\kappa$ changes the spreading by
\begin{equation}
1-\frac{j(\ell)}{\sinh\ell}=
\int_0^\ell\!\frac{\sinh t\,\sinh(\ell-t)}{\sinh\ell}\,\kappa(t)\,dt
+O(\|\kappa\|_1^2).
\label{eq:focuskernel}
\end{equation}
The kernel is positive, so positive curvature excess reduces the separation, and contributions of opposite sign can cancel.
The total magnitude $\|\kappa\|_1=\int_0^\ell|\kappa(t)|dt$ controls the remainder uniformly in the cutoff. \\

\noindent
The separation per unit angle is the Jacobi field $jN$, where $N$ is a parallel unit normal along the geodesic.
In units of $L$, the physical Gaussian curvature $\mathcal K$ contributes $L^2\mathcal K j=(-1+\kappa)j$ to the equation of geodesic deviation, giving
\begin{equation}
j''=(1-\kappa)j,\qquad j(0)=0,\qquad j'(0)=1 .
\label{eq:jacobi}
\end{equation}
On the reflection slice, the constraint gives $\kappa=8\pi G_N L^2\rho_\lambda$.
For example, if $0<\kappa<1$ is constant from launch to distance $t$, the solution is $j(t)=\sinh(\sqrt{1-\kappa}\,t)/\sqrt{1-\kappa}$.
Expanding at launch gives $j(t)=t+(1-\kappa)t^3/6+O(t^5)$, compared with $t+t^3/6+O(t^5)$ in vacuum.
In this example, the first difference from vacuum is cubic in the distance traveled. \\

\noindent
The reflection slice has dimensionless metric $h_\lambda/L^2$.
We denote the classical interval entropy by $S_\lambda=S_I(\lambda)$, its vacuum value by $S_0$, and its derivatives by $S_j=\partial_\lambda^j S_\lambda|_0$ for $j=1,2$.
For fixed endpoints $u<v<u+2\pi$, the distance $d_\lambda$ is the length of the shortest geodesic from $(\epsilon,u)$ to $(\epsilon,v)$ on a common radial cutoff surface.
We follow one smooth minimizing branch without conjugate points, meaning that no nonzero Jacobi field vanishes at both ends.
In the unit-circle boundary metric, $d_\lambda+2\ln\epsilon$ has a finite limit because the divergent term is independent of state and endpoints.
Assume this limit also exists after the mixed endpoint derivative and through two state derivatives.
The classical entropy relation gives $d_\lambda-d_0\to 6(S_\lambda-S_0)/c$ and $\partial_u\partial_v d_\lambda\to (6/c)\partial_u\partial_v S_\lambda$. \\

\noindent
To obtain the vacuum normalization at the same cutoff, let $r>0$ be its vacuum radial distance from the center in units of $L$, and put $a_r=\sinh r$ and $\alpha=(v-u)/2\in(0,\pi)$.
The hyperbolic law of cosines gives $\cosh d_0=\cosh^2 r-\sinh^2 r\cos(2\alpha)=1+2a_r^2\sin^2\alpha$, so $\sinh(d_0/2)=a_r\sin\alpha$.
At fixed $r$, differentiation gives $\partial_\alpha d_0=2a_r\cos\alpha/\sqrt{1+a_r^2\sin^2\alpha}$.
Since $d_0$ depends on the angles through $\alpha$ alone, its mixed derivative is
\begin{equation}
\partial_u\partial_v d_0=-\tfrac14\partial_\alpha^2d_0
=\frac{a_r(1+a_r^2)\sin\alpha}
{2(1+a_r^2\sin^2\alpha)^{3/2}} .
\label{eq:focusing_vacuum_mixed_distance}
\end{equation}
For fixed $\alpha$, the limit $r\to+\infty$ gives $1/(2\sin^2\alpha)$ and hence $\partial_u\partial_v S_0=c/(12\sin^2\alpha)$ by the entropy prescription. \\

\noindent
Orient the geodesic from $u$ to $v$, keep $u$ fixed, and slide $v$ along the cutoff circle.
At $a\in\{u,v\}$, write $U_a$ for the cutoff tangent in the direction of increasing angle, $T_a$ for the unit geodesic tangent, and $b_a=\langle U_a,N_a\rangle$ for the transverse component in $h_\lambda/L^2$.
A unit increase of $v$ displaces the far end sideways by $b_v$, while the near end stays fixed.
By Eq.~\eqref{eq:jacobi}, the transverse displacement at distance $s$ is $b_v j(s)/j(d_\lambda)$, since this solution has the prescribed values at both ends and the absence of conjugate points ensures a nonzero denominator.
Its initial slope $b_v/j(d_\lambda)$ is the rotation of the tangent at $u$ toward $N_u$.
The first variation of length gives $\partial_u d_\lambda=-\langle U_u,T_u\rangle$, where only $T_u$ rotates when $v$ moves, giving
\begin{equation}
\partial_u\partial_v d_\lambda=-\frac{b_u b_v}{j(d_\lambda)}.
\label{eq:focusing_mixed_distance}
\end{equation}
Smaller $j(d_\lambda)$ requires a larger rotation at $u$ to reach the same displaced endpoint at $v$.
To evaluate the endpoint factors, decompose the tangent to the cutoff into components along and across the geodesic, so $b_a^2=|U_a|^2-\langle U_a,T_a\rangle^2$.
In the common unit-circle boundary metric, $|U_a|=\epsilon^{-1}[1+o(1)]$ as $\epsilon\downarrow0$.
The geodesic meets the cutoff asymptotically orthogonally, so $\langle U_a,T_a\rangle/|U_a|\to0$.
It enters at $u$ and exits at $v$, so its consistently oriented normal has opposite projections onto the increasing-angle tangents there.
Consequently, $-b_u b_v=\epsilon^{-2}[1+o(1)]$, with the same leading coefficient at fixed state and in the vacuum. \\

\noindent
For the ratio to the vacuum spreading, $\mathfrak j(s)=j(s)/\sinh s$, variation of constants gives
\begin{align}
\mathfrak j(s)&=1-\int_0^s B_s(t)\kappa_\lambda(t)\mathfrak j(t)\,dt,\nonumber\\
B_s(t)&=\frac{\sinh(s-t)\sinh t}{\sinh s}
\leq\tfrac12\quad(0<t<s).
\end{align}
The bound follows from $B_s(t)=[\coth t+\coth(s-t)]^{-1}$, with $\coth x=\cosh x/\sinh x$ for real $x\ne0$.
We assume $\int_0^{d_\lambda}|\kappa_\lambda|ds=O(\lambda^2)$ uniformly in the cutoff.
In the $n$th iteration, for $n\ge1$, the integration points occur in order along the path.
Bounding the kernels by $1/2$ and using one of the $n!$ equivalent orderings gives $(\|\kappa_\lambda\|_1/2)^n/n!$.
Summing gives $\sup_{0\le s\le d_\lambda}|\mathfrak j(s)-1|\le\exp(\|\kappa_\lambda\|_1/2)-1=O(\lambda^2)$.
The integral already contains one curvature factor of order $O(\lambda^2)$ in $L^1$, so replacing $\mathfrak j$ by one costs $O(\lambda^4)$ uniformly in the cutoff.
As the endpoints recede, the kernel $B_{d_\lambda}$ tends to $1/2$ at each fixed interior point.
To compare geodesics of different lengths, place the arclength origin at an interior point in a fixed compact bulk region and extend each cutoff profile by zero beyond its ends.
As $\lambda\to0$ and $\epsilon\downarrow0$, assume that $\kappa_\lambda/\lambda^2$ converges in $L^1$ along this coordinate to $(8\pi G_N/L)e$ on the complete vacuum geodesic.
Local $L^1$ convergence suffices if the integral of the absolute rescaled curvature in both tails is uniformly small far from the origin.
Since $\exp(d_\lambda)/\sinh d_\lambda\to2$, the endpoint and deviation formulas give the finite residual
\begin{align}
\mathcal A_I(\lambda)
&\equiv \exp[6(S_\lambda-S_0)/c]
\frac{\partial_u\partial_v S_\lambda}{\partial_u\partial_v S_0}
=\lim_{\epsilon\downarrow0}\frac{\exp(d_\lambda)\partial_u\partial_v d_\lambda}
{\exp(d_0)\partial_u\partial_v d_0}
\nonumber\\
&=1+\frac{4\pi G_N}{L}\lambda^2\int_{\gamma_I}e\,d\sigma
+o(\lambda^2).
\end{align}
Here, $e=L^3\rho_2$, where $\rho_\lambda=\lambda^2\rho_2+o(\lambda^2)$, and $d\sigma$ is vacuum arclength divided by $L$.
In particular, $(\ln \mathcal A_I)'(0)=0$, so $\partial_u\partial_v S_1/\partial_u\partial_v S_0=-6S_1/c$.
To take the next derivative, write $S_{j,uv}=\partial_u\partial_v S_j$ for $j=0,1,2$.
At fixed endpoints, the logarithm in Eq.~\eqref{eq:ratio} is the sum of a term linear in $S_\lambda-S_0$ and the logarithm of $S_{\lambda,uv}/S_{0,uv}$, where $S_{\lambda,uv}=\partial_u\partial_v S_\lambda$.
Its second derivative at the vacuum is
\begin{equation}
(\ln\mathcal A_I)''(0)=\frac{6}{c}S_2+
\frac{S_{2,uv}}{S_{0,uv}}-
\left(\frac{S_{1,uv}}{S_{0,uv}}\right)^2 .
\label{eq:focusing_log_entropy_hessian}
\end{equation}
Substituting $S_{1,uv}/S_{0,uv}=-6S_1/c$ and the vacuum normalization from Eq.~\eqref{eq:focusing_vacuum_mixed_distance} gives
\begin{equation}
(\ln \mathcal A_I)''(0)=\frac3c\underbrace{\left[-\mathcal D S_2-
\frac{12}{c}S_1^2\right]}_{\displaystyle\mathcal Q_I}
=\frac{8\pi G_N}{L}\int_{\gamma_I} e\,d\sigma ,
\end{equation}
where $\mathcal D=-4\sin^2\alpha\,\partial_u\partial_v-2$.
Thus, $c=3L/(2G_N)$ gives $\mathcal Q_I=4\pi\int_{\gamma_I}e\,d\sigma$ directly from the focusing of spatial geodesics. \\

\noindent
The shrinking quartet in Eq.~\eqref{eq:ptolemy_focus_limit} gives the boundary measurement at fixed state, and Eq.~\eqref{eq:finite_quartet_response} gives its state Hessian.
The geodesic-deviation calculation supplies the physical meaning of that Hessian. It identifies which path probes the matter when the perturbation has finite amplitude.

\subsection{Finite-Amplitude Bounds along Perturbed Geodesics}
\label{app:finitefocus}

\noindent
At finite amplitude, follow the same smooth minimizing geodesic branch as the endpoints of a proper interval $I=(u,v)$, $u<v<u+2\pi$, vary on the reflection slice.
Its length gives the classical entropy $S$ after division by $4G_N$, using the same boundary metric and cutoff as in vacuum.
Assume the cutoff geodesics have no conjugate points and their renormalized length and mixed endpoint derivative converge.
We compare the entropy observable with the integrated curvature along this geodesic,
\begin{align}
\mathcal A_I&=\exp[6(S-S_0)/c]\,
\frac{\partial_u\partial_v S}{\partial_u\partial_v S_0},
&\tau_I&=\frac12\int_\gamma\kappa\,d\sigma,
\label{eq:focusing_definitions}
\end{align}
where $d\sigma=d\ell/L$ is arclength on the \emph{actual} geodesic in units of $L$ and $\kappa=1+L^2\mathcal K=8\pi G_NL^2\rho$ is the excess of its ambient Gaussian curvature $\mathcal K$ above the vacuum value, in dimensionless units.
We require $0<\mathcal A_I<\infty$.
At fixed state, center the cutoff curvature profiles at an interior point, extend them by zero past their endpoints, and assume they converge in $L^1(\mathbb R)$ to the profile on $\gamma$.
At finite amplitude, each successive stretch of curvature acts on a separation already altered upstream.
This dependence on the earlier path gives the following bounds, with the lower bound requiring $\kappa\ge0$ along the cutoff geodesics.
\begin{equation}
\tau_I\le\ln\mathcal A_I\quad\text{for signed }\kappa,\qquad
1-\mathcal A_I^{-1}\le\tau_I\quad\text{if }\kappa\ge0 .
\label{eq:focusing_bracket}
\end{equation}
For example, with nonnegative curvature excess, a reading $\mathcal A_I=2$ puts the half-integral $\tau_I$ between $1/2$ and $\ln 2$.
If the excess can have either sign, only the upper bound survives.
A reading below one still implies a negative integral, whereas a reading above one leaves its sign open. \\

\noindent
On a cutoff geodesic of length $d$, the arclength $s\in[0,d]$ starts at its initial end.
Its transverse Jacobi field $j$ satisfies $j''=(1-\kappa)j$, $j(0)=0$, $j'(0)=1$.
The absence of conjugate points gives $j(s)>0$ for $s>0$.
The ratio $\mathfrak j=j/\sinh s$ has continuous value $\mathfrak j(0)=1$.
Variation of constants gives
\begin{equation}
1-\mathfrak j(d)=\int_0^d B_d(t)\kappa(t)\mathfrak j(t)\,dt,
\qquad B_d(t)=\frac{\sinh t\,\sinh(d-t)}{\sinh d}.
\end{equation}
If $\kappa\ge0$, positivity of $j$ and of the integral kernel implies $0<\mathfrak j(s)\le1$ at every $s$, hence $1-\mathfrak j(d)\le\int_0^d B_d\kappa\,dt$.
For either sign, let $q=-(\ln\mathfrak j)'$ measure the fractional loss of separation per unit distance relative to vacuum.
Inserting $j=\sinh(s)\mathfrak j$ in the Jacobi equation gives $\mathfrak j''+2\coth(s)\mathfrak j'=-\kappa\mathfrak j$, or $q'+2\coth(s)q=\kappa+q^2$, with $q=O(s)$ at launch.
Multiplying by $\sinh^2s$ makes the left side a total derivative, so two integrations give
\begin{equation}
-\ln \mathfrak j(d)=\int_0^d B_d(t)\,[\kappa(t)+q(t)^2]\,dt
\ \ge\ \int_0^d B_d(t)\kappa(t)\,dt .
\end{equation}
Indeed, $\sinh^2t\int_t^d ds/\sinh^2s=B_d(t)$.
To translate spreading back to entropy, use the mixed distance derivative in Eq.~\eqref{eq:focusing_mixed_distance}.
Only the transverse endpoint projections $b_u,b_v$ enter, and their product has the same leading cutoff behavior as the vacuum product $b_{0,u}b_{0,v}$.
With $d_0$ the vacuum distance between the same endpoints, substituting $j(d)=\sinh(d)\mathfrak j(d)$ and $j_0(d_0)=\sinh d_0$ gives
\begin{equation}
\exp(d-d_0)\frac{\partial_u\partial_v d}{\partial_u\partial_v d_0}
=\frac{b_u b_v}{b_{0,u}b_{0,v}}\,
\frac{1-\exp(-2d_0)}{1-\exp(-2d)}\,\frac{1}{\mathfrak j(d)} .
\label{eq:cutoff_endpoint_normalization}
\end{equation}
As $\epsilon\downarrow0$, the two distances diverge, while their difference tends to $6(S-S_0)/c$ and the ratio of mixed derivatives tends to $S_{uv}/S_{0,uv}$.
Thus, the left side tends to $\mathcal A_I$ in Eq.~\eqref{eq:focusing_definitions}, and both factors preceding $1/\mathfrak j(d)$ on the right tend to one.
Finally, $0\le B_d\le1/2$ and $B_d\to1/2$ at each fixed interior point, so the assumed convergence gives $\int_0^d B_d\kappa\,dt\to\tau_I$.
Discarding the nonnegative $q^2$ term before taking the limit proves the upper bound for either curvature sign.
For continuous nonnegative density, $\mathcal A_I\ge1$, with equality precisely when $\rho$ vanishes everywhere on this geodesic.
A vacuum reading on one geodesic still allows matter elsewhere on the slice.
These bounds use the standard Jacobi equation \cite{Wald:1984rg} projected onto the unit normal in two dimensions.

\subsubsection{Exact Focusing Remainder for Signed Curvature}
\label{app:detailC9a}

\noindent
Measure dimensionless arclength $x\in\mathbb R$ from a fixed interior point of the oriented geodesic.
As the cutoff endpoints recede, the stated $L^1$ convergence gives a limit on each compact $x$ interval for $\mathfrak j=j/\sinh s$,
\begin{equation}
\mathfrak j(x)=1-\frac12\int_{-\infty}^{x}
[1-\exp[-2(x-t)]]\kappa(t)\mathfrak j(t)\,dt .
\label{eq:app_completefocus}
\end{equation}
At a fixed interior point $x$, the cutoff kernel approaches $[1-\exp(-2(x-t))]/2$ for $t<x$.
In iteration $n\ge0$, each kernel is bounded by $1/2$ and the $n$ curvature factors are ordered, giving an absolute bound $(\|\kappa\|_1/2)^n/n!$, with $\|\kappa\|_1=\int_{\mathbb R}|\kappa|\,dx$.
Strong $L^1$ convergence bounds the cutoff norms uniformly and gives convergence of each fixed product in $L^1(\mathbb R^n)$.
One can therefore pass to the kernel limit term by term, and the factorial bound controls the tail in Eq.~\eqref{eq:app_completefocus}.
At the receding final endpoint, the kernel instead approaches $1/2$, giving $\mathfrak j(+\infty)=1-\tfrac12\int_{\mathbb R}\kappa\mathfrak j\,dx=\mathcal A_I^{-1}$.
Thus, $j_{\rm in}(x)=\exp(x)\mathfrak j(x)$ solves $j_{\rm in}''=(1-\kappa)j_{\rm in}$, and $\exp(-x)j_{\rm in}(x)$ tends to one at the initial vertex and to $\mathcal A_I^{-1}$ at the final vertex. \\

\noindent
The bounded solution $\mathfrak j$ tends to one as $x\to-\infty$ and is nonnegative as a limit of positive cutoff solutions.
At any interior zero, nonnegativity would require $\mathfrak j'=0$.
Uniqueness in $\mathfrak j''+2\mathfrak j'=-\kappa \mathfrak j$ would then force $\mathfrak j\equiv0$.
Thus, $\mathfrak j>0$ and, because both endpoint limits are positive, $\mathfrak j$ is bounded away from zero.
Differentiation gives
\begin{equation}
\mathfrak j'(x)=-\int_{-\infty}^{x} \exp[-2(x-t)]\kappa(t)\mathfrak j(t)\,dt .
\end{equation}
Set $F=\kappa\mathfrak j\in L^1(\mathbb R)$, using boundedness of $\mathfrak j$.
The derivative formula gives $|\mathfrak j'(x)|\le\int_{-\infty}^x|F(t)|\,dt$, which tends to zero at the initial end.
Reversing the order of integration gives $\|\mathfrak j'\|_1\le\|F\|_1/2$.
At the final end, split at $a<x$ to obtain $|\mathfrak j'(x)|\le\exp[-2(x-a)]\|F\|_1+\int_a^\infty|F(t)|\,dt$, and send $x$ then $a$ to infinity.
Since $\mathfrak j$ is bounded away from zero, $q=-\mathfrak j'/\mathfrak j$ is bounded and integrable, $q^2$ is integrable, and $q$ vanishes at both ends.
On $[a,b]$, $a<b$, integrating its almost-everywhere equation $q'+2q=\kappa+q^2$ using absolute continuity gives
\[
\ln\frac{\mathfrak j(a)}{\mathfrak j(b)}-\frac12\int_a^b\kappa(x)\,dx
=\frac12\int_a^b q(x)^2\,dx-\frac{q(b)-q(a)}{2}.
\]
As $a\to-\infty$ and $b\to+\infty$, the last endpoint term vanishes, while $\mathfrak j(a)\to1$ and $\mathfrak j(b)\to\mathcal A_I^{-1}$, giving
\begin{equation}
\ln\mathcal A_I-\tau_I=\frac12\int_{\mathbb R}q(x)^2\,dx\ge0 .
\label{eq:app_focus_deficit}
\end{equation}
Changes in the fractional spreading rate $q$ contribute positively through $q^2$ whether they speed up or slow down the separation.
Equality forces $q=0$ everywhere by continuity, and its equation then gives $\kappa=0$ almost everywhere.
Conversely, this curvature profile gives $\mathfrak j=\mathcal A_I=1$ by Eq.~\eqref{eq:app_completefocus}.
Thus, a nonzero signed profile with vanishing curvature integral still gives $\mathcal A_I>1$.
For the lower comparison, assume $\kappa\ge0$ almost everywhere, so Eq.~\eqref{eq:app_completefocus} gives $0<\mathfrak j\le1$ and $1-\mathcal A_I^{-1}=\tfrac12\int_{\mathbb R}\kappa\mathfrak j\,dx$.
Subtracting this last expression from $\tau_I$ and substituting the same integral equation at $x$ gives
\[
\begin{aligned}
\tau_I-(1-\mathcal A_I^{-1})
&=\frac12\int_{\mathbb R}\kappa(x)[1-\mathfrak j(x)]\,dx\\
&=\frac14\int_{\mathbb R} dx\int_{-\infty}^{x}dt\,
[1-\exp(-2(x-t))]\,\kappa(x)\kappa(t)\mathfrak j(t).
\end{aligned}
\]
For a nonzero nonnegative profile, the set where $\kappa>0$ has positive measure, and the kernel is positive on the ordered half of the pairs from this set.
The difference is therefore strictly positive.
Bounding the kernel and $\mathfrak j$ by one gives the upper estimate $(\int_{\mathbb R}\kappa\,dx)^2/8=\tau_I^2/2$, so the lower bound agrees with $\tau_I$ to first order. \\

\noindent
If $m=\|\kappa\|_1$ is small, Eq.~\eqref{eq:app_completefocus} gives $\mathfrak j=1+O(m)$ uniformly.
Its linear spreading rate is $q_1(x)=\int_{-\infty}^x\exp[-2(x-t)]\kappa(t)\,dt$.
The one-sided exponential has $L^2$ norm $1/2$, so its convolution with $|\kappa|$ has norm at most $m/2$, where $\|f\|_2^2=\int_{\mathbb R}|f|^2dx$.
The exact rate contains the additional factor $\mathfrak j(t)/\mathfrak j(x)=1+O(m)$ in this integral, giving $\|q-q_1\|_2=O(m^2)$.
When $q_1$ is squared and integrated, the product of its two kernels integrates to $\exp(-2|s-t|)/4$ over $x\ge\max(s,t)$.
This gives the first nonlinear correction,
\begin{equation}
\ln\mathcal A_I-\tau_I
=\frac18\int_{\mathbb R^2}
\exp(-2|s-t|)\kappa(s)\kappa(t)\,ds\,dt+O(m^3).
\label{eq:app_chordquadraticcorrection}
\end{equation}
Although the integrand can have either sign for a signed profile, its double integral has a definite sign.
With Fourier convention $\widehat\kappa(\xi)=\int_{\mathbb R}\kappa(x)\exp(-\mathrm i\xi x)\,dx$ for real $\xi$, integration over the two half-lines gives
\[
\int_{\mathbb R}\exp(-2|x|)\exp(-\mathrm i\xi x)\,dx
=\frac{1}{2+\mathrm i\xi}+\frac{1}{2-\mathrm i\xi}
=\frac{4}{4+\xi^2}.
\]
This multiplier is integrable, and $\kappa$ is real and absolutely integrable, so Fourier inversion of the kernel and absolute convergence give
\[
\frac18\int_{\mathbb R^2}\exp(-2|s-t|)\kappa(s)\kappa(t)\,ds\,dt
=\frac1{4\pi}\int_{\mathbb R}\frac{|\widehat\kappa(\xi)|^2}{4+\xi^2}\,d\xi .
\]
The integral on the right is strictly positive if the continuous transform $\widehat\kappa$ is nonzero on an interval of frequencies.
Consequently, a vanishing integral forces $\widehat\kappa=0$ everywhere and Fourier uniqueness then gives $\kappa=0$ almost everywhere. \\

\noindent
For a scalar family with $\kappa_\lambda/\lambda^2\to(8\pi G_N/L)e$ in $L^1$ along the vacuum geodesic, the Fourier transforms of the rescaled curvatures converge uniformly.
The integrable multiplier therefore gives
\begin{equation}
\lim_{\lambda\to0}\frac{\ln\mathcal A_I(\lambda)-\tau_I(\lambda)}{\lambda^4}
=\frac{1}{4\pi}\left(\frac{8\pi G_N}{L}\right)^2
\int_{\mathbb R}\frac{|\widehat e(\xi)|^2}{4+\xi^2}\,d\xi .
\label{eq:fourth-order-focusing}
\end{equation}
Here, $\widehat e$ is the Fourier transform along the chord in dimensionless arclength.
The quadratic response measures only the integral $\widehat e(0)$, which can vanish for a signed density.
The fourth-order difference in Eq.~\eqref{eq:fourth-order-focusing} is positive for any density nonzero on a set of positive arclength, since all frequencies carry positive weight.
It compares the entropy reading with the actual $\tau_I(\lambda)$, including higher-order changes in path and matter.

\subsection{Endpoint Limit and Finite-Width Errors}
\label{app:quartet-resolution}

\noindent
For the narrow quartets of Sec.~\ref{sec:endpoint-limit}, one can reach the chord observable by taking the endpoint limit in a given state or, under the stated regularity assumptions, by taking the state derivatives at finite width first.
We derive both forms and estimate the effect of measurement error when the endpoint pairs have finite separation. \\

\noindent
Take the cutoff $\epsilon$ to zero while the four endpoints are distinct, so that the regulator resolves both short arcs before they shrink.
At fixed endpoints, the finite quantity
\[
Y(a,b)=\lim_{\epsilon\downarrow0}\epsilon D(a,b)
=2\sin[(b-a)/2]\,\exp\!\left[3(S(a,b)-S_0(a,b))/c\right]
\]
has the normalization $\lim_{\ell\downarrow0}Y(a,a+\ell)/\ell=1$ provided the vacuum-subtracted entropy tends to zero locally uniformly for a short arc.
Number the endpoints in the displayed cyclic order by $x_1=u$, $x_2=u+h$, $x_3=v$, and $x_4=v+k$.
With $Y_{ij}=Y(x_i,x_j)$ for $1\le i<j\le4$, the entropy difference gives $Y_{13}Y_{24}/(Y_{14}Y_{23})=\exp(3I_c/c)$.
The two short chords give $hk/(Y_{12}Y_{34})\to1$, and the two long chords both tend to $Y(u,v)$.
Since $I_c=O(hk)$, expanding the exponential leaves $3I_c/(c hk)$ at leading order in the last factor of
\begin{equation}
\lim_{h,k\downarrow0}\Pi
=\lim_{h,k\downarrow0}
\left[Y_{14}Y_{23}\,\frac{hk}{Y_{12}Y_{34}}\,
\frac{\exp(3I_c/c)-1}{hk}\right]
=\frac{3}{c}Y(u,v)^2 S_{uv} .
\label{eq:ptolemy_focus_limit}
\end{equation}
Using the vacuum sine law in this expression gives Eq.~\eqref{eq:ratio}. \\

\noindent
There is a direct way to see why precisely the mixed derivative survives.
The two products of long chords agree if either pair of endpoints coincides, so integrating their mixed derivative over the rectangle recovers their difference.
Here, subscripts denote endpoint derivatives on the fixed-state branch,
\[
\begin{aligned}
Y_{13}Y_{24}-Y_{14}Y_{23}
={}&\int_0^h da\int_0^k db\,
\bigl[Y(u,v)Y_{uv}(u+a,v+b)\\
&\hspace{82pt}{}-Y_u(u+a,v)Y_v(u,v+b)\bigr].
\end{aligned}
\]
Dividing by $hk$ and taking the two widths independently to zero gives $Y Y_{uv}-Y_uY_v=Y^2\partial_u\partial_v\ln Y$, with all factors evaluated at $(u,v)$.
In the common cutoff convention, $\partial_u\partial_v\ln Y=3S_{uv}/c$, as follows from the definition of $Y$ and the vacuum sine law. \\

\noindent
A smooth local change of cutoff replaces $Y(a,b)$ by $q(a)q(b)Y(a,b)$ with $q>0$.
The mixed derivative of $\ln Y$ is unchanged because it annihilates $\ln q(u)+\ln q(v)$.
The two short chords cancel the remaining factor $q(u)^2q(v)^2$ from the numerator of Eq.~\eqref{eq:ptolemy_focus_limit}. \\

\noindent
For a smooth orientation-preserving vacuum reparametrization $f$ common to both chiral sectors, let $\delta_f=[f(v)-f(u)]/2$.
The two factors of Eq.~\eqref{eq:ratio} in this family are
\[
\frac{\sin^2\delta_f}{\sin^2\alpha\,f'(u)f'(v)}
\quad\text{and}\quad
\frac{f'(u)f'(v)\sin^2\alpha}{\sin^2\delta_f},
\]
as follows by taking the mixed derivative of its interval entropy \cite{Calabrese:2004eu,Beach:2016ocq,deBoer:2016pqk}.
A reparametrization changes apparent chord length and sensitivity to endpoint displacements by reciprocal factors.
Their product is one for every proper arc and every amplitude on this vacuum branch. \\

\noindent
For $j=1,2$, let $A_j,B_j,C_j$ sum $S_j=\partial_\lambda^j S|_0$ over the pairs $(13,24)$, $(14,23)$, $(12,34)$, respectively.
Use $A,B,C$ for the undifferentiated sums and $\Delta$ for subtraction of their vacuum values.
The vacuum Ptolemy identity gives
\[
\Pi=\eta_Q^{-1}
\big\{\exp[(3/c)(\Delta A-\Delta C)]
-(1-\eta_Q)\exp[(3/c)(\Delta B-\Delta C)]\big\}.
\]
For a vacuum descendant, $\Pi$ remains one even though its individual interval entropies vary.
The measured state has the same first-order entropies as the matched descendant, so $\Pi'(0)=0$ in the measured state as well.
Differentiating the two exponentials once therefore gives $A_1-C_1=(1-\eta_Q)(B_1-C_1)$.
Differentiating again produces a difference of squares, $(A_1-C_1)^2-(1-\eta_Q)(B_1-C_1)^2=-\eta_Q(1-\eta_Q)(B_1-C_1)^2$.
Writing $R$ for the logarithmic defect defined in Eq.~\eqref{eq:bpi}, one finds
\begin{equation}
\frac{R''(0)}{\eta_Q}
=\frac c3\Pi''(0)
=\frac{A_2-B_2}{\eta_Q}+B_2-C_2
-\frac3c(1-\eta_Q)(B_1-C_1)^2 .
\label{eq:finite_quartet_response}
\end{equation}
The prefactor follows as well by expanding the exact identity $\exp(3R/c)-1=[D_{12}D_{34}/(D_{14}D_{23}+D_{12}D_{34})](\Pi-1)$.
The bracket has vacuum value $\eta_Q$, while $\Pi-1$ starts at second order. \\

\noindent
The vacuum sine formula gives $\eta_Q/(hk)\to1/(4\sin^2\alpha)$, and Eq.~\eqref{eq:rectangular_entropy_average} gives $(A_2-B_2)/(hk)\to S_{2,uv}$.
Thus, the first term on the right tends to $4\sin^2\alpha\,S_{2,uv}$.
For the other two terms, assume that $S_1,S_2,S_{2,uv}$ are continuous near $(u,v)$ and that each $S_j$ tends uniformly to zero on short arcs near $u$ and $v$ for $j=1,2$.
Each $B_j$ then tends to twice the response of the surviving interval, $2S_j(u,v)$, and each $C_j$ tends to zero.
To estimate how quickly this happens, center the two pairs at $u$ and $v$ by setting $x_{1,2}=u\mp h$, $x_{3,4}=v\mp h$, with $0<h<\min(\alpha,\pi-\alpha)$.
Their cross ratio is $\eta_Q=\sin^2h/\sin^2\alpha$, their diagonals cross at $(\cos\alpha/\cos h)n_\theta$ in the Klein disk, and the two sums become
\[
A_j=S_j(\theta-h,\alpha)+S_j(\theta+h,\alpha),\qquad
B_j=S_j(\theta,\alpha-h)+S_j(\theta,\alpha+h),
\]
where the arguments of $S_j$ are now the center and half-opening.
The two intervals in $A_j$ slide left and right at fixed width, while those in $B_j$ expand and contract at fixed center.
For twice continuously differentiable data, each symmetric pair cancels its linear change and contributes $h^2$ times the corresponding second derivative above the common value $2S_j$.
Their difference divided by $h^2$ therefore tends to $(\partial_\theta^2-\partial_\alpha^2)S_j=4\partial_u\partial_v S_j$, since $\partial_u=(\partial_\theta-\partial_\alpha)/2$ and $\partial_v=(\partial_\theta+\partial_\alpha)/2$.
Keep the target $0<\alpha<\pi$ fixed, set $s=\sin\alpha$, and let $F_h=R''(0)/\eta_Q$ for this centered quartet.
Assume that $S_2$ has continuous endpoint derivatives through order four near the target, that $S_1$ has them through order two there, and that $S_j(w-h,w+h)=O(h^2)$ for $w=u,v$ and $j=1,2$.
Taylor expansion of the symmetric pairs, followed by $\sin^2 h=h^2+O(h^4)$, gives
\[
\begin{aligned}
A_2-B_2&=h^2(\partial_\theta^2-\partial_\alpha^2)S_2+O(h^4),\qquad
\eta_Q=\frac{h^2+O(h^4)}{s^2},\\
B_j-C_j&=2S_j+O(h^2),\qquad j=1,2.
\end{aligned}
\]
Pairing the displaced intervals cancels the odd derivatives, so the remainder $O(h^4)$ in $A_2-B_2$ contributes $O(h^2)$ after division by $\eta_Q$.
Consequently, Eq.~\eqref{eq:finite_quartet_response} gives $F_h=\mathcal Q_I+O(h^2)$ for the source in Eq.~\eqref{eq:entropysource}, with a constant depending on the target and the stated bounds. \\

\noindent
With noisy readings, reducing $h$ makes the geometric approximation more accurate but also divides a measured difference by a smaller cross ratio.
Suppose that on the six intervals the measured $S_j$ have absolute errors at most $\epsilon_j$ for $j=1,2$, and denote the resulting estimate of $F_h$ by $\widetilde F_h$.
In the linear part of Eq.~\eqref{eq:finite_quartet_response}, the absolute coefficients add to $2[1+(1-\eta_Q)+\eta_Q]/\eta_Q=4/\eta_Q$.
For the slope contrast $d_h=B_1-C_1$ and its measured value $\widetilde d_h$, the four slopes give $|\widetilde d_h-d_h|\leq4\epsilon_1$.
Consequently,
\[
|\widetilde F_h-F_h|\leq
\frac{4\epsilon_2}{\eta_Q}
+\frac{12}{c}(1-\eta_Q)\epsilon_1
(2|\widetilde d_h|+4\epsilon_1).
\]
For directly measured, bounded state derivatives and $0\le\epsilon_1\le1$, the geometric bias and reading errors at fixed target and central charge sum to $O(h^2+\epsilon_2/h^2+\epsilon_1)$.
Balancing the first two terms gives $h=\epsilon_2^{1/4}$ and total error $O(\sqrt{\epsilon_2}+\epsilon_1)$ for small $\epsilon_2>0$. \\

\noindent
For measurements of ordinary entropy $\widetilde S(J,\lambda)$, sample each of the six intervals $J$ on one branch at amplitudes $-t,0,t$, with $t>0$.
The error bound $s_E$ per reading includes any cutoff error relative to the continuum data at fixed endpoints (Sec.~\ref{sec:endpoint-limit}).
Assume $|\partial_\lambda^j S(J,\lambda)|\le K_j$ for $j=3,4$ and $|\lambda|\le t$, uniformly over the narrowing family.
The centered differences have coefficients $(-1,0,1)/(2t)$ and $(1,-2,1)/t^2$.
Each set sums to zero, eliminating common offsets such as the state-independent vacuum subtraction.
Their absolute coefficients sum to $1/t$ and $4/t^2$, respectively, giving the following error bounds after adding the Taylor remainders,
\[
\epsilon_1=\frac{s_E}{t}+\frac{K_3t^2}{6},\qquad
\epsilon_2=\frac{4s_E}{t^2}+\frac{K_4t^2}{12}.
\]
Denote by $\widetilde F_{h,t}$ the estimate from Eq.~\eqref{eq:finite_quartet_response} using these differences.
At fixed target interval, $1/\eta_Q=\sin^2\alpha/\sin^2 h=O(h^{-2})$, so combining the state errors with the finite-width error yields
\[
|\widetilde F_{h,t}-\mathcal Q_I|
=O\!\left(h^2+\frac{t^2+s_E/t^2}{h^2}
+t^2+\frac{s_E}{t}\right)
\]
whenever $h,t$ are sufficiently small and $\epsilon_1\le1$.
For reading error $s_E\to0^+$, choose the state step $t=s_E^{1/4}$ first, balancing truncation against Hessian noise.
Then choose the angular width $h=s_E^{1/8}$, balancing the resulting noise amplification against the bias from a finite quartet.
The absolute error is $O(s_E^{1/4})$ for $\mathcal Q_I$ and smaller by $4\pi$ for the chord integral in Eq.~\eqref{eq:chord_limit}. \\

\noindent
In Fig.~\ref{fig:bweight}(b), the successive rectangles show how the support of the quartet weight contracts onto a diameter in the chord limit.
Center its crossing in a Klein frame so that its half-widths satisfy $a_Q^2+b_Q^2=1$ and $b_Q^2=\eta_Q$.
Put $X=a_Qv_1$, $Y=b_Qv_2$, $d^2=1-X^2-Y^2$ and $C_*=1-\max(|v_1|,|v_2|)$, with $|v_1|,|v_2|<1$.
After division by $\eta_Q$, the measure for the logarithmic-defect kernel is $a_Q^2C_*\,dv_1dv_2/d^4$.
For $|e|\le C d^{2\nu}$ with $\nu>0$, which includes the assumed falloff with $\nu=\Delta>1$, the inequalities $C_*\le1-|v_1|$ and $d^2\ge a_Q^2(1-v_1^2)$ bound the integrand by $C'[1+(1-v_1^2)^{\nu-1}]$ for small $b_Q$.
This bound is integrable at both ends and independent of $b_Q$, so dominated convergence permits the limit inside the integral.
At fixed $v_1$, the transverse weight has a flat part and two linear tails, with integral $2|v_1|(1-|v_1|)+2\int_{|v_1|}^1(1-v_2)\,dv_2=1-v_1^2$.
Since $d^4\to(1-v_1^2)^2$, the remaining measure is $dv_1/(1-v_1^2)$, precisely the arclength along the diameter in units of $L$.
The diagonals of the centered quartet have half-opening $\alpha$ and centers $\theta-h$ and $\theta+h$.
With $n_\psi=(\cos\psi,\sin\psi)$, their crossing solves $y\cdot n_{\theta-h}=y\cdot n_{\theta+h}=\cos\alpha$.
Subtracting gives $y\cdot(-\sin\theta,\cos\theta)=0$, and adding gives $y=p_hn_\theta$ with $p_h=\cos\alpha/\cos h$.
For each proper arc, $p_h\to\cos\alpha$ lies strictly inside the disk.
In a common orientation, the boost $B_h$ taking the disk origin to that crossing obeys
\[
d(B_h y)=\frac{d(y)}{\Gamma_h(1+p_h n_\theta\cdot y)},
\qquad \Gamma_h=(1-p_h^2)^{-1/2}.
\]
The time component of the unit hyperboloid vector $(1,y)/d(y)$ becomes $\Gamma_h(1+p_h n_\theta\cdot y)/d(y)$.
Because $|p_h|$ stays bounded away from one, the denominator of $d(B_hy)$ is uniformly bounded above and below by positive constants.
The transformed density therefore satisfies the same falloff bound uniformly in $h$ and converges at each interior point.
The limiting boost maps the diameter to $\gamma_I$ without changing its arclength, yielding Eq.~\eqref{eq:chord_limit}.
For unequal widths, Eq.~\eqref{eq:finite_quartet_response} has the same rectangular limit by Eq.~\eqref{eq:rectangular_entropy_average}.

\section{Boundary Stress and Relative-Entropy Subtraction}
\label{app:gravity}

\noindent
A change in boundary stress shifts interval entropies even in a geometry containing no scalar matter.
To measure the scalar contribution, we first calculate the response of a pure-gravity state with the same first-order boundary stress.
For a reduced state $\varrho_I(\lambda)$ and its vacuum value $\sigma_I=\varrho_I(0)$, the relevant measures are the relative entropy and its second derivative,
\begin{equation}
S_{\rm rel}(\varrho\Vert\sigma)
=\tr[\varrho(\ln\varrho-\ln\sigma)],\qquad
\FF_I=\left.\partial_\lambda^2 S_{\rm rel}(\varrho_I(\lambda)\Vert\sigma_I)\right|_{\lambda=0}.
\end{equation}
Use the same ultraviolet regulator for both states.
With the vacuum modular Hamiltonian $K_I^{(0)}=-\ln\sigma_I$ held fixed, $S_{\rm rel}=\Delta\langle K_I^{(0)}\rangle-\Delta S_I$, where $S_I=-\tr(\varrho_I\ln\varrho_I)$ and $\Delta$ subtracts the vacuum value.
The entropy first law makes the linear terms cancel, leaving a quadratic coefficient determined by the first-order change of $\varrho_I$.
Each of the left- and right-moving sectors has central charge $c$, giving interval entropy coefficient $c/3$, and time reflection equates their profiles in $x+t$ and $x-t$.
The vacuum transformation laws and modular Hamiltonian determine the boundary-stress part of the Hessian \cite{Beach:2016ocq,Hollands:2019czd,Panebianco:2019plp,Casini:2011kv}.
Ref.~\cite{Beach:2016ocq} expresses it by pairing stress changes at two points of the interval.
Below, the stress is the full energy density, and the Hessian is twice the coefficient of $\lambda^2$.

\subsection{Relative-Entropy Hessian on Spatial Line}
\label{app:detailB1}

\noindent
On the spatial line, the vacuum modular Hamiltonian weights the energy at $x$ by a function that vanishes at either end of the interval and is largest at the midpoint.
For $I=[u,v]$ with length $\ell=v-u$, write this weight and the small change of boundary coordinate as
\begin{equation}
k_I(x)=\frac{(x-u)(v-x)}{\ell},\qquad f_\lambda(x)=x+\lambda s(x)+O(\lambda^2).
\end{equation}
A vacuum descendant is produced by a conformal transformation $f_\lambda$ that preserves orientation and acts on both chiral sectors.
Here, $\ell=v-u>0$ for finite endpoints $u<v$, $\epsilon>0$ is the short-distance cutoff, and a prime denotes differentiation with respect to the spatial coordinate.
We use the following normalization for the transformed stress, the interval entropy, and the vacuum modular Hamiltonian,
\begin{align}
\langle T_{00}(x)\rangle_f&=-\frac c{12\pi}\{f,x\},&
\{f,x\}&=\frac{f'''}{f'}-\frac32\left(\frac{f''}{f'}\right)^2,\label{eq:stress-schwarzian}\\
S_I(f)&=\frac c3\ln\frac{f(v)-f(u)}{\epsilon\sqrt{f'(u)f'(v)}},&
K_I^{(0)}&=2\pi\int_u^v k_I(x)T_{00}(x)\,dx+\mathrm{constant}.
\label{eq:app_lineexact}
\end{align}
In Eq.~\eqref{eq:stress-schwarzian}, the Schwarzian derivative $\{f,x\}$ gives the stress induced by the coordinate change.
Its linear part $\lambda s'''$ is insensitive to any quadratic polynomial added to $s$.
The order-$\lambda^2$ term in $f_\lambda$ multiplies the first variation of relative entropy, which vanishes in vacuum.
Taking $f_\lambda=x+\lambda s$ thus gives the required Hessian from the expansions
\begin{align}
\{f_\lambda,x\}&=\lambda s'''-\lambda^2\left(s's'''+\tfrac32(s'')^2\right)+O(\lambda^3),\\
\Delta S_I&=\frac c3\left[
\lambda\left(A_s-\frac{s'(u)+s'(v)}2\right)
+\lambda^2\left(-\frac{A_s^2}{2}+\frac{s'(u)^2+s'(v)^2}{4}\right)\right]+O(\lambda^3),\\
A_s&=\frac{s(v)-s(u)}{\ell}.
\end{align}
At first order, $\int_u^v k_I s'''=s'(u)+s'(v)-2A_s$ gives the first-law cancellation.
At second order, integration by parts gives
\begin{equation}
\int_u^v k_I s's'''\,dx
=-\int_u^v k_I(s'')^2\,dx
+\frac{s'(u)^2+s'(v)^2}{2}
-\frac1\ell\int_u^v(s')^2\,dx.
\end{equation}
The Hessian $\partial_\lambda^2(\Delta\langle K_I^{(0)}\rangle-\Delta S_I)|_0$ therefore equals
\begin{equation}
\GG_I[s]=\frac c6\int_u^v k_I(s'')^2\,dx
-\frac{c}{3\ell}\int_u^v(s')^2\,dx
+\frac c3\left(\frac{s(v)-s(u)}{\ell}\right)^2.
\label{eq:app_Gline}
\end{equation}
Here, $s'$ is the local change in scale of the spatial coordinate, and $A_s$ is its mean over the interval.
The last two terms subtract $c/3$ times the variance of that local change in scale.
To determine which shapes of $s'$ the positive derivative term detects, rescale the interval to $[-1,1]$ using $t=2(x-u)/\ell-1$ and write $f(t)=s'(u+\ell(t+1)/2)$ and $\bar f=\tfrac12\int_{-1}^{1}f(t)\,dt=A_s$.
Since $k_I=\ell(1-t^2)/4$ and $dx=\ell\,dt/2$, Eq.~\eqref{eq:app_Gline} becomes
\[
\GG_I[s]=\frac{c}{12}\int_{-1}^{1}
\big[(1-t^2) f'(t)^2-2(f(t)-\bar f)^2\big]dt .
\]
Here, $f'$ is the derivative with respect to $t$.
Expand the smooth real function $f$ in Legendre polynomials $P_n$, the degree-$n$ solutions of $-[(1-t^2)P_n']'=n(n+1)P_n$ with $P_n(1)=1$, for integers $n\geq0$.
These polynomials form a complete orthogonal basis on $[-1,1]$ with $\int_{-1}^{1}P_n^2\,dt=2/(2n+1)$.
Writing $a_n=(2n+1)\int_{-1}^{1} f P_n\,dt/2$ gives $a_0=\bar f$.
Orthogonality separates the variance into independent contributions from each mode.
For the derivative term, integration by parts gives the same modes with an extra factor $n(n+1)$ and no edge term because $1-t^2$ vanishes at both ends.
Subtracting the variance gives the coefficient $n(n+1)-2=(n-1)(n+2)$ for $n\ge1$, hence
\[
\GG_I[s]=\frac{c}{6}\sum_{n=2}^{\infty}
\frac{(n-1)(n+2)}{2n+1}\,a_n^2 .
\]
The mean subtraction removes $n=0$, and for $n=1$ the derivative and variance terms cancel because $n(n+1)=2$.
Every higher mode raises the Hessian.
Thus, a vanishing Hessian on this interval allows $s'$ to vary linearly and $s$ to be any quadratic polynomial there.
These displacements have $s'''=0$ and induce no first-order stress on the measured interval. \\

\noindent
The associated bilinear form pairs every quadratic polynomial with every smooth $s$ to zero.
Constants and linear functions follow from the fundamental theorem of calculus.
For the quadratic direction, the two required identities are
\begin{equation}
\int_u^v k_I s''\,dx=s(u)+s(v)-\frac2\ell\int_u^v s\,dx,
\qquad
\int_u^v xs'\,dx=vs(v)-us(u)-\int_u^v s\,dx.
\end{equation}
For a quadratic form $Q$, its polarization is the symmetric bilinear form $[Q(s+q)-Q(s)-Q(q)]/2$.
The terms in the polarized version of Eq.~\eqref{eq:app_Gline} then cancel.
Thus, $\GG_I[s+q]=\GG_I[s]$ for every quadratic polynomial $q$, including all cross terms.
The stress tangent
\begin{equation}
t_1(x)=-\frac c{12\pi}s'''(x)
\end{equation}
therefore determines $\GG_I$ without a choice of global conformal frame.

\subsection{Stress Kernel for Line Intervals}
\label{app:detailB2}

\noindent
The measured stress gives $s'''=-(12\pi/c)t_1$ and determines $s$ up to a quadratic polynomial, which contributes zero to the Hessian.
Choose the integration constants so that $s(u)=s'(u)=s''(u)=0$.
Three successive integrations from $u$ to $x$ give the weight $(x-y)^2/2$ at a point $y$, yielding
\begin{equation}
s(x)=-\frac{6\pi}{c}\int_u^x(x-y)^2t_1(y)\,dy.
\end{equation}
With this expression for $s$, an exchange of the finite integrations in Eq.~\eqref{eq:app_Gline} gives
\begin{equation}
\GG_I[t_1]=\frac{12\pi^2}{c\ell^2}
\int_u^v dx\int_u^v dy\,
[\min(x,y)-u]^2[v-\max(x,y)]^2t_1(x)t_1(y).
\label{eq:app_Gstressline}
\end{equation}
For example, after translating $u$ to zero, the required elementary identity for $0\leq a\leq b\leq\ell$ is
\begin{equation}
\frac16\int_b^\ell k_I(x)\,dx
-\frac1{3\ell}\int_b^\ell(x-a)(x-b)\,dx
+\frac{(\ell-a)^2(\ell-b)^2}{12\ell^2}
=\frac{a^2(\ell-b)^2}{12\ell^2}.
\end{equation}
Eq.~\eqref{eq:app_Gstressline} uses only the restriction of $t_1$ to $I$. \\

\noindent
To determine what a short interval measures when the stress changes sign near its center, fix $x_0$ and take $I=[x_0-\ell/2,x_0+\ell/2]$ with $\ell\downarrow0$.
Assume $t_1$ has four continuous derivatives on a neighborhood of $x_0$, and write $T_j=t_1^{(j)}(x_0)$ for $j=0,1,2$, including $T_0=t_1(x_0)$.
On the half-square $u\leq x\leq y\leq v$, set $X=(x-u)/\ell$ and $Y=(y-u)/\ell$.
The stress product is then integrated against $X^2(1-Y)^2$, with overall coefficient $12\pi^2\ell^4/c$ and a factor of two from the second half-square.
For nonnegative integers $a,b$, integrate first in $X$ to obtain the moments
\[
\begin{aligned}
M_{ab}
&=2\int_0^1 (1-Y)^2Y^b\left(\int_0^Y X^{2+a}\,dX\right)dY\\
&=\frac{2}{3+a}\int_0^1 (1-Y)^2Y^{3+a+b}\,dY
=\frac{4}{(3+a)(4+a+b)(5+a+b)(6+a+b)} .
\end{aligned}
\]
Expand $t_1(x)=T_0+\ell(X-1/2)T_1+\ell^2(X-1/2)^2T_2/2+O(\ell^3)$ and expand $t_1(y)$ at the same center.
The reflection $(X,Y)\mapsto(1-Y,1-X)$ leaves the integration weight fixed and reverses the sign of every odd total order in the Taylor expansion of the stress product.
The constant moment is $M_{00}=1/90$, while the moments of $(X-1/2)(Y-1/2)$ and $[(X-1/2)^2+(Y-1/2)^2]/2$ are $1/5040$ and $1/2520$, respectively.
Thus, in line coordinates,
\begin{equation}
\GG_I[t_1]=\frac{2\pi^2}{15c}\ell^4 T_0^2
+\frac{\pi^2}{420c}\ell^6\bigl(T_1^2+2T_0T_2\bigr)+O(\ell^8).
\label{eq:short-interval-stress}
\end{equation}
At a simple zero of stress, $T_0=0$ and $T_1\ne0$, so the $\ell^4$ term vanishes and the leading signal is $\pi^2\ell^6T_1^2/(420c)$.
A shrinking interval thus still detects the local stress gradient. \\

\noindent
Extending a prescribed stress to a global descendant involves additional conditions at infinity.
For example, suppose the full line profile $t_1$ is smooth and compactly supported, and choose the quadratic freedom so that $s$ vanishes to the left of its support.
Integrating $s'''=-(12\pi/c)t_1$ three times gives $s(x)=-(6\pi/c)\int_{-\infty}^x(x-y)^2t_1(y)\,dy$.
To the right of the support, this becomes
\begin{equation}
s(x)=-\frac{6\pi}{c}\bigl(x^2\mathfrak m_0-2x\mathfrak m_1+\mathfrak m_2\bigr),
\qquad \mathfrak m_j=\int_{\mathbb R} y^j t_1(y)\,dy,\qquad j=0,1,2 .
\end{equation}
Consequently, all three moments must vanish if the generator is to be zero beyond both ends of the stress support, and their vanishing is sufficient for the explicit integral above to have compact support.
The freedom to add a quadratic polynomial shifts both exterior polynomials by the same amount, leaving their difference intact.

\subsection{Circle Intervals and Global Conformal Freedom}
\label{app:detailB3}

\noindent
On the unit-radius cylinder, the vacuum energy density is $-c/(24\pi)$.
Here, $x$ measures angle in radians and $\tau$ is dimensionless cylinder time.
An orientation-preserving change of angle can be described without a jump at $2\pi$ by using a smooth real function with $f'(x)>0$ and $f(x+2\pi)=f(x)+2\pi$.
The finite transformation reads
\begin{align}
T_{\tau\tau,f}(x)&=-\frac c{24\pi}[f'(x)]^2-\frac c{12\pi}\{f,x\},\\
S_I(f)&=\frac c3\ln
\frac{2\sin((f(v)-f(u))/2)}{\epsilon\sqrt{f'(u)f'(v)}}.
\end{align}
The infinitesimal stress relation is $t_1=-c(s'''+s')/(12\pi)$.
For $u=\theta-\alpha$, $v=\theta+\alpha$ and $k=[\cos(x-\theta)-\cos\alpha]/\sin\alpha$, a circle interval has center $\theta$ modulo $2\pi$ and half-opening $0<\alpha<\pi$.
The second-order expansion of the sine in $S_I(f)$ uses $(\ln\sin z)'=\cot z$ and $(\ln\sin z)''=-1/\sin^2 z$.
The endpoint terms from this expansion and the stress terms from the modular Hamiltonian together give
\begin{equation}
\GG_I[s]=\frac c6\int_u^v
\left[k(s'')^2-(2k+\cot\alpha)(s')^2\right]dx
+\frac{c[s(v)-s(u)]^2}{12\sin^2\alpha}.
\label{eq:app_Gcircle}
\end{equation}
The identities $k(u)=k(v)=0$, $k'(u)=1$, $k'(v)=-1$, and $k''=-k-\cot\alpha$ fix the endpoint terms.
In particular,
\begin{equation}
\int_u^v k s's'''\,dx
=-\int_u^v k(s'')^2\,dx+\frac{s'(u)^2+s'(v)^2}{2}
-\frac12\int_u^v(k+\cot\alpha)(s')^2\,dx.
\end{equation}
This identity expresses the dependence on the endpoint derivatives explicitly. \\

\noindent
A vanishing quadratic reading on one interval and vanishing readings on every interval impose different conditions.
For example, a smooth vector field supported outside the chosen interval is invisible there, but another interval can intersect its support.
The common nullspace consists of directions invisible to all intervals and includes $1$, $\sin x$, and $\cos x$.
The Green function argument in Appendix~\ref{app:stressgreen} identifies them as the full common nullspace of all intervals.
For a sine or cosine direction $q$, the equation $q''=-q$ simplifies the bilinear term obtained by polarizing Eq.~\eqref{eq:app_Gcircle}.
After integration by parts, the integral part, with its overall factor removed, is $\int_u^v[k'q+k''q']s'\,dx$.
The bracket is constant and equals $-[q(v)-q(u)]/(2\sin^2\alpha)$.
It cancels the endpoint bilinear term.
The constant mode vanishes immediately. \\

\noindent
The Fourier coefficient of the first stress variation at integer index $n\in\mathbb Z$ is
\begin{equation}
t_{1,n}=\frac1{2\pi}\int_0^{2\pi}t_1(x)\exp(-\mathrm{i} n x)\,dx.
\end{equation}
A periodic vacuum-orbit tangent has $t_{1,0}=t_{1,1}=t_{1,-1}=0$.
For $|n|\geq2$, inversion of the stress relation gives
\begin{equation}
s_n=\frac{12\pi}{c}\frac{t_{1,n}}{\mathrm{i} n(n^2-1)},\qquad
s(x)=\sum_{\substack{n\in\mathbb Z\\ |n|\geq2}}s_n\exp(\mathrm{i} n x).
\end{equation}
Here, $\mathrm i^2=-1$, and a real $s$ has $s_{-n}=\overline{s_n}$, with the overline denoting complex conjugation.
The three omitted modes generate transformations preserving the vacuum and leave $\GG_I$ unchanged.
A first-order mass shifts the zero mode, and singular holonomy can change how a local solution returns around the circle.
Such deformations lie outside this smooth vacuum orbit. \\

\noindent
Choose the angular origin so that the pole of $z=\tan(x/2)$ lies outside the interval, where $dz/dx=(1+z^2)/2>0$ on the chosen lift.
The state derivative removes the state-independent Schwarzian term from the stress transformation, leaving $t_{1,\mathbb R}(z)=4t_{1,S^1}(x)/(1+z^2)^2$.
For a smooth circle stress, the transformed profile decays as $O(|z|^{-4})$ as $|z|\to\infty$, so all three line moments in Appendix~\ref{app:detailB2} converge absolutely.
On the chart $-\pi<x<\pi$, changing variables with $dz=(1+z^2)dx/2$ gives
\begin{equation}
(\mathfrak m_0,\mathfrak m_1,\mathfrak m_2)=\int_{-\pi}^{\pi}
\bigl(1+\cos x,\,\sin x,\,1-\cos x\bigr)\,t_{1,S^1}(x)\,dx .
\end{equation}
Taking the sum and difference of the first and third entries shows that vanishing of the three line moments is equivalent to $t_{1,0}=t_{1,1}=t_{1,-1}=0$ on the circle.
The Fourier inverse above then supplies a periodic generator up to the three vacuum symmetries.
These three moment conditions concern a stress profile prescribed on the entire circle. \\

\noindent
Suppose the stress is given on a neighborhood of a closed proper arc $[u,v]$.
Inside that neighborhood, choose an angular chart $J$ of length less than $2\pi$ containing the arc.
With $s_{\rm loc}(u)=s_{\rm loc}'(u)=s_{\rm loc}''(u)=0$, the solution on $J$ is
\[
s_{\rm loc}(x)=-\frac{12\pi}{c}\int_u^x
[1-\cos(x-y)]\,t_1(y)\,dy .
\]
At $x=y$, the kernel and its first derivative vanish, and its second derivative equals one.
Three differentiations then give $s_{\rm loc}'''+s_{\rm loc}'=-(12\pi/c)t_1$, with the source supplied by the upper endpoint.
Multiply $s_{\rm loc}$ by a smooth cutoff supported inside $J$ and equal to one near $[u,v]$, then extend by zero around the circle.
This smooth periodic vector field generates an increasing circle map for sufficiently small $|\lambda|$ and reproduces the measured stress throughout $I$.
The extension determines the stress elsewhere and supplies the three vanishing global moments. Hence, a single proper arc places no additional moment condition on its smooth local data. \\

\noindent
We now apply the stereographic coordinate to the kernel itself.
The endpoint identity
\begin{equation}
z(x)-z(u)=\frac{\sin[(x-u)/2]}{\cos(x/2)\cos(u/2)}
\end{equation}
and its $(v,y)$ counterpart convert the squared differences in Eq.~\eqref{eq:app_Gstressline} into squared sines.
The cosine factors cancel against the Jacobians, the two stress factors, and $[z(v)-z(u)]^{-2}$, giving
\begin{equation}
\GG_I[t_1]=\frac{48\pi^2}{c\sin^2\alpha}
\int_u^v dx\int_u^v dy\,
\sin^2\!\frac{\min(x,y)-u}{2}\,
\sin^2\!\frac{v-\max(x,y)}2\,t_1(x)t_1(y).
\end{equation}
Both integrations run over the same lifted interval.
A smooth lift gives the same stress integral when the interval crosses a coordinate seam. \\

\noindent
One can also read off the first effect of the unit-circle geometry in the short-interval expansion.
Set $\ell=2\alpha$ and compare the circle expression to Eq.~\eqref{eq:app_Gstressline} evaluated on the same angular segment with the same local stress function.
In the variables $X,Y$ of Appendix~\ref{app:detailB2}, the ratio of circle to line kernels for $0<X\leq Y<1$ is
\[
\frac{\ell^2}{4\sin^2(\ell/2)}
\left[\frac{\sin(\ell X/2)}{\ell X/2}
\frac{\sin(\ell(1-Y)/2)}{\ell(1-Y)/2}\right]^2
=1+\frac{\ell^2}{12}\big[1-X^2-(1-Y)^2\big]+O(\ell^4),
\]
with the sine quotients extended by continuity at the endpoints.
The correction is symmetric under $(X,Y)\mapsto(1-Y,1-X)$, and its polynomial in brackets has moment $M_{00}-2M_{20}=2/315$.
Consequently, the circle result equals the right-hand side of Eq.~\eqref{eq:short-interval-stress} with an additional term $2\pi^2\ell^6T_0^2/(315c)$ at the order shown.
For a simple zero of the stress, the line and circle have the same leading coefficient $\pi^2T_1^2/(420c)$.

\subsection{Circle Stress Kernel from Second-Order Equation}
\label{app:stressgreen}

\noindent
The finite descendant entropy also gives a direct derivation of the circle stress kernel.
For the common circle reparametrization $f$ of the two chiralities, we use
\begin{equation}
D_f(x,y)=\frac{2\sin[(f(y)-f(x))/2]}{\sqrt{f'(x)f'(y)}},
\qquad q=\frac{6\pi}{c},\qquad T_f=T_{\tau\tau,f}.
\end{equation}
The interval entropy is $S_I(f)=(c/3)\ln[D_f(u,v)/\epsilon]$ with a fixed boundary cutoff $\epsilon$.
At fixed $x$, the finite circle stress transformation gives the differential equation in $y$
\begin{equation}
(-\partial_y^2+qT_f(y))D_f(x,y)=0,\qquad
D_f(x,x)=0,\qquad \left.\partial_yD_f(x,y)\right|_{y=x}=1 .
\end{equation}
To obtain the differential equation, put $A(y)=[f'(y)]^{-1/2}$ and set $h(y)=A(y)H(f(y))$, where $H(z)=\cos(z/2)$ or $2\sin(z/2)$.
In $h''$, the coefficient of $H'(f(y))$ is $2A'f'+Af''=0$.
The remaining terms use $H''(z)=-H(z)/4$ and $A''/A=-\{f,y\}/2$, where $\{f,y\}=f'''/f'-\tfrac32(f''/f')^2$.
Thus, both functions $p(y)=A(y)\cos(f(y)/2)$ and $r(y)=2A(y)\sin(f(y)/2)$ satisfy $h''=qT_fh$, including at the zeros of $h$.
Their Wronskian, the determinant $p r'-p'r$, equals $A^2 f'=1$.
The sine subtraction formula gives $D_f(x,y)=p(x)r(y)-r(x)p(y)$.
At $y=x$, this determinant vanishes, and its derivative with respect to $y$ is the unit Wronskian, which fixes the normalization of the initial-value solution.
On the proper interval $(u,v)$, the inverse of $-\partial_x^2+qT_f(x)$ with both endpoint values fixed to zero has the following Green function, the kernel of its inverse on functions obeying these homogeneous endpoint conditions.
\begin{equation}
G_f(x,y)=\frac{D_f(u,x_<)D_f(x_>,v)}{D_f(u,v)},\qquad
x_<=\min(x,y),\quad x_>=\max(x,y).
\label{eq:stressgreen}
\end{equation}
The denominator is nonzero for a proper interval and an orientation-preserving $f$.
The two factors guarantee the endpoint conditions.
For $L_f(x)=D_f(u,x)$ and $R_f(x)=D_f(x,v)$, the common second-order equation without a first-derivative term makes $L_fR_f'-L_f'R_f$ constant.
Its value at $u$ is $-D_f(u,v)$ because $L_f(u)=0$ and $L_f'(u)=1$.
Dividing by $D_f(u,v)$ gives a slope jump $\partial_xG_f(y^+,y)-\partial_xG_f(y^-,y)=-1$ from left to right, so $-\partial_x^2G_f$ supplies the required unit delta function. \\

\noindent
At fixed initial conditions, the linear variation of the differential equation has source $q\,\delta T_f(x)D_f(u,x)$ in its second-derivative form.
The solution $D_f(x,v)$ propagates that source from $x$ to the final endpoint, giving $\delta D_f(u,v)=q\int_u^vD_f(x,v)\,\delta T_f(x)D_f(u,x)\,dx$.
The ratio of this expression to $D_f(u,v)$ is $\delta\ln D_f=q\int_u^v G_f(x,x)\delta T_f(x)dx$.
For the second variation, the inverse-operator identity $\delta\mathcal B^{-1}=-\mathcal B^{-1}(\delta\mathcal B)\mathcal B^{-1}$ gives
\begin{equation}
\delta G_f(x,y)=-q\int_u^v G_f(x,z)\delta T_f(z)G_f(z,y)\,dz .
\end{equation}
At the vacuum configuration $f(x)=x$, Eq.~\eqref{eq:stressgreen} reduces to
\begin{equation}
G_0(x,y)=
\frac{2\sin[(x_<-u)/2]\sin[(v-x_>)/2]}{\sin\alpha},
\qquad G_0(x,x)=k_I(x).
\end{equation}
We denote the two stress derivatives by $t_j=\partial_\lambda^j T_f|_0$, $j=1,2$.
The preceding variations now read
\begin{align}
\left.\partial_\lambda^2 S_I(f)\right|_0
&=2\pi\int_u^v k_I(x)t_2(x)\,dx
-\frac{12\pi^2}{c}\int_u^v\!dx\int_u^v\!dy\,
G_0(x,y)^2t_1(x)t_1(y).
\end{align}
Subtracting this entropy variation from the second modular-energy variation cancels the entire term containing $t_2$.
The remainder is $\GG_I[t_1]=(12\pi^2/c)\int_u^v dx\int_u^v dy\,G_0(x,y)^2t_1(x)t_1(y)$, giving the circle prefactor $48\pi^2/(c\sin^2\alpha)$ on substituting $G_0$. \\

\noindent
The sign and the null directions of this stress kernel can be read off even for a stress tangent that changes sign on $I$.
The vacuum Dirichlet operator is $\mathcal B_0=-\partial_x^2-1/4$ on the interval of length $\ell=2\alpha<2\pi$.
Its normalized eigenfunctions and eigenvalues are
\[
\psi_n(x)=\sqrt{2/\ell}\,\sin[n\pi(x-u)/\ell],\qquad
d_n=(n\pi/\ell)^2-1/4>0,\quad n=1,2,\ldots .
\]
The series $G_0(x,y)=\sum_{n=1}^{\infty}\psi_n(x)\psi_n(y)/d_n$ converges uniformly because the eigenfunctions are uniformly bounded and $\sum_{n=1}^\infty d_n^{-1}<\infty$.
Uniform convergence permits termwise integration of the product of two such series.
For real $t_1$, the $x$ and $y$ integrals in each term are equal, giving the square in
\[
\GG_I[t_1]=\frac{12\pi^2}{c}\sum_{m,n=1}^{\infty}
\frac{1}{d_m d_n}\left|\int_u^v t_1(x)\psi_m(x)\psi_n(x)\,dx\right|^2 .
\]
A stress that changes sign still gives a nonnegative Hessian because every term in the series is a square divided by positive eigenvalues.
If the sum vanishes, the $m=1$ terms alone force all sine coefficients of $t_1\psi_1$ to vanish.
Completeness implies $t_1\psi_1=0$ almost everywhere, and the lowest sine mode is strictly positive inside the interval.
A continuous $t_1$ must therefore vanish throughout $I$, which is also sufficient for $\GG_I=0$.
The relation $s'''+s'=-(12\pi/c)t_1$ therefore identifies the null directions on one interval as those vector fields whose \emph{restriction} to that interval is a linear combination of $1$, $\sin x$, and $\cos x$.
The same three functions exhaust the common global nullspace when every interval is included.

\subsection{Canonical-Energy Separation and Pure-Gravity Example}
\label{app:detailB5}

\noindent
The background scalar vanishes, $\phi_0=0$, so the first coupling between a metric change and scalar matter has the form $h\phi^2$.
It enters at third order in the perturbations.
At quadratic order, the conserved charge known as canonical energy therefore separates into two terms,
\begin{equation}
\FF_I=\mathcal E_I^{\rm grav}(h_1,h_1)+\mathcal E_I^\phi(\phi_1,\phi_1).
\end{equation}
The first-order metric $h_1$ and scalar $\phi_1$ obey independent linearized equations.
The charge $\mathcal E_I$ generates the vacuum modular flow of the bulk wedge, and its metric part is evaluated in coordinates that fix the extremal surface to first order.
Changing to these coordinates leaves $\phi_1$ unchanged because the background scalar has zero derivative.
The matched descendant has the same $h_1$ and no scalar, so subtracting its Hessian $\GG_I[t_1]$ removes the metric term even if the second-order stresses differ.
Appendix~\ref{app:tomography} uses the remaining scalar charge.
The canonical-energy relation and the principle of subtracting a matched gravitational contribution are already present in Ref.~\cite{Lashkari:2015hha}. \\

\noindent
The following pure-gravity family quantifies the effect of the subtraction.
We choose a smooth, compactly supported vector field $s$ on the line with $s(x)=x^3$ on a neighborhood of the intervals under consideration.
For sufficiently small $\lambda$ of either sign, $f_\lambda=x+\lambda s$ is an orientation-preserving diffeomorphism and defines a pure vacuum-descendant state.
A translation of $x^3$ changes it by a quadratic polynomial, so Eq.~\eqref{eq:app_Gline} gives
\begin{equation}
\FF_{[x_0-R,x_0+R]}=\GG_{[x_0-R,x_0+R]}=\frac{8c}{15}R^4,
\qquad
\left[R^2(\partial_R^2-\partial_{x_0}^2)-2\right]\FF
=\frac{16c}{3}R^4.
\end{equation}
For $x_0=0$, one can see the coefficient directly from $s'=3x^2$, $s''=6x$, $k_I=(R^2-x^2)/(2R)$ and $A_s=R^2$ in Eq.~\eqref{eq:app_Gline}.
The stress representation explains the same coefficient using only a constant interval stress, since $t_1=-c/(2\pi)$ on this interval.
Write $a=x-u$ and $b=y-u$ on the half-square $u\leq x\leq y\leq v$, with $\ell=v-u=2R$.
Adding the other half by symmetry gives the unnormalized kernel integral
\[
2\int_0^\ell db\,(\ell-b)^2\int_0^b a^2\,da
=\frac{2}{3}\int_0^\ell b^3(\ell-b)^2\,db
=\frac{\ell^6}{90}.
\]
Multiplication by $12\pi^2 t_1^2/(c\ell^2)$ in Eq.~\eqref{eq:app_Gstressline} gives $c\ell^4/30=8cR^4/15$.
A pure-gravity deformation thus has a positive Hessian and a positive source if one uses $\FF$ alone.
Its matched Hessian is identical, giving zero after subtraction, as required by the constant curvature and unchanged polygon volume on this branch.
Appendix~\ref{app:cft} considers a state with both stress and scalar excitations.

\subsection{Local Source from Interval Entropies}
\label{app:localentropy}

\noindent
At fixed interval and cutoff, write $S_j(u,v)=\left.\partial_\lambda^jS_I(\lambda)\right|_0$ for $j=1,2$.
Each derivative removes the state-independent vacuum divergence, and the Taylor coefficient is $S_j/j!$.
With $u=\theta-\alpha$, $v=\theta+\alpha$, and $0<v-u<2\pi$, differentiating with the opposite endpoint fixed gives
\begin{equation}
\DD=-4\sin^2\alpha\,\partial_u\partial_v-2 .
\end{equation}
We first eliminate the modular energy and then the gravitational Fisher term, the second variation of relative entropy in the matched vacuum descendant.
Complementary arcs and a perturbation containing both scalar and stress contributions illustrate what remains in the measured entropy.

\subsubsection{Homogeneous Equation for Modular Energy}
\label{app:detailC7a}

\noindent
For a smooth circle energy-density variation $h(x)$, the change in vacuum modular energy is the linear functional \cite{Casini:2011kv}
\begin{align}
M_h(u,v)&=2\pi\int_u^v k(x;u,v)h(x)\,dx,\nonumber\\
k(x;u,v)&=\frac{2\sin[(x-u)/2]\sin[(v-x)/2]}{\sin\alpha}.
\end{align}
For $u<x<v$, one has
\begin{equation}
\partial_v k=\frac{\sin^2[(x-u)/2]}{\sin^2\alpha},
\qquad
\partial_u\partial_v k=-\frac{k}{2\sin^2\alpha}.
\end{equation}
Changing an interval endpoint normally changes both the integrand and the range of integration.
At the upper end, the endpoint term vanishes because $k(v;u,v)=0$, and at the lower end of the second differentiation it vanishes because $\partial_v k(u;u,v)=0$.
Thus, $\DD M_h=0$ for any stress profile $h$ in this formula.
Applying $\DD$ to the second state derivative of $S_{\rm rel}=\Delta\langle K_I^{(0)}\rangle-\Delta S_I$ eliminates the modular-energy term even when the second-order stress is unknown, leaving $\DD\FF_I=-\DD S_2$.
At first order, the first law gives $S_1=M_{t_1}$, an integral of the stress over the interval.
Slide the right endpoint once more while holding the left fixed.
Differentiating $\sin^2\alpha\,\partial_v S_1$ with respect to $v$ isolates $2\pi\sin^2\alpha\,t_1(v)$ at the moving endpoint, giving
\begin{equation}
2\pi t_1(v)=(\partial_v^2+\cot\alpha\,\partial_v)S_1(u,v).
\end{equation}
The corresponding expression at $u$ is $2\pi t_1(u)=(\partial_u^2-\cot\alpha\,\partial_u)S_1(u,v)$.
One can also recover the stress from first-state derivatives on shrinking intervals without differentiating at either endpoint.
Indeed, the nonnegative modular weight is even in $x-\theta$ and has integral
\[
\int_{\theta-\alpha}^{\theta+\alpha} k(x;\theta-\alpha,\theta+\alpha)\,dx
=\frac{1}{\sin\alpha}\int_{-\alpha}^{\alpha}(\cos z-\cos\alpha)\,dz
=2(1-\alpha\cot\alpha).
\]
Hence, $S_1(\theta-\alpha,\theta+\alpha)/[4\pi(1-\alpha\cot\alpha)]$ is a weighted average of $t_1$ on this interval and tends to $t_1(\theta)$ as $\alpha\downarrow0$ for a continuous stress tangent.
For twice continuously differentiable $t_1$, the odd Taylor term integrates to zero and the difference between this average and $t_1(\theta)$ is at most $\tfrac12\alpha^2\sup_{|x-\theta|\le\alpha}|t_1''(x)|$.
Using $1-\alpha\cot\alpha=\alpha^2/3+O(\alpha^4)$ gives
\[
t_1(\theta)=\frac{3}{4\pi}\lim_{\alpha\downarrow0}
\frac{S_1(\theta-\alpha,\theta+\alpha)}{\alpha^2}.
\]
In the smooth vacuum-descendant sector, the reconstructed stress has vanishing Fourier coefficients at $n=0,\pm1$, as required by the stress transformation in Appendix~\ref{app:detailB3}.
These formulas recover the stress when it is wanted separately.
For the chord measurement, its contribution can be removed directly from the measured entropy slope $S_1$.

\subsubsection{Gravitational Fisher Subtraction}
\label{app:detailC7b}

\noindent
A vacuum descendant generated by the same increasing circle map $f_\lambda$ in both chiral sectors has interval entropy $S^f$ and vacuum reference value $S^0$.
The mixed endpoint derivative of the transformed entropy in Appendix~\ref{app:gravity} is
\begin{align}
\partial_u\partial_v S^f
&=\frac{c}{12}\frac{f'_\lambda(u)f'_\lambda(v)}
{\sin^2[(f_\lambda(v)-f_\lambda(u))/2]}\nonumber\\
&=\frac{c}{12\sin^2\alpha}
\exp\!\left[-\frac{6}{c}(S^f-S^0)\right].
\label{eq:app_entliouville}
\end{align}
The cutoff cancels between the two entropies on the right.
This is the interval form of the vacuum-family Liouville equation \cite{deBoer:2016pqk}.
Its first state derivative gives $\DD S_1^f=0$.
Its second derivative at $f_0(x)=x$ is
\begin{equation}
\partial_u\partial_v S^f_2
=-\frac{S^f_2}{2\sin^2\alpha}
+\frac{3(S^f_1)^2}{c\sin^2\alpha}.
\end{equation}
Multiplying the mixed derivative by $-4\sin^2\alpha$ and subtracting $2S^f_2$ cancels the term linear in $S^f_2$, leaving $\DD S^f_2=-12(S^f_1)^2/c$.
The first law gives $S^f_1=S_1$ for a descendant with the same first-order stress.
Its Hessian $\GG_I[t_1]$ is the second modular-energy variation minus $S^f_2$, and the former is annihilated by $\DD$ according to Appendix~\ref{app:detailC7a}.
Consequently,
\begin{align}
\DD\GG_I[t_1]&=\frac{12}{c}S_1^2,\nonumber\\
\mathcal Q_I&\equiv-\DD S_2-\frac{12}{c}S_1^2
=\DD(\FF_I-\GG_I[t_1]).
\label{eq:app_entropysource}
\end{align}
This cancellation uses boundary identities and a matching first-order tangent to the smooth vacuum orbit.
The actual state can contain scalar matter, whose contribution is identified as $\mathcal Q_I=4\pi\XX e$ by the leading semiclassical canonical-energy relation under the hypotheses of Appendix~\ref{app:tomography}. \\

\noindent
One can see directly why the double stress integral in Appendix~\ref{app:stressgreen} produces the square of a single interval measurement.
Use $G_0$ from Eq.~\eqref{eq:stressgreen} in the vacuum, and hold the interior integration points $x,y$ fixed while varying $u,v$.
Differentiating the two sine factors and $1/\sin\alpha$ gives the symmetric formulas
\begin{align*}
\partial_u G_0(x,y)&=-\frac{\sin[(v-x)/2]\sin[(v-y)/2]}{\sin^2\alpha},\\
\partial_v G_0(x,y)&=\frac{\sin[(x-u)/2]\sin[(y-u)/2]}{\sin^2\alpha}.
\end{align*}
They imply $4\sin^2\alpha\,(\partial_uG_0)(\partial_vG_0)=-k_I(x)k_I(y)$.
In $\ln G_0$, only $-\ln\sin\alpha$ depends on both endpoints, so $\partial_u\partial_v\ln G_0=-1/(4\sin^2\alpha)$.
Using $\partial_u\partial_v G_0^2=2G_0^2\partial_u\partial_v\ln G_0+4(\partial_uG_0)(\partial_vG_0)$ now yields
\[
\DD G_0(x,y)^2=4k_I(x)k_I(y),\qquad u<x,y<v .
\]
In the stress integral, the endpoint terms vanish because $G_0^2$ has a double zero whenever either integration point reaches an endpoint.
Consequently, $\DD\GG_I=(48\pi^2/c)[\int_u^v k_I(x)t_1(x)\,dx]^2=(12/c)S_1^2$ by the interval first law.
This calculation shows how endpoint differentiation removes the kernel's dependence on the ordering of $x$ and $y$. \\

\noindent
A single endpoint derivative also determines how the gravitational Fisher term changes when the interval grows.
Keep the same smooth real stress profile on a neighborhood of the moving interval, and define $a_u(x)=\sin[(x-u)/2]/\sin\alpha$ and $a_v(x)=\sin[(v-x)/2]/\sin\alpha$.
Differentiating with respect to $v$, one factor of $G_0$ in $G_0^2$ differentiates to $a_u(x)a_u(y)$, and the two endpoint terms from the moving integration limits vanish.
The derivative with respect to $u$ instead gives $-a_v(x)a_v(y)$ for the differentiated factor.
Writing $\langle h,g\rangle_I=\int_u^v h(x)g(x)\,dx$ for real functions, we obtain
\begin{equation}
\begin{aligned}
\partial_v\GG_I[t_1]&=\frac{24\pi^2}{c}
\langle a_u t_1,\mathcal B_0^{-1}(a_u t_1)\rangle_I,\\
-\partial_u\GG_I[t_1]&=\frac{24\pi^2}{c}
\langle a_v t_1,\mathcal B_0^{-1}(a_v t_1)\rangle_I .
\end{aligned}
\label{eq:stress-fisher-endpoint-growth}
\end{equation}
For the vacuum Dirichlet operator $\mathcal B_0$ of Appendix~\ref{app:stressgreen}, expansion in its normalized eigenfunctions gives $\langle h,\mathcal B_0^{-1}h\rangle_I=\sum_{n=1}^{\infty}|\langle h,\psi_n\rangle_I|^2/d_n$.
Every $d_n$ is positive for $0<v-u<2\pi$, and both $a_u$ and $a_v$ are positive in the open interval.
Thus, either outward derivative in Eq.~\eqref{eq:stress-fisher-endpoint-growth} is strictly positive whenever $t_1$ is nonzero somewhere in the interval, even if the stress changes sign.
When $t_1$ vanishes throughout the interval, both derivatives are zero. \\

\noindent
The coefficient of the square of the total $S_1$ uses equality of the two chiral variations on the reflection slice.
To see where this enters, write $S^f=S^f_{\rm L}+S^f_{\rm R}$ for a vacuum descendant with independently varied left and right maps, retaining equal central charges $c$.
For $a\in\{\mathrm L,\mathrm R\}$, each chiral entropy has half the logarithmic coefficient of the common-map expression.
Its mixed derivative therefore has vacuum prefactor $c/(24\sin^2\alpha)$ and exponential $\exp[-12(S^f_a-S^0_a)/c]$ in place of the corresponding factors in Eq.~\eqref{eq:app_entliouville}.
Writing $S_{1,a}=\partial_\lambda S^f_a|_0$, the second derivatives of these two exponentials give
\[
\DD\GG_I
=\frac{24}{c}\big(S_{1,\rm L}^2+S_{1,\rm R}^2\big)
=\frac{12}{c}\big[(S_{1,\rm L}+S_{1,\rm R})^2+(S_{1,\rm L}-S_{1,\rm R})^2\big].
\]
On the reflection slice used here, $S_{1,\rm L}=S_{1,\rm R}=S_1/2$, recovering Eq.~\eqref{eq:app_entropysource}.
With unequal chiral perturbations, the same equal-time total $S_1$ leaves the squared difference undetermined.

\subsubsection{Complementary Arcs and Mixed State Variations}
\label{app:detailC7c}

\noindent
The complement $I^c$ is specified by half-opening $\pi-\alpha$ and center $\theta+\pi$ modulo $2\pi$.
Purity gives the same ordinary entropy to the two arcs, and their common geodesic gives the same chord reading $\mathcal Q_{I^c}=\mathcal Q_I$.
Their relative entropies can differ because the scalar contribution before applying $\DD$ samples all the energy on the selected side of the chord.
The weights on the two sides are $\max(w_I,0)$ and $\max(-w_I,0)$.
Their difference is $w_I$, so the difference of the two weighted integrals is $4\pi\int_{\mathbb H^2} w_I e\,d\mu$, which vanishes after applying $\DD$ because $\DD w_I=0$.
Data at all centers and $0<\alpha\leq\pi/2$ therefore suffice for reconstruction.
For $p=\cos\alpha>0$, the short arcs give $J(p,\theta)=\mathcal Q_I/\sqrt{1-p^2}$ for $I=(\theta-\alpha,\theta+\alpha)$.
Complementary arcs fix $J(-p,\theta)=J(p,\theta+\pi)$.
Only after completing the data in this way and extending by zero for $|p|\geq1$ is the full-line Hilbert filter applied. \\

\noindent
The state variation in Eq.~\eqref{eq:cftmixedhalf} contains both a real scalar excitation and a conformal deformation.
The scalar has zero first-order stress, so only the deformation $f_\lambda(x)=x+\lambda\xi\sin(2x)$ contributes to $S_1$, with $\xi$ the dimensionless conformal amplitude.
The first derivative of $S^f$ gives $S_1=(2c\xi/3)\sin^2\alpha\cos(2\theta)$.
At $(\alpha,\theta)=(\pi/2,0)$, this is $2c\xi/3$, giving $\DD\GG=16c\xi^2/3$, even though the undifferentiated Fisher contribution is $\GG=8c\xi^2/5$.
For the $\Delta=3$ cosine mode with scalar amplitude $a$ and $\beta^2=\pi a^2/2$, a signed coordinate $t\in(-1,1)$ along this diameter gives
\begin{equation}
\XX e=\tfrac12a^2\int_{-1}^1(9t^2+3)(1-t^2)^2\,dt=\frac{16a^2}{7}.
\end{equation}
At this interval, $-\DD S_2=128\beta^2/7+16c\xi^2/3$.
The measured slope supplies $12S_1^2/c=16c\xi^2/3$, exactly the conformal term to subtract.
The chord reading is therefore $\mathcal Q_I=128\beta^2/7$ for every $\xi$, including a pure conformal perturbation with $\beta=0$.

\section{Energy Tomography and Polygon Reconstruction}
\label{app:tomography}

\noindent
Which intervals distinguish a localized change in energy from one spread along a geodesic?
Varying the opening at a fixed center scans parallel chords, while varying the center rotates the scan.
We derive the chord integral from interval entropies and show how to combine the scans for one polygon.
We then quantify the effects of incomplete angular coverage, finite sampling, and reading errors.
Appendix~\ref{app:polygon-window-details} gives the explicit line weights for polygon edges and explains how complementary arcs can be averaged.

\subsection{Canonical Energy and Normalization}
\label{app:detailC1}

\noindent
The equality between the relative-entropy Hessian and canonical energy is an input from semiclassical holography \cite{Lashkari:2015hha}.
The subtraction $\widehat\FF_I=\FF_I-\GG_I$ of Appendix~\ref{app:gravity} removes the contribution of the first-order metric perturbation, leaving the canonical energy of the linearized scalar field.
The region $r_I^{(0)}$ lies between $I$ and its geodesic $\gamma_I$ on the vacuum reflection slice with future unit normal $n$.
The modular Killing vector $\xi_I$ generates the bulk isometry that implements the vacuum modular flow and fixes $\gamma_I$.
The scalar stress tensor has the expansion $T_{\mu\nu}=\lambda^2 T_{\mu\nu}^{[2]}+o(\lambda^2)$.
Its ordinary Killing energy is the integral of $T_{\mu\nu}\xi_I^\mu n^\nu$ over $r_I^{(0)}$ with the two spacetime indices contracted.
Twice its quadratic coefficient gives
\begin{equation}
\widehat\FF_I=2\int_{r_I^{(0)}}
\sum_{\mu,\nu=0}^2T_{\mu\nu}^{[2]}\xi_I^\mu n^\nu\,dA_0.
\end{equation}
Thus, the state Hessian contains $2T^{[2]}$ because the stress expansion uses the coefficient of $\lambda^2$.
The vector $\xi_I$ has lapse (normal component) $-n\cdot\xi_I=2\pi Lw_I$ on the time-zero slice.
Its spatial components vanish there, so only the normal energy density contributes to the charge.
The normal-normal contraction is $\sum_{\mu,\nu=0}^{2}T_{\mu\nu}^{[2]}n^\mu n^\nu=\rho_2$.
With $dA_0=L^2d\mu$ and $e=L^3\rho_2$, the integral becomes
\begin{equation}
\widehat\FF_I=4\pi\int_{w_I>0} w_I e\,d\mu .
\label{eq:weightedenergy}
\end{equation}
The factor of two is also consistent with $S_{\rm rel}=\lambda^2\FF_I/2+o(\lambda^2)$.
For a coherent free scalar, the displacement changes the mean field and its modular energy.
The fluctuations about that mean retain their vacuum covariance, so the entropy of a fixed bulk region is unchanged.
Source-free boundary conditions and the stated falloff remove the asymptotic surface term, and the modular Killing vector vanishes on the RT surface.

\subsection{Distributional Derivation of Kinematic Source}
\label{app:detailC2}

\noindent
The interval half-opening determines $p=\cos\alpha\in(-1,1)$ and $q=\sin\alpha>0$.
Its center determines the unit vectors $n_\theta=(\cos\theta,\sin\theta)$ and $n_\theta^\perp=(-\sin\theta,\cos\theta)$, with $0\leq\theta<2\pi$.
The Klein disk uses Euclidean coordinates $y=(y_1,y_2)$, $r=|y|<1$, in which hyperbolic geodesics are chords.
One way to fix the normalization of the lapse is to realize the unit hyperbolic plane as the sheet $-X_0^2+X_1^2+X_2^2=-1$, $X_0>0$.
Its embedding is $X=(1,y)/\sqrt{1-r^2}$.
The unit spacelike vector $a_I=(p,n_\theta)/q$ in this auxiliary Minkowski space is normal to the chord plane $X\cdot a_I=0$.
The scalar $w=X\cdot a_I$ has a positive sign toward $I$ and a unit normal derivative on the chord, the normalization associated with a modular boost of rapidity $2\pi$ per unit flow parameter.
Thus, $r_I^{(0)}=\{w>0\}$ and $-n\cdot\xi_I=2\pi Lw$ there.
A fixed point $y$ has components $x=y\cdot n_\theta$ and $z=y\cdot n_\theta^\perp$.
The lapse is
\begin{equation}
w=\frac{x-p}{q\sqrt{1-r^2}},\qquad r^2=x^2+z^2.
\end{equation}
Here, $\DD=q^2(\partial_\alpha^2-\partial_\theta^2)-2$ acts on interval parameters at fixed $y$.
Since $\partial_\theta^2 x=-x$, $\partial_\alpha^2[(x-p)/q]=2(x-p)/q^3-x/q$ gives $\DD w=0$.
On the level set $w=0$, $x=p$ and
\begin{equation}
\partial_\alpha w=\frac1{\sqrt{1-r^2}},\qquad
\partial_\theta w=\frac{z}{q\sqrt{1-r^2}},\qquad
q^2\left[(\partial_\alpha w)^2-(\partial_\theta w)^2\right]=1.
\end{equation}
As an interval varies, a fixed bulk point enters the region of Eq.~\eqref{eq:weightedenergy} when its chord crosses that point.
Its weight $w_+=\max(w,0)=w\Theta(w)$ is continuous at the crossing, but switches there from identically zero to the smooth function $w$.
The derivative of the Heaviside step $\Theta$ is the Dirac delta $\delta$.
For $j\in\{\alpha,\theta\}$, the first derivative is $\partial_jw_+=\Theta(w)\partial_jw$, since $w\delta(w)=0$.
A second derivative gives $\partial_j^2w_+=\Theta(w)\partial_j^2w+\delta(w)(\partial_jw)^2$, so
\begin{equation}
\DD w_+=\delta(w).
\label{eq:app_distribution}
\end{equation}
To see which line measure the delta function selects, project $a_I$ into the tangent plane of the hyperboloid.
This projection is $a_I+wX$, with squared norm $1+w^2$, so $|\nabla w|=1$ on the chord.
The one-dimensional normalization of the Dirac delta is $\int_{\mathbb R}\delta(a) f(a)\,da=f(0)$ for a smooth test function $f$.
In local coordinates given by distance along and distance across the chord, integration against $\delta(w)$ removes the across-chord coordinate.
The conversion from transverse distance to $w$ contributes $1/|\nabla w|$, the Jacobian in the coarea formula.
Since this factor is one at $w=0$, Eq.~\eqref{eq:app_distribution} leaves precisely the arclength integral
\begin{equation}
\DD\widehat\FF_I=4\pi\int_{\mathbb H^2}e\delta(w)\,d\mu
=4\pi\int_{\gamma_I}e\,d\sigma=4\pi\XX e.
\end{equation}
Each pair $(\theta,\alpha)$ labels a geodesic on one spatial slice. Hence, the differentiations in $\DD$ compare intervals at the same physical time.
The Lorentzian metric $(d\theta^2-d\alpha^2)/\sin^2\alpha$ on this space of geodesics is the two-dimensional de Sitter metric of unit radius.
Its scalar wave operator $\Box_{\rm dS}$ satisfies $\DD=-(\Box_{\rm dS}+2)$ in the convention of the kinematic Einstein equation \cite{Czech:2016tqr}.
The intervals $(\alpha,\theta)$ and $(\pi-\alpha,\theta+\pi)$ select opposite wedges adjacent to the same chord.
Their $\DD\widehat\FF$ agree because the line integral discards the orientation.
Relative-entropy energy tomography predates the present application \cite{Lin:2014hva}. \\

\noindent
In the Poincar\'e half-plane $\bar h=(dx^2+dz^2)/z^2$, $x\in\mathbb R$ and $z>0$, with a boundary interval centered at $x_0$ and of radius $R>0$, the same statement reads
\begin{equation}
w=\frac{R^2-(x-x_0)^2-z^2}{2Rz},\qquad
\DD=R^2(\partial_R^2-\partial_{x_0}^2)-2.
\end{equation}
The locus $w=0$ is the semicircle $(x-x_0)^2+z^2=R^2$ orthogonal to the boundary.
The region beneath that semicircle has $w>0$ and lies in the interval wedge.
Thus, the delta distribution selects the same geodesic as in the disk description.

\subsection{Reduction to Euclidean Radon Transform}
\label{app:radoninversion}

\noindent
To recover the density on the reflection slice, measure $S_j(\alpha,\theta)=\partial_\lambda^jS_I|_{\lambda=0}$, $j=1,2$, at fixed endpoints $I=(\theta-\alpha,\theta+\alpha)$ for $0<\alpha<\pi$ and $0\leq\theta<2\pi$.
Form $\mathcal Q_I=-\DD S_2-12 S_1^2/c$ from Eq.~\eqref{eq:entropysource} and divide by $\sin\alpha$ to get $J(p,\theta)$ at $p=\cos\alpha$.
For example, at $\theta=0$ the chord is the vertical line $y_1=\cos\alpha$, which moves from the right side of the disk to the left as $\alpha$ increases from zero to $\pi$.
Rotating the center rotates this parallel family.
To reconstruct at $y$, first filter each family's readings in offset and then average their values on the lines through $y$, selected by $p=y\cdot n_\theta$, giving
\begin{equation}
g(y)=\frac1{16\pi^2}\int_0^{2\pi}(\Lambda J)(y\cdot n_\theta,\theta)\,d\theta .
\label{eq:backproj}
\end{equation}
The factor $1/(16\pi^2)$ uses the full angular range, so each unoriented chord is counted twice. \\

\noindent
For $y=pn_\theta+t n_\theta^\perp$ with $|p|<1$, the disk restricts the Euclidean arclength coordinate to $|t|<\sqrt{1-p^2}$.
Writing $r=|y|$, the unit-curvature Klein metric has squared line element $|dy|^2/(1-r^2)+(y\cdot dy)^2/(1-r^2)^2$.
Holding $p$ and $\theta$ fixed gives $|dy|^2=dt^2$ and $y\cdot dy=t\,dt$.
Thus, $d\sigma^2=(1-p^2)dt^2/(1-p^2-t^2)^2$, which separates the chord factor $\sqrt{1-p^2}$ from the position factor $1/(1-r^2)$.
The hyperbolic line integral therefore reduces to a Euclidean Radon transform,
\begin{equation}
\frac{\XX e(p,\theta)}{\sqrt{1-p^2}}=\RR_Eg(p,\theta),
\qquad  g(y)=\frac{e(y)}{1-|y|^2}\quad (|y|<1).
\label{eq:app_Radon}
\end{equation}
The zero extension of $g$ off the disk gives zero line integrals for $|p|\ge1$.
Since $\mathcal Q_I=4\pi\XX e$, the measured function is $J=4\pi\RR_Eg$.
The complementary interval sends $(p,\theta)$ to $(-p,\theta+\pi)$ modulo $2\pi$ and gives the same $J$ at the reversed normal. \\

\noindent
Our convention for the one-dimensional Fourier transform is
\begin{equation}
\widetilde H(\sigma,\theta)=\int_{\mathbb R}\exp(-\mathrm{i}\sigma p)H(p,\theta)\,dp.
\end{equation}
The filter $\Lambda$ multiplies this transform by $|\sigma|$.
The sign function $\operatorname{sgn}(x)$ equals $1$, $0$, or $-1$ according as $x>0$, $x=0$, or $x<0$.
For real $p$, the Hilbert transform of a sufficiently smooth decaying function $B$ is
\begin{equation}
(\HH B)(p)=\frac1\pi\lim_{\epsilon\downarrow0}
\int_{|p-q|>\epsilon}\frac{B(q)}{p-q}\,dq.
\end{equation}
The integration variable $q$ ranges over $\mathbb R$ in this formula.
The symmetric exclusion defines the Cauchy principal value.
The multipliers of $\HH$ and $\partial_p$ are $-\mathrm{i}\,\operatorname{sgn}(\sigma)$ and $\mathrm{i}\sigma$ for the real Fourier frequency $\sigma$, respectively.
Their product is $|\sigma|$, so $\Lambda=\HH\partial_p$.
The Fourier slice identity follows by substituting $y=pn_\theta+t n_\theta^\perp$ in the Fourier transform of $\RR_Eg$.
With $\widehat g(\xi)=\int_{\mathbb R^2}g(y)\exp(-\mathrm{i}\xi\cdot y)\,d^2y$, this change of variables reads
\[
\widetilde{\RR_Eg}(\sigma,\theta)
=\int_{\mathbb R^2}g(pn_\theta+t n_\theta^\perp)\exp(-\mathrm{i}\sigma p)\,dp\,dt
=\widehat g(\sigma n_\theta).
\]
The offset frequency $\sigma\in\mathbb R$ probes the bulk wave vector $\xi=\sigma n_\theta\in\mathbb R^2$.
Sweeping all offsets at one orientation probes a single line in this frequency plane, and rotating the scan through half a turn covers the plane.
Identical complete readings therefore imply identical transforms and hence equal integrable densities almost everywhere, or everywhere in the interior for continuous densities.
With $0\le\theta<\pi$, the Jacobian is $d^2\xi=|\sigma|\,d\sigma\,d\theta$.
The factor $|\sigma|$ supplies the filter, and the inverse transform in $\sigma$ supplies $2\pi$ against the two-dimensional prefactor $(2\pi)^{-2}$, giving \cite{Facchi:2009nls}
\begin{equation}
g(y)=\frac1{2\pi}\int_0^\pi
\Lambda\RR_Eg(y\cdot n_\theta,\theta)\,d\theta
=\frac1{4\pi}\int_0^{2\pi}
\Lambda\RR_Eg(y\cdot n_\theta,\theta)\,d\theta.
\end{equation}
To recover $g(y)$, sum the filtered readings on the lines passing through $y$.
A full circle counts every line twice, so using $J=4\pi\RR_Eg$ gives the normalization $1/(16\pi^2)$ in Eq.~\eqref{eq:backproj}.
For the classes in Eq.~\eqref{eq:app_sourceclasses}, the zero extension of $g$ belongs to $H^2(\mathbb R^2)$, meaning its weak derivatives through order two are square integrable.
Inversion gives its continuous representative in the interior, and the prescribed boundary falloff makes the polygon integral finite. \\

\noindent
The physical quantity recovered pointwise is $e(y)=(1-|y|^2)g(y)$, the coefficient in $T_{\mu\nu}n^\mu n^\nu=\lambda^2e/L^3+o(\lambda^2)$.
Eq.~\eqref{eq:energy-density} gives this coefficient as a quadratic combination of the initial field, its spatial gradient, and its normal derivative.
Determining that combination fixes the proper-energy integral on each polygon, while reconstructing the individual scalar data and their later evolution requires further information. \\

\noindent
Finer offset resolution alone cannot recover spatial variations invisible in every measured direction.
Suppose every $-1<p<1$ is measured at just $m\ge1$ distinct orientations $\theta_j$ modulo $\pi$, $1\le j\le m$.
Choose a smooth real $\psi$ supported on a compact set inside the disk and change the rescaled density by
\[
\delta g(y)=\left[\prod_{j=1}^{m}(n_{\theta_j}^{\perp}\cdot\nabla_y)\right]\psi(y).
\]
The derivative along each measured line is a derivative with respect to its position coordinate $t$.
Because the derivatives commute, choose the direction $n_{\theta_j}^\perp$ last.
Integration along $y=pn_{\theta_j}+t n_{\theta_j}^\perp$ then leaves only the endpoint values of the preceding compactly supported function, both zero.
Thus, $\delta J(p,\theta_j)=0$ at every measured offset.
For example, choose $\psi$ to agree on a small disk with $(a\cdot y)^m$, where the constant vector $a$ satisfies $a\cdot n_{\theta_j}^{\perp}\ne0$ for every $j$.
There, $\delta g=m!\prod_{j=1}^{m}(a\cdot n_{\theta_j}^{\perp})\ne0$, proving that these chord readings leave a genuine ambiguity in the density. \\

\noindent
Even a polygon integral can detect such a change.
For example, measure only the normal $n_0=(1,0)$ and take a target polygon whose interior intersects $\{y_2>0\}$.
Choose a nonnegative, nonzero $\psi$ supported in that intersection, so $\delta g=\partial_{y_2}\psi$ and $\delta e=(1-r^2)\delta g$.
Since $e=(1-r^2)g$ and $d\mu=d^2y/(1-r^2)^{3/2}$, integration by parts gives
\[
\int_{P_0}\delta e\,d\mu
=\int_{P_0}\frac{\partial_{y_2}\psi}{\sqrt{1-r^2}}\,d^2y
=-\int_{P_0}\frac{y_2\psi}{(1-r^2)^{3/2}}\,d^2y<0 .
\]
The same effect is visible if a narrow positive bump in $g$ moves vertically toward the horizontal diameter while staying above it and inside the polygon.
Every vertical scan retains the same line integral, yet the bump's contribution to the polygon decreases because $1/\sqrt{1-r^2}$ decreases at fixed $y_1$.
An ambiguity persists even with measurements in any finite number of orientations and any target polygon $P_0$ with nonempty interior. \\

\noindent
To see this, write $v_j=n_{\theta_j}^\perp$ for $1\le j\le m$, let $D_0$ be the identity, and set $D_j=(v_j\cdot\nabla_y)D_{j-1}$.
With $h(y)=(1-|y|^2)^{-1/2}$ and $\psi$ supported strictly inside $P_0$, the change $\delta g=D_m\psi$ gives
\begin{equation}
\delta T_P=\int_{P_0} h(y)D_m\psi(y)\,d^2y
=(-1)^m\int_{P_0}\psi(y)D_mh(y)\,d^2y .
\label{eq:finite-orientations-polygon}
\end{equation}
Compact support inside $P_0$ eliminates all boundary terms when the derivatives move to $h$.
To change the polygon integral, it suffices to choose $\psi\ge0$ where $D_mh$ has a fixed nonzero sign.
Writing $D_jh=H_j(y)(1-|y|^2)^{-j-1/2}$ gives $H_0=1$ and
\[
H_{j+1}=(1-|y|^2)(v_{j+1}\cdot\nabla_y H_j)
+(2j+1)(v_{j+1}\cdot y)H_j,\qquad 0\le j<m.
\]
Thus, $H_m$ is a polynomial, and at any unit vector $u$ its value is $H_m(u)=\prod_{j=1}^m[(2j-1)(v_j\cdot u)]$.
Only finitely many unit vectors are perpendicular to one of the $v_j$, so $H_m$ is a nonzero polynomial.
A polynomial that vanishes on an open disk has all its Taylor coefficients zero at its center, so $H_m$ cannot vanish throughout the interior of $P_0$.
On a sufficiently small disk about a point where $H_m\ne0$, the function $D_mh$ has a fixed strict sign.
A nonnegative, nonzero smooth $\psi$ supported there makes Eq.~\eqref{eq:finite-orientations-polygon} nonzero while leaving all the measured chord readings unchanged. \\

\noindent
Two nonnegative scalar energy profiles can realize this signed difference.
Choose a smooth cutoff $b$ supported inside $P_0$ that is one on the support of $\delta e=(1-|y|^2)D_m\psi$.
Set the initial field $\phi_1$ to zero and choose the two dimensionless momenta $\sqrt{L}\,\pi_\phi^{(1)}=b+\delta e/2$ and $\sqrt{L}\,\pi_\phi^{(2)}=b-\delta e/2$.
By Eq.~\eqref{eq:energy-density}, $e^{(a)}=L(\pi_\phi^{(a)})^2/2\ge0$ for $a=1,2$, and the difference of the two squares gives $e^{(1)}-e^{(2)}=b\,\delta e=\delta e$.
Their chord observables therefore agree at all the prescribed orientations, although their polygon area Hessians differ by $(24\pi L^2/c)\delta T_P$.
Under the volume comparison of Appendix~\ref{app:globalmaximum}, this also gives the difference of their volume Hessians.
Sec.~\ref{sec:square-calibration} treats a restricted scalar-mode family in which fewer measurements suffice. \\

\noindent
The requirement that every measured interval probe the same density imposes relations between chords of different orientations.
For $g$ integrable on the disk and any integer $k\geq0$, define the angular moment of the entropy source by
\[
\mathcal M_k(\theta)=\int_0^\pi(\cos\alpha)^k\,
\mathcal Q_I(\alpha,\theta)\,d\alpha,
\qquad I=(\theta-\alpha,\theta+\alpha).
\]
Here, the power of the offset is evaluated at $p=\cos\alpha$.
Since $J=\mathcal Q_I/\sin\alpha$, the factor $\sin\alpha$ in $|dp|$ cancels its denominator.
Using $J=4\pi\RR_Eg$ and integrating first along each line yields
\[
\mathcal M_k(\theta)
=\int_{-1}^{1}p^k J(p,\theta)\,dp
=4\pi\int_{|y|<1}(y_1\cos\theta+y_2\sin\theta)^k g(y)\,d^2y .
\]
The offset of the unique line through $y$ with normal $n_\theta$ is $y\cdot n_\theta$.
Since its magnitude is at most one on the disk, absolute integrability of $g$ justifies exchanging the two integrals.
The entire angular dependence of the $k$th moment is therefore contained in $(y_1\cos\theta+y_2\sin\theta)^k$, a homogeneous polynomial of degree $k$ in $\cos\theta$ and $\sin\theta$.
For $k=0$, integrating over offsets counts each point once in every orientation, so this moment must be constant.
For $k=1$, the integral is the projection of $4\pi\int_{|y|<1} y g(y)\,d^2y$ onto $n_\theta$ and hence contains only the first angular harmonic.
For example, the first moments of three scans must satisfy $\mathcal M_1(\pi/4)=[\mathcal M_1(0)+\mathcal M_1(\pi/2)]/\sqrt2$, since the diagonal projection is the sum of the two axial projections divided by $\sqrt2$.
More generally, expressing the sines and cosines in terms of $\exp(\pm\mathrm{i}\theta)$ shows that the angular Fourier coefficient of index $n$ vanishes for $|n|>k$ or when $k-n$ is odd.
These are necessary consistency tests for the unfiltered entropy source and require no positivity assumption on the scalar density. \\

\noindent
The same relations can be expressed directly in terms of the measured entropy derivatives, without differentiating the endpoint data.
Write $S_j=\partial_\lambda^j S|_{0}$ for $j=1,2$ and, for any $2\pi$-periodic function $F(\alpha,\theta)$, define its angular Fourier coefficient by
\[
F_n(\alpha)=\frac{1}{2\pi}\int_0^{2\pi}
\exp(-\mathrm{i} n\theta)F(\alpha,\theta)\,d\theta,
\qquad n\in\mathbb Z .
\]
Square the measured slope $S_1$ before taking $(S_1^2)_n$, since different angular harmonics mix in the product.
Write $\mathcal M_{k,n}$ for the coefficient of $\mathcal M_k$ and $u_k(\alpha)=\sin^2\alpha\cos^k\alpha$ for integer $k\ge0$.
Assume the derivatives in Eq.~\eqref{eq:entropysource} exist in the interior, $S_1,S_2$ are bounded, and $\sin^2\alpha\,\partial_\alpha S_2\to0$ uniformly in $\theta$ at both ends.
Periodicity transfers $\partial_\theta^2$ to the exponential, giving $-n^2$.
Two integrations in $\alpha$ leave $[u_k'S_{2,n}-u_k\partial_\alpha S_{2,n}]_0^\pi=0$, since $u_k$ has double zeros and the derivative of $S_2$ obeys the stated limit.
The moment is therefore
\begin{equation}
\mathcal M_{k,n}=\int_0^\pi
\left\{[2\cos^k\alpha-u_k''(\alpha)-n^2u_k(\alpha)]S_{2,n}(\alpha)
-\frac{12}{c}\cos^k\alpha\,(S_1^2)_n(\alpha)\right\}\,d\alpha .
\label{eq:raw-entropy-angular-moments}
\end{equation}
Thus, whenever $|n|>k$ or $k-n$ is odd, the measured entropy derivatives satisfy one integral relation with zero on the left. \\

\noindent
For $k=0$, $u_0''(\alpha)=2\cos(2\alpha)$, and the coefficient of $S_{2,n}$ reduces to $(4-n^2)\sin^2\alpha$.
For $n=0$, the surviving moment measures $4\pi\int_{|y|<1}g(y)\,d^2y=4\pi\int_{\mathbb H^2}\sqrt{1-|y|^2}\,e(y)\,d\mu$.
The square root follows from $g=e/(1-|y|^2)$ and $d\mu=d^2y/(1-|y|^2)^{3/2}$.
Appendix~\ref{app:disk_energy_sumrule} obtains the full unweighted proper energy by using a different opening-angle weight.
For $n=2$ or $n=-2$, rotational independence of $\mathcal M_0$ forces $\mathcal M_{0,n}=0$, and the $S_{2,n}$ term drops out identically.
In these two harmonics, the condition reduces to $\int_0^\pi (S_1^2)_n(\alpha)\,d\alpha=0$ and involves only the measured first-order response. \\

\noindent
To obtain the polygon response, substitute $e=(1-r^2)g$ and $d\mu=d^2y/(1-r^2)^{3/2}$ into the area law of Sec.~\ref{sec:polygon-area-law}.
This gives $A''(0)/L^2=(24\pi/c)\int_{P_0}g(y)(1-r^2)^{-1/2}\,d^2y$.
When the reflection slice realizes the maximal volume in the prescribed homology class, $V''(0)=A''(0)$ by Appendix~\ref{app:globalmaximum}.
Substituting Eq.~\eqref{eq:backproj} multiplies $24\pi/c$ by $1/(16\pi^2)$ and yields
\begin{equation}
\frac{V''(0)}{L^2}
=\frac{3}{2\pi c}\int_{P_0}\frac{d^2y}{\sqrt{1-r^2}}
\int_0^{2\pi}d\theta\,
\left.\HH\partial_p\left[
\frac{\DD\widehat\FF(p,\theta)}{\sqrt{1-p^2}}\right]\right|_{p=y\cdot n_\theta}.
\label{eq:app_explicitresponse}
\end{equation}
Both the integration domain and the kernel use the known vacuum geometry.
The formula applies to data in the range of the scalar transform, under the differentiability and falloff conditions above.
For exact continuum data, one can reconstruct $e$ once and integrate it over each target polygon.
For a single polygon, Eq.~\eqref{eq:app_polygonfilter} in Appendix~\ref{app:detailC4} puts its geometry into a known weight and computes the response directly.
Because the filter multiplies offset-frequency errors by $|\sigma|$, finite-precision reconstruction requires the smoothing and error estimates in Appendices~\ref{app:ordinary_entropy_weights}--\ref{app:continuum_stability}.

\subsection{Polygon Kernel in Interval Parameters}
\label{app:detailC4}

\noindent
For a single polygon, its sides fix how much of each line contributes to the integral.
In Eq.~\eqref{eq:app_explicitresponse}, the filtered reading is constant along a line, so it factors out of the integral over the segment inside the polygon.
For $|p|<1$, these two factors are
\begin{align}
B_\theta(p)&=\HH\partial_p\!
\left[\frac{\DD\widehat\FF(p,\theta)}{\sqrt{1-p^2}}\right],\\
W_P(p,\theta)&=\int_{\{t:\;pn_\theta+t n_\theta^\perp\in P_0\}}
\frac{dt}{\sqrt{1-p^2-t^2}} .
\label{eq:app_weightdef}
\end{align}
First, extend the divided data by zero outside the disk's offset range and apply the Hilbert filter to obtain $B_\theta$.
The polygon weight $W_P$ in Eq.~\eqref{eq:app_weightdef} integrates $\mathbf1_{P_0}(y)/\sqrt{1-|y|^2}$ along each line, with the integrand zero outside the disk.
At fixed $\theta$, rotation sends $d^2y$ to $dp\,dt$, where $p$ selects the line and $t$ runs along it.
Integrating first over $t$ in Eq.~\eqref{eq:app_explicitresponse} gives
\begin{equation}
\frac{V''(0)}{L^2}=\frac{3}{2\pi c}
\int_0^{2\pi}\!d\theta\int_{-1}^{1}\!dp\,
W_P(p,\theta)B_\theta(p).
\label{eq:app_polygonfilter}
\end{equation}
At each orientation, first filter the whole scan in $p$, then weight it by the polygon.
A line missing the polygon has zero weight in the final sum, but its raw reading enters the filter at other offsets. \\

\noindent
The RT pairing determines the sides of each convex Klein component.
The component is the intersection of the oriented half-planes $y\cdot m_j\leq b_j$ with unit outward normals $m_j$ and signed offsets $b_j$, $1\leq j\leq J$, where $J$ counts the component's sides.
The ideal endpoints of each side determine its chord and its inside half-plane uniquely once the component is specified.
At fixed $(p,\theta)$, the disk requires $-\sqrt{1-p^2}<t<\sqrt{1-p^2}$.
Each side imposes the additional inequality
\begin{equation}
(m_j\cdot n_\theta^\perp)t\leq b_j-p(m_j\cdot n_\theta).
\end{equation}
Positive and negative coefficients set upper and lower bounds on $t$, respectively.
A zero coefficient leaves all $t$ admissible if the right side is nonnegative and leaves none otherwise.
If the resulting chord segment is nonempty with endpoints $t_-\leq t_+$, its contribution is
\begin{equation}
W_{P,\mathrm{component}}(p,\theta)
=\arcsin\!\frac{t_+}{\sqrt{1-p^2}}
-\arcsin\!\frac{t_-}{\sqrt{1-p^2}}.
\end{equation}
Here, $\arcsin$ is the inverse sine on $[-1,1]$ with values in $[-\pi/2,\pi/2]$.
An empty intersection contributes zero.
For a union of components, split the chord intersection into disjoint segments and add their contributions.
The total is bounded by the weight of the whole chord, $\arcsin 1-\arcsin(-1)=\pi$, so $0\le W_P\le\pi$.
Taking the entire disk as the target and integrating over the line orientation gives the full-disk sum rule below.
The values assigned to lines that coincide with an edge do not affect Eq.~\eqref{eq:app_polygonfilter}. \\

\noindent
Both classes in Eq.~\eqref{eq:app_sourceclasses} give a zero extension of $g$ with Lipschitz first derivatives.
For the second class, $g=(1-r^2)^2a_0$ and its first derivatives vanish on the unit circle, and its piecewise second derivatives are bounded.
For the first class, compact support removes the boundary terms.
Since $\DD\widehat\FF/\sqrt{1-p^2}=4\pi\RR_Eg$, differentiating its chord integral over $-1\le t\le1$ gives a derivative supported in $[-1,1]$ and uniformly Lipschitz in $p$.
Near the principal-value pole, subtract its value at the pole to bound the difference quotient by the Lipschitz constant.
For $|p-q|\ge1$, the absolute integral of the derivative bounds the remaining contribution.
Thus, $B_\theta$ is uniformly bounded on $|p|\le1$, and $\int_{|y|<1}(1-|y|^2)^{-1/2}d^2y=2\pi$ justifies exchanging integrals in Eq.~\eqref{eq:app_polygonfilter}. \\

\noindent
The ordinary polygon integral also converges for both lowest scalar phases at every $\Delta>1$, including fractional dimensions outside Eq.~\eqref{eq:app_sourceclasses}.
Their densities in Appendix~\ref{app:kg} are combinations of $(1-r^2)^\Delta$ and $(1-r^2)^{\Delta+1}$.
Eq.~\eqref{eq:powerchords} therefore expresses $J$ as a combination of $(1-p^2)_+^{\Delta-1/2}$ and $(1-p^2)_+^{\Delta+1/2}$, with $x_+=\max(x,0)$.
Both vanish continuously at $p=\pm1$, so differentiating their zero extensions produces no endpoint delta functions.
Near each endpoint, $|\partial_pJ|^2=O((1-|p|)^{2\Delta-3})$ is integrable for $\Delta>1$.
The Hilbert multiplier has unit modulus away from zero, so $\Lambda J=\HH\partial_pJ$ is also square integrable, and its Poisson regularization converges in $L^2$.
Since $0\le W_P\le\pi$ on the finite offset and angular domain, the pairing converges absolutely to the same value as in Appendix~\ref{app:ordinary_entropy_weights}. \\

\noindent
The lowest momentum mode of a massless scalar provides a case in which the angular average can be evaluated for every bulk point.
At scaling dimension $\Delta=2$, its quadratic density is $e(r)=E(1-r^2)^3$, with $E=e(0)=2a^2$ in the amplitude convention of Appendix~\ref{app:chordphasegeneral}.
Thus, $g(r)=E(1-r^2)^2$ for $r<1$ and is zero elsewhere.
For a chord of Euclidean half-length $s=\sqrt{1-p^2}$, its line integral is $E\int_{-s}^{s}(s^2-t^2)^2dt=16Es^5/15$.
The measured data are consequently $J(p)=(64\pi E/15)(1-p^2)_+^{5/2}$, independent of $\theta$. \\

\noindent
The filter can be computed on $|p|<1$ from $w(q)=\sqrt{(1-q^2)_+}$.
With $q=\cos u$, the numerator of the Hilbert integral is $\sin^2 u=(1-p^2)+(p-\cos u)(p+\cos u)$.
The substitution $v=\tan(u/2)$ gives $\operatorname{pv}\int_0^\pi du/(p-\cos u)=0$, hence $(\HH w)(p)=p$.
The moments $\int_{-1}^1w(q)dq=\pi/2$ and $\int_{-1}^1q^2 w(q)dq=\pi/8$ fix the remaining terms.
Polynomial division in the Hilbert integral gives $\HH[qw](p)=p^2-1/2$ and $\HH[q^3w](p)=p^4-p^2/2-1/8$.
Since $J'(q)=-(64\pi E/3)(q-q^3)w(q)$, the filtered data on the interior are
\[
(\Lambda J)(p)=8\pi E\left(1-4p^2+\frac83p^4\right),\qquad |p|<1 .
\]
The remaining angular average involves only the zeroth, second, and fourth powers of the chord offset. \\

\noindent
At radius $r<1$, a rotation lets us write $p=y\cdot n_\theta=r\cos\theta$.
The averages of $p^2$ and $p^4$ over $0\le\theta<2\pi$ are $r^2/2$ and $3r^4/8$, respectively.
The mean of $\Lambda J$ is therefore $8\pi E(1-r^2)^2$, and Eq.~\eqref{eq:backproj} recovers $g(r)=E(1-r^2)^2$.
In particular, the filtered data are negative at $p^2=1/2$ even though $e(r)>0$ throughout the disk.
The angular integral restores the positive density, giving $V''(0)/L^2=(24\pi E/c)\int_{P_0}(1-r^2)^{3/2}d^2y$ for a finite-area polygon whenever the reflection area realizes its volume. \\

\noindent
For the $\Delta=3/2$ field phase, this filter can be written in elementary functions.
For $|p|<1$, $J_+=3\pi\beta^2(-1/8+5p^2/4-9p^4/8)$ and $J_+=0$ outside.
For a polynomial $P$, the symmetric Cauchy principal value satisfies
\[
\operatorname{pv}\!\int_{-1}^1\!\frac{P(t)}{p-t}\,dt
=P(p)\ln\frac{1+p}{1-p}
+\int_{-1}^1\!\frac{P(t)-P(p)}{p-t}\,dt.
\]
The second integrand is a polynomial.
With $P(t)=(5t-9t^3)/2$, this gives
\[
(\Lambda J_+)(p)=3\beta^2\!\left[
9p^2-2+\frac{5p-9p^3}{2}\ln\frac{1+p}{1-p}\right].
\]
The logarithmic endpoint singularities are integrable.
At the center, Eq.~\eqref{eq:backproj} samples $p=0$ in every direction and gives $e_+(0)=-3\beta^2/(4\pi)=-3a^2/8$.
The positive semicircle entropy Hessian and the negative central density therefore follow from the same boundary data. \\

\noindent
The same backprojection recovers the entire field-phase profile directly.
For $0<r<1$, we write $z=\sqrt{1-r^2}$, $k=r/(1+z)$, and $p=r\cos\theta$ after rotating the point to the horizontal axis.
Since $r=2k/(1+k^2)$, the ratios $1\pm r\cos\theta=|1\pm k\exp(\mathrm{i}\theta)|^2/(1+k^2)$ give the uniformly convergent series
\[
\ln\frac{1+r\cos\theta}{1-r\cos\theta}
=4\sum_{j=0}^{\infty}\frac{k^{2j+1}}{2j+1}\cos[(2j+1)\theta].
\]
Here, $\langle\cdot\rangle$ denotes the angular average $(2\pi)^{-1}\int_0^{2\pi}(\cdot)\,d\theta$.
With $U=\ln[(1+p)/(1-p)]$, elementary cosine orthogonality and $4\cos^3\theta=3\cos\theta+\cos3\theta$ yield
\[
\langle p^2\rangle=r^2/2,\qquad
\langle pU\rangle=2rk,\qquad
\langle p^3U\rangle=\tfrac12r^3(3k+k^3/3).
\]
Thus, $\langle\Lambda J_+\rangle/(3\beta^2)=9r^2/2-2+5rk-9r^3(3k+k^3/3)/4$.
The identities $rk=1-z$ and $r^2=1-z^2$ reduce this expression to $2z(3r^2-1)$.
Eq.~\eqref{eq:backproj} now gives $g(r)=3\beta^2(3r^2-1)z/(4\pi)$ and $e_+(r)=3a^2(3r^2-1)(1-r^2)^{3/2}/8$.
The formulas extend continuously to $r=0$.

\subsubsection{Sum Rule for Integrated Normal Energy}
\label{app:disk_energy_sumrule}

\noindent
For the entire disk, angular averaging gives every point its proper area weight.
Consequently, if $\int_{\mathbb H^2}|e|\,d\mu<\infty$, a positive-weight average of the chord readings $\mathcal Q_I=4\pi(\XX e)(I)$ gives the total normal energy.
If the entropy variations also satisfy the endpoint conditions specified below, this energy can be written directly in terms of $S_1$ and $S_2$, giving
\begin{equation}
\int_{\mathbb H^2}e\,d\mu
=\frac{1}{8\pi^2}\int_0^{2\pi}\!d\theta\int_0^\pi
\frac{\mathcal Q_I}{\sin^2\alpha}\,d\alpha
=\frac{1}{4\pi^2}\int_0^{2\pi}\!d\theta\int_0^\pi
\frac{S_2-6S_1^2/c}{\sin^2\alpha}\,d\alpha .
\label{eq:whole_disk_sumrule}
\end{equation}
Indeed, $p=\cos\alpha$ and $y=pn_\theta+t n_\theta^\perp$ give $dp\,dt=d^2y$.
At a fixed center, the Klein arclength formula yields
\[
\int_0^\pi\frac{(\XX e)(I)}{\sin^2\alpha}\,d\alpha
=\int_{|y|<1}\frac{e(y)\,d^2y}{(1-|y|^2)[1-(y\cdot n_\theta)^2]} .
\]
For the angular integral, rotate $y$ onto the horizontal axis, set $r=|y|$, use fourfold symmetry, and substitute $t=\tan\theta$ in the first quadrant.
This gives $\int_0^{2\pi}[1-(y\cdot n_\theta)^2]^{-1}d\theta=4\int_0^\infty dt/(t^2+1-r^2)=2\pi/\sqrt{1-r^2}$.
Consequently, the integral of $(\XX e)(I)/\sin^2\alpha$ over $\theta,\alpha$ equals $2\pi\int e\,d\mu$.
The same calculation for $|e|$ justifies changing integration order for signed densities under the stated absolute-integrability hypothesis. \\

\noindent
The polygon window gives a second way to see why this full-disk weight is positive.
For the whole unit disk $\mathbb D$, the chord endpoints are $t_\pm=\pm\sqrt{1-p^2}$, so Eq.~\eqref{eq:app_weightdef} gives $W_{\mathbb D}(p,\theta)=\pi\mathbf 1_{(-1,1)}(p)$ almost everywhere.
Its zero extension matters because the jump at each end of the offset interval contributes to the filter.
Let $\Lambda_\delta$ be the regularized filter of Appendix~\ref{app:ordinary_entropy_weights}, with multiplier $|\sigma|\exp(-\delta|\sigma|)$ for smoothing scale $\delta>0$.
Its convolution kernel is the derivative of $z/[\pi(\delta^2+z^2)]$ with respect to $z$.
Integration over the offsets from $-1$ to $1$ therefore gives, for $|p|<1$,
\[
(\Lambda_\delta W_{\mathbb D})(p,\theta)
=\frac{1+p}{\delta^2+(1+p)^2}
+\frac{1-p}{\delta^2+(1-p)^2}
\ \xrightarrow[\delta\downarrow0]{}\ \frac{2}{1-p^2}.
\]
Both terms are positive, and their sum for every $\delta>0$ is bounded by its limit.
The calculation above applied to $|e|$ bounds $\int_0^{2\pi}d\theta\int_{-1}^1 |J(p,\theta)|\,dp/(1-p^2)$ by $8\pi^2\int_{\mathbb H^2}|e|\,d\mu$.
Dominated convergence thus permits the limit in the pairing $(16\pi^2)^{-1}\int_0^{2\pi}d\theta\int_{-1}^1 J(\Lambda_\delta W_{\mathbb D})\,dp$.
Substituting $p=\cos\alpha$ and $J=\mathcal Q_I/\sin\alpha$ gives exactly the first expression in Eq.~\eqref{eq:whole_disk_sumrule}.
This argument works even though the full-disk position weight $1/\sqrt{1-r^2}$ has a divergent square norm, so the finite-polygon square-norm estimate does not directly apply. \\

\noindent
For the last equality in Eq.~\eqref{eq:whole_disk_sumrule}, we assume the absolute integrability of $S_2/\sin^2\alpha$ and $S_1^2/\sin^2\alpha$ in the measure $d\theta\,d\alpha$ and that $\partial_\alpha S_2\to0$ uniformly at $\alpha=0,\pi$.
On each compact subinterval of $0<\alpha<\pi$, the entropy variations have the two continuous endpoint derivatives used in Eq.~\eqref{eq:entropysource}.
Indeed, Eq.~\eqref{eq:entropysource} reads
\[
\frac{\mathcal Q_I}{\sin^2\alpha}
=(\partial_\theta^2-\partial_\alpha^2)S_2+
\frac{2S_2-12S_1^2/c}{\sin^2\alpha}.
\]
The angular derivative integrates to zero by periodicity.
On $\delta<\alpha<\pi-\delta$, the other derivative integrates to $[\partial_\alpha S_2]_{\pi-\delta}^{\delta}$, which vanishes uniformly as $\delta\downarrow0$.
For the lowest scalar modes with $\Delta>1$, Eq.~\eqref{eq:cftelementaryS} gives $S_2=O(\alpha^2)$ and $\partial_\alpha S_2=O(\alpha)$ at a small arc.
Purity gives the corresponding estimates at the other endpoint.
If $S_1=0$ and every ordinary interval entropy Hessian is positive, the full-disk sum has a positive integrand after the derivative terms have been integrated away.
The integrated normal energy is therefore positive at quadratic order.
The same entropies can give a negative response for a smaller polygon because its reconstruction uses the weight for that polygon. \\

\noindent
The factor $1/\sin^2\alpha$ in Eq.~\eqref{eq:whole_disk_sumrule} makes the angular cutoff relevant when the entropies have finite precision.
Write $Z_I=S_2-6S_1^2/c$ for the entropy combination before weighting, and let $E_{\mathbb D}=\int_{\mathbb H^2}e\,d\mu$.
Assume that $|Z_I|\le M\sin^2\alpha$ uniformly in $\theta$ for $0<\min(\alpha,\pi-\alpha)<\rho_0$, with fixed $M>0$ and $0<\rho_0<\pi/2$.
This condition holds, for example, when both variations vanish quadratically at the two ends of the opening range.
On the retained range $\rho\le\alpha\le\pi-\rho$, let $|\widetilde S_j-S_j|\le\sigma_j$ for $j=1,2$, where $0<\rho<\rho_0$.
Factoring the difference of the two squares bounds $|\widetilde Z_I-Z_I|$ by $\epsilon=\sigma_2+(6/c)\sigma_1(2\|\widetilde S_1\|_\infty+\sigma_1)$, where $\widetilde Z_I=\widetilde S_2-6\widetilde S_1^2/c$ and the supremum is over the retained intervals.
The two omitted strips have total width $2\rho$, while the integral of $1/\sin^2\alpha$ over the retained openings is $2\cot\rho$.
Integrating over the full $2\pi$ range of centers therefore gives the estimate
\begin{equation}
\widehat E_\rho=\frac{1}{4\pi^2}\int_0^{2\pi}\!d\theta
\int_\rho^{\pi-\rho}\!\frac{\widetilde Z_I}{\sin^2\alpha}\,d\alpha,
\qquad
|\widehat E_\rho-E_{\mathbb D}|\le\frac{M\rho+\epsilon\cot\rho}{\pi}.
\label{eq:disk-energy-finite-precision}
\end{equation}
If the data have a common error bound $\epsilon>0$ at the chosen openings and $\sqrt{\epsilon/M}<\rho_0$, then $\rho=\sqrt{\epsilon/M}$ and $\cot\rho\le1/\rho$ bound the total error by $2\sqrt{M\epsilon}/\pi$.
The term $M\rho/\pi$ in Eq.~\eqref{eq:disk-energy-finite-precision} bounds the missing true signal, and the term $\epsilon\cot\rho/\pi$ bounds the error on the measured arcs. \\

\noindent
At $\Delta=3/2$, the field-phase integrand is $(3a^2/8)(3r^2-1)d^2y$, so its full-disk integral is $3\pi a^2/16$ and its equal-arc triangle integral is $-9\sqrt3a^2/128$.
The boundary evaluation uses the integral in Eq.~\eqref{eq:positive32}.
The half-angle substitution gives $\int_0^\pi[1+\sqrt{1-z^2}\sin\alpha]^{-1}d\alpha=2\arcsin|z|/|z|$ for $0<|z|\le1$, with its continuous limit at zero.
Consequently, $\int_0^\pi S_{2,+}\,d\alpha/(\beta^2\sin^2\alpha)=12\int_0^1 z\arcsin z\,dz-3\pi/4=3\pi/4$.
The sum rule gives $3\beta^2/8=3\pi a^2/16$ for the integrated normal energy.
A state-independent nonnegative weight on ordinary $S_2$ cannot reproduce the response of that triangle on the whole class of scalar states.
Eq.~\eqref{eq:whole_disk_sumrule} concerns a finite integral of the normal density.
A volume comparison for the entire asymptotically AdS slice also requires a specified regulator and vacuum subtraction.

\subsection{Poisson Filter for Ordinary Entropy Variations}
\label{app:ordinary_entropy_weights}

\noindent
At finite resolution, the polygon still specifies the target through its known weight, while smoothing controls how finely the density is resolved near its sides.
Moving the endpoint derivatives to this weight leaves only the measured state derivatives $S_1,S_2$.
The Poisson construction converges for square-integrable zero-extended density, while the Gaussian construction below bounds the bias under H\"older continuity.
The Poisson filter $\Lambda_\delta$ multiplies each frequency by $|\sigma|\exp(-\delta|\sigma|)$ for $\delta>0$.
Since the inverse transform of $\exp(-\delta|\sigma|)$ is $\delta/[\pi(\delta^2+s^2)]$ for $s\in\mathbb R$, differentiating in $\delta$ gives the kernel
\begin{equation}
k_\delta(s)=-\partial_\delta\frac{\delta}{\pi(\delta^2+s^2)}
=\frac{\delta^2-s^2}{\pi(\delta^2+s^2)^2}.
\end{equation}
Because this convolution kernel is real and even, the filter in Eq.~\eqref{eq:app_polygonfilter} can act on the polygon weight inside the pairing.
The known function $h_P(y)=\mathbf1_{P_0}(y)/\sqrt{1-|y|^2}$ equals $1/\sqrt{1-|y|^2}$ on the polygon and zero elsewhere, including outside the disk.
Its line integrals $W_P=\RR_Eh_P$ determine the two polygon weights,
\begin{align}
H_\delta(\alpha,\theta)
&=(\Lambda_\delta W_P)(\cos\alpha,\theta)
=\int_{P_0}\frac{k_\delta(\cos\alpha-y\cdot n_\theta)}
{\sqrt{1-|y|^2}}\,d^2y,\notag\\
B_\delta(\alpha,\theta)
&=2H_\delta+
(\partial_\theta^2-\partial_\alpha^2)
(\sin^2\alpha\, H_\delta).
\label{eq:polygon_entropy_weights}
\end{align}
At fixed $\delta>0$, bounded kernel derivatives and integrable $h_P$ make both weights smooth through $\alpha=0,\pi$.
They give the regularized volume response
\begin{equation}
T_\delta=\frac{3}{2\pi c}\int_0^{2\pi}\!d\theta\int_0^\pi\!d\alpha\,
\left[B_\delta S_2-\frac{12}{c}H_\delta S_1^2\right],
\qquad \frac{V''(0)}{L^2}=\lim_{\delta\downarrow0}T_\delta.
\label{eq:polygon_entropy_average}
\end{equation}
Changing from the line offset $p=\cos\alpha$ to the interval half-opening $\alpha$ gives $|dp|=\sin\alpha\,d\alpha$, canceling the denominator $\sqrt{1-p^2}$ of the normalized source.
The remaining integrand is $H_\delta\mathcal Q_I$, with $\mathcal Q_I=\sin^2\alpha(\partial_\theta^2-\partial_\alpha^2)S_2+2S_2-12S_1^2/c$.
Integrating each second derivative by parts twice yields the known function $\sin^2\alpha\,H_\delta$ and gives Eq.~\eqref{eq:polygon_entropy_average}.
Periodic interval centers cancel the $\theta$ boundary terms.
At $\alpha=0,\pi$, both $\sin^2\alpha\,H_\delta$ and its first opening-angle derivative vanish.
It suffices that $S_1,S_2$ be bounded and $\sin^2\alpha\,\partial_\alpha S_2\to0$ uniformly in $\theta$ at both ends, as in the two pure scalar examples.
The measured quantities are thus the two state derivatives of ordinary interval entropy, with no endpoint differences between measurements. \\

\noindent
Extend $g=e/(1-|y|^2)$ by zero outside the disk, so $J(p,\theta)=\mathcal Q_I/\sqrt{1-p^2}=4\pi\RR_E g$ for each chord.
By the Fourier slice identity, the offset frequency $\sigma$ measures the bulk wave vector $\boldsymbol\xi=\sigma n_\theta$.
Multiplying the data by $\exp(-\delta|\sigma|)$ in frequency space therefore multiplies each bulk component by $\exp(-\delta|\boldsymbol\xi|)$, equivalent to averaging with the following positive kernel of unit integral on the Klein plane,
\[
P_\delta^{(2)}(y)=\frac{\delta}{2\pi(\delta^2+|y|^2)^{3/2}},
\qquad y\in\mathbb R^2 .
\]
Along a line of offset $p$, put $A=\delta^2+p^2>0$.
The identity $\partial_t[t/(A\sqrt{A+t^2})]=(A+t^2)^{-3/2}$ gives
\[
(\RR_E P_\delta^{(2)})(p,\theta)
=\frac{\delta}{2\pi}\int_{\mathbb R}\frac{dt}{(A+t^2)^{3/2}}
=\frac{\delta}{\pi(\delta^2+p^2)} .
\]
The Fourier transform in offset is $\exp(-\delta|\sigma|)$, so the Fourier slice identity gives $g_\delta=P_\delta^{(2)}*g$, with $*$ denoting convolution on $\mathbb R^2$.
Consequently, $T_\delta=(24\pi/c)\int_{\mathbb R^2}h_P g_\delta\,d^2y$ integrates the averaged density over the original polygon. \\

\noindent
Smoothing allows matter just outside the polygon to affect its reading.
Suppose a bounded $g$ is supported at Euclidean distance at least $\ell>0$ from $P_0$, so the exact polygon energy vanishes.
At every $y\in P_0$, only displacements of length at least $\ell$ can contribute.
Their total kernel mass is $\int_\ell^\infty\delta r(\delta^2+r^2)^{-3/2}dr=\delta/\sqrt{\delta^2+\ell^2}$, giving
\[
|T_\delta|\leq\frac{24\pi}{c}\|h_P\|_1\|g\|_\infty
\frac{\delta}{\sqrt{\delta^2+\ell^2}} .
\]
Here, $\|h_P\|_1$ uses Euclidean area, and the bound tends to zero at fixed separation as $\delta\downarrow0$.
For general $g\in L^2(\mathbb R^2)$, the Fourier multiplier tends to one and is bounded by one, so Fourier norm conservation gives $\|g_\delta-g\|_2\to0$.
Since $\|h_P\|_2^2\le\int_{P_0}d\mu<\infty$, Cauchy--Schwarz gives convergence for this particular polygon even when the source crosses its sides.
For the source classes in Eq.~\eqref{eq:app_sourceclasses}, applying $1-\exp(-t)\le t$ for $t\ge0$ in Fourier space gives
\begin{equation}
\left|T_\delta-\frac{V''(0)}{L^2}\right|
\leq\frac{24\pi}{c}\,\delta\,\|h_P\|_2\,\|\nabla g\|_2,
\end{equation}
where $\nabla$ is the Euclidean gradient.
For exact continuum data, a bound on $\|\nabla g\|_2$ thus gives a numerical bias bound at a chosen resolution.

\subsection{Gaussian Filter and Errors in Ordinary Entropies}
\label{app:gaussianweights}

\noindent
Even with exact interval readings, Gaussian smoothing changes the polygon integral by averaging the density over a finite neighborhood.
This smoothing bias and the error from imperfect readings have different bounds.
Use the Fourier conventions of Appendix~\ref{app:radoninversion} and set $s=\sin\alpha$, $p=\cos\alpha$, and
\[
g_\varepsilon^{\rm heat}(y)=\frac{\exp[-|y|^2/(4\varepsilon)]}{4\pi\varepsilon},\qquad
k_\varepsilon(z)=\frac1{2\pi}\int_{\mathbb R}|\sigma|\exp(-\varepsilon\sigma^2)
\cos(\sigma z)\,d\sigma,\qquad \varepsilon>0 .
\]
The normalization $1/(2\pi)$ in the one-dimensional inverse transform makes $k_\varepsilon$ the convolution kernel for the filtered Gaussian with multiplier $|\sigma|\exp(-\varepsilon\sigma^2)$.
The elementary sine transform of the Gaussian yields
\begin{equation}
k_\varepsilon(z)=\frac{1-2u\mathcal D_{\rm Daw}(u)}{2\pi\varepsilon},
\qquad u=\frac{z}{2\sqrt\varepsilon},\qquad
\mathcal D_{\rm Daw}(u)=\exp(-u^2)\int_0^u\exp(v^2)\,dv .
\end{equation}
To see the sine-transform identity, put $f(u)=\int_0^\infty\exp(-v^2)\sin(2uv)\,dv$.
Integration by parts gives $f'(u)=1-2uf(u)$ and $f(0)=0$, so $f=\mathcal D_{\rm Daw}$, Dawson's integral.
The cosine transform above is proportional to $f'(u)$. \\

\noindent
The Gaussian kernel is real and even, so convolution can act on the polygon window to give a known weight $H_\varepsilon$.
Two integrations by parts move the endpoint derivatives from the measured entropy onto this weight, giving the operator $\mathcal B$ below.
\begin{align}
H_\varepsilon(\alpha,\theta)
&=\int_{-1}^1 W_P(\zeta,\theta)k_\varepsilon(\zeta-p)\,d\zeta
=\int_{P_0}\frac{k_\varepsilon(y\cdot n_\theta-p)}{\sqrt{1-|y|^2}}\,d^2y ,
\nonumber\\
\mathcal B H
&=2H+s^2 H_{\theta\theta}-(s^2 H)_{\alpha\alpha}.
\label{eq:kinematic_adjoint}
\end{align}
At fixed $\varepsilon$, the kernel and its required derivatives are bounded, and periodicity cancels the angular boundary terms.
To see what the opening-angle endpoints require, put $H=H_\varepsilon$, choose $0<\rho<\pi/2$, and integrate twice on $[\rho,\pi-\rho]$, obtaining
\[
-\int_\rho^{\pi-\rho}s^2 H\,\partial_\alpha^2 S_2\,d\alpha
=-\int_\rho^{\pi-\rho}S_2\,\partial_\alpha^2(s^2H)\,d\alpha
+\left[S_2\,\partial_\alpha(s^2H)-s^2H\,\partial_\alpha S_2\right]_{\alpha=\rho}^{\alpha=\pi-\rho}.
\]
At distance $b=\min(\alpha,\pi-\alpha)$ from an endpoint, boundedness of $H$ and its derivatives gives $s^2H=O(b^2)$ and $\partial_\alpha(s^2H)=O(b)$ uniformly in $\theta$.
The first boundary term therefore vanishes for bounded $S_2$, and the second vanishes if $s^2\partial_\alpha S_2\to0$ uniformly as $b\downarrow0$.
Once these conditions hold for the true entropy, only values of $S_1$ and $S_2$ enter the measured integral.
Using $G=\RR_Eg$, $W_P=\RR_E h_P$, and $J=4\pi G$ now gives three equivalent expressions,
\begin{align}
\mathcal V_\varepsilon
&=\frac{24\pi}{c}\int_{\mathbb R^2} h_P(y)
(g_\varepsilon^{\rm heat}*g)(y)\,d^2y\nonumber\\
&=\frac{6}{\pi c}\int_0^\pi\!d\theta\int_{\mathbb R}\!d\sigma\,
|\sigma|\exp(-\varepsilon\sigma^2)
\widetilde G(\sigma,\theta)\overline{\widetilde W_P(\sigma,\theta)}
\nonumber\\
&=\frac{3}{2\pi c}\int_0^{2\pi}\!d\theta\int_0^\pi\!d\alpha
\left[S_2\mathcal B H_\varepsilon-\frac{12}{c}S_1^2 H_\varepsilon\right].
\label{eq:volumedual}
\end{align}
The star denotes convolution on the Euclidean plane, so $(a*b)(y)=\int_{\mathbb R^2}a(y-y')b(y')\,d^2y'$.
The first equality follows by two-dimensional Fourier inversion.
In the second line, signed frequency and half the angular range cover the Fourier plane once, with Jacobian $|\sigma|$.
For the last line, the full angular range covers each unoriented chord twice and $|dp|=s\,d\alpha$ cancels the denominator in $G=I_\gamma/s$.
As $\varepsilon\downarrow0$, a continuous zero extension $g=O[(1-|y|^2)^{\Delta-1}]$ with $\Delta>1$ converges uniformly under this smoothing.
Since $h_P$ is integrable, $\mathcal V_\varepsilon\to V''(0)/L^2$ under the volume hypotheses. \\

\noindent
Suppose $\|\widetilde S_j-S_j\|_\infty\le\sigma_j$ for $j=1,2$, with $\widetilde{\mathcal V}_\varepsilon$ formed using these data.
Then, $|\widetilde S_1^2-S_1^2|\le \sigma_{\rm sq}=\sigma_1(2\|\widetilde S_1\|_\infty+\sigma_1)$.
If $|g(y)-g(y')|\le H_\nu |y-y'|^\nu$ on the full plane for $0<\nu\le1$, subtraction in the last and first lines of Eq.~\eqref{eq:volumedual}, respectively, gives
\begin{align}
|\widetilde{\mathcal V}_\varepsilon-\mathcal V_\varepsilon|
&\le\frac{3}{2\pi c}\left[\sigma_2\|\mathcal B H_\varepsilon\|_1+
\frac{12}{c}\sigma_{\rm sq}\|H_\varepsilon\|_1\right],
\label{eq:reconstruction_noise}\\
\left|\mathcal V_\varepsilon-\frac{V''(0)}{L^2}\right|
&\le\frac{24\pi}{c}\,\|h_P\|_1 H_\nu(4\varepsilon)^{\nu/2}.
\label{eq:smoothingbias}
\end{align}
The first line propagates reading errors through the absolute weights of this polygon at fixed resolution.
The second compares its exact and smoothed integrals.
A displacement of length $r$ changes the density by at most $H_\nu r^\nu$.
The Gaussian has $\langle r^2\rangle=4\varepsilon$, so Jensen's inequality gives $\langle r^\nu\rangle\le(4\varepsilon)^{\nu/2}$, and integration over the target supplies its Euclidean norm $\|h_P\|_1$. \\

\noindent
One can estimate the effective filter order by differentiating before doing the polygon integral.
Set $\zeta=y\cdot n_\theta$, $t=y\cdot n_\theta^\perp$, $d=\sqrt{1-|y|^2}$, $z=\zeta-p$, and temporarily write $k=k_\varepsilon(z)$, with primes denoting derivatives in $z$.
The identities $\zeta_\theta=t$, $t_\theta=-\zeta$, and $p_\alpha=-s$ give
\begin{equation}
\mathcal B H_\varepsilon
=s^2\int_{P_0}\big[4k-(\zeta+5p)k'+(t^2-s^2)k''\big]\frac{d^2y}{d}.
\label{eq:gaussian_differentiated_weight}
\end{equation}
This formula also quantifies the loss from omitting short arcs in the last line of Eq.~\eqref{eq:volumedual}.
For fixed $\varepsilon>0$, the expression inside the integral is uniformly bounded, and $\int_{P_0}d^2y/d$ is finite, giving $\mathcal B H_\varepsilon=O(b^2)$ as $b\downarrow0$.
If $S_j=O(b^2)$ uniformly for $j=1,2$, both terms in the entropy pairing are $O(b^4)$, since $H_\varepsilon$ is bounded.
The integrated error from omitting both strips $b<\rho$ is therefore $O(\rho^5)$ as $\rho\downarrow0$ at fixed smoothing.
For pure states, complementarity relates the two strips.
Controlling the dependence on $\varepsilon$ additionally requires an estimate for the term with $k''$ near the disk boundary.
Using $t^2-s^2=-d^2-(\zeta+p)z$ reduces it to $-d k''$ and a multiple of $zk''/d$.
For the first term, integration by parts in the direction $n_\theta$ gives
\[
-\int_{P_0}d\,k''\,d^2y
=-\int_{\partial P_0}d\,k'(n_{\rm out}\cdot n_\theta)\,d\ell_E
-\int_{P_0}\frac{\zeta}{d}k'\,d^2y .
\]
Here, $n_{\rm out}$ is the Euclidean outward normal and $d\ell_E$ is Euclidean arclength.
For an ideal vertex $v$, put $x=1-y\cdot v$ and $t_v=y\cdot v^\perp$, where $v^\perp$ is a unit vector orthogonal to $v$ in the Klein plane.
The two straight sides meeting at $v$ have equations $t_v=m_1x$ and $t_v=m_2x$ with finite slopes $m_1<m_2$ after ordering.
In the wedge between these sides, $d^2=2x-x^2-t_v^2=2x+O(x^2)$ uniformly as $x\downarrow0$.
Thus, the area integral of $1/d$ in $0<x<\tau$ is $O(\int_0^\tau x^{1/2}dx)=O(\tau^{3/2})$.
The cut $x=\tau$ has length $(m_2-m_1)\tau$, and its flux is bounded by $O(\tau^{3/2})\|k_\varepsilon'\|_\infty$.
For fixed $\varepsilon>0$, these estimates justify removing the cuts before varying the smoothing parameter.
A convex inscribed polygon has Euclidean perimeter at most $2\pi$, $\|h_P\|_1\le2\pi$, and $W_P\le\pi$.
For a finite union, the argument uses its total Euclidean boundary length.
For each $y$, the $\alpha$ integral in Eq.~\eqref{eq:gaussian_differentiated_weight} uses $s^2 d\alpha\le |dp|$.
Meanwhile, $|H_\varepsilon|\le\pi\|k_\varepsilon\|_{L^1(\mathbb R)}$ follows directly from the window in Eq.~\eqref{eq:kinematic_adjoint}.
Consequently, for constants $C_P,C'_P$ depending only on the polygon,
\begin{align}
\|H_\varepsilon\|_1&\le C_P\|k_\varepsilon\|_{L^1(\mathbb R)},\nonumber\\
\|\mathcal B H_\varepsilon\|_1&\le C'_P
(\|k_\varepsilon\|_{L^1(\mathbb R)}+
\|k_\varepsilon'\|_{L^1(\mathbb R)}
+\|zk_\varepsilon''\|_{L^1(\mathbb R)}) .
\end{align}
Scaling gives $k_\varepsilon(z)=\varepsilon^{-1}k_1(z/\sqrt\varepsilon)$, and integration by parts in the cosine transform on either half-line gives $k_1^{(j)}(z)=O(|z|^{-2-j})$ for $j=0,1,2$.
Hence, $\|H_\varepsilon\|_1=O(\varepsilon^{-1/2})$ and $\|\mathcal B H_\varepsilon\|_1=O(\varepsilon^{-1})$ for $0<\varepsilon\le1$.
The weight carrying the endpoint derivatives grows faster, so sufficient conditions for vanishing measurement error are $\sigma_2=o(\varepsilon)$ and $\sigma_{\rm sq}/c=o(\sqrt\varepsilon)$.
With $\delta=\max(\sigma_2,\sigma_{\rm sq}/c)$ at fixed $c$, geometry, and H\"older bound, balance the noise $\delta/\varepsilon$ with the bias $\varepsilon^{\nu/2}$.
This gives $\varepsilon\asymp\delta^{2/(2+\nu)}$ and polygon error $O(\delta^{\nu/(2+\nu)})$, or $O(\delta^{1/3})$ for a Lipschitz density.
Here, $f\asymp g$ means that $f/g$ is bounded above and below by positive constants near zero. \\

\noindent
If the shortest arcs cannot be resolved, let $\widetilde{\mathcal V}_{\varepsilon,\rho}$ use measured data only for $\rho\le\alpha\le\pi-\rho$ in the last line of Eq.~\eqref{eq:volumedual}.
Assume the omitted true data satisfy $|S_j(\alpha,\theta)|\le M_jb^2$ for $j=1,2$ and $0<b<\rho_0$, with fixed $M_1,M_2$ and $0<\rho_0<\pi/2$.
Kernel scaling gives $\|k_\varepsilon\|_\infty=O(\varepsilon^{-1})$, while $\|k_\varepsilon'\|_\infty$ and $\|z k_\varepsilon''\|_\infty$ are both $O(\varepsilon^{-3/2})$.
In Eq.~\eqref{eq:gaussian_differentiated_weight}, the integration by parts just proved reduces the remaining $k_\varepsilon''$ term to integrals of $k_\varepsilon'$ with bounded total weights in the interior and along the sides.
Consequently, $|\mathcal B H_\varepsilon|\le C_P s^2\varepsilon^{-3/2}$, and the convolution with $W_P\le\pi$ gives $|H_\varepsilon|\le C_P\varepsilon^{-1/2}$ for $0<\varepsilon\le1$.
On either omitted strip, $s\le b$ makes $|S_2\mathcal B H_\varepsilon|$ at most $C_P M_2 b^4\varepsilon^{-3/2}$, and $|S_1^2H_\varepsilon|$ is at most $C_P M_1^2 b^4\varepsilon^{-1/2}$.
Integrating $b^4$ from $0$ to $\rho<\rho_0$ gives $\rho^5/5$, while restricting the noise integral in Eq.~\eqref{eq:reconstruction_noise} only decreases its absolute-error bound.
At fixed $c$ and fixed geometry and regularity constants, the triangle inequality therefore yields
\[
\left|\widetilde{\mathcal V}_{\varepsilon,\rho}-\frac{V''(0)}{L^2}\right|
\le C\left[\varepsilon^{\nu/2}
+\frac{\sigma_2}{\varepsilon}
+\frac{\sigma_{\rm sq}}{c\sqrt{\varepsilon}}
+\frac{\rho^5}{\varepsilon^{3/2}}\right].
\]
Choosing $\rho=O(\varepsilon^{(3+\nu)/10})$ keeps the loss from missing openings within the smoothing bias.
This rate assumes a uniform error bound throughout the measured region, which for a finite grid also requires control of variation between samples.

\subsection{Error Bound for Continuum Chord Data}
\label{app:continuum_stability}

\noindent
To estimate an integrated density, we can split the inversion filter $|\sigma|$ equally between the data and the known polygon weight.
Write $\delta J$ for the error in $J(p,\theta)=\mathcal Q_I/\sqrt{1-p^2}$ at $p=\cos\alpha$, extended by zero outside $|p|<1$.
Each half-filter $\Lambda^{1/2}$ multiplies the Fourier transform in $p$ by $|\sigma|^{1/2}$.
The square-integral norms $\|\cdot\|_{p,\theta}$ and $\|\cdot\|_y$ use $dp\,d\theta$ on $\mathbb R\times[0,2\pi)$ and $d^2y$ on the plane, respectively.
The Fourier slice identity and Plancherel's identity (equality of norms in position and Fourier space with the stated normalization) give
\begin{equation}
\|\Lambda^{1/2}\RR_E h\|_{p,\theta}^2=4\pi\|h\|_y^2,
\end{equation}
because $(\sigma,\theta)\mapsto\sigma n_\theta$ covers the Fourier plane twice, with Jacobian $|\sigma|$.
The identity holds in particular for $h_P=\mathbf1_{P_0}/\sqrt{1-r^2}$, defined as zero off the polygon.
Its norm is bounded by the vacuum area, since $\|h_P\|_y^2=\int_{P_0}d^2y/(1-r^2)\leq\int_{P_0}d\mu=\pi(N_v-2\chi)$.
To evaluate the first integral for a given polygon, set $u=r^2$ and $q(u)=-\log(1-u)/(2u)$ for $0<u<1$, with $q(0)=1/2$ by continuity.
The radial vector field $F(y)=yq(|y|^2)$ satisfies $\nabla\cdot F=2\,d[uq(u)]/du=1/(1-u)$.
On side $j$, write $y=b_jm_j+t m_j^\perp$, where $m_j$ is its outward unit normal, $m_j^\perp$ is a unit tangent, $b_j$ is its signed offset and $|t|<s_j=\sqrt{1-b_j^2}$.
Since $F\cdot m_j=b_jq(b_j^2+t^2)$ and the side length element is $dt$ in increasing $t$, the divergence theorem gives
\[
\|h_P\|_y^2=\sum_{j=1}^{N_v} b_j\int_{-s_j}^{s_j}q(b_j^2+t^2)\,dt ,
\]
where the sum counts sides over all components and a diameter contributes zero.
To justify the boundary integral at an ideal vertex $v$, truncate the polygon at $y\cdot v=1-\varepsilon$ with $\varepsilon>0$.
On this cut, the two incident straight sides bound a segment of length $O(\varepsilon)$, so its tangential coordinate relative to $v$ is $O(\varepsilon)$ and $1-|y|^2=2\varepsilon+O(\varepsilon^2)$.
Consequently, the flux through the cut is $O(\varepsilon|\log\varepsilon|)$ and vanishes as $\varepsilon\to0^+$.
At the same time, the side integrals converge because $1-b_j^2-t^2=(s_j-t)(s_j+t)$ vanishes linearly at either endpoint.
In Eq.~\eqref{eq:app_polygonfilter}, the data error is paired with the known window $W_P=\RR_E h_P$ and the filter supplies one factor of $|\sigma|$.
Placing one factor of $|\sigma|^{1/2}$ on each function allows Cauchy--Schwarz to bound their pairing in terms of the corresponding square-integral norms,
\begin{equation}
\left|\delta\!\left(\frac{V''(0)}{L^2}\right)\right|
\leq\frac{3}{c\sqrt\pi}\|h_P\|_y\|\Lambda^{1/2}\delta J\|_{p,\theta}
\leq\frac{3\sqrt{N_v-2\chi}}{c}\|\Lambda^{1/2}\delta J\|_{p,\theta}.
\label{eq:continuum_stability}
\end{equation}
Eq.~\eqref{eq:continuum_stability} applies when the error is already known for normalized chord readings, even if they correspond to no exact density.
The half-filter weights rapid oscillations in offset by $|\sigma|^{1/2}$.
For errors specified on sampled entropies, Eq.~\eqref{eq:grid-entropy-error} in Appendix~\ref{app:detailC6a} instead uses continuity between samples and places endpoint derivatives on known weights.

\subsubsection{Conditional Bound for Sampled Entropy Derivatives}
\label{app:detailC6a}

\noindent
The preceding estimate requires an error bound on $J$, which already contains endpoint derivatives.
For measurements of $S_1,S_2$, Eq.~\eqref{eq:polygon_entropy_average} instead places those derivatives on known weights.
The true $S_1,S_2$ are bounded and satisfy the endpoint conditions of Appendix~\ref{app:ordinary_entropy_weights}.
The readings themselves need only have bounded absolute error, $\|\widetilde S_j-S_j\|_\infty\le\varepsilon$ for $j=1,2$, over $0<\alpha<\pi$ and all centers $\theta$ modulo $2\pi$.
Substituting these readings into Eq.~\eqref{eq:polygon_entropy_average} gives $\widetilde T_\delta$. \\

\noindent
For $0<\delta\le1$ and integer $m=0,1,2$, Fourier inversion of $|\sigma|\exp(-\delta|\sigma|)$ gives
\[
\|k_\delta^{(m)}\|_\infty
\le \frac{1}{\pi}\int_0^\infty \sigma^{m+1}\exp(-\delta\sigma)d\sigma
=\frac{(m+1)!}{\pi\delta^{m+2}} .
\]
For $s=\sin\alpha$, the differentiated weight is $B_\delta=4s^2H_\delta-4s\cos\alpha\,\partial_\alpha H_\delta +s^2(\partial_\theta^2-\partial_\alpha^2)H_\delta$.
The first and second angular derivatives of the argument $\cos\alpha-y\cdot n_\theta$ have absolute value at most one.
The chain rule in Eq.~\eqref{eq:polygon_entropy_weights} consequently gives
\[
\|H_\delta\|_\infty\le\frac{\|h_P\|_1}{\pi\delta^2},
\qquad
\|B_\delta\|_\infty\le
\|h_P\|_1\big(4\|k_\delta\|_\infty+4\|k_\delta'\|_\infty+2\|k_\delta''\|_\infty\big)
\le\frac{24\|h_P\|_1}{\pi\delta^4},
\]
where $\|h_P\|_1=\int_{\mathbb R^2}|h_P(y)|\,d^2y\le2\pi$ because the integral over the full disk is $2\pi$.
To see the derivative constants, set $p=\cos\alpha$, $s=\sin\alpha$, and $\zeta=y\cdot n_\theta$.
The coefficient of $k_\delta'$ after applying $\mathcal B$ to $k_\delta(p-\zeta)$ is $s^2(\zeta+5p)$, whose magnitude is at most $1+10/(3\sqrt3)<4$.
The corresponding coefficient of $k_\delta''$ has magnitude at most two.
The parameter cylinder has measure $2\pi^2$, and the error in $S_1^2$ is at most $(2\|S_1\|_\infty+\varepsilon)\varepsilon$.
Hence,
\[
|\widetilde T_\delta-T_\delta|
\le\frac{3\|h_P\|_1}{c}
\left[\frac{24}{\delta^4}+
\frac{12(2\|S_1\|_\infty+\varepsilon)}{c\,\delta^2}\right]\varepsilon .
\]
The inverse powers of $\delta$ quantify the amplification of measurement errors at finer resolution. \\

\noindent
The bias has a complementary bound when $g$ has positive Sobolev regularity.
For $0<\nu\le1$, we denote by $|\nabla|^\nu g$ the Fourier multiplier $|\boldsymbol\xi|^\nu$ applied to $g$.
The inequality $1-\exp(-t)\le t^\nu$ for $t\ge0$ in the two-dimensional Fourier reconstruction gives
\[
\left|T_\delta-\frac{V''(0)}{L^2}\right|
\le\frac{24\pi}{c}\|h_P\|_2\,
\||\nabla|^\nu g\|_2\,\delta^\nu .
\]
For $0<\varepsilon\le1$, equating the powers in the noise term $O(\varepsilon\delta^{-4})$ and the bias $O(\delta^\nu)$ gives $\delta=\varepsilon^{1/(4+\nu)}$ and error $O(\varepsilon^{\nu/(4+\nu)})$ at fixed norms.
The exponent is $1/5$ for $g\in H^1$, meaning that $g$ and its first weak derivatives are square integrable on $\mathbb R^2$.
For finite samples, assume that $S_1,S_2$ are Lipschitz continuous on the closed parameter cylinder with common constant $M$, so their variation is bounded by $M$ times the distance in the product of opening-angle and unit-circle metrics.
If each interval is within distance $h$ of a sample whose reading errs by at most $\eta$, using its nearest sample gives uniform error $\varepsilon\le Mh+\eta$ by the triangle inequality.
The same nearest-sample construction supplies uniform input to the Gaussian estimates in Eqs.~\eqref{eq:reconstruction_noise} and~\eqref{eq:smoothingbias} under their H\"older hypothesis on the rescaled density.
A numerical value for $M$ is additional input to the estimate.
Appendix~\ref{app:finite} exhibits the finite-data ambiguity when smooth profiles have no prescribed common bound. \\

\noindent
For samples $I_k$, $1\le k\le N$, let $C_k$ contain the intervals nearest to $I_k$, breaking ties by the lowest index, and let $h$ be the greatest distance to an assigned sample.
At $I_k$, use one branch and cutoff to measure the entropy at $-t,0,t$ with absolute error at most $s_E$ per reading.
Denote the centered first and second differences, with denominators $2t$ and $t^2$, by $y_{1,k}$ and $y_{2,k}$.
Assume uniform bounds $K_j$ on the absolute state derivatives of orders $j=3,4$ over $|\lambda|\le t$ at every sample.
Sec.~\ref{sec:signed-energy-bound} bounds the measurement and truncation errors there.
Using each value throughout its cell adds at most $Mh$ by Lipschitz continuity, giving the uniform bounds
\begin{equation}
\sigma_1\leq U_1=Mh+\frac{s_E}{t}+\frac{K_3 t^2}{6},\qquad
\sigma_2\leq U_2=Mh+\frac{4s_E}{t^2}+\frac{K_4 t^2}{12}.
\label{eq:grid-state-errors}
\end{equation}
Taking $t=s_E^{1/4}$ for sufficiently small $s_E>0$ makes both bounds $O(h+\sqrt{s_E})$ at fixed $M,K_3,K_4$. \\

\noindent
For Gaussian smoothing at scale $\varepsilon>0$, integrate the geometric weights in Eq.~\eqref{eq:volumedual} over each sampling cell $C_k$.
The two cell weights are $b_{k,\varepsilon}=\int_{C_k}\mathcal B H_\varepsilon\,d\theta\,d\alpha$ and $w_{k,\varepsilon}=\int_{C_k}H_\varepsilon\,d\theta\,d\alpha$.
Treating the measured derivatives as constant in each cell lets them factor out and gives the finite sum
\begin{equation}
\widehat{\mathcal V}_\varepsilon
=\frac{3}{2\pi c}\sum_{k=1}^{N}
\left(b_{k,\varepsilon}y_{2,k}
-\frac{12}{c}w_{k,\varepsilon}y_{1,k}^{\,2}\right).
\label{eq:grid-entropy-volume}
\end{equation}
Each cell contributes three entropy readings at different amplitudes, from which $y_{1,k}$ and $y_{2,k}$ are calculated.
All angular derivatives act on the polygon weights.
For bounded $S_1$, factoring a difference of squares gives $|S_1^2-y_{1,k}^2|\le(2\|S_1\|_\infty+\sigma_1)\sigma_1=O(h+\sqrt{s_E})$.
For a zero-extended density $g$ that is H\"older continuous of order $0<\nu\le1$, Eqs.~\eqref{eq:reconstruction_noise} and~\eqref{eq:smoothingbias} bound the leading noise and bias by terms proportional to $(h+\sqrt{s_E})/\varepsilon$ and $\varepsilon^{\nu/2}$.
Equating these orders at fixed $c$, polygon, and regularity bounds selects $\varepsilon\asymp(h+\sqrt{s_E})^{2/(2+\nu)}$ and gives
\begin{equation}
\left|\widehat{\mathcal V}_\varepsilon-\frac{V''(0)}{L^2}\right|
=O\!\left((h+\sqrt{s_E})^{\nu/(2+\nu)}\right)
\qquad (h,s_E\to0^+).
\label{eq:grid-entropy-error}
\end{equation}
This rate assumes exact cell weights and the volume comparison hypotheses.
The grid spacing $h$ measures the gaps between sampled intervals, while $s_E$ bounds the error in each entropy reading.
Once $h$ falls below $\sqrt{s_E}$, improving reading precision has a greater effect on this bound, up to fixed constants.
Errors in the numerical weights enter separately.
For $1\leq k\leq N$, let $\eta_{b,k}$ and $\eta_{w,k}$ bound the absolute errors of the two numerical cell weights at the chosen $\varepsilon$, and denote by $\widehat{\mathcal V}_\varepsilon^{\rm num}$ the result of using those weights in Eq.~\eqref{eq:grid-entropy-volume}.
Write $d\omega=d\theta\,d\alpha$ for the parameter measure.
At every point of $C_k$, Eq.~\eqref{eq:grid-state-errors} gives $|S_2-y_{2,k}|\leq U_2$ and $|S_1^2-y_{1,k}^2|\leq U_1(2|y_{1,k}|+U_1)$, since the difference of two squares factors.
Subtracting the continuum pairing in Eq.~\eqref{eq:volumedual} cell by cell and adding the smoothing bias in Eq.~\eqref{eq:smoothingbias} yields
\begin{align}
\left|\widehat{\mathcal V}_\varepsilon^{\rm num}-\frac{V''(0)}{L^2}\right|
&\leq \frac{24\pi}{c}\|h_P\|_1H_\nu(4\varepsilon)^{\nu/2} \nonumber\\
&\quad+\frac{3}{2\pi c}\sum_{k=1}^{N}\left[
U_2\int_{C_k}|\mathcal B H_\varepsilon|\,d\omega
+\frac{12}{c}U_1(2|y_{1,k}|+U_1)\int_{C_k}|H_\varepsilon|\,d\omega \right.\nonumber\\
&\hspace{105pt}\left.{}+\eta_{b,k}|y_{2,k}|+\frac{12}{c}\eta_{w,k}y_{1,k}^2
\right].
\label{eq:grid-error-budget}
\end{align}
In Eq.~\eqref{eq:grid-error-budget}, errors can change sign within a cell; hence the integrals of absolute weights.
Errors in the weights themselves multiply individual recorded values.
Eq.~\eqref{eq:grid-entropy-error} retains its rate if these summed numerical errors have the same order or smaller.
Let $E$ be the right side of Eq.~\eqref{eq:grid-error-budget} evaluated with valid bounds on regularity, reading errors, and numerical errors.
The resulting interval $[\widehat{\mathcal V}_\varepsilon^{\rm num}-E,\widehat{\mathcal V}_\varepsilon^{\rm num}+E]$ determines the sign of $V''(0)$ whenever it excludes zero.

\subsection{Polygon Window and Complementary Chords}
\label{app:polygon-window-details}

\noindent
For the polygon integral in Eq.~\eqref{eq:boundarywindow}, $s=\sqrt{1-p^2}$ is the half-length of the Euclidean chord at signed offset $p$ and $t$ is Euclidean arclength from its midpoint.
The window $W_P$ is the integral of $1/\sqrt{s^2-t^2}$ over its intersection with the polygon.
For a convex component, let $\delta_j=x_{j+1}-x_j$, $\beta_j=x_j+\delta_j/2$ for cyclically ordered vertices, with $x_{m+1}=x_1+2\pi$ for an $m$-gon.
The edge half-planes require
\[
t\sin(\beta_j-\theta)\le\cos(\delta_j/2)-p\cos(\beta_j-\theta),
\qquad j=1,\ldots,m .
\]
Their intersection with $-s\le t\le s$ gives the integration limits $t_-,t_+$, including when an individual gap exceeds $\pi$.
Since $\int dt/\sqrt{s^2-t^2}=\arcsin(t/s)$, the contribution of a nonempty segment is $[\arcsin(t/s)]_{t_-}^{t_+}$, with the inverse sine taking values in $[-\pi/2,\pi/2]$.
Empty segments give zero, and disjoint segments add.
In particular, $0\le W_P\le\pi$.
At fixed $\theta$, the map $(p,t)\mapsto pn_\theta+tn_\theta^\perp$ is a rotation with Jacobian one.
Multiplying the reconstruction of $g$ by $h_P$ and integrating over $y$ therefore integrates $h_P$ along each line while holding $(\Lambda J)(p,\theta)$ fixed.
This gives the weight and normalization in Eq.~\eqref{eq:boundarywindow}. \\

\noindent
In Eq.~\eqref{eq:boundarywindow}, complementary arcs label the same chord by $(p,\theta)$ and $\mathcal C(p,\theta)=(-p,\theta+\pi\bmod 2\pi)$, where $p\in\mathbb R$ and $0\le\theta<2\pi$.
Reversal of the position coordinate $t$ preserves $W_P$ and, by Eq.~\eqref{eq:chord_data_conventions}, the exact reading $J$.
Average the two measured values as $J_{\rm av}=(\widetilde J+\widetilde J\circ\mathcal C)/2$.
The even multiplier $|\sigma|$ commutes with offset reversal, so applying $\mathcal C$ to the second half of Eq.~\eqref{eq:boundarywindow} gives
\[
\int_0^{2\pi}\!d\theta\int_{-1}^1\!dp\,W_P(\Lambda\widetilde J)
=2\int_0^\pi\!d\theta\int_{-1}^1\!dp\,W_P(\Lambda J_{\rm av}) .
\]
The average cancels equal and opposite errors but retains an offset common to both readings.
Eq.~\eqref{eq:continuum_stability} uses an independent bound on chord errors to bound the polygon error. \\

\noindent
Nearby entropy readings can have small errors whose differences are amplified by endpoint differentiation.
A Gaussian multiplier $\exp(-\varepsilon\sigma^2)$ lets Eq.~\eqref{eq:volumedual} place those derivatives on known polygon weights.
Suppose the zero extension obeys $|g(y)-g(y')|\le H_\nu|y-y'|^\nu$ throughout $\mathbb R^2$, with $H_\nu<\infty$ and $0<\nu\le1$ (H\"older continuity).
For uniform errors $O(\delta)$ in $S_2$ and $S_1^2/c$, taking $\varepsilon\asymp\delta^{2/(2+\nu)}$ gives polygon error $O(\delta^{\nu/(2+\nu)})$ at fixed $c$ and geometry.
Appendix~\ref{app:gaussianweights} derives the weights, and Appendix~\ref{app:continuum_stability} gives bounds for chord readings and sampled entropies.

\section{Interval Response from Vacuum CFT Two-Point Function}
\label{app:cft}

\noindent
What distinguishes the field and momentum preparations when measurements are confined to one interval?
Both excite the same pairs of eigenstates of the vacuum modular Hamiltonian $-\ln\sigma_I$, but the ket and bra contributions add with different phases.
The relative-entropy Hessian is quadratic in the first change of the reduced state, so both responses follow from one vacuum two-point function.
Their squared weights differ by a constant at each frequency, giving the difference from one moment of that function for every $\Delta>0$.
Rotational symmetry also fixes the second-order modular energy needed for the individual ordinary entropies.
We connect these boundary responses to geodesic energy by normalizing one bulk quantum and isolating the scalar contribution in states that also perturb the boundary stress.

\subsection{Field and Momentum Phases of Primary Excitations}
\label{app:detailD1}

\noindent
Normalize the Hermitian spin-zero primary of dimension $\Delta>0$ by $\langle\mathcal O(z)\mathcal O(0)\rangle=|z|^{-2\Delta}$ on the plane.
In radial quantization, $|\mathcal O\rangle=\mathcal O(0)|0\rangle$ then has unit norm and is orthogonal to the vacuum.
On a unit-radius cylinder, the following normalized states realize the two tangents,
\begin{equation}
|\psi_{\lambda,\varsigma}\rangle=
\frac{|0\rangle+\lambda\beta b_\varsigma|\mathcal O\rangle}
{\sqrt{1+\lambda^2\beta^2}},\qquad
b_+=1,\quad b_-=\mathrm{i},\quad \varsigma\in\{+,-\},\quad \lambda,\beta\in\mathbb R .
\label{eq:cfttangent}
\end{equation}
At the observation time, the $+$ wave is momentarily at rest, and the $-$ wave passes through zero.
For each phase, Eq.~\eqref{eq:cfttangent} and its coherent completion have the same first derivative, also after tracing out the complement.
This derivative fixes the relative-entropy Hessian because relative entropy is stationary at the vacuum. At the same time, the ordinary entropy Hessian also needs the modular-energy Hessian.
We differentiate at $\lambda=0$ and fixed central charge $c$ before taking $c\to\infty$ at fixed $\beta^2/c$. \\

\noindent
Rotations make the answer independent of the interval center.
For $I=(-\alpha,\alpha)$ with $0<\alpha<\pi$, the vacuum density matrix is $\sigma=\exp(-2\pi K)/Z$, where $K$ generates translations of Euclidean modular time and $Z=\tr \exp(-2\pi K)$ normalizes $\tr\sigma$ to one.
Thus, the modular Hamiltonian $-\ln\sigma$ is $2\pi K+\ln Z$.
The conformal equivalence between the vacuum in an interval and a thermal state in modular coordinates is described in Ref.~\cite{Casini:2011kv}.
The plane coordinate $z=\exp(\tau_E+\mathrm{i}\varphi)$ depends on Euclidean cylinder time $\tau_E$ and angle $\varphi$.
The logarithmic map from the plane to the modular cylinder is
\begin{equation}
\zeta=\ln\!\left[-e^{-\mathrm{i}\alpha}
\frac{z-e^{\mathrm{i}\alpha}}{z-e^{-\mathrm{i}\alpha}}\right].
\label{eq:cftmodularmap}
\end{equation}
Our logarithm branch takes the interval interior to $\operatorname{Im}\zeta=0$ on one bank of the cut, with a jump of $2\pi\mathrm{i}$ to the other bank.
At $z=0$ and $z=\infty$, the argument of the logarithm is $-\exp(\mathrm{i}\alpha)$ and $-\exp(-\mathrm{i}\alpha)$, respectively.
The two arguments have unit modulus, so the insertions share a spatial position on the modular cylinder. At the same time, their phases put them at times $\pi+\alpha$ and $\pi-\alpha$ modulo $2\pi$.
As the interval widens, the two times move apart symmetrically about the midpoint $\pi$.
At either insertion, the map rescales lengths by $2\sin\alpha$ in a local coordinate ($z$ near zero, $1/z$ near infinity), so each primary contributes $d_\alpha=(2\sin\alpha)^\Delta$.
An insertion at time $\tau_{\rm m}$ divides a trip around the thermal circle into segments of duration $\tau_{\rm m}$ and $2\pi-\tau_{\rm m}$, each evolved by $K$.
Placing the perturbation in the ket or in the bra interchanges the two durations and gives
\begin{align}
\delta\rho_\varsigma
&=\frac{\beta d_\alpha}{Z}
\big[b_\varsigma e^{-(\pi-\alpha)K}\mathcal O e^{-(\pi+\alpha)K}
+b_\varsigma^* e^{-(\pi+\alpha)K}\mathcal O e^{-(\pi-\alpha)K}\big],
\label{eq:cftdeltarho}
\end{align}
where $\delta\rho_\varsigma=\partial_\lambda\varrho_{I,\varsigma}|_0$.
With the same short-distance regulator for the reference and excited states, choose an orthonormal basis $\{|j\rangle:j\in\mathcal J\}$ that diagonalizes $K$ on the support of $\sigma$.
The reference probability is $p_j=\exp(-2\pi E_j)/Z>0$ for each retained eigenvalue $E_j$.
An insertion connects two basis states with matrix element $\mathcal O_{jk}=\langle j|\mathcal O|k\rangle$ and modular frequency $\omega_{jk}=E_j-E_k$.
This frequency measures evolution generated by the interval operator $K$, while the wave frequency $\Delta$ measures evolution in global cylinder time.
The parts of the two propagation factors of duration $\pi$ supply $\sqrt{p_jp_k}$, and their remaining durations $\pm\alpha$ supply $\exp(\pm\alpha\omega_{jk})$.
After removing $\beta d_\alpha$, the ket and bra coefficients are therefore $\sqrt{p_jp_k}\,b_\varsigma \exp(\alpha\omega_{jk})$ and $\sqrt{p_jp_k}\,b_\varsigma^* \exp(-\alpha\omega_{jk})$.
Their sum gives
\begin{equation}
(\delta\rho_\varsigma)_{jk}
=2\beta d_\alpha\sqrt{p_jp_k}\,h_\varsigma(\alpha\omega_{jk})\mathcal O_{jk},
\qquad h_+(x)=\cosh x,\quad h_-(x)=\mathrm{i}\sinh x .
\label{eq:cftmatrix}
\end{equation}
In the field phase, equal real ket and bra coefficients add the two exponentials, producing the even function $\cosh x$.
In the momentum phase, conjugate coefficients $+\mathrm{i}$ and $-\mathrm{i}$ produce their difference, the odd function $\mathrm{i}\sinh x$.
For either phase, $h_\varsigma(-x)=h_\varsigma(x)^*$ for real $x$.
Reversing a transition therefore conjugates its matrix element in Eq.~\eqref{eq:cftmatrix}, since $\omega_{kj}=-\omega_{jk}$ and $\mathcal O_{kj}=\mathcal O_{jk}^*$.
At equal modular energy, the momentum contribution vanishes because $h_-(0)=0$.
Its tangent has zero trace, while the field tangent has trace $2\beta d_\alpha\sum_{j\in\mathcal J}p_j\mathcal O_{jj}=2\beta d_\alpha\tr(\sigma\mathcal O)$.
In the continuum, this expectation comes from the vacuum one-point function on the plane, which vanishes because a translation-invariant constant cannot scale with nonzero dimension $\Delta$.
If a regulator leaves a residual one-point term, set $X=\partial_\lambda\widetilde\varrho_\lambda|_0$ with unnormalized state $\widetilde\varrho_0=\sigma$.
Differentiating $\widetilde\varrho_\lambda/\tr\widetilde\varrho_\lambda$ subtracts $\sigma\tr X$ from $X$ and restores the traceless tangent required below.

\subsection{Spectral Integral for Relative-Entropy Hessian}
\label{app:detailD2}

\noindent
The first derivative of $S(\varrho\Vert\sigma)=\tr[\varrho(\ln\varrho-\ln\sigma)]$ along a normalized family is $\tr[\partial_\lambda\varrho(\ln\varrho-\ln\sigma)]$, which vanishes at $\varrho=\sigma$.
The leading change in distinguishability is therefore quadratic, $S(\varrho_{I,\varsigma}(\lambda)\Vert\sigma_I)=\lambda^2\FF_{I,\varsigma}/2+o(\lambda^2)$.
Differentiating once more gives $\FF_{I,\varsigma}=\tr[\delta\rho_\varsigma D\ln_\sigma[\delta\rho_\varsigma]]$, where $D\ln_\sigma[X]$ denotes the linear change in the matrix logarithm under $X$.
A diagonal probability perturbation has logarithmic weight $1/p_j$, while coherence between $j$ and $k$ involves both reference probabilities.
To obtain both weights at once, use $\ln x=\int_0^\infty[(1+\xi)^{-1}-(x+\xi)^{-1}]\,d\xi$ for $x>0$.
Varying the inverse matrix puts $(\sigma+\xi\mathbf1)^{-1}$ on each side of $\delta\rho_\varsigma$.
Taking the trace then pairs each transition with its reverse, producing the squared magnitude of its matrix element and the two denominators $p_j+\xi$ and $p_k+\xi$.
The resulting quadratic form is called the Bogoliubov--Kubo--Mori form,
\begin{align}
\FF_{I,\varsigma}
&=\int_0^\infty\!d\xi\;
\tr\!\big[\delta\rho_\varsigma(\sigma+\xi\mathbf 1)^{-1}
\delta\rho_\varsigma(\sigma+\xi\mathbf 1)^{-1}\big]
=\sum_{j,k\in\mathcal J}|(\delta\rho_\varsigma)_{jk}|^2\,m(p_j,p_k),
\label{eq:cftbkm}\\
m(p_j,p_k)&=\int_0^\infty\frac{d\xi}{(p_j+\xi)(p_k+\xi)}
=\frac{\ln p_j-\ln p_k}{p_j-p_k}
=\frac{\pi\omega_{jk}}{\sqrt{p_jp_k}\sinh(\pi\omega_{jk})}.
\label{eq:bkm_sum}
\end{align}
The quotient is the slope between two points on the increasing function $\ln p$, so each transition contributes nonnegatively.
For equal probabilities, the slope becomes the derivative $1/p_j$, which covers transitions within a degenerate eigenspace.
The underlying trace identity also applies when the state and its variation fail to commute.
For a regulated state $\varrho$ with positive eigenvalues and a Hermitian variation $X$, the inverse-matrix representation gives
\[
\tr\!\left(\varrho\,D\ln_{\varrho}[X]\right)
=\tr\!\left[X\int_0^\infty \varrho(\varrho+\xi\mathbf 1)^{-2}\,d\xi\right]
=\tr X .
\]
Cyclicity moves $X$ to the left, and each eigenvalue $p>0$ of $\varrho$ contributes $\int_0^\infty p(p+\xi)^{-2}\,d\xi=1$ to the operator inside the trace.
Taking $X=\partial_\lambda\varrho$ gives $\tr X=0$ and hence the stated first derivative.
On differentiating once more at $\varrho=\sigma$, the term proportional to $\partial_\lambda^2\varrho$ multiplies $\ln\varrho-\ln\sigma=0$.
Thus, two normalized completions with the same reduced tangent give the same relative-entropy Hessian, including the coherent completion and Eq.~\eqref{eq:cfttangent}.
Vacuum CFT modular correlators for perturbative relative entropy have been developed in Refs.~\cite{Lashkari:2015dia,Faulkner:2017tkh,Sarosi:2017rsq}. \\

\noindent
After inserting the matrix elements of the tangent, all dependence on the interval width and on the phase enters through a common prefactor and the frequency $\omega_{jk}$ of each transition.
We can therefore sum the squared operator matrix elements at fixed frequency once for both preparations, defining the midpoint spectral measure
\begin{equation}
\mu_\Delta(\omega)=\sum_{j,k\in\mathcal J}\sqrt{p_jp_k}
|\mathcal O_{jk}|^2\delta(\omega-\omega_{jk}) ,
\end{equation}
where $\delta$ is the Dirac distribution, characterized by $\int_{\mathbb R} f(\omega)\delta(\omega-\omega_0)\,d\omega=f(\omega_0)$ for smooth test functions $f$ and real $\omega_0$.
To determine this measure without diagonalizing the reduced state, evolve the operator as $\mathcal O(t)=\exp(\mathrm{i}Kt)\mathcal O\exp(-\mathrm{i}Kt)$ for real modular time $t$.
In $\tr[\sigma^{1/2}\mathcal O(t)\sigma^{1/2}\mathcal O(0)]$, each pair contributes its spectral weight $\sqrt{p_jp_k}|\mathcal O_{jk}|^2$ multiplied by $\exp(\mathrm{i}\omega_{jk}t)$.
The correlation function is thus the Fourier transform of $\mu_\Delta$.
To evaluate it, start with Euclidean time separation $\tau_{\rm m}$ in the strip $0<\operatorname{Re}\tau_{\rm m}<2\pi$, where propagation around the thermal circle gives
\[
C(\tau_{\rm m})=Z^{-1}\tr\!\left[e^{-(2\pi-\tau_{\rm m})K}\mathcal O e^{-\tau_{\rm m} K}\mathcal O\right]
=\sum_{j,k\in\mathcal J}\sqrt{p_jp_k}\,
e^{(\tau_{\rm m}-\pi)\omega_{jk}}|\mathcal O_{jk}|^2 .
\]
At $\tau_{\rm m}=\pi$, both propagation segments have length $\pi$, and each supplies one square root of a reference probability.
Continuing to $\tau_{\rm m}=\pi+\mathrm{i}t$ retains their product $\sqrt{p_jp_k}$ and inserts the phase $\exp(\mathrm{i}\omega_{jk}t)$.
For real $0<\tau_{\rm m}<2\pi$, the map $w=\exp\zeta$ puts the equal-position operators at $1$ and $\exp(\mathrm{i}\tau_{\rm m})$ on the plane, a distance $2\sin(\tau_{\rm m}/2)$ apart.
Their scale factors $|w|^\Delta$ are both one, so the normalized plane two-point function gives $C(\tau_{\rm m})=[2\sin(\tau_{\rm m}/2)]^{-2\Delta}$.
Writing $\tau_{\rm m}=x+\mathrm{i}y$ gives $\operatorname{Re}\sin(\tau_{\rm m}/2)=\sin(x/2)\cosh(y/2)>0$ for $0<x<2\pi$ and real $y$.
Its power therefore continues from the positive real sine without encountering zero or crossing the logarithm cut, even for noninteger $\Delta$.
At zero real modular time, the correlation function sums all transition weights, giving $C(\pi)=2^{-2\Delta}$.
Since $\mathcal O$ is Hermitian, each transition and its reverse have equal weight and opposite frequency, so the measure is even.
At $\tau_{\rm m}=\pi+\mathrm{i}t$, the continued correlator is therefore
\begin{equation}
\int_{\mathbb R}e^{\mathrm{i}\omega t}\mu_\Delta(\omega)\,d\omega
=[2\cosh(t/2)]^{-2\Delta},\qquad t\in\mathbb R .
\end{equation}
For the inverse Fourier transform, set $u=\exp(t)$, so $dt=du/u$ and $2\cosh(t/2)=(1+u)/\sqrt u$.
The frequency factor becomes $u^{-\mathrm{i}\omega}$, giving
\begin{align}
\mu_\Delta(\omega)&=\frac{1}{2\pi}\int_0^\infty
\frac{u^{\Delta-\mathrm{i}\omega-1}}{(1+u)^{2\Delta}}\,du
=\frac{|\Gamma(\Delta+\mathrm{i}\omega)|^2}{2\pi\Gamma(2\Delta)} .
\label{eq:cftspectrum}
\end{align}
Here, the Euler gamma function is $\Gamma(z)=\int_0^\infty v^{z-1}\exp(-v)dv$ for $\operatorname{Re}z>0$.
We used Euler's beta integral $\int_0^\infty u^{x-1}(1+u)^{-x-y}du=\Gamma(x)\Gamma(y)/\Gamma(x+y)$ for $\operatorname{Re}x,\operatorname{Re}y>0$.
To return from the two-point function to the entropy derivative, insert the squared tangent $4\beta^2d_\alpha^2p_jp_k|h_\varsigma|^2|\mathcal O_{jk}|^2$ in Eq.~\eqref{eq:cftbkm}.
The logarithmic weight contains $1/\sqrt{p_jp_k}$, leaving exactly the factor $\sqrt{p_jp_k}|\mathcal O_{jk}|^2$ already determined by the correlator.
The weighted sum over transitions therefore becomes a single frequency integral for the continuum relative-entropy Hessian,
\begin{equation}
\FF_{I,\varsigma}=4\pi\beta^2d_\alpha^2\int_{-\infty}^{\infty}
\mu_\Delta(\omega)\frac{\omega |h_\varsigma(\alpha\omega)|^2}{\sinh(\pi\omega)}\,d\omega .
\label{eq:cftgeneralintegral}
\end{equation}
There is no singularity at zero frequency, because $\omega/\sinh(\pi\omega)\to1/\pi$ as $\omega\to0$.
At large $|\omega|$ and fixed $\Delta>0$, Stirling's formula gives $|\Gamma(\Delta+\mathrm{i}\omega)|^2\sim2\pi|\omega|^{2\Delta-1}\exp(-\pi|\omega|)$.
The two phase factors grow with the interval opening. However, the full integrand is bounded by a constant times $(1+|\omega|)^{2\Delta}\exp[-2(\pi-\alpha)|\omega|]$ at large frequency.
For every $0<\alpha<\pi$, the positive gap $\pi-\alpha$ makes the high-frequency tail integrable.
If that gap is bounded below, the exponential still dominates the powers of $|\omega|$ introduced by differentiating in $\alpha$.

\subsection{Exact Phase Difference at Arbitrary Scaling Dimension}
\label{app:phasegap}

\noindent
The two phase weights can be subtracted before integration because Eq.~\eqref{eq:cftmatrix} gives $|h_+(x)|^2-|h_-(x)|^2=\cosh^2 x-\sinh^2 x=1$ at every frequency.
All opening dependence is then in $d_\alpha^2$, and the remaining moment of the midpoint correlator has Fourier kernel
\begin{equation}
\frac{\omega}{\sinh(\pi\omega)}
=\frac{1}{4\pi}\int_{\mathbb R}dt\,
e^{-\mathrm{i}\omega t}\operatorname{sech}^2(t/2),
\qquad \operatorname{sech}x=1/\cosh x .
\label{eq:phasefilter}
\end{equation}
Indeed, setting $v=\exp(t)$ in this Fourier integral gives $4\int_0^\infty v^{-\mathrm{i}\omega}(1+v)^{-2}dv =4\Gamma(1-\mathrm{i}\omega)\Gamma(1+\mathrm{i}\omega) =4\pi\omega/\sinh(\pi\omega)$, including the value at zero by continuity.
The integrable factors permit exchanging the $t$ and $\omega$ integrals.
The midpoint correlator then contributes $[2\cosh(t/2)]^{-2\Delta}$, and $d_\alpha^2=(2\sin\alpha)^{2\Delta}$ cancels the corresponding power of two.
Eq.~\eqref{eq:cftgeneralintegral} gives
\begin{align}
\FF_{I,+}-\FF_{I,-}
&=\beta^2\sin^{2\Delta}\!\alpha
\int_{\mathbb R}\operatorname{sech}^{2\Delta+2}(t/2)\,dt
\nonumber\\
&=2\beta^2\sin^{2\Delta}\!\alpha
\int_{-1}^1 (1-z^2)^\Delta\,dz
\nonumber\\
&=\frac{2\sqrt\pi\,\Gamma(\Delta+1)}
{\Gamma(\Delta+3/2)}\,\beta^2\sin^{2\Delta}\!\alpha .
\label{eq:phasegapall}
\end{align}
In the second line, $z=\tanh(t/2)$.
In the last line, $z^2$ is the integration variable on the positive half-interval, so the integral equals Euler's beta integral with arguments $1/2$ and $\Delta+1$.
At $\Delta=3$, integrating $(1-z^2)^3$ gives the coefficient $64/35$ directly.
For nonzero $\beta$, the ordinary entropy Hessian is larger in the momentum phase at every proper opening.
The difference depends on the opening only through $\sin\alpha$, as required for the equal ordinary entropies of an arc and its complement in a pure state.
It vanishes as $\alpha^{2\Delta}$ on a shrinking arc and as $(\pi-\alpha)^{2\Delta}$ on an arc with a shrinking complement.
These are limits through proper arcs, since at $\alpha=\pi$ the reference state of the full system has rank one and the logarithm no longer has the same support.
For finite openings, the same spectral measure determines the ordering of the two relative-entropy Hessians and how their ratio changes with interval size.
The shrinking-interval coefficient then follows from its second frequency moment.
For $\beta\ne0$, set $w_\Delta(\omega)=\mu_\Delta(\omega)\omega/\sinh(\pi\omega)$, with its value at $\omega=0$ defined by continuity.
Dividing Eq.~\eqref{eq:cftgeneralintegral} for the momentum phase by the phase difference cancels the common factor $4\pi\beta^2d_\alpha^2$ and gives
\[
\frac{\FF_{I,-}}{\FF_{I,+}-\FF_{I,-}}
=\frac{\int_{\mathbb R}w_\Delta(\omega)\sinh^2(\alpha\omega)\,d\omega}
{\int_{\mathbb R}w_\Delta(\omega)\,d\omega}.
\]
Call the ratio $A_\Delta(\alpha)$.
Its denominator normalizes $w_\Delta$ to a probability weight over frequencies, so the entire dependence on opening is the weighted mean of $\sinh^2(\alpha\omega)$.
The weight is positive for every real $\omega$, and differentiating the mean gives
\[
A_\Delta'(\alpha)
=\frac{\int_{\mathbb R}w_\Delta(\omega)\,\omega\sinh(2\alpha\omega)\,d\omega}
{\int_{\mathbb R}w_\Delta(\omega)\,d\omega}>0,
\qquad 0<\alpha<\pi .
\]
For $\omega\ne0$ and $\alpha>0$, the numerator integrand is positive.
The exponential bound in Appendix~\ref{app:detailD2} dominates the extra frequency factor on each compact subinterval of $0<\alpha<\pi$.
It follows that $\FF_{I,-}/\FF_{I,+}=A_\Delta/(1+A_\Delta)$ increases strictly with the half-opening.
This comparison concerns the two relative-entropy responses, each formed with the vacuum on that same arc.
For small $\alpha$, the coefficient of $\alpha^2$ in $A_\Delta$ is the normalized second moment of $w_\Delta$.
The exponential frequency bound below Eq.~\eqref{eq:cftgeneralintegral} controls the integrated Taylor remainder by $O(\alpha^4)$ as $\alpha\to0^+$.
To compute the moment, multiply the Fourier kernel in Eq.~\eqref{eq:phasefilter} by $\omega^2$.
Two integrations by parts replace $k(t)=\operatorname{sech}^2(t/2)$ by $-k''(t)=-k(t)+3k(t)^2/2$, with vanishing boundary terms.
The midpoint correlator contributes the same factor $[2\cosh(t/2)]^{-2\Delta}$ to both moments.
Writing $J_\nu=\int_{\mathbb R}\operatorname{sech}^{\nu}(t/2)\,dt$ for $\nu>0$, integration of the derivative of $\tanh(t/2)\operatorname{sech}^{\nu}(t/2)$ gives $J_{\nu+2}/J_\nu=\nu/(\nu+1)$.
Consequently,
\[
\frac{\int_{\mathbb R}\omega^2w_\Delta(\omega)\,d\omega}
{\int_{\mathbb R}w_\Delta(\omega)\,d\omega}
=-1+\frac32\frac{J_{2\Delta+4}}{J_{2\Delta+2}}
=\frac{\Delta}{2\Delta+3}.
\]
On a short interval, the real field preparation contributes already at order $\alpha^{2\Delta}$ to the relative-entropy Hessian.
For the momentum preparation, $\sinh^2(\alpha\omega)$ supplies two extra powers of $\alpha$, so its Hessian is $\Delta\alpha^2/(2\Delta+3)+O(\alpha^4)$ times the phase difference in Eq.~\eqref{eq:phasegapall}.
For $\Delta=3$, this gives the coefficient $64\beta^2/105$ multiplying $\alpha^8$. \\

\noindent
The relative-entropy calculation also determines an ordinary entropy response once the interval modular energy is supplied.
Subtract the vacuum eigenvalue $E_0$ of the cylinder Hamiltonian $H$ before differentiating.
Since $(H-E_0)|0\rangle=0$, only the squared first-order excitation contributes to quadratic energy, giving $\partial_\lambda^2\langle H\rangle|_0=2\Delta\beta^2$ for every smooth normalized completion here.
Rotations spread this energy uniformly around the circle, so $\partial_\lambda^2\langle T_{\tau\tau}(x)\rangle|_0=\Delta\beta^2/\pi$ for $x$ modulo $2\pi$.
On $u=-\alpha<x<v=\alpha$, the vacuum modular Hamiltonian $K_I^{(0)}=-\ln\sigma_I$ weights the density by $2\pi k_I(x)$ with $k_I(x)=(\cos x-\cos\alpha)/\sin\alpha$.
The integral $\int_u^v k_I(x)\,dx=2(1-\alpha\cot\alpha)$ therefore gives the same modular-energy Hessian $4\Delta\beta^2(1-\alpha\cot\alpha)$ in both phases.
For $S_{2,\pm}=\partial_\lambda^2 S(\varrho_{I,\pm})|_0$, differentiating the exact identity $S(\varrho_I\Vert\sigma_I)=\Delta\langle K_I^{(0)}\rangle-\Delta S_I$ gives $S_{2,\pm}=M_I''-\FF_{I,\pm}$ and hence $S_{2,-}-S_{2,+}=\FF_{I,+}-\FF_{I,-}$.
Here, $\Delta$ before an expectation value or an entropy denotes subtraction of its vacuum value.
In principle, $S_{2,\pm}$ can be inferred from the limit of $[S_I(\lambda)+S_I(-\lambda)-2S_I(0)]/\lambda^2$ as $\lambda\to0$, using ordinary entropies in the chosen preparation.
These boundary statements hold for finite central charge and require no bulk description.

\subsection{Interval Entropy for Both Phases}
\label{app:elementaryentropy}

\noindent
To obtain the two individual entropy Hessians, we need their mean and their difference.
Their mean is the common modular-energy Hessian minus the mean of the two relative-entropy Hessians.
We divide by $\beta^2$ for $\beta\ne0$, since both responses vanish at $\beta=0$.
Begin on a minor arc with $0<\alpha<\pi/2$, $s=\sin\alpha$, $p=\cos\alpha>0$, and time kernel $k(t)=\operatorname{sech}^2(t/2)$.
The identity $|h_\pm(\alpha\omega)|^2=[\cosh(2\alpha\omega)\pm1]/2$ separates the two contributions to Eq.~\eqref{eq:cftgeneralintegral}.
The Fourier kernel for the term containing $\cosh(2\alpha\omega)$ is the average of $k(t+2\mathrm{i}\alpha)$ and $k(t-2\mathrm{i}\alpha)$ through the identity
\begin{equation}
\frac{\omega\cosh(2\alpha\omega)}{\sinh(\pi\omega)}
=\frac1{4\pi}\int_{\mathbb R} e^{-\mathrm{i}\omega t}
\operatorname{Re} k(t+2\mathrm{i}\alpha)\,dt .
\end{equation}
Indeed, shifting the two contours back to the real axis in Eq.~\eqref{eq:phasefilter} gives the factors $\exp(\mp2\alpha\omega)$.
The kernel is analytic for $|\operatorname{Im}t|<\pi$ and decays exponentially at either end of each shifted contour.
The nearest poles lie at $t=\pm\mathrm{i}\pi$, where $\cosh(t/2)=0$.
The shifts $\pm2\mathrm{i}\alpha$ stay inside them for $2\alpha<\pi$.
Its integrability and the finite total midpoint spectral measure allow the frequency integral to be performed first, giving $[2\cosh(t/2)]^{-2\Delta}$. \\

\noindent
The substitution $u=\tanh(t/2)$ maps the real axis to $(-1,1)$.
Since $\cosh(t/2+\mathrm{i}\alpha)=\cosh(t/2)(p+\mathrm{i} s u)$, the relative-entropy Hessians become
\begin{equation}
\frac{\FF_{I,\pm}}{\beta^2}
=s^{2\Delta}\int_{-1}^{1}(1-u^2)^\Delta
\left[\operatorname{Re}\frac{1}{(p+\mathrm{i} s u)^2}\ \pm1\right]du .
\label{eq:cftelementaryF}
\end{equation}
The common real kernel gives the average of the two responses.
In their difference, it cancels, leaving twice the integral of $(1-u^2)^\Delta$ with the prefactor shown.
For the common part, $\operatorname{Re}(p+\mathrm{i} s u)^{-2}=(p^2-s^2u^2)/(p^2+s^2u^2)^2$ is the derivative of $u/(p^2+s^2u^2)$.
The boundary term in integration by parts vanishes because $(1-u^2)^\Delta$ vanishes at $u=\pm1$.
With $z=s^2(1-u^2)$, so $1-z=p^2+s^2u^2$, the result is
\begin{equation}
\frac{\FF_{I,+}+\FF_{I,-}}{2\beta^2}
=2\Delta s^2\int_{-1}^1\frac{u^2 z^{\Delta-1}}{1-z}\,du .
\end{equation}
For the rotationally invariant completions, the boundary energy-density Hessian is spatially uniform and fixed by the total energy Hessian $2\Delta\beta^2$.
Its integral against the vacuum modular Hamiltonian therefore has the same value for both phases.
This common modular-energy Hessian has the equivalent integral form $M_I''/\beta^2=2\Delta s^2\int_{-1}^1u^2/(1-z)\,du =4\Delta(1-\alpha\cot\alpha)$.
The last equality uses $s^2u^2/(1-z)=1-p^2/(p^2+s^2u^2)$ and $\arctan(s/p)=\alpha$ on the stated short arc.
Thus, for the rotationally invariant pure completions, the \emph{ordinary} entropy Hessians $S_{2,\pm}=M_I''-\FF_{I,\pm}$ are
\begin{equation}
\frac{S_{2,\pm}}{\beta^2}
=2\Delta s^2\int_{-1}^1 u^2\frac{1-z^{\Delta-1}}{1-z}\,du
\ \mp\ s^{2\Delta}\int_{-1}^1(1-u^2)^\Delta\,du .
\label{eq:cftelementaryS}
\end{equation}
At $z=1$, the numerator and denominator both vanish, and the ratio of their derivatives gives the finite value $\Delta-1$.
As $u$ approaches either endpoint, $z=s^2(1-u^2)$ vanishes linearly, so the power $z^{\Delta-1}$ is integrable for $\Delta>0$.
The mean relative-entropy Hessian after integration by parts also gives the half-circle limit directly.
For $s\uparrow1$, its integrand divided by $2\Delta$ is
\[
\frac{s^2u^2}{p^2+s^2u^2}\,s^{2\Delta-2}(1-u^2)^{\Delta-1},
\qquad p^2=1-s^2.
\]
The fraction belongs to $[0,1]$ and tends to one for every $u\ne0$ in $(-1,1)$.
For $s\geq1/2$, the remaining factor is bounded by a constant times $(1-u^2)^{\Delta-1}$, which is integrable for every $\Delta>0$.
Dominated convergence gives the mean $2\Delta\beta^2 I_{\Delta-1}$, where $I_\nu=\int_{-1}^{1}(1-u^2)^\nu\,du$ for $\nu>-1$.
Integration of $\partial_u[u(1-u^2)^\Delta]$ gives $2\Delta I_{\Delta-1}=(2\Delta+1)I_\Delta$.
Adding or subtracting half of Eq.~\eqref{eq:phasegapall} now gives both semicircle Hessians,
\[
\frac{\FF_{I(\pi/2),+}}{\beta^2}=2(\Delta+1)I_\Delta,
\qquad
\frac{\FF_{I(\pi/2),-}}{\beta^2}=2\Delta I_\Delta ,
\]
where $I(\pi/2)$ denotes a semicircle.
As the interval approaches a semicircle, the two poles of the original real kernel in Eq.~\eqref{eq:cftelementaryF} approach $u=0$, which explains why integration by parts precedes that limit.
For a larger arc with $\pi/2<\alpha<\pi$, its complement has half-opening $\pi-\alpha$ and the same $s=\sin\alpha$.
Purity equates their ordinary entropies at every amplitude and hence also their second derivatives, extending Eq.~\eqref{eq:cftelementaryS} to these arcs without shifting a contour past a pole.
Their relative entropies also involve the modular energy, whose weight differs on the two regions. \\

\noindent
The spectral integral also determines the relative-entropy Hessian on the longer arc directly.
Write $I(\gamma)$ for an arc of half-opening $0<\gamma<\pi$ in the rotationally invariant family and set $x=2\alpha-\pi$.
For each $\varsigma\in\{+,-\}$, subtraction of the two factors $|h_\varsigma|^2$ in Eq.~\eqref{eq:cftgeneralintegral} gives
\[
|h_\varsigma(\alpha\omega)|^2-|h_\varsigma((\pi-\alpha)\omega)|^2
=\tfrac12\big[\cosh(2\alpha\omega)-\cosh(2(\pi-\alpha)\omega)\big]
=\sinh(\pi\omega)\sinh(x\omega).
\]
The first factor cancels the denominator of the spectral kernel, and $d_{\pi-\alpha}=d_\alpha$.
For real $|x|<\pi$, the midpoint correlator at Euclidean separation $\pi+x$ gives
\[
\int_{\mathbb R}\mu_\Delta(\omega)\cosh(x\omega)\,d\omega
=C(\pi+x)=[2\cos(x/2)]^{-2\Delta}.
\]
Here, the evenness of $\mu_\Delta$ replaces $\exp(x\omega)$ with its even part.
The exponential decay in Eq.~\eqref{eq:cftspectrum} allows differentiation with respect to $x$ within this open interval.
Its derivative is the remaining spectral moment, namely $\Delta\tan(x/2)C(\pi+x)$.
Since $d_\alpha^2C(2\alpha)=1$ and $\tan(x/2)=-\cot\alpha$, the result for both phases is
\[
\FF_{I(\alpha),\varsigma}-\FF_{I(\pi-\alpha),\varsigma}
=-4\pi\Delta\beta^2\cot\alpha
=M_{I(\alpha)}''-M_{I(\pi-\alpha)}'' .
\]
For a pure state, the ordinary entropies of an arc and its complement agree, so their modular-energy difference must be matched by their relative-entropy difference.
The relation above exhibits that cancellation before any limiting procedure.
For an arc approaching the full circle, write $\delta=\pi-\alpha$ and let $\delta\to0^+$ at fixed $\Delta>0$ and $\beta\ne0$.
On the small complement, the modular-energy Hessian is $O(\delta^2)$ and each relative-entropy Hessian is $O(\delta^{2\Delta})$ by Appendix~\ref{app:phasegap}.
Their difference tends to zero there, so purity implies that the same ordinary entropy Hessian tends to zero on the large arc. \\

\noindent
Both relative-entropy Hessians on the larger arc receive a common contribution from modular frequencies of order $1/\delta$.
For $\omega\to+\infty$, the spectral density in Eq.~\eqref{eq:cftspectrum}, the factor $1/\sinh(\pi\omega)$, and the factor $|h_\varsigma(\alpha\omega)|^2$ contribute exponentials $\exp(-\pi\omega)$, $\exp(-\pi\omega)$, and $\exp(2\alpha\omega)$, respectively.
Including their prefactors gives
\[
\mu_\Delta(\omega)\frac{\omega|h_\varsigma(\alpha\omega)|^2}{\sinh(\pi\omega)}
=\frac{\omega^{2\Delta}\exp(-2\delta\omega)}{2\Gamma(2\Delta)}[1+o(1)] .
\]
The relative remainder tends to zero uniformly for $\pi/2\leq\alpha<\pi$, since $\exp(-2\alpha\omega)\leq\exp(-\pi\omega)$ controls the subleading terms in either phase factor.
The integral over any fixed bounded frequency range contributes only $O(\delta^{2\Delta})$ to Eq.~\eqref{eq:cftgeneralintegral} because of $d_\alpha^2=(2\sin\delta)^{2\Delta}$.
On the remaining two tails, evenness supplies a factor of two, and the substitution $v=2\delta\omega$ gives $\int_0^\infty\omega^{2\Delta}\exp(-2\delta\omega)\,d\omega=\Gamma(2\Delta+1)/(2\delta)^{2\Delta+1}$.
The frequency range contributing to the tails grows as $1/\delta$, making the tail integral of order $\delta^{-(2\Delta+1)}$.
Multiplying by the vanishing insertion factor $d_\alpha^2\sim(2\delta)^{2\Delta}$ leaves a simple $1/\delta$ pole common to both phases, with its coefficient fixed by $\Gamma(2\Delta+1)=2\Delta\Gamma(2\Delta)$ in Eq.~\eqref{eq:cftgeneralintegral}.
Dividing the exact phase difference in Eq.~\eqref{eq:phasegapall} by this common leading field-phase Hessian gives
\begin{equation}
\FF_{I(\pi-\delta),\varsigma}\sim\frac{4\pi\Delta\beta^2}{\delta},
\qquad
1-\frac{\FF_{I(\pi-\delta),-}}{\FF_{I(\pi-\delta),+}}
\sim\frac{\Gamma(\Delta)}{2\sqrt{\pi}\,\Gamma(\Delta+3/2)}\,\delta^{2\Delta+1}.
\label{eq:cft-near-full-phases}
\end{equation}
Here, $\sim$ means that the quotient of the two sides tends to one as $\delta\to0^+$ through proper arcs. \\

\noindent
For $\Delta=3$, the quotient reduces to $1+z$.
The three elementary integrals on $[-1,1]$ are $\int_{-1}^1u^2du=2/3$, $\int_{-1}^1u^2(1-u^2)du=4/15$ and $I_3=32/35$.
It follows that
\begin{align}
\frac{S_{2,\pm}}{\beta^2}
&=4s^2+\frac85s^4\mp\frac{32}{35}s^6,\nonumber\\
\frac{\FF_{I,\pm}}{\beta^2}
&=12(1-\alpha\cot\alpha)-4s^2-\frac85s^4
\pm\frac{32}{35}s^6 .
\label{eq:cftclosed}
\end{align}
At a half circle, $M_I''/\beta^2=12$.
In $(+,-)$ order, $S_{2,\pm}/\beta^2=(164,228)/35$ and $\FF_{I,\pm}/\beta^2=(256,192)/35$.
For every proper arc, the real tangent has the larger relative-entropy Hessian and the smaller ordinary entropy Hessian, because the phase integral in Eq.~\eqref{eq:cftelementaryS} is strictly positive.
For a shrinking arc, the respective relative-entropy Hessians divided by $\beta^2$ start at $64\alpha^6/35$ and $64\alpha^8/105$.

\subsection{CFT Entropy Hessians at Scaling Dimension Three Halves}
\label{app:cft_three_halves}

\noindent
At $\Delta=3/2$, each interval entropy can be evaluated in elementary functions, making its dependence on the opening explicit.
With $B=\beta^2$, the midpoint spectrum in Eq.~\eqref{eq:cftspectrum} simplifies by the identities $\Gamma(z+1)=z\Gamma(z)$ and $\Gamma(z)\Gamma(1-z)=\pi/\sin(\pi z)$ evaluated at $z=1/2+\mathrm{i}\omega$,
\begin{equation}
\mu_{3/2}(\omega)=\frac{\omega^2+1/4}{4\cosh(\pi\omega)},
\qquad \omega\in\mathbb R .
\end{equation}
The factor $\omega^2+1/4$ has a simple origin in the time dependence of the two-point function.
At $\Delta=3/2$, the midpoint correlator is the cube of the function for $\Delta=1/2$.
For the function $g(\xi)=[2\cosh(\xi/2)]^{-1}$ on $\xi\in\mathbb R$, direct differentiation gives
\begin{equation}
2g^3=(1/4-\partial_\xi^2)g,\qquad
\widehat{g^3}(\omega)=\frac{\pi(\omega^2+1/4)}{2\cosh(\pi\omega)}.
\label{eq:thermal_spectral_density}
\end{equation}
With $\widehat g(\omega)=\int_{\mathbb R}g(\xi)\exp(-\mathrm{i}\omega\xi)\,d\xi$, the identity $\widehat g=\pi/\cosh(\pi\omega)$ follows by setting $t=\exp\xi$ and evaluating Euler's beta integral with arguments $1/2\pm\mathrm{i}\omega$.
Thus, $\widehat{g^3}/(2\pi)=\mu_{3/2}$ agrees with the gamma-function result.
For $s=\sin\alpha$, the factors $h_\pm$ defined in Eq.~\eqref{eq:cftmatrix} obey $|h_\pm(\alpha\omega)|^2=[\cosh(2\alpha\omega)\pm1]/2$.
The factor of two in the reduced-state tangent has already been squared in the prefactor $4\pi B d_\alpha^2$ of Eq.~\eqref{eq:cftgeneralintegral}.
At this dimension, $d_\alpha^2=(2s)^3$ and the product of the spectral density with the universal frequency kernel is
\[
\mu_{3/2}(\omega)\frac{\omega}{\sinh(\pi\omega)}
=\frac{\omega(\omega^2+1/4)}{2\sinh(2\pi\omega)} ,
\]
with its continuous value at $\omega=0$.
Multiplication by $|h_\pm|^2$ and the prefactor gives
\begin{equation}
\frac{\FF_{I,\pm}}{8\pi B s^3}
=\int_{\mathbb R}\frac{\omega(\omega^2+1/4)
[\cosh(2\alpha\omega)\pm1]}{\sinh(2\pi\omega)}\,\mathrm d\omega .
\label{eq:thermal_frequency_integral}
\end{equation}
To perform both integrals at once, introduce a variable $x$ in $\sinh(x\omega)$.
One or three derivatives in $x$ produce $\omega\cosh(x\omega)$ or $\omega^3\cosh(x\omega)$, exactly the two powers of frequency that occur here.
For real $x$ with $|x|<2\pi$, the generating integral $J(x)$ converges and its integrand is even in $\omega$.
On $\omega>0$, the expansion $1/\sinh(2\pi\omega)=2\sum_{n=0}^\infty\exp[-(4n+2)\pi\omega]$ reduces the integral to Laplace integrals.
Each term uses $\int_0^\infty\exp(-b\omega)\sinh(x\omega)\,d\omega=x/(b^2-x^2)$ with $b=(4n+2)\pi>|x|$, giving
\begin{equation}
J(x)=\int_{\mathbb R}\frac{\sinh(x\omega)}{\sinh(2\pi\omega)}\,\mathrm d\omega
=4x\sum_{n=0}^\infty\frac{1}{(4n+2)^2\pi^2-x^2}
=\tfrac12\tan(x/4).
\label{eq:thermal_elementary_integral}
\end{equation}
For the last identity, use the cosine product $\cos(x/4)=\prod_{n=0}^\infty[1-x^2/((4n+2)^2\pi^2)]$.
On compact subsets of $|x|<2\pi$, the logarithmic derivatives of the factors converge uniformly, since their magnitudes are bounded by a constant times $(1+n)^{-2}$.
Taking the logarithmic derivative gives $-\tfrac14\tan(x/4)=-2x\sum_{n=0}^\infty[(4n+2)^2\pi^2-x^2]^{-1}$, fixing both the argument and the normalization of $J(x)$.
On any closed subinterval of $|x|<2\pi$, exponential decay at large frequency permits all three derivatives under the integral.
Acting on the elementary result $J(x)=\tfrac12\tan(x/4)$ gives $(\partial_x^3+\partial_x/4)J(x)=3/[64\cos^4(x/4)]$.
For the term with $\cosh(2\alpha\omega)$ in Eq.~\eqref{eq:thermal_frequency_integral}, we set $x=2\alpha$, and for the constant term we set $x=0$.
This gives the two phase responses
\begin{align}
\frac{\FF_{I,\pm}}{B}
&=\frac{3\pi s^3}{8}\left[\frac1{\cos^4(\alpha/2)}\pm1\right]
\label{eq:app_field_fisher32}\\
&=3\pi\left[\tan(\alpha/2)-\frac{s}{2}\pm\frac{s^3}{8}\right],
\qquad 0<\alpha<\pi .
\label{eq:primary_fisher}
\end{align}
The identity $s^3/[8\cos^4(\alpha/2)]=\tan(\alpha/2)-s/2$ follows from the half-angle formula.
The same analytic strip covers intervals smaller than, equal to, or larger than a semicircle.
For the same arc and $B>0$, dividing the minus response by the plus response in Eq.~\eqref{eq:app_field_fisher32} gives the elementary ratio
\[
\frac{\FF_{I,-}}{\FF_{I,+}}
=\frac{1-\cos^4(\alpha/2)}{1+\cos^4(\alpha/2)} .
\]
As $\alpha\to0^+$, the ratio is $\alpha^2/4+O(\alpha^4)$, whereas it tends to one as $\alpha\to\pi^-$ through proper arcs.
The latter limit expresses equality of the leading divergences of the two relative-entropy Hessians, even though the finite-opening phase gap is strictly positive. \\

\noindent
For the ordinary entropy, the common modular-energy Hessian gives $S_{2,\pm}/\beta^2=6(1-\alpha\cot\alpha)-\FF_{I,\pm}/\beta^2$.
For completeness, the positive representation in Eq.~\eqref{eq:positive32} follows directly from this closed expression.
We denote its integral by $J_{\rm ent}(s)=\int_{-1}^1x^2/[1+s\sqrt{1-x^2}]\,dx$ for $0<s\le1$.
The substitution $x=\sin v$ gives the integrand $(\cos v-\cos^3v)/(1+s\cos v)$ for $-\pi/2<v<\pi/2$.
Polynomial division gives, for $0<s<1$,
\begin{align}
J_{\rm ent}(s)
&=\int_{-\pi/2}^{\pi/2}
\left[-\frac{\cos^2v}{s}+\frac{\cos v}{s^2}
+\frac1s-\frac1{s^3}+\frac{1-s^2}{s^3(1+s\cos v)}\right]dv\nonumber\\
&=\frac{\pi}{2s}+\frac{2}{s^2}-\frac{\pi}{s^3}
+\frac{2\sqrt{1-s^2}\,\arccos s}{s^3}.
\end{align}
Here, the integral of $(1+s\cos v)^{-1}$ over the indicated interval equals $2\arccos(s)/\sqrt{1-s^2}$ by the substitution $w=\tan(v/2)$.
For all $0<\alpha<\pi$, $\sqrt{1-s^2}\arccos(s)=\cos\alpha\,(\pi/2-\alpha)$.
The elementary identities therefore give $S_{2,+}/\beta^2=3s^2J_{\rm ent}(s)-3\pi s^3/8$, including $s=1$ by continuity.

\subsection{Canonical Mode Normalization and Bulk Energy Densities}
\label{app:kg}

\noindent
To convert the boundary amplitude $\beta$ into a bulk field amplitude, we choose a single-trace primary $\mathcal O$ whose unit-normalized state describes one bulk scalar quantum at leading order.
Its lowest positive-frequency wave in global AdS$_3$ is $u_0=N\operatorname{sech}^{\Delta}\chi\,\exp(-\mathrm{i}\Delta\tau)$, and fixing its conserved norm fixes the remaining constant $N$.
On the global $\tau=0$ slice, $dA=L^2\sinh\chi\,d\chi\,d\varphi$, $0\leq\varphi<2\pi$, and $n=(L\cosh\chi)^{-1}\partial_\tau$ is the future unit normal.
The conserved Klein--Gordon inner product is $(u,v)_{\rm KG}=\mathrm{i}\int_{\tau=0}[u^*(nv)-v(nu^*)]dA$, where $nv$ denotes the normal derivative.
Since $nu_0=-\mathrm{i}\Delta u_0/(L\cosh\chi)$, the angular integral gives
\begin{equation}
(u_0,u_0)_{\rm KG}=4\pi\Delta L N^2
\int_0^\infty\tanh\chi\,\operatorname{sech}^{2\Delta}\chi\,d\chi=2\pi L N^2 .
\end{equation}
Differentiating $\operatorname{sech}^{2\Delta}\chi$ gives the radial integral $1/(2\Delta)$, so unit norm fixes $N=(2\pi L)^{-1/2}$.
The linearized mean field is $\beta(b_\varsigma u_0+b_\varsigma^*u_0^*)$, since $u_0$ multiplies the annihilation operator and $u_0^*$ the creation operator.
At linear order, the real combination $u_0+u_0^*$ gives the cosine wave, and $\mathrm{i}(u_0-u_0^*)$ gives the sine wave.
Both have amplitude $A=2\beta/\sqrt{2\pi L}$.
Hence,
\begin{equation}
a=A\sqrt L,\qquad \beta^2=\pi a^2/2 .
\label{eq:cftamplitude}
\end{equation}
The cylinder energy above the vacuum is $\lambda^2\Delta\beta^2+o(\lambda^2)$ in both states because $H-E_0$ annihilates the vacuum.
For rotationally invariant completions, this fixes the same uniform second-order boundary energy density for both phases. \\

\noindent
To exhibit the pure coherent completion, write $\widehat a,\widehat a^\dagger$ for the oscillator operators and $|n\rangle=(\widehat a^\dagger)^n|0\rangle/\sqrt{n!}$, $n=0,1,2,\ldots$.
The number states obey $\widehat a|0\rangle=0$ and $[\widehat a,\widehat a^\dagger]=1$.
These operators are distinct from the dimensionless classical amplitude $a=A\sqrt L$.
For the field tangent, $|1\rangle$ is dual to $|\mathcal O\rangle$ and
\begin{equation}
\exp[\lambda\beta(\widehat a^\dagger-\widehat a)]|0\rangle
=(1-\lambda^2 B/2)|0\rangle+\lambda\beta|1\rangle
+\frac{\lambda^2 B}{\sqrt2}|2\rangle+O(\lambda^3).
\label{eq:coherent_completion}
\end{equation}
For either phase, put $z_\varsigma=\lambda\beta b_\varsigma$ and displace the oscillator by $\exp(z_\varsigma\widehat a^\dagger-z_\varsigma^*\widehat a)$.
Since the commutator of the creation and annihilation exponents is the scalar $|z_\varsigma|^2$, ordering the displacement and using the vacuum annihilation condition gives $\exp(-|z_\varsigma|^2/2)\exp(z_\varsigma\widehat a^\dagger)|0\rangle$.
The coefficients of $|1\rangle$ and $|2\rangle$ in this expansion are proportional to $z_\varsigma$ and $z_\varsigma^2/\sqrt2$, respectively, which gives Eq.~\eqref{eq:coherent_completion} for $b_+=1$. \\

\noindent
The cylinder Hamiltonian measures the oscillator occupation, whereas local quadratic observables can also connect number states differing by two quanta.
For a coherent state, $\widehat a$ has eigenvalue $z_\varsigma$, so the occupation is $\langle\widehat a^\dagger\widehat a\rangle=|z_\varsigma|^2$ and the pair amplitude is $\langle\widehat a^2\rangle=z_\varsigma^2$.
For the vacuum-plus-one-particle state of Eq.~\eqref{eq:cfttangent}, the occupation is $|z_\varsigma|^2/(1+|z_\varsigma|^2)$, the same to quadratic order.
Its pair amplitude vanishes because annihilating two quanta kills both terms in the state.
For the mode $\widehat\phi_0=u_0\widehat a+u_0^*\widehat a^\dagger$, vacuum normal ordering places creation operators to the left and gives
\[
\langle :\widehat\phi_0^2:\rangle
=2|u_0|^2\langle\widehat a^\dagger\widehat a\rangle
+2\operatorname{Re}\!\left[u_0^2\langle\widehat a^2\rangle\right].
\]
Both completions therefore deposit the same energy to order $\lambda^2$, with occupation $\lambda^2 B$.
Only the coherent state has the pair term in the local expectation value, with opposite signs for the two phases because $b_+^2=1$ and $b_-^2=-1$.
In this state, each normal-ordered product of the free field and its derivatives factorizes into a product of their mean values.
This factorization is the step that permits using a classical wave to calculate the local stress.
The boundary entropy Hessian follows already from the reduced-state data, since the common reduced tangent fixes the relative-entropy Hessian and rotational symmetry fixes the modular-energy Hessian. \\

\noindent
At $\Delta=3$, the even result, expressed through $p=\cos\alpha$, is
\begin{equation}
\FF_{I,+}
=\frac{2\pi a^2}{35}(64+39p^2+10p^4-8p^6)
-\frac{6\pi a^2p\arccos p}{\sqrt{1-p^2}},\qquad -1<p<1 .
\label{eq:cftbulkmatch}
\end{equation}
The inverse cosine takes values in $(0,\pi)$.
The expectation of the stress tensor has zero first variation for either primary tangent since its vacuum two-point function with $\mathcal O$ vanishes.
The angle-independent functions in Eq.~\eqref{eq:cftclosed} give the following sources under $\DD=(1-p^2)[(1-p^2)\partial_p^2-p\partial_p]-2$, namely
\begin{equation}
\frac{\DD\FF_{I,+}}{4\pi}
=\frac{16a^2}{35}(5+9p^2)(1-p^2)^3,\qquad
\frac{\DD\FF_{I,-}}{4\pi}
=\frac{144a^2}{35}(1-p^2)^4 .
\label{eq:cftbothtransforms}
\end{equation}
The first expression gives the integrated energy on a chord for the field at maximum displacement.
For the momentum phase, the field and its spatial gradient vanish on the slice, leaving only the kinetic energy from $n\phi_1=(3A/L)\operatorname{sech}^4\chi$.
Squaring this derivative gives $e_-(r)=(9a^2/2)(1-r^2)^4$ in the Klein radius $r=\tanh\chi$.
The Euclidean half-length of the chord is $q=\sqrt{1-p^2}$.
The Klein chord at signed distance coordinate $p$ has $r^2=p^2+t^2$ and length element $q\,dt/(q^2-t^2)$, $-q<t<q$.
Direct integration gives $\XX e_-=(9a^2/2)q^8\int_{-1}^1(1-v^2)^3dv=(144a^2/35)q^8$, in agreement with Eq.~\eqref{eq:cftbothtransforms}.
The one-particle norm therefore fixes the strength of the density independently of the interval size.
The pointwise reconstruction in Appendix~\ref{app:detailE2} and the direct metric calculation in Appendix~\ref{app:directarea} give the two corresponding volumes.

\subsection{Geodesic Phase Difference at Arbitrary Scaling Dimension}
\label{app:chordphasegeneral}

\noindent
To compare chord energies at general mass, take the standard normalizable falloff $\Delta=1+\sqrt{1+M_\phi^2L^2}>1$ \cite{Klebanov:1999tb}, so $M_\phi^2L^2=\Delta(\Delta-2)$.
At $\tau=0$, the cosine field has zero velocity. The sine field vanishes on the whole slice, so the momentum density $(n\phi_1)\partial_i\phi_1$ is zero for both, with $i=1,2$ labeling spatial coordinates.
Time reflection preserves the cosine data and, together with $\phi\mapsto-\phi$, the sine data.
For an even potential and reflection-invariant gravitational initial data, both thus permit a reflection-symmetric metric.
The boundary difference in Eq.~\eqref{eq:phasegapall} requires only the two tangents, the vacuum modular correlator, and pure rotationally invariant completions for the ordinary entropies. \\

\noindent
The radial variable $D=1-r^2=\operatorname{sech}^2\chi$ records the common falloff.
On the vacuum slice, the canonical scalar stress gives $2e_\pm=L^3[(n\phi_{1,\pm})^2+L^{-2}(\partial_\chi\phi_{1,\pm})^2+M_\phi^2\phi_{1,\pm}^2]$.
The cosine mode has $\partial_\chi\phi_{1,+}=-A\Delta r D^{\Delta/2}$ and the sine mode has $n\phi_{1,-}=(A\Delta/L)D^{(\Delta+1)/2}$.
The resulting densities are
\begin{equation}
e_+=\tfrac{a^2}{2}\big[\Delta^2r^2+\Delta(\Delta-2)\big]D^\Delta,\qquad
e_-=\tfrac{a^2}{2}\Delta^2D^{\Delta+1},\qquad
e_--e_+=a^2\Delta(1-\Delta r^2)D^\Delta .
\end{equation}
For $1<\Delta<2$, $e_+$ can be negative near the center.
Only the quadratic response, valid for signed profiles, enters this comparison.
On the chord with $p=\cos\alpha$, the rescaled longitudinal coordinate $x$ satisfies $t=sx$, with $s=\sqrt{1-p^2}$ and $-1<x<1$.
Then, $D=s^2(1-x^2)$ and $d\sigma=dx/(1-x^2)$.
The moment $I_\nu=\int_{-1}^1(1-x^2)^\nu dx$ is finite for $\nu>-1$.
For $\Delta>0$, integrating the derivative of $x(1-x^2)^\Delta$ from $-1$ to $1$ yields $I_{\Delta-1}=(2\Delta+1)I_\Delta/(2\Delta)$.
Therefore,
\begin{align}
(\XX e_- -\XX e_+)(I)
&=a^2\Delta s^{2\Delta}\big[(1-\Delta)I_{\Delta-1}+\Delta s^2 I_\Delta\big]\nonumber\\
&=a^2 I_\Delta s^{2\Delta}\left[\tfrac12(\Delta+1)-\Delta^2p^2\right].
\label{eq:alldeltabulkphase}
\end{align}
On the boundary, subtracting phase Hessians in Eq.~\eqref{eq:cftgeneralintegral} cancels the common term because $\cosh^2x-\sinh^2x=1$.
That frequency integral converges for each $0<\alpha<\pi$, so the subtraction for larger arcs requires no contour shift.
Eq.~\eqref{eq:phasegapall} and equality of the two modular energies then give $S_{2,-}-S_{2,+}=2\beta^2 I_\Delta s^{2\Delta}$.
The corresponding normalized endpoint source is
\begin{equation}
\frac{\mathcal Q_--\mathcal Q_+}{4\pi}=
a^2 I_\Delta s^{2\Delta}\left(\frac{\Delta+1}{2}-\Delta^2 p^2\right).
\label{eq:chordphase_main}
\end{equation}
Since $\partial_\alpha^2s^{k}=k(k-1)s^{k-2}\cos^2\alpha-ks^k$ for $k=2\Delta$, application of $-\DD/(4\pi)$ to the boundary difference gives exactly Eq.~\eqref{eq:alldeltabulkphase}, including its normalization.
For $\Delta>1$, the chord difference vanishes at $|p|=\sqrt{(\Delta+1)/2}/\Delta$, inside the local zero $r=1/\sqrt\Delta$ since their squared ratio is $(\Delta+1)/(2\Delta)<1$.
A chord tangent to the local zero samples an outer region favoring the field phase, so cancellation requires an inner segment as well.
All chord integrals converge directly on the open disk. \\

\noindent
The same comparison fixes each phase separately at general $\Delta>1$.
The normalized average is $H=(S_{2,-}+S_{2,+})/(2\beta^2)$, with $s=\sin\alpha$.
By Eq.~\eqref{eq:cftelementaryS},
\begin{equation}
H=2\Delta\big[T_0(s)-T_{\Delta-1}(s)\big],\qquad
T_b(s)=\sum_{j=0}^\infty A_{b+j}s^{2(b+j+1)},\qquad
A_v=\int_{-1}^1x^2(1-x^2)^v\,dx .
\end{equation}
Here, $0<s<1$, $b\in\{0,\Delta-1\}$, and $v>-1$.
The geometric expansion of $1/[1-s^2(1-x^2)]$ converges uniformly on every compact range of $s$ below one, even after two derivatives, because its ratio is at most $s^2$.
We can therefore act term by term with the radial operator $\DD=s^2\partial_\alpha^2-2$, which gives
\begin{equation}
\DD s^k=(k-2)(k+1)s^k-k^2 s^{k+2}.
\end{equation}
For $v>0$, integration by parts gives $A_v/A_{v-1}=2v/(2v+3)$.
For each power above the lowest, the two terms of the differential operator contribute from neighboring terms in the series.
Their coefficient is $2v(2v+3)A_v-(2v)^2 A_{v-1}=0$ by this recurrence.
The lowest power has no preceding term and survives with coefficient $2b(2b+3)A_b$ multiplying $s^{2b+2}$.
Using $A_{\Delta-1}=I_{\Delta-1}/(2\Delta+1)$ then gives
\begin{equation}
-\DD H=4\Delta(\Delta-1)I_{\Delta-1}s^{2\Delta}.
\label{eq:allDeltacommonCFT}
\end{equation}
Neither the series nor the coefficient cancellation requires $\Delta$ to be an integer.
For example, both $A_v$ identities follow directly upon setting $x^2=z$ on $0<x<1$ and evaluating Euler's beta integral.
The modular integral in Eq.~\eqref{eq:cftgeneralintegral} is smoothly differentiable in $\alpha$ on compact subsets of $(0,\pi)$, so the identity extends to $\alpha=\pi/2$ by continuity of its two derivatives.
For $\pi/2<\alpha<\pi$, purity gives the same $H$ at $\pi-\alpha$ and its second derivative is unchanged. \\

\noindent
For the independent bulk integrals, the two required densities are $e_++e_-=a^2\Delta(\Delta-1)(1-r^2)^\Delta$ and $e_-=a^2\Delta^2(1-r^2)^{\Delta+1}/2$.
With $r^2=1-s^2(1-x^2)$ and $d\sigma=dx/(1-x^2)$, they give
\begin{equation}
\XX(e_++e_-)=a^2\Delta(\Delta-1)I_{\Delta-1}s^{2\Delta},
\qquad \XX e_-=\tfrac12a^2\Delta^2 I_\Delta s^{2\Delta+2}.
\end{equation}
The first expression follows from Eq.~\eqref{eq:allDeltacommonCFT} on multiplying by $2\beta^2/(4\pi)=a^2/4$.
The boundary phase difference and $2\Delta I_{\Delta-1}=(2\Delta+1)I_\Delta$ give the second expression.
Inverting the power transform in Eq.~\eqref{eq:powerchords} sends $I_{\Delta-1}s^{2\Delta}$ to $(1-r^2)^\Delta$ and $I_\Delta s^{2\Delta+2}$ to $(1-r^2)^{\Delta+1}$.
The first profile fixes the density sum and the second fixes the momentum density, so subtraction recovers the field density for all $\Delta>1$.

\subsection{Simultaneous Stress and Scalar Perturbations}
\label{app:detailD8}

\noindent
At $\Delta=3$, add a boundary stress perturbation to the field-phase state of Eq.~\eqref{eq:cfttangent}.
Let the same conformal reparametrization $f_\lambda(\varphi)=\varphi+\lambda\xi\sin(2\varphi)$, $\xi\in\mathbb R$, act on the left- and right-moving sectors.
The reduced tangent now has both a stress contribution and a scalar contribution.
The cross term of Eq.~\eqref{eq:cftbkm} is a vacuum two-point function of a smeared stress tensor with $\mathcal O$, integrated over relative modular time.
Indeed, the spectral multiplier $\pi\omega/\sinh(\pi\omega)$ depends only on the modular-frequency difference, and its inverse Fourier transform translates one insertion by vacuum modular flow.
This flow is conformal for a vacuum interval.
The conformal Ward identity expresses an insertion of the stress tensor through infinitesimal conformal transformations of the other insertions.
For a single scalar insertion on the plane, it gives $\langle T(z)\mathcal O(w,\bar w)\rangle=0$, and the same holds for the antiholomorphic stress tensor, since $\langle\mathcal O\rangle=0$ on the plane.
The identity contribution in a conformal transformation of the stress tensor also multiplies $\langle\mathcal O\rangle$.
The mixed term consequently vanishes on every proper interval. \\

\noindent
The circle stress law gives $t_1(\varphi)=c\xi\cos(2\varphi)/(2\pi)$.
For $I=(-\pi/2,\pi/2)$, the reparametrization tangent vanishes at the two endpoints.
The circle quadratic form for $\GG_I$ therefore reduces to the integral below.
Since $\int_{-\pi/2}^{\pi/2}\cos\varphi\,d\varphi=2$ and $\int_{-\pi/2}^{\pi/2}\cos\varphi\cos(4\varphi)\,d\varphi=-2/15$, the stress contribution and the full response are
\begin{align}
\GG_I&=\frac{c\xi^2}{6}\int_{-\pi/2}^{\pi/2}
\cos\varphi\,[4-12\cos(4\varphi)]\,d\varphi
=\frac{8c\xi^2}{5},\\
\FF_{I,\mathrm{mixed}}
&=\frac{8c\xi^2}{5}+\frac{256\beta^2}{35}.
\label{eq:cftmixedhalf}
\end{align}
For general $I$, subtracting the stress quadratic form leaves $\FF_{I,+}$ of Eq.~\eqref{eq:cftclosed}.
On the bulk side, the same linearized boundary graviton changes neither the zero-background scalar tangent nor its quadratic density, and geodesic displacement changes its polygon integral only beyond second order.

\section{Scalar Energy and Volume Response of Convex Polygons}
\label{app:scalar}

\noindent
For equal-amplitude preparations, a polygon measures how the energy is distributed over the radii it encloses.
We find that a regular triangle shrinks even though every proper interval gains entropy, and use two interval widths to distinguish phase from amplitude.
At $\Delta=3$, we compute the polygon response both from chord data and by following the moving geodesic sides, then give conditions for this slice to attain maximal volume.
For the dimension-$3/2$ field phase, Appendices~\ref{app:angular-chord-derivation} and \ref{app:quartet-sign-derivation} derive the angular signal of a displaced mode and the sign change when four endpoints move in a fixed state.

\subsection{Partition Dependence of Lowest Scalar Modes at General Scaling Dimension}
\label{app:generalcross}

\noindent
The phase comparison uses any available lowest scalar mode whose standard normalizable boundary condition selects the larger root $\Delta=1+\sqrt{1+M_\phi^2L^2}>1$ of the mass--dimension relation in global AdS$_3$.
The bulk field falls as $\exp(-\Delta\chi)$ at large geodesic radius $\chi$.
For the lowest rotational mode, the complex solution is $u_0=\mathcal N_0\operatorname{sech}^{\Delta}\chi\,\exp(-\mathrm{i}\Delta\tau)$, with $\mathcal N_0=(2\pi L)^{-1/2}$ and $M_\phi^2L^2=\Delta(\Delta-2)$.
To derive the lowest frequency, write a rotationally invariant mode as $R(\chi)\exp(-\mathrm{i}\Omega\tau)$ with dimensionless frequency $\Omega>0$.
Set $P_{\rm rad}(\chi)=\sinh\chi\cosh\chi$ and $W_{\rm rad}(\chi)=\tanh\chi$, and let a prime denote $d/d\chi$.
The radial Klein--Gordon equation becomes
\[
-(P_{\rm rad}R')'+M_\phi^2L^2P_{\rm rad}R=\Omega^2 W_{\rm rad}R .
\]
Multiplying by $R^*$ and integrating over $0<\chi<\infty$ gives $\Omega^2\int_0^\infty W_{\rm rad}|R|^2\,d\chi=\int_0^\infty P_{\rm rad}(|R'|^2+M_\phi^2L^2|R|^2)\,d\chi$.
To isolate the lowest possible $\Omega$, expand the square below and integrate its cross term $\Delta P_{\rm rad}\tanh\chi\,\partial_\chi|R|^2$ by parts.
Using $(P_{\rm rad}\tanh\chi)'=2P_{\rm rad}$, $P_{\rm rad}\operatorname{sech}^2\chi=W_{\rm rad}$, and $M_\phi^2L^2=\Delta(\Delta-2)$ gives
\[
(\Omega^2-\Delta^2)\int_0^\infty W_{\rm rad}|R|^2\,d\chi
=\int_0^\infty P_{\rm rad}|R'+\Delta\tanh\chi\,R|^2\,d\chi .
\]
At the origin, $R$ is bounded and $R'=O(\chi)$ by smoothness.
At infinity, $R,R'=O(\exp(-\Delta\chi))$, so the boundary terms vanish for $\Delta>1$.
For a nonzero mode, its weighted norm on the left is positive, and the square on the right is nonnegative.
The smallest possible frequency is therefore $\Omega=\Delta$ if the square can vanish everywhere.
The regular profile $R=R(0)\operatorname{sech}^\Delta\chi$ does exactly that because $R'=-\Delta\tanh\chi\,R$. \\

\noindent
To compare polygons, we start with the phase-dependent densities, their normalization, and their geodesic integrals from Appendix~\ref{app:chordphasegeneral}.
The Euclidean density $(e_--e_+)/(1-r^2)$, extended by zero, belongs to $L^1\cap L^2$ on the plane. Hence, the Fourier slice identity determines it from its chord integrals even for noninteger $\Delta>1$.
Here, $L^p(\mathbb R^2)$ denotes the class of functions with finite integral of the $p$th power of their absolute value, for $p=1,2$ in this argument.
In polar Klein coordinates $d^2y=r\,dr\,d\varphi$, the density difference times the area measure is $a^2\Delta(1-\Delta r^2)(1-r^2)^{\Delta-3/2}r\,dr\,d\varphi$.
Along a ray, the momentum phase carries more energy near the center and the field phase carries more near the boundary.
To decide which contribution wins over the entire ray segment, set $w=r^2$, so $0\leq w<1$ and $r\,dr=dw/2$.
The difference integrated out to radius $R$ per unit polar angle is
\begin{equation}
H(R)=\tfrac12 a^2\Delta\int_0^{R^2}(1-\Delta w)(1-w)^{\Delta-3/2}dw .
\label{eq:raygap}
\end{equation}
The weight in the full polygon is $(1-r^2)^{\Delta-3/2}d^2y$.
It is positive and integrable, so the ordering for any polygon is the sign of $1-\Delta\langle r^2\rangle_P$, with the average taken with this weight.
The angular area density along a ray has the sign of $1-\Delta r^2$.
Its radial primitive $H(R)$ is positive for every $0<R\leq1$, since it first increases, then decreases, and its one-sided limit $H(1)\equiv\lim_{R\to1^-}H(R)$ at the ideal boundary is
\[
H(1)=\frac{a^2\Delta}{2}\left[\frac{1-\Delta}{\Delta-1/2}
+\frac{\Delta}{\Delta+1/2}\right]
=\frac{a^2\Delta}{4\Delta^2-1}.
\]
For a polygon $P$ containing the origin in its interior, every ray at angle $\varphi$ meets the boundary once at radius $R_P(\varphi)>0$.
The integral of the density difference over that ray is $H(R_P(\varphi))$ by Eq.~\eqref{eq:raygap}.
Its angular integral is therefore strictly positive at nonzero amplitude.
The area Hessians have this ordering by Eq.~\eqref{eq:geometricresponse}, and the same ordering holds for the volume Hessians when the reflection slice maximizes the volume as in Appendix~\ref{app:globalmaximum}.
Conversely, if the entire polygon lies in $r>1/\sqrt\Delta$ and $a\ne0$, the phase difference is negative at every interior point.
Under the same volume comparison, this gives $V_-''(0)<V_+''(0)$.
For example, suppose that all vertices lie in a boundary arc of half-opening $0<\delta<\arccos(1/\sqrt\Delta)$.
Let $n$ point from the origin to the midpoint of that arc.
Each vertex $v$ then satisfies $n\cdot v\ge\cos\delta$.
Every point in the polygon is a convex combination of its vertices and therefore satisfies $r\ge n\cdot y\ge\cos\delta>1/\sqrt\Delta$.
The open conditions on the polygon vertices also show that the two orderings persist for sufficiently small changes in the vertex angles. \\

\noindent
For a polygon that straddles the circle $r=1/\sqrt{\Delta}$ and misses the origin, the two sign tests leave the ordering undetermined.
One can nevertheless evaluate the difference from the endpoints alone when the mode, its center, and its amplitude are specified.
In the Euclidean Klein disk, set $F(y)=yH(r)/r^2$ for $r=|y|>0$, with $F(0)=0$.
Eq.~\eqref{eq:raygap} gives $H(r)/r^2\to a^2\Delta/2$ at the origin, so the field is regular there.
Its flux across a circle of radius $r$ is $2\pi H(r)$, the integrated phase difference inside that circle.
Indeed, a radial vector of magnitude $H(r)/r$ has Euclidean divergence $r^{-1}\partial_r H(r)$, which is the density difference per unit $d^2y$.
For each of the $N_v$ sides, let $2\alpha_j$ be the positive angular gap between consecutive vertices in cyclic order and let the outward unit normal point toward the midpoint of that boundary arc.
On side $j$, the normal coordinate is $p_j=\cos\alpha_j$ and the tangential coordinate $t$ runs from $-s_j$ to $s_j$, where $s_j=\sin\alpha_j$.
With $E_\pm(P)=\int_P e_\pm\,d\mu$, the divergence theorem gives
\begin{equation}
E_-(P)-E_+(P)
=\sum_{j=1}^{N_v}p_j\int_{-s_j}^{s_j}
\frac{H(\sqrt{p_j^2+t^2})}{p_j^2+t^2}\,dt .
\label{eq:general-phase-side-integral}
\end{equation}
To justify the formula at an ideal vertex, first scale $P$ by a factor $0<\rho<1$ about the disk center and then take $\rho\to1^-$.
The divergence is absolutely integrable on the disk for $\Delta>1$, and the vector field is bounded and continuous on the closed disk because $H(1)$ is finite.
The interior and side integrals therefore converge to those of $P$.
A diameter has $p_j=0$ and contributes zero flux, including through its midpoint by continuity.
For a gap larger than a semicircle, $p_j<0$ keeps the sign of the outward flux in Eq.~\eqref{eq:general-phase-side-integral}. \\

\noindent
The conserved energy for global time translations, also called the Killing energy, gives both phases the same value despite their different radial distributions.
On the vacuum slice $\Sigma_0$, $\partial_\tau=L\cosh\chi\,n$, so the coefficient of $\lambda^2$ in the scalar energy generated by dimensionless time translations is $\int_{\Sigma_0}\cosh\chi\,e_\pm\,d\mu$.
Here, $d\mu=dA_0/L^2$ and $e_\pm=L^3\rho_{2,\pm}$ supply precisely the three powers of $L$ in the area element and lapse.
With $w=r^2$, the lapse-weighted measure is $\cosh\chi\,d\mu=dw\,d\varphi/[2(1-w)^2]$ for $0\le w<1$ and $0\le\varphi<2\pi$.
For the momentum mode, $e_-=(a^2\Delta^2/2)(1-w)^{\Delta+1}$, so its energy coefficient is
\[
\int_{\Sigma_0}\cosh\chi\,e_-\,d\mu
=\frac{\pi a^2\Delta^2}{2}\int_0^1(1-w)^{\Delta-1}\,dw
=\frac{\pi a^2\Delta}{2}=\Delta\beta^2 .
\]
Subtracting the field-mode energy coefficient from the momentum-mode coefficient gives $\pi a^2\Delta\int_0^1(1-\Delta w)(1-w)^{\Delta-2}\,dw=\pi a^2\Delta[w(1-w)^{\Delta-1}]_0^1=0$ for $\Delta>1$.
Thus, the second derivative for either phase is $2\Delta\beta^2$, in the same normalization as Eq.~\eqref{eq:primary_energy}.
The polygon response in Eq.~\eqref{eq:app_shape} integrates the normal energy with the area measure alone. \\

\noindent
We can also start on the gravity side and derive the boundary entropy difference by following the length of one geodesic.
In areal radial gauge, the metric functions defined in Appendix~\ref{app:directarea} have expansions $f_\pm=1+x^2+\lambda^2 f_{2,\pm}+o(\lambda^2)$, with $x=\sinh\chi$ at vacuum.
Define $\delta e(r)=e_-(r)-e_+(r)$ at fixed Klein radius and $\delta e_{\rm ar}(x)=\delta e(x/\sqrt{1+x^2})$ at the corresponding vacuum areal coordinate.
Then, Eq.~\eqref{eq:genericcross} gives $\delta e_{\rm ar}(x)=a^2\Delta[1-(\Delta-1)x^2](1+x^2)^{-\Delta-1}$.
Consequently, $2x\,\delta e_{\rm ar}(x)=a^2\Delta\,\partial_x[x^2(1+x^2)^{-\Delta}]$.
Regularity fixes $f_{2,\pm}(0)=0$ because the ratio of circumference to proper radius approaches $2\pi$ at the origin.
Integrating the radial constraint with this condition gives
\begin{equation}
\delta f_2(x)\equiv f_{2,-}-f_{2,+}
=-\frac{16\pi G_N}{L}\int_0^x t\,\delta e_{\rm ar}(t)\,dt
=-\frac{8\pi G_N}{L}a^2\Delta\,x^2(1+x^2)^{-\Delta}.
\label{eq:generalmetricdiff}
\end{equation}
At this order, the radial metric coefficient is larger in the momentum phase for every $x>0$, even though its normal energy density is smaller at large radius.
The areal gauge fixes the angular metric coefficient.
To parametrize the vacuum geodesic, use arclength $\sigma$ in units of $L$ and write a dot for $d/d\sigma$.
In the dimensionless metric $dx^2/(1+x^2)+x^2d\varphi^2$, rotational symmetry conserves $x^2\dot\varphi$.
Choose the orientation so that $x^2\dot\varphi=b\geq0$.
The unit-speed condition then gives $\dot x^2=(1+x^2)(1-b^2/x^2)$, and the minimum radius on the geodesic is $x=b$.
An arc and its complement have the same geodesic, so the angle swept on one radial half is $\gamma=\min(\alpha,\pi-\alpha)$.
For $b>0$, dividing the angular velocity by the outgoing radial velocity and substituting $x=b/\sin v$ gives
\[
\gamma=\int_b^\infty
\frac{b\,dx}{x\sqrt{1+x^2}\sqrt{x^2-b^2}}
=\int_0^{\pi/2}\frac{\sin v\,dv}{\sqrt{b^2+\sin^2v}}
=\arcsin\!\left(\frac{1}{\sqrt{1+b^2}}\right).
\]
The last integral uses $u=\cos v$, hence $b=\cot\gamma=|\cot\alpha|$, with $b=0$ at a diameter by continuity. \\

\noindent
At order $\lambda^2$, the geodesic length changes by the metric perturbation integrated along the vacuum curve, since the fixed-endpoint geodesic is stationary.
Integrating over the full vacuum curve avoids a separate endpoint term at its turning point, where the two radial parametrizations meet.
In units of $L$, the length variation for either phase is $\lambda^2\int[-f_{2,\pm}/(2f_0^2)](dx/d\sigma)^2d\sigma+o(\lambda^2)$, where $f_0=1+x^2$.
The factor of two from the second derivative cancels $1/2$.
Both radial halves of the chord contribute equally.
On the outgoing half, $\dot x^2d\sigma=\dot x\,dx$, with $\dot x=\sqrt{1+x^2}\sqrt{x^2-b^2}/x$ for $x>b$.
Subtracting the field-phase entropy Hessian from the momentum-phase Hessian, setting $\delta S_2=S_{2,-}-S_{2,+}$, and including the RT factor $L/(4G_N)$ therefore gives, by Eq.~\eqref{eq:generalmetricdiff},
\[
\delta S_2=4\pi a^2\Delta\int_b^\infty
\frac{x\sqrt{x^2-b^2}}{(1+x^2)^{\Delta+3/2}}\,dx .
\]
The substitution $z=(x^2-b^2)/(1+b^2)$ reduces the last integral to $(1+b^2)^{-\Delta}/2$ times $\int_0^\infty z^{1/2}(1+z)^{-\Delta-3/2} dz$.
With $v=z/(1+z)$ and integration by parts, this integral is $\int_0^1v^{1/2}(1-v)^{\Delta-1}dv=I_\Delta/(2\Delta)$.
The identity $1/(1+b^2)=\sin^2\alpha$ therefore gives $\delta S_2=\pi a^2 I_\Delta\sin^{2\Delta}\alpha$, including the smooth limit $b\to0^+$ for a diameter.
In the same areal gauge, every fixed domain of positive area has a larger area response in the momentum phase, because the area element is proportional to $x/\sqrt f$.
The opposite ordering of the actual outer polygon consequently requires displacing the geodesic sides.

\subsection{Polygon Sums for Positive Interval-Entropy Variations}
\label{app:onephase}

\noindent
At $\Delta=3/2$, the density times hyperbolic area is polynomial in the Klein radius, so its polygon integral reduces to a sum over straight sides.
This makes it possible to use a measured interval response to predict the volume for other choices of endpoints, find its extrema, and distinguish changes in the metric from displacements of the sides. \\

\noindent
The $\Delta=3/2$ field mode has $M_\phi^2L^2=-3/4$, above the AdS$_3$ stability bound $-1$.
Its local normal energy density can be negative even though its cylinder energy is positive.
In Eq.~\eqref{eq:cftelementaryS}, $z=s^2(1-u^2)$, $s=\sin\alpha$, and $(1-z^{\Delta-1})/(1-z)=1/(1+\sqrt z)$ at this dimension.
The remaining moment is $I_{3/2}=3\pi/8$, as follows from $u=\sin t$ in $I_{3/2}=\int_{-1}^1(1-u^2)^{3/2}du$.
These substitutions give Eq.~\eqref{eq:positive32} for $0<s\le1$, with the same result on complementary arcs by purity.
Its first term is nonincreasing and its second strictly decreasing, so the minimum of $S_{2,+}/(\beta^2s^2)$ occurs at $s=1$.
There the integral of $1-\sqrt{1-u^2}$ from $-1$ to $1$ is $2-\pi/2$, yielding a strictly positive minimum. \\

\noindent
In Klein coordinates, we write $D=1-r^2$.
The general scalar densities give $e_+=\tfrac38 a^2(3r^2-1)D^{3/2}$ and $e_-=\tfrac98 a^2D^{5/2}$.
With $d\mu=D^{-3/2}d^2y$, each response is the Euclidean integral of a polynomial.
At a common amplitude $a$, the sum of the energy densities times $d\mu$ is $a^2[3(3r^2-1)+9(1-r^2)]d^2y/8$.
The terms proportional to $r^2$ cancel pointwise, so integration gives
\begin{equation}
E_+(P)+E_-(P)=\frac{3a^2}{4}\operatorname{Area}_{\rm E}(P),
\qquad \operatorname{Area}_{\rm E}(P)=\int_P d^2y .
\label{eq:equal-amplitude-energy-sum}
\end{equation}
Here, $\operatorname{Area}_{\rm E}(P)$ is the coordinate area in the Klein disk.
The polygon $P$ is the convex hull of $N_v\ge3$ distinct unit boundary points in cyclic order.
For $1\le j\le N_v$, the boundary arc from vertex $j$ to vertex $j+1$ has opening $2\alpha_j$, with $N_v+1$ identified with $1$.
Thus, $0<\alpha_j<\pi$ and $\sum_{j=1}^{N_v}\alpha_j=\pi$.
The outward unit normal of the corresponding straight side points to the midpoint of this arc, so its signed normal coordinate is $p_j=\cos\alpha_j$ even for a major arc.
If $t$ measures signed Euclidean distance along the side from its midpoint, then $r^2=p_j^2+t^2$, $-s_j\le t\le s_j$, where $s_j=\sin\alpha_j$. \\

\noindent
The Euclidean vector field $F(y)=(3r^2-2)y/4$ has the required outward flux.
The elementary divergence identities $\nabla\cdot y=2$ and $\nabla\cdot(r^2 y)=4r^2$ give $\nabla\cdot F=3r^2-1$.
The flux of $F$ through side $j$ is
\begin{align}
\int_{-s_j}^{s_j}\frac{p_j}{4}\,[3(p_j^2+t^2)-2]dt
&=\frac{p_js_j}{2}(3p_j^2+s_j^2-2)\nonumber\\
&=\frac{p_js_j}{2}(2p_j^2-1)
=\frac18\sin(4\alpha_j).
\end{align}
The sum of the outward fluxes gives
\begin{equation}
E_+(P)\equiv\int_Pe_+d\mu=\frac{3a^2}{64}\sum_{j=1}^{N_v}\sin(4\alpha_j).
\label{eq:polygon32}
\end{equation}
A side through the center contributes zero because its outward normal is perpendicular to $y$ all along that side.
For any choice of the boundary vertices, the positive proportionality factor $V_+''(0)/L^2=(24\pi/c)E_+(P)$ transfers the sign of the side sum directly to the volume response. \\

\noindent
For a triangle, the three half-openings equal the opposite Euclidean interior angles by the inscribed angle theorem.
The angle sum $\alpha_3=\pi-\alpha_1-\alpha_2$ gives
\[
\sin 4\alpha_1+\sin 4\alpha_2-\sin[4(\alpha_1+\alpha_2)]
=-4\sin(2\alpha_1)\sin(2\alpha_2)\sin(2\alpha_3).
\]
For the Euclidean triangle drawn with straight chords, each factor $\sin(2\alpha_j)$ is positive for an acute angle, zero for a right angle, and negative for an obtuse angle.
The unit circle passes through all three vertices, so its center lies inside an acute triangle, on the hypotenuse of a right triangle, and outside an obtuse triangle.
The circle center is also the center of the scalar mode, which explains why its position alone fixes the sign for a triangle.
The physical hyperbolic angles at the ideal vertices are all zero.
For equal arcs, the sum is $N_v\sin(4\pi/N_v)$.
At $N_v=3$, $\operatorname{Area}_{\rm E}(P)=3\sqrt3/4$, so $E_+=-9\sqrt3\,a^2/128$ and $E_-=81\sqrt3\,a^2/128$.
The regular square has $E_+=0$, and $E_+>0$ for every regular polygon with $N_v\ge5$.
The integral of the same $e_+\,d\mu$ over the full unit disk is $3\pi a^2/16>0$.

\subsubsection{Interval-Entropy and Triangle-Volume Sign Thresholds}
\label{sec:entropy-opening-threshold}
\noindent
For which masses does every boundary interval gain entropy while the regular triangle shrinks?
For a nonzero lowest field mode, these requirements amount to two sign tests, one on the ordinary entropy Hessian at every opening and one on the energy integrated over a single triangle.
For $1<\Delta\le 2$, we set $q=\Delta-1$, $s=\sin\alpha\in(0,1]$, and $b=s^2(1-u^2)$ with $-1\le u\le1$.
Eq.~\eqref{eq:cftelementaryS} gives the ordinary entropy Hessian in the form
\[
H_\Delta(s)\equiv\frac{S_{2,+}(\alpha)}{\beta^2s^2}
=2\Delta\int_{-1}^1u^2\frac{1-b^q}{1-b}\,du-I_\Delta s^{2q}.
\]
The quotient has value $q$ at $b=1$ by continuity.
Integrating the derivative of $x^q$ from $b$ to $1$ and putting $x=b+(1-b)t$ gives
\[
\frac{1-b^q}{1-b}=q\int_0^1[b+(1-b)t]^{q-1}\,dt,\qquad 0\le b<1 .
\]
For $0<q<1$, increasing $b$ increases the base for every $0<t<1$ and therefore strictly decreases the integrand.
At $q=1$, the quotient equals one throughout.
In both cases, it is bounded above by its value one at $b=0$, and $-I_\Delta s^{2q}$ decreases strictly for $0<s\le1$.
Thus, $H_\Delta(s)$ is continuous and strictly decreasing on the physical range, including the case $\Delta=2$.
For $s\to0^+$, the integrand is bounded by $2\Delta u^2$ and tends to that function, while the last term in $H_\Delta$ vanishes.
The short-arc limit is therefore $2\Delta\int_{-1}^1u^2\,du=4\Delta/3>0$.
At the other endpoint $s=1$, a semicircle gives the smallest value of $H_\Delta$.
Because $S_{2,+}=\beta^2s^2 H_\Delta(s)$ with a positive prefactor on every proper arc, a positive semicircle reading implies positive readings at all widths in the stated mass range.
At $s=1$, the denominator $1-b$ simplifies to $u^2$, and the moment recurrence $2\Delta I_{\Delta-1}=(2\Delta+1)I_\Delta$ gives
\[
H_\Delta(1)=2\Delta(2-I_{\Delta-1})-I_\Delta
=4\Delta-2(\Delta+1)I_\Delta .
\]
The integral $I_\Delta$ decreases strictly with $\Delta$, while $2\Delta/(\Delta+1)$ increases strictly.
At $\Delta=1$, the former exceeds the latter, whereas $I_{3/2}=3\pi/8<6/5$.
Consequently, a unique $\Delta_*\in(1,3/2)$ solves $I_{\Delta_*}=2\Delta_* /(\Delta_*+1)$.
Throughout $1<\Delta\le2$, positivity of the field-phase Hessian on \emph{every} proper interval is equivalent to $\Delta>\Delta_*$.
At $\Delta=\Delta_*$, only semicircles have zero Hessian.
For $1<\Delta<\Delta_*$, the positive short-arc limit and the negative semicircle value instead give a unique $s_*(\Delta)\in(0,1)$ with $H_\Delta(s_*)=0$.
Since $\beta^2s^2>0$, the ordinary entropy Hessian is positive for $\sin\alpha<s_*$, zero for $\sin\alpha=s_*$, and negative for $\sin\alpha>s_*$.
The two zero half-openings $\arcsin s_*$ and $\pi-\arcsin s_*$ are the same physical sign test on an arc and its complement, which have equal ordinary entropy by purity.
For known $\Delta_*<\Delta\le2$, the ratio of the momentum and field profiles of $S_2/(\beta^2 s^2)$ is $1+2I_\Delta s^{2\Delta-2}/H_\Delta(s)$ by Eq.~\eqref{eq:phasegap_main}.
It increases strictly, so two different values of $s$ also distinguish the two pure phases in this range. \\

\noindent
The regular ideal triangle $T$ has a negative field-phase response throughout $3/2\le\Delta\le8/5$.
We write $D=1-r^2$ and $k=\Delta-3/2$.
The formulas for $e_++e_-$ and $e_-$ in Appendix~\ref{app:kg} yield
\[
e_+\,d\mu=\frac{a^2\Delta^2}{2}D^k
\left(2-\frac{2}{\Delta}-D\right)d^2y .
\]
We denote the normalized Euclidean average with this weight by $\langle f\rangle_k=\int_T f D^k d^2y/\int_T D^k d^2y$.
Here, $D$ is largest at the mode center.
The sign of the polygon integral therefore depends on how strongly the weight samples this central region.
For the regular triangle, the three unit vertex vectors $v_i$, $1\le i\le3$, sum to zero.
In barycentric coordinates $y=\sum_{i=1}^3 t_i v_i$, uniform area measure gives $\langle t_i t_j\rangle_0=(1+\delta_{ij})/12$ for $1\le i,j\le3$ by integration on $t_1,t_2\ge0$, $t_1+t_2\le1$, $t_3=1-t_1-t_2$.
Here, $\delta_{ij}=1$ for $i=j$ and $0$ otherwise.
Therefore, $\langle r^2\rangle_0=(\sum_{i=1}^3 |v_i|^2+|\sum_{i=1}^3 v_i|^2)/12=1/4$ and $\langle D\rangle_0=3/4$.
For $k\ge0$, the functions $D$ and $D^k$ are nondecreasing functions of the same variable, so $\langle D^{k+1}\rangle_0\ge\langle D\rangle_0\langle D^k\rangle_0$, strictly for $k>0$ since $D$ varies over $T$.
Thus, $\langle D\rangle_k\ge3/4$, whereas $2-2/\Delta\le3/4$ on the stated interval.
At $\Delta=3/2$, the latter bound is strict, and at $\Delta=8/5$, the former is strict.
It follows that $E_+(T)<0$ throughout this interval, even though every $S_{2,+}$ is positive.
The corresponding masses obey $-3/4\le M_\phi^2 L^2\le-16/25$ under standard quantization.
This elementary bound gives a sufficient window.
The following monotonicity argument fixes its sharp endpoints. \\

\noindent
To find the mass at which the regular triangle changes from shrinking to growing, we first establish that its field-phase volume can change sign only once as $\Delta$ increases.
The argument works for any nondegenerate convex ideal polygon $P$ whose interior contains the mode center.
For $k=\Delta-3/2>-1/2$, we write $J_P(k)=\int_P D^k d^2y$ with $D=1-r^2$, and let $\langle f\rangle_k=J_P(k)^{-1}\int_P fD^k d^2y$.
The sign of its field-phase energy is the sign of
\[
f_P(\Delta)=2-\frac{2}{\Delta}-\frac{J_P(k+1)}{J_P(k)}
=2-\frac{2}{\Delta}-\langle D\rangle_k .
\]
Since $D=1-r^2$ is largest near the center, increasing $k$ weights the central part of the polygon more strongly.
To differentiate the resulting mean, one must vary both its numerator and its normalizing integral.
Differentiating $D^k$ inserts $\ln D$, so the quotient rule gives $d\langle D\rangle_k/dk=\operatorname{Cov}_k(D,\ln D)$, with $\operatorname{Cov}_k(X,Y)=\langle XY\rangle_k-\langle X\rangle_k\langle Y\rangle_k$.
The factor $D^k|\ln D|$ remains integrable for $k>-1$, which justifies differentiation under the integral.
At radius $r=\sqrt{1-D}$, let $w_P(D)$ be the fraction of the circle lying inside $P$.
With $y_\theta=r(\cos\theta,\sin\theta)$, this fraction is $w_P(D)=(2\pi)^{-1}\int_0^{2\pi}\mathbf1_P(y_\theta)\,d\theta$, where $\mathbf1_P$ equals one inside $P$ and zero outside.
Moving toward the origin keeps each included direction inside the convex polygon, so $w_P(D)$ is nondecreasing in $D$.
The full-disk probability density of $D$ after weighting by $D^k$ is $(k+1)D^k$ on $(0,1)$.
Polygon averaging reweights this distribution by $w_P(D)$.
For the decreasing function $g(D)=(1-D)(-\ln D)$, this gives $\langle g\rangle_k\le(k+1)\int_0^1D^k g(D)\,dD$.
Indeed, the covariance of $g$ and $w_P$ under the full-disk distribution is nonpositive, as follows by averaging $[g(D)-g(D')][w_P(D)-w_P(D')]\le0$ over independent $D,D'$.
Consequently,
\[
\operatorname{Cov}_k(D,\ln D)=\langle g\rangle_k
-\langle1-D\rangle_k\langle-\ln D\rangle_k
\le \frac{2k+3}{(k+1)(k+2)^2}<\frac{2}{\Delta^2}.
\]
The first bound uses $\int_0^1x^a(-\ln x)dx=(a+1)^{-2}$ for $a>-1$.
The strict bound is equivalent to $\Delta^3<(\Delta-1/2)(\Delta+1/2)^2$, whose difference is $(4\Delta^2-2\Delta-1)/8>0$ for $\Delta>1$.
The bound gives $df_P/d\Delta=2/\Delta^2-\operatorname{Cov}_k(D,\ln D)>0$, since $dk/d\Delta=1$.
Its limit as $\Delta\to1^+$ is $-\langle D\rangle_{-1/2}<0$ and its value at $\Delta=2$ is $1-\langle D\rangle_{1/2}>0$.
The opposite endpoint signs give at least one zero between $1$ and $2$, and strict increase excludes a second.
Thus, $\Delta_P$ separates negative and positive field-phase volume responses for this polygon. \\

\noindent
For the regular triangle, $T$, symmetry divides it into six sectors with $0\le\xi\le\pi/3$ and $0\le r\le(2\cos\xi)^{-1}$, where $\xi$ is the angle from the side normal.
Radial integration gives
\[
J_T(k)=\frac{3}{k+1}\int_0^{\pi/3}
\left[1-\left(1-\frac{1}{4\cos^2\xi}\right)^{k+1}\right]d\xi .
\]
The unique zero of $f_T$ is $\Delta_T\simeq 1.60914446$.
The strict inequalities $S_{2,+}(I)>0$ for every proper interval $I$ and $V_{+,T}''(0)<0$ therefore hold simultaneously exactly for $\Delta_*<\Delta<\Delta_T$ in standard quantization, with $\Delta_*\simeq1.46267391$ determined by the semicircle condition above.
At the lower endpoint, the semicircle entropy Hessian vanishes, and the triangle volume Hessian is negative.
At the upper endpoint, all proper interval entropy Hessians are positive, and the triangle volume Hessian is zero.

\subsubsection{Inferring Scalar Phase and Amplitude from Two Measurements}
\label{sec:two-width-proof}
\noindent
At $\Delta=3/2$, a single interval measures a combination of the phase and the amplitude of an excitation.
Suppose its center is known and only the lowest spatial mode is excited, in either the time-even field phase $(+)$ or the time-odd momentum phase $(-)$.
A second interval width can distinguish these possibilities without an independent amplitude measurement if the ratio of the two readings depends on the phase.
Let $\beta_+$ and $\beta_-$ denote the respective candidate boundary amplitudes, both nonzero, with $a_\pm^2=2\beta_\pm^2/\pi$.
For $s=\sin\alpha\in(0,1]$, division of the entropy Hessian by $\beta_\pm^2s^2$ leaves a known positive function $H_\pm(s)$.
Rotations account for independence of the interval center, and purity equates the entropies at complementary openings $\alpha$ and $\pi-\alpha$.
The field profile $H_+$ is strictly decreasing by Eq.~\eqref{eq:positive32}.
Since $I_{3/2}=3\pi/8$, Eq.~\eqref{eq:phasegap_main} gives $H_-(s)=H_+(s)+qs$, with $q=3\pi/4$. \\

\noindent
For two measurements with $0<s_1<s_2\le1$, we write $y_i=S_2(\alpha_i)/s_i^2$ and $h_i=H_+(s_i)$, where $s_i=\sin\alpha_i$ and $i=1,2$.
As the squared amplitude varies, the possible pairs $(y_1,y_2)$ lie on one of two rays through the origin, directed along $(h_1,h_2)$ for the field phase or $(h_1+qs_1,h_2+qs_2)$ for the momentum phase.
The profile vectors have determinant
\[
h_1(h_2+qs_2)-h_2(h_1+qs_1)
=q(h_1s_2-h_2s_1)>0 ,
\]
because $h_1>h_2>0$ and $s_2>s_1$.
The two predicted pairs therefore lie on rays of different slopes. Hence, the ratio $y_2/y_1$ selects the phase, and $\beta_+^2=y_1/h_1$ or $\beta_-^2=y_1/(h_1+qs_1)$ fixes its magnitude.
Both readings are insensitive to the sign of the amplitude.
Eq.~\eqref{eq:polygon32} gives every field-phase polygon response.
The momentum-phase energy is
\[
E_-(P)=\frac{3a_-^2}{4}\operatorname{Area}_{\rm E}(P)
-\frac{3a_-^2}{64}\sum_{j=1}^{N_v}\sin(4\alpha_j)
\]
by Eq.~\eqref{eq:equal-amplitude-energy-sum} evaluated at $a=a_-$.
The volume follows from Eq.~\eqref{eq:geometricresponse} in either case. \\

\noindent
Because both $H_+(s)$ and $H_-(s)$ are positive, adjusting the amplitude of either preparation can reproduce any one positive interval reading.
For a semicircle, Eq.~\eqref{eq:positive32} gives $H_+(1)=3(16-5\pi)/8$ and hence $H_-(1)=3(16-3\pi)/8$.
Choosing $\beta_-^2/\beta_+^2=(16-5\pi)/(16-3\pi)$ matches the semicircle Hessians but gives regular-triangle energies $-9\sqrt3\,a_+^2/128$ and $81\sqrt3\,a_-^2/128$.
The volumes inherit these opposite signs because the multiplier $24\pi/c$ in Eq.~\eqref{eq:geometricresponse} is positive.
The matched states have different cylinder-energy Hessians $E_{\rm cyl}''(0)=2\Delta\beta^2$.
If the energy is known independently, it fixes $\beta^2$, and the gap in Eq.~\eqref{eq:phasegap_main} lets one proper interval distinguish the phases. \\

\noindent
Two widths also suffice to predict the polygon volume for either definite phase.
Introduce the squared amplitude $B=\beta^2>0$ and its momentum part $M$, equal to zero in the field phase and to $B$ in the momentum phase.
Both cases obey the linear system $y_i=h_i B+q s_i M$ for $i=1,2$.
The combinations $s_2y_1-s_1y_2$ and $h_1y_2-h_2y_1$ eliminate $M$ and $B$, respectively.
With $d_{12}=h_1s_2-h_2s_1>0$, this gives $B=(s_2y_1-s_1y_2)/d_{12}$ and $M=(h_1y_2-h_2y_1)/(q d_{12})$.
For the counterclockwise vertices $\boldsymbol v_j=(\cos\varphi_j,\sin\varphi_j)$, the oriented triangle with vertices $0,\boldsymbol v_j,\boldsymbol v_{j+1}$ contributes $\tfrac12\det(\boldsymbol v_j,\boldsymbol v_{j+1})=\tfrac12\sin(2\alpha_j)$.
Here, $2\alpha_j$ is the positive angular gap and $\boldsymbol v_{N_v+1}=\boldsymbol v_1$.
Summing these signed areas gives $A_{\rm E}=\tfrac12\sum_{j=1}^{N_v}\sin(2\alpha_j)$, since overlapping exterior pieces cancel when the origin is outside $P$.
At equal amplitudes, the two energy measures add to $3a^2 d^2y/4$, and integration over $P$ gives $E_-(P)=3a^2A_{\rm E}/4-E_+(P)$.
The amplitude relation $a^2=2B/\pi$ and Eqs.~\eqref{eq:geometricresponse} and~\eqref{eq:polygon32} give the geometric factors $F=\sum_{j=1}^{N_v}\sin(4\alpha_j)$ and $16A_{\rm E}-F$ in $4cV''/(9L^2 B)$ for the two respective phases.
For either definite parity, the scaled volume is therefore $F(B-M)+(16A_{\rm E}-F)M=FB+KM$, where $K=16A_{\rm E}-2F$.
Substituting the two measured coefficients yields
\[
\frac{4c}{9}\frac{V''(0)}{L^2}
=\frac{(qs_2F-h_2K)y_1+(h_1K-qs_1F)y_2}{qd_{12}}.
\]
For a third interval with $s_3=\sin\alpha_3$ and $0<\alpha_3<\pi$, the predicted Hessian satisfies $S_2(\alpha_3)/s_3^2=B H_+(s_3)+q M s_3$ with the same $B$ and $M$.
In particular, the two pure phases require $M(B-M)=0$. \\

\noindent
Eqs.~\eqref{eq:square_entropy_ratio}--\eqref{eq:direct_square_volume} specialize the two-width measurement to a semicircle and a quarter circle.
The contrasts at small finite amplitude obey $\Pi_\square(\lambda)-1=(6/c)B k_\varsigma\lambda^2+o(\lambda^2)$, so the sign agrees with $k_\varsigma$ for sufficiently small nonzero $\lambda$.
The phase-specific constants also predict a further interval with full opening $\pi/3$.
Using $M''/B=6-\sqrt3\pi$ and $\tan(\pi/12)=2-\sqrt3$ gives
\[
S_{2,\varsigma}(\pi/6)
=\frac{d_\square}{k_\varsigma}
\left(6+2\sqrt3\,\pi-\frac{C_\varsigma\pi}{64}\right),
\qquad C_+=339,\quad C_-=333 .
\]
For either definite parity, the same square measurement fixes the normalization appearing in each term. \\

\noindent
To determine what the interval measurements reveal about an intermediate phase, consider a normalized pure-state family $|\psi_\lambda\rangle$ with $|\psi_0\rangle=|0\rangle$.
Its tangent at the vacuum is $\left.\partial_\lambda|\psi_\lambda\rangle\right|_{\lambda=0}=\beta\exp(\mathrm{i}\varphi)|\mathcal{O}\rangle$, where $\beta>0$ and $0\leq\varphi<2\pi$. 
We keep the spin-zero Hermitian primary, a pure rotationally invariant completion, and the same boundary observation time.
In Eq.~\eqref{eq:cftmatrix}, the two insertions give $h_\varphi(x)=\cos\varphi\,\cosh x+\mathrm{i}\sin\varphi\,\sinh x$, where $x=\alpha\omega$ and $\omega$ is real.
The field and momentum terms are real and purely imaginary, respectively, so squaring the magnitude removes their interference.
The relative-entropy Hessian in Eq.~\eqref{eq:cftbkm} is therefore a sum of the two unit-amplitude responses with weights $B_{\rm c}=\beta^2\cos^2\varphi$ and $B_{\rm s}=\beta^2\sin^2\varphi$.
Rotational symmetry and the cylinder-energy Hessian $2\Delta(B_{\rm c}+B_{\rm s})$ give the same linear combination for the modular-energy contribution and hence for the ordinary entropy Hessian. \\

\noindent
For an arbitrary pair $0<s_1<s_2\leq1$, put $f=B_{\rm s}/B=\sin^2\varphi$, with $B=B_{\rm c}+B_{\rm s}>0$.
The normalized readings $y_i=S_2(\alpha_i)/s_i^2$ then satisfy $y_i=B g_i(f)$, where $g_i(f)=h_i+qs_i f$ for $i=1,2$.
Here, $h_i=H_+(s_i)$ and $q=3\pi/4$, so $g_i(f)>0$ throughout $0\leq f\leq1$.
Taking the ratio eliminates $B$ and gives
\[
R(f)=\frac{y_2}{y_1}=\frac{h_2+qs_2 f}{h_1+qs_1 f},
\qquad
R'(f)=\frac{q d_{12}}{(h_1+qs_1 f)^2}>0 .
\]
Thus, exact positive readings fit a single-primary tangent precisely when $R(0)\leq y_2/y_1\leq R(1)$.
Writing $R=y_2/y_1$, the recovered fraction is $f=(Rh_1-h_2)/[q(s_2-Rs_1)]$, since $s_2-R(f)s_1=d_{12}/g_1(f)>0$.
Either width then gives $B=y_i/g_i(f)$.
A ratio at its lower endpoint selects the pure field preparation, and a ratio at its upper endpoint selects the pure momentum preparation.
A ratio strictly between them requires both quadratures within this single-mode family. \\

\noindent
With uncertain readings, we can constrain the phase without dividing by a noisy measurement near zero.
For $i=1,2$, divide the error limits on $S_2(\alpha_i)$ by $s_i^2$ to obtain $[l_i,u_i]$ for $y_i$.
A fraction $f$ fits both readings precisely when the amplitude intervals $[l_i/g_i(f),u_i/g_i(f)]$ overlap at some $B>0$.
This requires $u_1,u_2>0$, after which the remaining conditions are
\[
l_1 g_2(f)\leq u_2 g_1(f),\qquad
l_2 g_1(f)\leq u_1 g_2(f).
\]
Both inequalities are linear in $f$, so their intersection with $[0,1]$ gives the compatible fractions, possibly none.
If the error bounds allow every pair in the stated rectangle, every retained fraction is realized by an amplitude in the overlap. \\

\noindent
For the square, the two measured quantities are the semicircle response $Y_{\rm h}=S_2(\pi/2)$ and its excess over the quarter-circle response $d_\square=Y_{\rm h}-S_2(\pi/4)$.
They obey $Y_{\rm h}=U_+B_{\rm c}+U_-B_{\rm s}$ and $d_\square=k_+B_{\rm c}+k_-B_{\rm s}$.
Since $U_\pm>0$ and $k_+<0<k_-$, the determinant $\mathcal D_\square=U_+k_--U_-k_+$ is positive and the two readings determine the squared quadrature amplitudes,
\[
B_{\rm c}=\frac{k_-Y_{\rm h}-U_-d_\square}{\mathcal D_\square},
\qquad
B_{\rm s}=\frac{U_+d_\square-k_+Y_{\rm h}}{\mathcal D_\square}.
\]
An admissible single-primary tangent has $B_{\rm c},B_{\rm s}\ge0$, and its overall amplitude obeys $\beta^2=B_{\rm c}+B_{\rm s}$.
For nonzero $\beta$, the fraction $B_{\rm s}/\beta^2=\sin^2\varphi$ is accessible, while the relative signs of the quadratures drop out of these entropy Hessians.
In particular, $d_\square=0$ at $\sin^2\varphi=-k_+/(k_--k_+)=(36-25\sqrt2)/(8-2\sqrt2)$ although both readings are positive.
The polygon-volume map in Eq.~\eqref{eq:square_to_polygon} was derived for definite parity.
For an intermediate phase, the two interval readings still determine the squared quadrature amplitudes. However, the simultaneous field and velocity can carry radial momentum.
Indeed, the coherent mean field $\phi_1=A\operatorname{sech}^{\Delta}\!\chi\cos(\Delta\tau-\varphi)$, with amplitude $A$ and phase $\varphi$, gives a quadratic normal--radial stress at $\tau=0$ of
\[
\left.\lim_{\lambda\to0}\frac{T_{\mu\nu}n^\mu(\partial_\chi)^\nu}{\lambda^2}\right|_{\tau=0}
=(n\phi_1)\,\partial_\chi\phi_1
=-\frac{A^2\Delta^2}{L}\tanh\chi\,
\operatorname{sech}^{2\Delta+1}\!\chi\,\sin\varphi\cos\varphi .
\]
Here, $n=(L\cosh\chi)^{-1}\partial_\tau$ is orthogonal to $\partial_\chi$, so the metric term in $T_{\mu\nu}$ drops out.
For two nonzero quadratures, this flux is nonzero at $\chi>0$.
The gravitational momentum constraint requires zero normal--tangential stress on a reflection slice, whose extrinsic curvature vanishes.
The boundary phase inference thus extends beyond the range of the reflection-slice volume formula at that observation time.

\subsubsection{Square Kernel and Triangle Moments}
\noindent
The square contrast and the square's volume sample the same density with different spatial weights.
To find the contrast, rotate the mode-centered Klein coordinates $(x,y)$ so that the square $Q$ has sides $x,y=\pm1/\sqrt2$.
Its quartet weight is $K_Q=C/d$, with $d=\sqrt{1-x^2-y^2}$, $m=\max(|x|,|y|)$ and $C=1-\sqrt2\,m$.
The two terms in the scalar density then require the geometric integrals
\[
I=\int_Q(C/d)\,dx\,dy,\qquad J=\int_Q Cd\,dx\,dy .
\]
The bulk field- and momentum-phase measures are $(3a^2/8)(2-3d^2)dx\,dy$ and $(9a^2/8)d^2 dx\,dy$, respectively.
Since $\beta^2=\pi a^2/2$, the finite-quartet energy formula gives $c\Pi_\square''(0)/(12\beta^2)=(3/4)(2I-3J)$ in the field phase and $(9/4)J$ in the momentum phase. \\

\noindent
The eight octants of the square give identical contributions.
In the octant $0\le y\le x\le l=1/\sqrt2$, the primitive of $1/d$ in $y$ gives $A(x)=\arcsin[x/\sqrt{1-x^2}]$ at the upper endpoint.
The corresponding integral of $d$ is $\{x\sqrt{1-2x^2}+(1-x^2)A(x)\}/2$.
With $L_n=\int_0^l x^nA(x)dx$ for $n=0,1,2,3$, it follows that
\[
I=8(L_0-\sqrt2 L_1),\qquad
J=4(1/6-\pi/32+L_0-\sqrt2 L_1-L_2+\sqrt2 L_3).
\]
Here, $1/6-\pi/32=\int_0^l(x-\sqrt2 x^2)\sqrt{1-2x^2}\,dx$.
The derivative $A'(x)=[(1-x^2)\sqrt{1-2x^2}]^{-1}$ and integration by parts yield $(n+1)L_n=l^{n+1}\pi/2-H_{n+1}$, where
\[
H_j=\int_0^l\frac{x^j\,dx}{(1-x^2)\sqrt{1-2x^2}},\qquad 0\le j\le4.
\]
The substitution $x=\sin t/\sqrt2$ gives $H_0=\pi/2$, $H_1=\pi/4$, and $H_{j+2}=H_j-2^{-(j+1)/2}\int_0^{\pi/2}\sin^j t\,dt$ for $j=0,1,2$.
Thus,
\[
I=\pi(3\sqrt2-4),\qquad J=\frac{\pi}{24}(23\sqrt2-28).
\]
The resulting coefficients are
\[
\frac34(2I-3J)=\frac{3\pi}{32}(25\sqrt2-36)=k_+,\qquad
\frac94J=\frac{3\pi}{32}(23\sqrt2-28)=k_- .
\]
These surface integrals give the same signed square curvatures as the boundary modular-energy and relative-entropy calculation. \\

\noindent
The outer panels of Fig.~\ref{fig:triangle-signs} compare triangles of equal vacuum area $\pi L^2$ using the same interval data.
Let $T_{\rm c}$ be the centered regular triangle and $T_{\rm o}$ the exterior triangle with gaps $\pi/3,\pi/3,4\pi/3$.
Their shape factors are
\[
\begin{array}{c|cc}
P & F(P) & A_{\rm E}(P)\\ \hline
T_{\rm c} & -3\sqrt3/2 &3\sqrt3/4\\
T_{\rm o} &  3\sqrt3/2 &\sqrt3/4 .
\end{array}
\]
For example, $\sum_{j=1}^{3}\sin(2x_j)=2\sin(2\pi/3)+\sin(8\pi/3)=3\sqrt3/2$ for $T_{\rm o}$, where $x_j$ is the full angular gap for $j=1,2,3$.
For either definite phase, $S_2(\pi/2)-S_2(\pi/4)$ fixes the common squared amplitude $\beta^2$ by Eq.~\eqref{eq:square_readout_main}.
For the field phase, the two triangles have the tabulated shape factors $F$.
For the momentum phase, the corresponding factors $16A_{\rm E}-F$ are $27\sqrt3/2$ and $5\sqrt3/2$. \\

\noindent
Which feature of a triangle makes its field-phase response change sign?
For a nondegenerate convex polygon $P$, define $m_2(P)=A_{\rm E}(P)^{-1}\int_P|y|^2d^2y$, the mean squared radius of a point chosen uniformly in the Euclidean area inside the Klein polygon.
Because the densities with respect to this area are linear in $r^2$, their integrals depend only on the area and this one average,
\[
E_+(P)=\frac{3a^2A_{\rm E}(P)}8[3m_2(P)-1],
\qquad E_-(P)=\frac{9a^2A_{\rm E}(P)}8[1-m_2(P)] .
\]
Thus, $F=8A_{\rm E}(3m_2-1)$ and $16A_{\rm E}-F=24A_{\rm E}(1-m_2)$.
The field-phase sign changes at $m_2=1/3$, whereas the momentum response is positive for every polygon of nonzero area, since $r<1$ in its interior.
For equal amplitudes, the momentum response exceeds the field response precisely when $m_2<2/3$. \\

\noindent
The radius moment $m_2$ can be found directly from the three unit vertex vectors $v_i$ of an ideal triangle.
In barycentric coordinates, a point is the weighted average $y=\sum_{i=1}^3t_i v_i$ with $t_i\ge0$ and $\sum_{i=1}^3t_i=1$.
Averaging $|y|^2=\sum_{i,j=1}^3t_i t_j v_i\cdot v_j$ with $\langle t_i t_j\rangle=(1+\delta_{ij})/12$ gives $m_2=(3+|v_1+v_2+v_3|^2)/12$.
The vector sum vanishes by symmetry for the centered regular triangle $T_{\rm c}$.
For $T_{\rm o}$, rotate the middle vertex to angle zero and the other two to $\pm\pi/3$, so the sum points along the middle vertex and has length $1+2\cos(\pi/3)=2$.
The two mean squared radii are therefore $1/4$ and $7/12$, on opposite sides of the field-phase threshold $1/3$.
Substitution in the density integral gives
\[
\big(E_+(T_{\rm c}),E_+(T_{\rm o})\big)
=\frac{9\sqrt3 a^2}{128}(-1,1),\qquad
\big(E_-(T_{\rm c}),E_-(T_{\rm o})\big)
=\frac{3\sqrt3 a^2}{128}(27,5).
\]
Multiplication by $24\pi/c$ gives $V''/L^2$.
The amplitude relation $\pi a^2=2\beta^2$ and Eq.~\eqref{eq:square_phase_amplitude} recover the square-readout prediction for each triangle.
In particular, the ratio $V''(T_{\rm o})/V''(T_{\rm c})$ is $-1$ in the field phase and $5/27$ in the momentum phase for every nonzero amplitude.

\subsubsection{Extremal Partitions for Fixed Field Modes}
\label{app:polygonextremes}

\noindent
The field-phase semicircle susceptibility is $\mathcal S=S_{2,+}(\pi/2)=3(16-5\pi)\beta^2/8>0$ for $\Delta=3/2$ and $\beta\ne0$.
For a convex ideal polygon $P$ with $N=N_v\ge3$ distinct vertices, its successive angular gaps are $x_j=2\alpha_j>0$ with $\sum_{j=1}^N x_j=2\pi$.
Eq.~\eqref{eq:semicircle_calibration} identifies the normalized volume measurement with a dimensionless function of the gaps,
\begin{equation}
F(P)\equiv\frac{(16-5\pi)c}{6\mathcal S}\frac{V_+''(0)}{L^2}
=\sum_{j=1}^N\sin(2x_j).
\label{eq:extremal_partition}
\end{equation}
A new boundary vertex divides an old gap $x+y<2\pi$ into two gaps $x,y>0$.
The sides of the resulting polygon enclose the old polygon and an additional ideal triangle.
The corresponding increment of $F$ is
\[
F_{\rm new}-F_{\rm old}
=\sin(2x)+\sin(2y)-\sin[2(x+y)]
=4\sin x\,\sin y\,\sin(x+y).
\]
The operation increases the vacuum area by $\pi L^2$ regardless of the sign of this \emph{quadratic} change.
If $x+y<\pi$, all three sine factors are positive.
Every refinement of a polygon whose interior contains the mode center therefore increases its quadratic volume response.
For a major gap $\pi<x+y<2\pi$, the response decreases when both new gaps are smaller than $\pi$.
In that case, the added triangle itself contains the mode center. \\

\noindent
To obtain a lower bound for all polygons, we remove vertices in an order that never increases the response.
For $N\ge4$, the $N$ adjacent-pair sums $x_j+x_{j+1}$ have average $4\pi/N\le\pi$, since each gap occurs twice and $x_{N+1}=x_1$.
At least one pair therefore has total opening at most $\pi$.
For this pair, the three sine factors in the insertion formula are nonnegative, so removing the vertex between the two gaps leaves $F$ unchanged or reduces it.
Repeating the construction gives a triangle with response at most that of the original polygon, so the triangular minimum is a lower bound for every number of vertices.
For a triangle, $x_3=2\pi-x_1-x_2$ gives $F=-4\sin x_1\sin x_2\sin x_3$.
The signs have a geometric interpretation.
The inward half-plane of side $j$ is $n_j\cdot y\le p_j$ in the Klein disk, with $n_j$ the outward Euclidean unit normal and $p_j=\cos(x_j/2)$.
Since $\sin x_j=2p_j\sin(x_j/2)$ and at most one gap can exceed $\pi$, a triangle has $F<0$ exactly when all $p_j>0$ and the mode center $y=0$ lies strictly inside it.
A side through the center gives $F=0$, and a triangle missing the center has $F>0$.
For an exterior triangle, the two smaller gaps $x,y$ satisfy $x+y<\pi$.
The third gap is $2\pi-x-y$, so $F=4\sin x\sin y\sin z$ with $z=\pi-x-y>0$.
The strict concavity of $\ln\sin$ on $(0,\pi)$ bounds this product by $4\sin^3(\pi/3)=3\sqrt3/2$, with equality only at $x=y=z=\pi/3$.
Thus, the positive maximum among triangles occurs for cyclic gaps $(\pi/3,\pi/3,4\pi/3)$, up to rotation and cyclic relabeling.
It has the same magnitude as the negative response of the equal-gap triangle in Fig.~\ref{fig:triangle-signs}. \\

\noindent
To determine what a negative polygon response implies about the mode center, divide the polygon into triangles by nonintersecting diagonals.
Eq.~\eqref{eq:geometricresponse} adds their density integrals, since their shared sides have zero area.
If the center lies outside the polygon, it lies outside each triangle, and every response is positive.
If the center lies on a side, that side is a diameter, and a fan triangulation from one endpoint contains exactly one triangle with the diameter as a side.
That triangle contributes zero, and every other triangle contributes positively.
Thus, a polygon whose boundary passes through the center has zero response for $N=3$ and positive response for $N>3$.
A negative response therefore places the mode center strictly in the interior.
To bound how negative the response can be, note that for $0<x_j<\pi$ the second derivative of $\ln\sin x$ is $-1/\sin^2 x<0$.
Jensen's inequality gives
\[
\prod_{j=1}^3\sin x_j
\leq \sin^3\!\left(\frac{x_1+x_2+x_3}{3}\right)
= \frac{3\sqrt3}{8},
\]
with equality only for $x_1=x_2=x_3=2\pi/3$.
For $N>4$, at least one adjacent pair has sum strictly below $\pi$, so the first removal can be chosen to decrease $F$ strictly.
For $N=4$, the only other possibility is that every pair sum is $\pi$, in which case the two pairs of opposite vertices are antipodal and $F=0$.
Thus, over all finite polygons, the regular triangle, up to rotation, is the unique minimizer.
For each fixed $N>3$, its value is still the infimum.
To approach it, replace one gap $2\pi/3$ in the regular triangle with $N-3$ gaps of size $\epsilon$ and one gap of size $2\pi/3-(N-3)\epsilon$.
For $0<\epsilon<2\pi/[3(N-3)]$, the $N$ vertices are distinct, while continuity of the sine sum gives the triangle value as $\epsilon\downarrow0$. \\

\noindent
At the opposite end, write $x_{\max}=\max_{1\leq j\leq N}x_j$ for the largest gap.
Integrating $0\leq1-\cos u\leq u^2/2$ from $0$ to $t>0$ gives $0<t-\sin t\leq t^3/6$.
Since $\sum_{j=1}^N x_j=2\pi$, summing this inequality at $t=2x_j$ gives
\[
0<4\pi-F(P)\leq\frac{4}{3}\sum_{j=1}^N x_j^3
\leq\frac{8\pi}{3}x_{\max}^2 .
\]
Thus, any sequence with largest gap tending to zero approaches $4\pi$, while a finite polygon always has a strictly smaller response.
For a regular $N$-gon, all gaps equal $2\pi/N$, so $F=N\sin(4\pi/N)$.
In Fig.~\ref{fig:signs}, increasing $N$ from three to five takes $4\pi/N$ from $4\pi/3$ through $\pi$ to $4\pi/5$, giving the three displayed signs.
As $N\to\infty$ through integers, the regular sequence has $x_{\max}=2\pi/N\to0$, so together with the triangular minimum it gives the sharp bounds
\begin{equation}
-\frac{3\sqrt3}{2}\ \le F(P)<4\pi .
\end{equation}
These bounds concern the quadratic response of the specified field mode.
An ordinary entropy difference at small amplitude obeys $S_+(\pi/2,\lambda)-S_0(\pi/2)=\mathcal S\lambda^2/2+o(\lambda^2)$.
Consequently, Eq.~\eqref{eq:semicircle_calibration} also predicts the corresponding vacuum-subtracted polygon volume at this order directly from that difference.

\subsubsection{Conformal Factor and Geodesic Motion in Field Phase}
\label{app:fieldflux32}

\noindent
An outward side displacement adds a strip of area, and an inward displacement removes one, giving the boundary term in Eq.~\eqref{eq:edgefluxmain}.
We evaluate this term in two radial coordinates to see which part of the result is coordinate independent.
Near vacuum side $j$, $1\le j\le N_v$, let $\sigma\in\mathbb R$ be unit hyperbolic arclength and $\ell$ the outward normal distance.
To construct these Fermi coordinates, follow the side to arclength $\sigma$ and move a distance $\ell$ along its normal geodesic.
The vacuum metric is $d\ell^2+\cosh^2\ell\,d\sigma^2$.
For the perturbed side $\ell=\lambda^2\zeta_j(\sigma)+o(\lambda^2)$, the length integrand in units of $L$ is
\[
e^{\lambda^2\psi(\ell,\sigma)+o(\lambda^2)}
\sqrt{\cosh^2\ell+(\partial_\sigma\ell)^2}.
\]
In its Euler--Lagrange equation, the derivative with respect to $\partial_\sigma\ell$ is $\lambda^2\zeta_j'+o(\lambda^2)$ and the derivative with respect to $\ell$ is $\lambda^2(\zeta_j+\partial_{\boldsymbol n}\psi)+o(\lambda^2)$.
Here, $\partial_{\boldsymbol n}\psi=\partial_\ell\psi(0,\sigma)$.
The linear equation and its solution at fixed ideal endpoints are
\[
\zeta_j''-\zeta_j=\partial_{\boldsymbol n}\psi,\qquad
\zeta_j(\sigma)=-\tfrac12\int_{-\infty}^{\infty}
e^{-|\sigma-v|}\partial_{\boldsymbol n}\psi(0,v)\,dv .
\]
A solution that grows as $\exp(\sigma)$ or $\exp(-\sigma)$ would displace an ideal endpoint and is excluded there.
At the source position $\sigma=v$, the jump of one in the kernel's first derivative ensures that its second derivative supplies a unit point source.
For the radial solution below, the coordinate along the chord is $t=\sin\alpha_j\tanh\sigma$ and $z=\sin\alpha_j\operatorname{sech}\sigma$.
With $p_j=\cos\alpha_j$, we have $\partial_{\boldsymbol n}\psi\,d\sigma=-p_j\psi'(z)dt$ and $\psi'(z)=O(z)$, so the forcing is integrable even near the ideal endpoints.
The finite integral of the forcing lets us integrate the displacement along the full side.
Each source at $v$ contributes the same integrated kernel, $-1$, giving $\int_\mathbb R\zeta_j\,d\sigma=-\int_\mathbb R\partial_{\boldsymbol n}\psi\,d\sigma$.
At order $\lambda^2$, the change of area is the change of the area element plus the area swept out by the sides.
Its second state derivative is $V''(0)/L^2=4\int_{P_0}\psi\,d\mu+2\sum_{j=1}^{N_v}\int_\mathbb R\zeta_j\,d\sigma$, reproducing Eq.~\eqref{eq:edgefluxmain}.
The areal metric $h_\lambda/L^2=dx^2/f_\lambda(x)+x^2d\varphi^2$ fixes the conformal-gauge decomposition of Eq.~\eqref{eq:edgefluxmain} for this mode, where $\varphi$ is the azimuthal angle.
With $\chi$ the vacuum radial coordinate and $z=\operatorname{sech}\chi$, the areal radius in the conformal metric is $x=\exp(\lambda^2\psi+o(\lambda^2))\sinh\chi$.
Equality of the radial line elements gives $f_\lambda(x)=(dx/d\chi)^2\exp(-2\lambda^2\psi+o(\lambda^2))$.
Its coefficient at order $\lambda^2$, with $f_\lambda=f_0+\lambda^2 f_2+o(\lambda^2)$ at fixed areal radius, therefore satisfies
\[
\frac{d}{d\chi}\bigl(\psi\operatorname{sech}\chi\bigr)
=\frac{f_2(\sinh\chi)}{2\sinh\chi\cosh^2\chi}.
\]
For $\Delta=3/2$, Eq.~\eqref{eq:app_metricprofiles} gives $f_2=-\mu_a(1-2z+z^3)$ and $\mu_a=9\pi a^2/c$ for the field phase.
The boundary condition $\psi\to0$ at infinity selects
\begin{align}
\psi(z)&=\frac{\mu_a}{2z}\int_0^z\frac{(1-2v+v^3)v^2}{1-v^2}\,dv\nonumber\\
&=\frac{\mu_a}{2z}\left[\ln(1+z)-z+\frac{z^2}{2}-\frac{z^4}{4}\right].
\label{eq:conformalfield32}
\end{align}
At the center, $\psi(1)=\mu_a(\ln 2-3/4)/2<0$, while near the ideal boundary $\psi(z)\sim\mu_a z^2/6>0$.
The ordinary entropy Hessian in Eq.~\eqref{eq:positive32} measures an integral of $\psi$ along an entire chord so that it can be positive for every opening even though the metric contracts locally at the center.
For the polygon area in Eq.~\eqref{eq:edgefluxmain}, $\psi$ controls the change in the interior area element, and its outward derivative controls the displacement of each side.
Along side $j$, take constant normal coordinate $p_j$ and tangential coordinate $t$, giving $z=\sqrt{1-p_j^2-t^2}$.
For radial $\psi$, the unit hyperbolic normal derivative times edge arclength becomes
\[
\partial_{\boldsymbol n}\psi\,d\sigma
=\sqrt{1-r^2}\,p_j(\partial_r\psi/r)\,dt
=-p_j\psi'(z)\,dt .
\]
The first equality follows from the Klein metric in Appendix~\ref{app:radoninversion}; the prime in the last expression denotes differentiation with respect to $z$.
For the regular triangle, each of its three sectors covers $-\pi/3\le\varphi\le\pi/3$ relative to the side normal, with side radius $R(\varphi)=1/(2\cos\varphi)$.
Since $d\mu=-z^{-2}dz\,d\varphi$ when integrating outward, the two contributions in order are
\begin{align}
4\!\int_{P_0}\psi\,d\mu
&=12\!\int_{-\pi/3}^{\pi/3}\!\!d\varphi
\int_{\sqrt{1-R(\varphi)^2}}^1 \frac{\psi(z)}{z^2}\,dz
\simeq 0.217735\,a^2/c,\nonumber\\
-2\!\int_{\partial P_0}\partial_{\boldsymbol n}\psi\,d\sigma
&=3\!\int_{-\sqrt3/2}^{\sqrt3/2}\!\!\psi'\bigl(\sqrt{3/4-t^2}\bigr)\,dt
\simeq -9.400094\,a^2/c .
\end{align}
Their sum equals the invariant response $-27\sqrt3\,\pi a^2/(16c)$ from Eq.~\eqref{eq:polygon32}. \\

\noindent
The total area is independent of the radial coordinate, but changing coordinates changes which interior points are held fixed when the state varies.
To see how this redistributes area between the metric term and the displacement of the sides, label points of fixed areal radius $x$ by $\bar\chi=\operatorname{arsinh}x$.
Expanding $\sinh\bar\chi=\exp(\lambda^2\psi+o(\lambda^2))\sinh\chi$ at fixed $\bar\chi$ gives $\chi=\bar\chi-\lambda^2\psi\tanh\bar\chi+o(\lambda^2)$, with $\psi$ evaluated at the vacuum point.
Let $X=-\psi\tanh\bar\chi\,\partial_{\bar\chi}$ generate this map from areal to conformal coordinates, and take its divergence and scalar products in the unit vacuum metric.
Pulling back the area form adds $(\operatorname{div}X)d\mu$ to its $\lambda^2$ coefficient, so the area-element contribution to the Hessian changes by $2\int_{P_0}\operatorname{div}X\,d\mu$.
The graph of the same side in areal coordinates is obtained with the inverse map, so its outward displacement coefficient changes from $\zeta_j$ to $\zeta_j-X\cdot\boldsymbol n$.
The change of its contribution to the Hessian is therefore $-2\int_{\gamma_j(0)}X\cdot\boldsymbol n\,d\sigma$.
The divergence theorem cancels these changes after summing over the sides.
The ideal-vertex cutoff arcs have vanishing length by Appendix~\ref{app:detailA5}, and $X$ is bounded there. \\

\noindent
For the regular triangle, this cancellation even allows the area-element term to change sign.
At fixed $x$, the radial area density is proportional to $1/\sqrt{f_\lambda}$, so its contribution to the dimensionless area Hessian is $-\int_{P_0}(f_2/f_0)\,d\mu$.
Putting $z_*(\varphi)=\sqrt{1-1/(4\cos^2\varphi)}$ on the side and using $f_0=z^{-2}$ gives
\begin{equation}
\begin{aligned}
-\int_{P_0}\frac{f_2}{f_0}\,d\mu
&=3\mu_a\int_{-\pi/3}^{\pi/3}
\left[\frac14-z_*+z_*^2-\frac{z_*^4}{4}\right]d\varphi\\
&=3\mu_a\left(\frac{\pi}{6}-\frac{5\sqrt3}{16}\right).
\end{aligned}
\label{eq:triangle-areal-metric-term}
\end{equation}
For the term linear in $z_*$, the substitution $\tan\varphi=\sqrt3\sin v$ on the positive half of the sector gives $\int_{-\pi/3}^{\pi/3}z_*\,d\varphi=3\int_0^{\pi/2}\cos^2v\,dv/(1+3\sin^2v)=\pi/2$.
The even powers follow by integrating $\sec^2\varphi$ and $\sec^4\varphi$ with $t=\tan\varphi$.
With $\mu_a=9\pi a^2/c$, Eq.~\eqref{eq:triangle-areal-metric-term} gives approximately $-1.498577\,a^2/c$ from the area element.
Subtracting this from the invariant Hessian gives $3\mu_a(\sqrt3/4-\pi/6)\simeq-7.683783\,a^2/c$ from the moving sides in the same areal coordinates.

\subsection{Scalar Initial Data at Scaling Dimension Three}
\label{app:detailE1}

\noindent
At $\Delta=3$, the scalar mass squared is positive, and both preparations have nonnegative local energy density.
Their polygon volume responses can still change order because moving the boundary endpoints changes which radii contribute.
In global AdS$_3$, let $\tau$ be dimensionless time, $\chi\geq0$ the dimensionless radial distance, and $\varphi\in[0,2\pi)$ the azimuth.
The vacuum metric is
\begin{equation}
ds^2=L^2[-\cosh^2\chi\,d\tau^2+d\chi^2+\sinh^2\chi\,d\varphi^2].
\end{equation}
We consider a real scalar with $M_\phi^2L^2=3$.
The scaling dimension $\Delta$ of its normalizable boundary mode obeys $\Delta(\Delta-2)=M_\phi^2L^2$, which gives $\Delta=3$.
For a radial mode $U(\chi)\exp(-\mathrm{i}\Delta\tau)$, the Klein--Gordon equation reduces to
\begin{equation}
\frac{\partial_\chi(\sinh\chi\cosh\chi\,\partial_\chi U)}{\sinh\chi\cosh\chi}
+\left(\frac{\Delta^2}{\cosh^2\chi}-M_\phi^2L^2\right)U=0.
\end{equation}
The regular solution $U=(\cosh\chi)^{-\Delta}$ obeys $U'=-\Delta\tanh\chi\,U$.
The radial derivative term thus equals $(-2\Delta+\Delta^2\tanh^2\chi)U$.
The identity $\tanh^2\chi+\operatorname{sech}^2\chi=1$ leaves the mass relation above.
The two quadratures, or independent time phases of this oscillator, are proportional to the real part and minus the imaginary part of $U\exp(-\mathrm{i}\Delta\tau)$.
Our sign convention is
\begin{equation}
\phi_{1,+}=A\operatorname{sech}^3\chi\cos(3\tau),\qquad
\phi_{1,-}=A\operatorname{sech}^3\chi\sin(3\tau),\qquad a=A\sqrt{L}.
\end{equation}
Here, $\operatorname{sech}\chi=1/\cosh\chi$, and the two profiles are even and odd under $\tau\mapsto-\tau$.
For the odd profile, time reversal changes the field's sign everywhere, leaving the quadratic stress unchanged.
Time reversal combined with $\phi\mapsto-\phi$ then preserves the scalar data, so both quadratures admit a reflection-symmetric metric.
On the reflection slice, the unit normal $n=(L\cosh\chi)^{-1}\partial_\tau$ converts the coordinate derivative into the proper velocity.
The even field has $n\phi_{1,+}=0$ at $\tau=0$, leaving its spatial gradient and mass as energy sources.
The odd field vanishes throughout the slice, so its spatial gradient also vanishes there, while $n\phi_{1,-}=(3A/L)\operatorname{sech}^4\chi$ supplies kinetic energy.
Putting these data in $2\rho_2=(n\phi_1)^2+L^{-2}(\partial_\chi\phi_1)^2+M_\phi^2\phi_1^2$ and using $e=L^3\rho_2$ gives the following densities in the Klein radius $r=\tanh\chi\in[0,1)$,
\begin{align}
e_{+}(r)&=\tfrac12a^2(9r^2+3)(1-r^2)^3,&
e_{-}(r)&=\tfrac92a^2(1-r^2)^4,\quad
g_j(r)=\frac{e_j(r)}{1-r^2},\quad j\in\{+,-\}.
\label{eq:app_scalardensity}
\end{align}
Both source profiles obey Eq.~\eqref{eq:app_sourceclasses}.
In Fig.~\ref{fig:matter}(c), the horizontal coordinate is the Klein radius $r$, and the vertical coordinate is $e_\pm(r)/a^2$ for $a\ne0$.
At $r=0$, Eq.~\eqref{eq:app_scalardensity} gives intercepts $3/2$ and $9/2$ for the field and momentum phases, respectively.
For $r<1$, equating the two densities cancels $(1-r^2)^3$ and gives $9r^2+3=9(1-r^2)$, so the curves cross at $r=1/\sqrt3$.
The equal-arc triangle in panel (a) samples both sides of this crossing, while the clustered triangle in panel (b) lies entirely at $r\geq\cos(\pi/4)=1/\sqrt2$ and samples only $e_+>e_-$ in its interior.

\subsection{Interval Data and Radial Inversion}
\label{app:detailE2}

\noindent
For a radial source, a chord reading depends only on its offset from the center.
Let $p=\cos\alpha\in(-1,1)$ be the signed offset and $s$ the signed Euclidean coordinate along the chord from its midpoint, so $r^2=p^2+s^2$ and $|s|<\sqrt{1-p^2}$.
The hyperbolic line element $\sqrt{1-p^2}\,ds/(1-r^2)$ separates an offset factor from a radial factor, giving the chord integral in terms of $g_j=e_j/(1-r^2)$,
\begin{equation}
(\XX e_j)(p)=\sqrt{1-p^2}\int_{-\sqrt{1-p^2}}^{\sqrt{1-p^2}}g_j\big(\sqrt{p^2+s^2}\big)\,ds .
\end{equation}
The substitution $s=t\sqrt{1-p^2}$ turns the integrands into polynomials on $-1<t<1$.
Their moments $\int_{-1}^1 t^{2k}dt=2/(2k+1)$ for $k=0,1,\ldots$ give
\begin{equation}
(\XX e_{+})(p)=\frac{16a^2}{35}(5+9p^2)(1-p^2)^3,\qquad
(\XX e_{-})(p)=\frac{144a^2}{35}(1-p^2)^4 .
\label{eq:app_bothchords}
\end{equation}
For nonzero amplitude and $|p|<1$, dividing out the common positive factor shows that the two chord energies coincide when $5+9p^2=9(1-p^2)$, or $p^2=2/9$.
The momentum phase gives the larger chord energy at smaller offsets and the smaller energy at larger offsets.
The local densities instead coincide at $r^2=1/3$, since a chord through that circle also samples the material between the circle and the outer boundary.
For the even phase, integrating the modular weight transverse to the chord normal gives
\begin{align}
\FF_{+}(p)&=\frac{32\pi a^2}{5\sqrt{1-p^2}}
\int_p^1(v-p)(3v^2+2)(1-v^2)^{3/2}\,dv\nonumber\\
&=\frac{2\pi a^2}{35}(64+39p^2+10p^4-8p^6)
-\frac{6\pi a^2p\arccos p}{\sqrt{1-p^2}} .
\label{eq:app_scalarF}
\end{align}
Here, $v$ is the coordinate along the chord normal and $\arccos p\in(0,\pi)$.
For the odd case, the weighted integral is
\begin{equation}
\FF_{-}(p)=\frac{96\pi a^2}{5\sqrt{1-p^2}}\int_p^1(v-p)(1-v^2)^{5/2}\,dv
=\FF_{+}(p)-\frac{32\pi a^2}{35}(1-p^2)^3 .
\end{equation}
The difference $\FF_+-\FF_-$ contains the integral of $-(v-p)\partial_v[v(1-v^2)^{5/2}]$.
Both boundary terms vanish, and the remaining integral $\int_p^1v(1-v^2)^{5/2}dv=(1-p^2)^{7/2}/7$ gives the stated sign and coefficient.
The interval operator acting on rotationally invariant functions is $\DD=(1-p^2)^2\partial_p^2-p(1-p^2)\partial_p-2$.
Each expression satisfies $\DD\FF_j=4\pi\XX e_j$. \\

\noindent
To recover the radial density from the interval response, form $H_j(p)=\DD\FF_j(p)/(4\pi\sqrt{1-p^2})$.
For $0<p<1$, the chord starts at radius $p$ at its midpoint and reaches radius one at either end.
Its two halves contribute equally, and $ds=r\,dr/\sqrt{r^2-p^2}$ on the positive half gives $H_j(p)=2\int_p^1 rg_j(r)/\sqrt{r^2-p^2}\,dr$.
Setting $u=p^2$ and $v=r^2$ absorbs the factor $2r\,dr$ into $dv$ and gives the normalized chord profile
\[
H_j(\sqrt u)=\int_u^1\frac{g_j(\sqrt v)}{\sqrt{v-u}}\,dv .
\]
A second integral over offsets $t<u<1$ converts the chord data into the cumulative density outside radius $\sqrt t$.
Multiplying by $(u-t)^{-1/2}$ and exchanging the order of integration leaves the inner integral $\int_t^v[(u-t)(v-u)]^{-1/2}\,du$ for $0<t<v<1$.
Setting $u=t+(v-t)x$ makes that integral $\int_0^1[x(1-x)]^{-1/2}dx=\pi$, independent of $t$ and $v$.
Thus, $\int_t^1 H_j(\sqrt u)(u-t)^{-1/2}du=\pi\int_t^1 g_j(\sqrt v)dv$.
The derivative of the right-hand side is $-\pi g_j(\sqrt t)$, so moving the lower endpoint isolates the unknown density at one radius.
On the data side, write $u=t+w$ before differentiating to keep the integrable singularity at $w=0$.
Since $H_j(1)=0$, the moving upper endpoint contributes zero, and differentiating the remaining integrand gives
\begin{equation}
g_j(r)=-\frac{1}{\pi}\int_r^1\frac{H_j'(p)}{\sqrt{p^2-r^2}}\,dp,\qquad
e_j(r)=(1-r^2)g_j(r).
\end{equation}
The boundary functions with $\beta^2=\pi a^2/2$ give
\begin{equation}
H'_{+}=-\tfrac{16}{5}a^2p(1+9p^2)(1-p^2)^{3/2},\qquad
H'_{-}=-\tfrac{144}{5}a^2p(1-p^2)^{5/2}.
\end{equation}
The substitution $p^2=r^2+(1-r^2)\sin^2 v$, $0\leq v\leq\pi/2$, reduces the integrals to even powers of $\cos v$.
For the odd phase, $p\,dp/\sqrt{p^2-r^2}=\sqrt{1-r^2}\cos v\,dv$ and $1-p^2=(1-r^2)\cos^2v$.
The integral $\int_0^{\pi/2}\cos^6 v\,dv=5\pi/32$ then gives $g_-=(9a^2/2)(1-r^2)^3$.
The even phase gives $g_+=a^2(9r^2+3)(1-r^2)^2/2$ by the same substitution.
Each density follows from $e_j=(1-r^2)g_j$.

\subsection{Interval Filters for Both Scalar Phases}
\label{app:detailE3}

\noindent
For the dimension-three example, we can also obtain polygon energies by filtering the chord readings with respect to offset.
For $j\in\{+,-\}$, the input $H_j(p)$ is extended by zero for $|p|\geq1$ and is given inside the disk by Eq.~\eqref{eq:cftbothtransforms},
\[
H_+(p)=\tfrac{16a^2}{35}(5+9p^2)(1-p^2)^{5/2},\qquad
H_-(p)=\tfrac{144a^2}{35}(1-p^2)^{7/2}.
\]
The filter $\Lambda=\HH\partial_p$ acts conveniently on the polynomials $T_n(\cos x)=\cos(nx)$ and $U_n(\cos x)=\sin[(n+1)x]/\sin x$ for $0<x<\pi$ and integers $n\geq0$.
These are the Chebyshev polynomials of the first and second kinds.
On $|p|<1$, the Hilbert transform of $\sqrt{1-p^2}\,U_n(p)$, extended by zero before transformation, is $T_{n+1}(p)$.
For $n=0$, the substitution $q=\cos x$ leaves $p$ plus a term proportional to $\operatorname{PV}\int_0^\pi(p-\cos x)^{-1}dx$.
The latter vanishes.
With $w=\tan(x/2)$ and $w_*=\sqrt{(1-p)/(1+p)}$, its primitive is proportional to $\ln |(w-w_*)/(w+w_*)|$.
Its endpoint values at $0$ and $+\infty$ both vanish, and its two logarithmic divergences at $w=w_*$ cancel in the principal value.
For $n=1$, the identities $U_1(q)=2q$ and $q/(p-q)=p/(p-q)-1$ give $2p^2-1$.
The recurrence $U_{n+1}(q)=2qU_n(q)-U_{n-1}(q)$ extends these first two transforms to higher degrees because $\int_{-1}^1\sqrt{1-q^2}\,U_n(q)\,dq=0$ for $n\geq1$.
Both chord profiles reach zero at $p=\pm1$, so extending them by zero and differentiating adds no delta function there.
After division by $a^2\sqrt{1-p^2}$, each profile is an even polynomial of degree at most six, which accounts for the four terms in
\[
\frac{H_j(p)}{a^2\sqrt{1-p^2}}=\sum_{k=0}^{3}c_{j,k}U_{2k}(p),\quad j\in\{+,-\}.
\]
The even polynomials on the two sides of this equation have coefficients
\[
(c_{j,0},c_{j,1},c_{j,2},c_{j,3})=
\begin{cases}
(7/4,-13/20,-1/20,9/140),&j=+,\\
(9/4,-27/20,9/20,-9/140),&j=-.
\end{cases}
\]
Applying the Hilbert transform replaces each weighted $U_{2k}$ by $T_{2k+1}$.
The identity $T_n'(p)=nU_{n-1}(p)$ for integers $n\ge1$ then gives
\begin{align}
(\Lambda H_+)(p)&=\tfrac{a^2}{5}(144p^6-200p^4+30p^2+15),\\
(\Lambda H_-)(p)&=-\tfrac{9a^2}{5}(16p^6-40p^4+30p^2-5).
\end{align}
Only offsets $|p|<1$ enter the polygon integral.
In Eq.~\eqref{eq:app_polygonfilter}, the filtered input is $B_\theta=4\pi\Lambda H_j$.
Equivalently, backprojection averages the displayed polynomials over $p=r\cos\theta$, multiplying the full angular integral by $(1-r^2)/(4\pi)$.
The averages of $\cos^{2k}\theta$ at $k=0,1,2,3$ are $1,1/2,3/8,5/16$.
They return both densities in Eq.~\eqref{eq:app_scalardensity}, providing a direct evaluation of the reconstruction from the CFT functions.

\subsection{Polygon Integrals of Reconstructed Energy Densities}
\label{app:detailE4}

\noindent
To integrate the reconstructed density over the regular triangle, divide it into three congruent sectors.
In each, the angle from the side normal satisfies $-\pi/3\le\xi\le\pi/3$, and the side $r\cos\xi=1/2$ sets the upper radial limit $R(\xi)=1/(2\cos\xi)$.
Integrating from zero to $R$ and writing $U=1-R^2$ gives
\begin{align}
E_{+}(R)&=a^2\big(39/70-6U^{5/2}/5+9U^{7/2}/14\big),&
E_{-}(R)&=(9a^2/14)(1-U^{7/2}),
\end{align}
where $E_j(R)=\int_0^R e_j(r)r(1-r^2)^{-3/2}dr$.
The angular moments are $U_\nu=\int_{-\pi/3}^{\pi/3}[1-R(\xi)^2]^\nu d\xi$ for $\nu>-1$.
A convenient evaluation sets $t=\tan\xi$, followed by $t=\sqrt{3}\sin v$ on the positive half of the integral.
This gives
\begin{equation}
U_\nu=2\sqrt{3}(3/4)^\nu\int_0^{\pi/2}\frac{\cos^{2\nu+1}v}{1+3\sin^2v}\,dv,\qquad
U_{5/2}=\frac{53\pi}{256},\quad U_{7/2}=\frac{289\pi}{2048}.
\end{equation}
The auxiliary integrals $D_m=\int_0^{\pi/2}\cos^{2m}v\,dv/(1+3\sin^2v)$ are needed at nonnegative integers $m$.
Because $1+3\sin^2v=4-3\cos^2v$,
\begin{equation}
D_0=\pi/4,\qquad 3D_m=4D_{m-1}-\frac{\pi}{2}\frac{(2m-2)!}{4^{m-1}[(m-1)!]^2},\quad m=1,2,\ldots .
\end{equation}
The change of variable $w=\tan v$ gives $D_0=\int_0^\infty (1+4w^2)^{-1}dw=\pi/4$.
For an integer $n\ge0$, $n!=\prod_{k=1}^{n}k$ with $0!=1$.
The factorials in the recurrence come from the even cosine moments.
The moments $U_{5/2}$ and $U_{7/2}$ use $D_3$ and $D_4$, respectively.
The resulting triangle integrals are
\begin{equation}
\int_{P_0}e_{+}\,d\mu=\frac{91911\pi a^2}{143360},\qquad
\int_{P_0}e_{-}\,d\mu=\frac{29061\pi a^2}{28672}.
\label{eq:app_bothenergies}
\end{equation}
For each phase, the coefficient of $\lambda^2$ in $V/L^2$ is $12\pi/c$ times its entry in Eq.~\eqref{eq:app_bothenergies}.

\subsection{Dependence on Boundary Partition at Scaling Dimension Three}
\label{app:partition_ordering}

\noindent
For unequal boundary arcs, subtract the caps outside the polygon from the whole disk.
Take either phase of the dimension-three mode and a convex ideal polygon $P$ with $N=N_v\geq3$ distinct cyclically ordered vertices.
For side $j=1,\ldots,N$, the intervening arc has opening $2\alpha_j$, with $0<\alpha_j<\pi$ and $\sum_{j=1}^N\alpha_j=\pi$.
The caps between the sides and these arcs have disjoint interiors and cover the exterior.
In each cap, use outward normal coordinate $x$ and transverse coordinate $t$.
The cap has $x\geq p_j=\cos\alpha_j$ and $t^2<1-x^2$.
The densities in the Klein disk satisfy
\begin{equation}
e_-\,d\mu=\tfrac92a^2(1-r^2)^{5/2}\,dx\,dt,
\qquad (e_++e_-)\,d\mu=6a^2(1-r^2)^{3/2}\,dx\,dt .
\end{equation}
At fixed $x$, the substitution $t=\sqrt{1-x^2}\sin v$, $-\pi/2\leq v\leq\pi/2$, gives the transverse integrals of $(1-r^2)^{5/2}$ and $(1-r^2)^{3/2}$ as $5\pi(1-x^2)^3/16$ and $3\pi(1-x^2)^2/8$, respectively.
For $m\in\{2,3\}$, the cap integral and its polygon remainder are
\begin{equation}
J_m(p)=\int_p^1(1-x^2)^m\,dx,\qquad
K_m(P)=J_m(-1)-\sum_{j=1}^N J_m(p_j).
\end{equation}
The full disk integral minus these cap integrals gives
\begin{equation}
E_-(P)=\frac{45\pi a^2}{32}K_3(P),\qquad
E_+(P)=\frac{9\pi a^2}{4}K_2(P)-E_-(P),
\qquad E_\pm(P)=\int_P e_\pm\,d\mu .
\label{eq:r2_cap_energies}
\end{equation}
With $[V_\pm(\lambda)-V(0)]/L^2=C_\pm\pi^2a^2\lambda^2/c+o(\lambda^2)$, Eq.~\eqref{eq:geometricresponse} gives $C_\pm=12E_\pm/(\pi a^2)$ and hence
\begin{equation}
C_-=\tfrac{135}{8}K_3(P),\qquad C_++C_-=27K_2(P).
\label{eq:triangleanswer}
\end{equation}
For three equal arcs, $p_j=1/2$ for all $j$.
The two polynomial integrals give $K_2=353/480$ and $K_3=3229/4480$.
Since $V_\pm''(0)/L^2=24\pi E_\pm(P)/c$, their ratio is $48435/30637\simeq1.581$. \\

\noindent
The local difference $e_--e_+=3a^2(1-3r^2)(1-r^2)^3$ favors the momentum phase near the center and the field phase farther out.
For a polygon containing the center, each ray traverses a single segment from $r=0$ to its side, so the relevant quantity is the accumulated difference per unit angle,
\begin{equation}
A(R)=\int_0^R\frac{(e_--e_+)r\,dr}{(1-r^2)^{3/2}}
=a^2\left[\frac{3}{35}+\frac65(1-R^2)^{5/2}
-\frac97(1-R^2)^{7/2}\right].
\end{equation}
For $a\ne0$, $A$ increases from zero to its maximum at $R=1/\sqrt3$, then decreases to $A(1)=3a^2/35$.
Because even this endpoint value is positive, a ray of any nonzero length starting at the origin collects more energy in the momentum phase.
A convex polygon with the origin in its interior contains precisely one such segment $[0,R(\varphi)]$ per direction, giving $E_-(P)-E_+(P)=\int_0^{2\pi}A(R(\varphi))\,d\varphi>0$. \\

\noindent
To reverse the ordering, place all vertices in an arc of half-opening $\delta<\pi/2$.
The convex hull then obeys $r\geq\cos\delta$, which bounds the contribution from the center.
The pointwise ratio $e_-/e_+=3(1-r^2)/(1+3r^2)$ decreases with $r$, so its value at $\cos\delta$ bounds its weighted average over the polygon.
Consequently, $0<V_-''(0)/V_+''(0)\leq3\sin^2\delta/(1+3\cos^2\delta)$, which is below unity for $\delta<\arccos(1/\sqrt3)$.
For endpoints $-\pi/4,0,\pi/4$, the upper bound is $3/5$.
More precisely, $K_m=J_m(\cos(\pi/4))-2J_m(\cos(\pi/8))$, by $J_m(-p)=J_m(-1)-J_m(p)$, and Eq.~\eqref{eq:r2_cap_energies} gives the ratio $0.32649913\ldots$.
Thus, changing only the partition reverses the ordering for the same two states in triangles of equal vacuum area $\pi L^2$.

\subsection{Direct Area Calculation with Moving Geodesic Sides}
\label{app:directarea}

\noindent
The same polygon response can be computed geometrically by following each geodesic as the metric changes, keeping its boundary endpoints fixed.
This calculation applies to either lowest-mode phase in standard quantization for real $\Delta>1$ and any convex ideal polygon.
With rotational symmetry and a regular center, use the circumference $2\pi Lx$ to define radius $x$.
In vacuum, $x=\sinh\chi$, and the reflection-slice metric is
\begin{equation}
h_\lambda=L^2\big[dx^2/f_\lambda(x)+x^2d\varphi^2\big],\qquad f_0(x)=1+x^2.
\end{equation}
A radial proper distance element is $L\,dx/\sqrt{f_\lambda}$, so the Gaussian curvature is $-f_\lambda'/(2L^2 x)$.
The slice Hamiltonian constraint with $f_\lambda(0)=1$ gives
\begin{equation}
\begin{aligned}
f_\lambda(x)&=1+x^2-16\pi G_NL^2\int_0^x t\rho_\lambda(t)\,dt\\
&=f_0(x)-\mu_a\lambda^2 k_\pm(x)+o(\lambda^2),\qquad
\mu_a=\frac{6\pi a^2\Delta}{c}.
\end{aligned}
\end{equation}
The variable $z=(1+x^2)^{-1/2}$ decreases from one at the center to zero at infinity.
The densities satisfy $e_-=(a^2\Delta^2/2)z^{2\Delta+2}$ and $e_++e_-=a^2\Delta(\Delta-1)z^{2\Delta}$.
Integration using $x\,dx=-z^{-3}dz$ yields
\begin{equation}
k_-=1-z^{2\Delta},\qquad k_++k_-=2(1-z^{2\Delta-2}).
\label{eq:app_metricprofiles}
\end{equation}
At $\Delta=3$, these give $k_-=1-z^6$ and $k_+=1-2z^4+z^6$.
Their common limit $1$ at infinity gives the same quadratic mass. However, geodesics with a finite turning radius encounter different radial metrics. \\

\noindent
Join each boundary vertex to the center by a radial geodesic, assigning each side $j$, $1\le j\le N$, a sector of opening $2\gamma_j$ with $\gamma_j=\min(\alpha_j,\pi-\alpha_j)$.
A sector adjacent to a short arc contributes with a plus sign, while one adjacent to a major arc is subtracted, as encoded by $\sigma_j=\operatorname{sgn}(\cos\alpha_j)$.
To see why the signed sum gives the area even when the center is outside the polygon, write the dimensionless area form as the exterior derivative of $A_\lambda(x)\,d\varphi$, where $A_\lambda(x)=\int_0^x t\,dt/\sqrt{f_\lambda(t)}$.
The one-form is regular at the center because $A_\lambda(x)\sim x^2/2$, and its integral along a radial segment vanishes because $\varphi$ is constant.
Stokes' theorem therefore reduces each sector's area to the oriented integral along its geodesic side.
An antipodal pair gives a radial side and zero contribution on the branch through the vacuum. \\

\noindent
A nondiametral side with short half-opening $\gamma$ has minimum radius $b>0$.
At that point, its unit tangent is purely angular.
Rotational symmetry conserves $x^2|d\varphi/ds|$ along the geodesic.
Its turning-point value is $b$, with $s$ denoting arclength in units of $L$.
The unit-speed condition gives $|d\varphi/dx|=b/[x\sqrt{x^2-b^2}\sqrt{f_\lambda(x)}]$ for $x>b$.
The change $x=b/\sin t$, $0<t\leq\pi/2$, gives $|d\varphi|=dt/\sqrt{f_\lambda(b/\sin t)}$ in order of decreasing $x$.
The integral of $A_\lambda|d\varphi|$ over one half-side is its unsigned half-sector area $U_\lambda(\gamma)$ in units of $L^2$.
The opening integral pins the two vertices to their boundary positions. In contrast, the area integral follows the geodesic selected by that condition.
For a fixed half-opening, the two integrals obey
\begin{align}
\gamma&=\int_0^{\pi/2}\frac{dt}{\sqrt{f_\lambda(b/\sin t)}},\label{eq:cap-fixed-opening}\\
U_\lambda(\gamma)+\gamma&=\int_0^{\pi/2} B_\lambda(b/\sin t)\,dt,
\qquad B_\lambda(x)=\frac{1+A_\lambda(x)}{\sqrt{f_\lambda(x)}}.
\label{eq:app_directarea}
\end{align}
The boundary opening is prescribed, so a metric change must generally move the turning radius $b_\lambda$ in Eq.~\eqref{eq:cap-fixed-opening}.
In vacuum, $\gamma=\arctan(1/b_0)$ gives $d\gamma/db_0=-1/(1+b_0^2)\ne0$ for $0<\gamma<\pi/2$, fixing the small compensating change uniquely.
For either phase $\varsigma\in\{+,-\}$, put $b_{\lambda,\varsigma}=b_0+\mu_a\lambda^2 b_{2,\varsigma}+o(\lambda^2)$ and $z_0(t)=\sin t/\sqrt{b_0^2+\sin^2 t}$ for $0\leq t\leq\pi/2$.
At fixed $b_0$, the coefficient of $\mu_a\lambda^2$ in $1/\sqrt{f_\lambda(b_0/\sin t)}$ is $z_0(t)^3 k_\varsigma(z_0(t))/2$, with $k_\varsigma$ expressed in the variable $z$ as in Eq.~\eqref{eq:app_metricprofiles}.
The value at $t=0$ is understood by continuity from $t>0$.
At fixed endpoints, the total angular change is zero, giving
\begin{equation}
0=-\frac{b_{2,\varsigma}}{1+b_0^2}
+\frac12\int_0^{\pi/2} z_0(t)^3 k_\varsigma(z_0(t))\,dt .
\label{eq:cap-turning-shift}
\end{equation}
For $0<z<1$, $k_-(z)-k_+(z)=2z^{2\Delta-2}(1-z^2)>0$.
At equal nonzero amplitude, the turning point therefore lies on a circle of larger circumference in the momentum phase to this order. \\

\noindent
Keeping the endpoints fixed imposes an opening-angle condition on the turning radius, so varying the metric changes the integration domain and the area element.
The radial constraint lets us combine these two changes before we expand the area.
The constraint gives $f_\lambda'=2x(1-8\pi G_NL^2\rho_\lambda)$ and the area primitive gives $A_\lambda'=x/\sqrt{f_\lambda}$.
Hence, $(\sqrt{f_\lambda})'=(1-8\pi G_NL^2\rho_\lambda)A_\lambda'$.
The common initial value $\sqrt{f_\lambda(0)}=1+A_\lambda(0)=1$ fixes the integral identity
\begin{equation}
B_\lambda(x)=1+\frac{8\pi G_NL^2}{\sqrt{f_\lambda(x)}}
\int_0^x\frac{u\,\rho_\lambda(u)}{\sqrt{f_\lambda(u)}}\,du .
\label{eq:radial_area_primitive}
\end{equation}
In vacuum, $B_0=1$, so its integral over the angle equals $\pi/2$ for every turning radius.
Its derivative with respect to $b$ vanishes, removing the explicit turning-point shift at $b=b_0$ from Eq.~\eqref{eq:app_directarea}.
The remaining part of $B_\lambda$ already contains a density of order $\lambda^2$, so a shift of order $\lambda^2$ in its path affects only higher orders.
Consequently, forming $U_\lambda+\gamma$ and evaluating the density term on the vacuum chord accounts for the fixed-opening displacement in Eq.~\eqref{eq:cap-turning-shift}.
In vacuum, $U_0(\gamma)=\pi/2-\gamma$, and the oriented sum gives $2\sum_{j=1}^N(\pi/2-\alpha_j)=(N-2)\pi$ whether or not the center is inside. \\

\noindent
At fixed $x$, the expansion of $B_\lambda$ reads $1+\mu_a\lambda^2 B_{2,\pm}(z)+o(\lambda^2)$.
At order $\mu_a\lambda^2$, the change in the primitive is $\tfrac12\int_0^x u k((1+u^2)^{-1/2})(1+u^2)^{-3/2}du=\tfrac12\int_z^1 k(v)\,dv$.
The change in $1/\sqrt f$ at the same order is $z^3 k(z)/2$.
Their product rule, with $1+A_0=1/z$, gives $B_{2,\pm}=(z/2)\int_z^1 k(v)\,dv+z^2k(z)/2$ for the selected profile.
For $k(z)=1-z^m$, it is $m(z-z^{m+2})/[2(m+1)]$, for every real $m>0$.
Consequently,
\begin{equation}
B_{2,-}=\frac{\Delta}{2\Delta+1}(z-z^{2\Delta+2}),\qquad
B_{2,+}+B_{2,-}=\frac{2(\Delta-1)}{2\Delta-1}(z-z^{2\Delta}).
\end{equation}
We assume the integrated mass remainder in $f_\lambda$ is uniformly $o(\lambda^2)$ in $x$.
Under this assumption, the remainder in Eq.~\eqref{eq:app_directarea} is integrable uniformly in $\lambda$.
The angular moments are $\mathcal J_\ell(b)=\int_0^{\pi/2}z(t;b)^\ell\,dt$ with $z(t;b)=\sin t/\sqrt{b^2+\sin^2t}$ and real $\ell>0$.
We have $\mathcal J_1(b_0)=\gamma$.
The two powers needed here are $\ell=2n$ with $n\in\{\Delta,\Delta+1\}$, where $n$ can be noninteger.
The variable $q=b/\sqrt{1+b^2}$ equals $\cos\gamma$ at $b=b_0$.
The derivative of $\mathcal J_{2n}$ takes the following form after the substitution $u=\cot t$.
\begin{align}
\frac{d\mathcal J_{2n}}{db}&=-2nb\int_0^\infty\frac{du}{(1+b^2+b^2u^2)^{n+1}}
=-\frac{d_n}{(1+b^2)^{n+1/2}},\nonumber\\
d_n&=2n\int_0^{\pi/2}\cos^{2n}w\,dw=n I_{n-1/2},
\qquad d_3=\frac{15\pi}{16},\quad d_4=\frac{35\pi}{32} .
\end{align}
The second equality uses $bu=\sqrt{1+b^2}\tan w$, where $I_s=\int_{-1}^1(1-t^2)^sdt$ for $s>-1$.
Integration by parts gives $\int_0^{\pi/2}\cos^{2n}w\,dw=(2n-1)(2n)^{-1}\int_0^{\pi/2}\cos^{2n-2}w\,dw$.
Since $dq/db=(1+b^2)^{-3/2}$ and $\lim_{b\to+\infty}\mathcal J_{2n}(b)=0$, integrating in $q$ yields $\mathcal J_{2n}(b)=d_n J_{n-1}(q)$ where $J_m(p)=\int_p^1(1-t^2)^m dt$ and $K_m(P)=J_m(-1)-\sum_{j=1}^N J_m(p_j)$, with $p_j=\cos\alpha_j$.
In particular, $d_nJ_{n-1}(-1)=\pi$ by symmetry and $\mathcal J_{2n}(0)=\pi/2$.
Thus, the orientation identity $\sigma_j[\gamma_j-d_nJ_{n-1}(|p_j|)]=\alpha_j-d_nJ_{n-1}(p_j)$ also holds for a major arc.
For $p_j<0$, it follows from $J_m(-p_j)=J_m(-1)-J_m(p_j)$, and at $p_j=0$ both sides vanish. \\

\noindent
The signed sector sum $V/L^2=2\sum_{j=1}^N\sigma_jU_\lambda(\gamma_j)$ gives the coefficient of $\lambda^2$ in $V_-/L^2$ as $2\mu_a\Delta d_{\Delta+1}K_\Delta/(2\Delta+1)$.
The coefficient of $\lambda^2$ in $(V_++V_-)/L^2$ is $4\mu_a(\Delta-1)d_\Delta K_{\Delta-1}/(2\Delta-1)$.
By the same integration by parts, $d_n=(2n-1)I_{n-3/2}/2$.
The definition $V_\pm/L^2=(N-2)\pi+(12\pi/c)E_\pm\lambda^2+o(\lambda^2)$ then gives
\begin{equation}
E_-=\frac{a^2\Delta^2}{2} I_{\Delta-1/2}K_{\Delta},\qquad
E_++E_-=a^2\Delta(\Delta-1)I_{\Delta-3/2}K_{\Delta-1}.
\end{equation}
These are independently the integrals of the normal densities over the vacuum polygon.
Indeed, $e_-\,d\mu=(a^2\Delta^2/2)(1-r^2)^{\Delta-1/2}d^2y$ and $(e_++e_-)d\mu=a^2\Delta(\Delta-1)(1-r^2)^{\Delta-3/2}d^2y$.
The transverse Klein integral at fixed normal coordinate $p$ is $\int_{-\sqrt{1-p^2}}^{\sqrt{1-p^2}}(1-p^2-t^2)^q dt=I_q(1-p^2)^{q+1/2}$ for $q>-1$.
The cap subtraction then gives the two displayed expressions.
At $\Delta=3$, the dimensionless coefficients $C_\pm$ are defined by $V_\pm/L^2=(N-2)\pi+C_\pm\pi^2a^2\lambda^2/c+o(\lambda^2)$.
Then, the preceding expressions yield
\begin{equation}
C_-(P)=\frac{135}{8}K_3(P),\qquad C_+(P)+C_-(P)=27K_2(P).
\label{eq:app_directboth}
\end{equation}
For either phase, $\varsigma\in\{+,-\}$, the angular integral is $H_{\varsigma}(\gamma)=\int_0^{\pi/2}B_{2,\varsigma}(z(t;\cot\gamma))\,dt$.
The vertices at $-\pi/4,0,\pi/4$ give two short arcs of opening $\pi/4$ and one major arc of opening $3\pi/2$.
Their signed sum is $C_{\varsigma}=(18/\pi)[4H_{\varsigma}(\pi/8)-2H_{\varsigma}(\pi/4)]$.
Numerically, $(C_+,C_-)=(0.51829086,0.16922152)$ in this triangle.
Three equal arcs give $(275733/35840,87183/7168)$.

\subsection{Finite Causal Regions and Global Volume Maximality}
\label{app:scalarmax}

\noindent
The direct calculation gives the area on the reflection slice.
To identify it with the maximum over all admissible fillings in the positive-mass example, we bound both the time extent of the polygon's causal development and the scalar amplitude there.
Global optical coordinates obey $r=\sin\zeta$, where $0\leq\zeta\leq\pi/2$ on the conformal completion.
The vacuum metric is
\begin{equation}
g_0=L^2\sec^2\zeta\left(-d\tau^2+\gamma_{\rm opt}\right),
\qquad
\gamma_{\rm opt}=d\zeta^2+\sin^2\zeta\,d\varphi^2.
\label{eq:app_opticalmetric}
\end{equation}
For a light ray, the spacetime line element vanishes, so dividing by the positive conformal factor in Eq.~\eqref{eq:app_opticalmetric} gives $d\tau^2=\gamma_{\rm opt}(dx,dx)$.
In these coordinates, vacuum light travels one unit of angular distance on the closed unit hemisphere $\mathcal B$ per unit of $\tau$.
To bound how long a point can remain causally separated from the exterior $Z_\lambda=\mathcal B\setminus\operatorname{int}P_\lambda$, we need the largest distance from a point to that exterior.
This optical inradius is
\begin{equation}
R_\lambda^{\rm opt}
=\sup_{x\in\mathcal B}d_{\rm opt}(x,Z_\lambda).
\end{equation}
For the regular ideal triangle, the exterior consists of three caps $n_j\cdot y\geq1/2$, one for each outward unit normal $n_j$, $1\le j\le3$.
Choose $n_j$ as the pole on the optical hemisphere extended to the full unit sphere.
The latitude of the point is then $\arcsin(n_j\cdot y)$, and the edge of cap $j$ has latitude $\pi/6$.
From a point inside the triangle, the shortest route to this cap follows a meridian. It has length equal to the latitude difference $\pi/6-\arcsin(n_j\cdot y)$.
Since the three normals sum to zero, at least one projection $n_j\cdot y$ is nonnegative, placing at least one cap within distance $\pi/6$.
At the center, all three projections vanish, and the bound is attained,
\begin{equation}
R_0^{\rm opt}=\frac{\pi}{6}.
\label{eq:app_triangleinradius}
\end{equation}
This radius sets the vacuum scale used to bound the perturbed causal region. \\

\noindent
We assume the exact family admits the common regular conformal development of Appendix~\ref{app:causalclass} on the closed slab $\mathcal W=[-\pi/4,\pi/4]\times\mathcal B$ and obeys
\begin{equation}
\widetilde g_\lambda=\frac{\cos^2\zeta}{L^2}g_\lambda
\longrightarrow-d\tau^2+\gamma_{\rm opt}
\quad\hbox{uniformly on }\mathcal W.
\label{eq:app_conformalcontinuity}
\end{equation}
Here, uniform convergence refers to $\lambda\to0$ from either side.
We take $\tau$ to be a Cauchy temporal coordinate of this conformal development, so the two end slices are ambient Cauchy slices.
We also assume $Z_\lambda\to Z_0$ in the Hausdorff distance induced by $\gamma_{\rm opt}$.
For nonempty closed sets $A,B\subset\mathcal B$, the Hausdorff distance is $d_{\rm H}(A,B)=\max\{\sup_{x\in A}d_{\rm opt}(x,B),\sup_{y\in B}d_{\rm opt}(y,A)\}$, with $d_{\rm opt}(x,B)=\inf_{z\in B}d_{\rm opt}(x,z)$.
The latter convergence is a uniform geometric regularity condition on the fixed RT phase, stronger than almost-everywhere convergence of polygon indicators.
If each point of either set lies within distance $d_{\rm H}$ of the other set, the triangle inequality bounds the difference of the two distance functions everywhere by $d_{\rm H}$.
Taking their maxima gives
\begin{equation}
|R_\lambda^{\rm opt}-R_0^{\rm opt}|
\leq d_{\rm H}(Z_\lambda,Z_0)\longrightarrow0.
\end{equation}
Hence, controlling the largest motion of the boundary also controls the largest optical travel time across the polygon in the reference metric. \\

\noindent
For sufficiently small $|\lambda|$, every vector $\partial_\tau+(3/4)X$ with $|X|_{\rm opt}\leq1$ is timelike on $\mathcal W$.
For example, an operator-norm bound $\|\widetilde g_\lambda-\widetilde g_0\|<7/25$, measured with $d\tau^2+\gamma_{\rm opt}$, implies
\begin{equation}
\widetilde g_\lambda\left(\partial_\tau+\frac34X,
\partial_\tau+\frac34X\right)
\leq-\frac7{16}+\frac{25}{16}
\|\widetilde g_\lambda-\widetilde g_0\|<0.
\end{equation}
Eq.~\eqref{eq:app_triangleinradius} also gives $R_\lambda^{\rm opt}<3\pi/16$ for sufficiently small $|\lambda|$.
Thus, a timelike curve from $\tau=0$ outside $P_\lambda$ reaches each point at $\tau=\pi/4$.
Its optical path has length less than $3\pi/16$.
The speed bound $3/4$ allows that path to fit in time $\pi/4$, with a vertical timelike segment occupying any unused time.
The strict margin permits the starting point to lie in the open complement, away from the edge.
An inextendible past continuation of this curve avoids $P_\lambda$.
Thus, the entire $\tau=\pi/4$ slice is outside $D^+(P_\lambda)$.
Any later point is also outside.
A past causal curve either reaches that slice and can be continued along an avoiding curve, or already fails to meet $P_\lambda$.
The past argument is identical.
We conclude
\begin{equation}
\mathcal N_\lambda\subset\{|\tau|<\pi/4\}.
\label{eq:app_slabtrapping}
\end{equation}
The causal cone bound applies throughout the slab, including points beyond a caustic of the normal congruence.
For another fixed polygon, a slab $[-T,T]$ and a speed $0<v<1$ with $vT>R_0^{\rm opt}$ give the same strict margin under the corresponding uniform metric and set convergence assumptions. \\

\noindent
If $\Psi_\lambda=\sqrt{8\pi G_N}\phi_\lambda$ tends uniformly to zero on $\mathcal W$, Eq.~\eqref{eq:app_slabtrapping} controls it on the entire polygon development.
Since $M_\phi^2L^2=3$,
\begin{equation}
q_\lambda=\frac32
\|\Psi_\lambda\|_{L^\infty(\mathcal N_\lambda)}^2<1
\end{equation}
for sufficiently small $|\lambda|$.
The global comparison therefore applies to every admissible filling in that neighborhood. \\

\noindent
For a completion differentiable in the uniform norm, $\phi_\lambda=\lambda\phi_1+o(\lambda)$ on $\mathcal W$.
Both linear modes obey $|\phi_1|\leq |A|$ throughout the slab.
For the even mode, equality holds at the center on the reflection slice, which remains inside the triangle.
Hence,
\begin{equation}
q_\lambda\leq18\pi\frac{a^2}{c}\lambda^2+o(\lambda^2).
\label{eq:app_modesmallness}
\end{equation}
For the even mode, the point where $|\phi_1|=|A|$ lies in the triangle, so the leading coefficient attains the upper bound.
To turn the estimate into a criterion at a specified amplitude, suppose the remainder satisfies $\|\phi_\lambda-\lambda\phi_1\|_\infty\leq C|A|\lambda^2$ with a specified constant $C$.
The triangle inequality then gives $\|\phi_\lambda\|_\infty\leq |A\lambda|(1+C|\lambda|)$, and a sufficient condition is
\begin{equation}
18\pi\frac{a^2}{c}\lambda^2(1+C|\lambda|)^2<1.
\end{equation}
Eq.~\eqref{eq:app_modesmallness} suffices for the susceptibility because the derivative is taken at zero, where the uniform smallness condition holds on some open interval of amplitudes.
A numerical bound on the size of that interval requires control of the remainder as above.

\subsection{Radial Inversion and Displaced Chords}
\label{app:angular-chord-derivation}

\noindent
For a fixed scalar preparation, how does an interval measurement change as its endpoints move?
A quartet of endpoints selects both a region and a spatial weight within that region.
Since the scalar stress begins at order $\lambda^2$, the polygon has zero first derivative of area.
If its positive and negative proper energies balance, the volume Hessian $V_+''(0)$ also vanishes under the comparison of Appendix~\ref{app:geometry}.
The weighted entropy reading can nevertheless have either sign.
We compute the dependence on chord orientation and exhibit both signs for the field phase of the dimension-$3/2$ mode, with $B=\beta^2=\pi a^2/2>0$.
Eqs.~\eqref{eq:primary_modular_energy} and~\eqref{eq:closed_fisher_main} give the ordinary entropy Hessian per unit $B$,
\begin{equation}
s_2(\alpha)=\frac{S_{2,+}(\alpha)}{B}
=6-6(\alpha-\pi/2)\cot\alpha-3\pi(1/s-s/2+s^3/8),
\qquad s=\sin\alpha .
\label{eq:entropy_calibration}
\end{equation}
Replacing the arc by its complement sends $\alpha$ to $\pi-\alpha$.
The sine is unchanged, and the two factors in $(\alpha-\pi/2)\cot\alpha$ both change sign, giving the equal entropies required by purity.
For the bulk mode $\phi_{1,+}=A D^{3/4}$ with $D=1-r^2$ at $\tau=0$, the radial derivative is $\partial_\chi\phi_{1,+}=-(3/2)r\phi_{1,+}$ and the normal momentum vanishes.
Inserting $M_\phi^2L^2=-3/4$ in the local energy formula $e_+=(L^3/2)[L^{-2}(\partial_\chi\phi_{1,+})^2+M_\phi^2\phi_{1,+}^2]$ gives $e_+=3a^2(3r^2-1)D^{3/2}/8$, where $a=A\sqrt L$ and $r^2=X^2+Y^2$.
Multiplication by $d\mu=D^{-3/2}\,dX\,dY$ gives Eq.~\eqref{eq:modedensity}. \\

\noindent
Here the input is the interval entropy Hessian as a function of the endpoints, so nearby intervals supply derivatives with respect to width and orientation.
For a mode at the origin, rotations leave the chord reading unchanged.
It depends only on the signed offset $p=\cos\alpha$, with $-1<p<1$ and $p=0$ for a diameter.
Write $d=\sqrt D$, $s=\sqrt{1-p^2}$ and $I_\gamma=(\XX e_+)(p)$ for the hyperbolic line integral of the density.
The Klein arclength is $\mathrm d\sigma=s\,\mathrm dt/D$, where $t$ is Euclidean arclength, so $G(p)=I_\gamma/s$ integrates $e_+/D$ in the Euclidean disk.
For the centered mode, the measured Hessian per unit squared amplitude is $s_2(\alpha)$ and Eq.~\eqref{eq:raycurvature} reduces to $4\pi I_\gamma/B=2s_2-s^2\partial_\alpha^2s_2$.
At offset $p$, the closest point of the chord has radius $|p|$, so the integral samples only radii $r\geq|p|$.
Measuring arclength $t$ from that closest point gives $r^2=p^2+t^2$ and hence $|dt|=r\,dr/\sqrt{r^2-p^2}$ on each half of the chord.
The factor of two from the two halves yields $G(p)=2\int_{|p|}^1 (e_+/d^2)\,r\,dr/\sqrt{r^2-p^2}$, the Abel transform of the radial density.
The inverse reconstructs $e_+/d^2$ from the variation of $G$ with offset in Eq.~\eqref{eq:mode_abel}.
For the last equality, differentiation gives $\partial_pG=(3B/8)p(5-9p^2)$.
Then, $t=\sqrt{p^2-r^2}$ gives $\int_0^d(5-9r^2-9t^2)dt=2d(1-3r^2)$.
This expression agrees with the canonical density in Eq.~\eqref{eq:modedensity} at every radius.
Fig.~\ref{fig:bweight}(c) plots the measured line profile $32G(p)/(3B)$ against the signed chord offset $p$.
For $B>0$, its interior zeros are $p=\pm1/3$, whereas the local density changes sign at $r=1/\sqrt3$.
A chord with $1/3<|p|<1/\sqrt3$ therefore passes through negative density but has a positive line integral because its positive outer segments outweigh its negative central segment.
In polygon variables $\delta_j=2\alpha_j=x_{j+1}-x_j$ with $x_{N_v+1}=x_1+2\pi$, the field-phase specialization of Eq.~\eqref{eq:square_to_polygon} gives Eq.~\eqref{eq:modepolygon}.
The area integral in the latter equation follows by multiplying Eq.~\eqref{eq:modedensity} by $24\pi/B$.
For four vertices, $\sum_{j=1}^{4}\delta_j=2\pi$ and two applications of the sine sum-to-product formula give
\[
\sum_{j=1}^4\sin(2\delta_j)
=4\sin(\delta_1+\delta_2)\sin(\delta_2+\delta_3)\sin(\delta_1+\delta_3).
\]
The first two factors test whether either diagonal passes through the center of the scalar mode.
For diagonal $13$, a diameter has endpoints half a turn apart, so $\delta_1+\delta_2=\pi$.
For diagonal $24$, the corresponding condition is $\delta_2+\delta_3=\pi$.
Either condition makes the integrated energy vanish for this radial profile, even though the density varies across the polygon.
For the third factor, write the direction of a chord joining angular positions $x_a$ and $x_b$ as $(x_a+x_b)/2+\pi/2$ modulo $\pi$, which follows by subtracting their unit-circle endpoint vectors.
The difference between the directions of the two diagonals is $(\delta_1+\delta_3)/2$ modulo $\pi$, so perpendicular diagonals in the Euclidean Klein chart give zero.
The example in Fig.~\ref{fig:squarecertificate}(c) has diagonals $X=7/25$ and $Y=7/25$.
For the trapezoid $Q_{\rm trap}$ with vertices $(\pm4/5,3/5)$ and $(\pm3/5,4/5)$, all four doubled gaps have sine $336/625$, so the sum is $1344/625$ and $cV_+''/(L^2B)=3024/625$. \\

\noindent
Now fix the endpoint coordinates in the observer's disk and displace the scalar center.
Use the AdS isometry whose action on the reflection slice is $T_v(X,Y)=(X-v,Y/\gamma)/(1-vX)$, where $|v|<1$ and $\gamma=(1-v^2)^{-1/2}$.
The displaced state assigns to each point $y$ the centered scalar field at $T_v y$ and therefore has its center at $(v,0)$.
Equivalently, one can calculate the reading for the fixed physical chord by evaluating its image in the centered state.
The parameter $v$ specifies the displacement, while $\lambda$ varies the excitation amplitude at that fixed displacement.
Thus, the image half-opening $\alpha'$ and the endpoint cutoff factors are constant when taking state derivatives, giving $S'=0$ and $S''=Bs_2(\alpha')$. \\

\noindent
To find the image interval, start with the chord $n_\theta\cdot y=p$, where $n_\theta=(\cos\theta,\sin\theta)$ and $-1<p<1$.
Write $y'=(X',Y')=T_v y$ and $D_v=\gamma(1-vX)$.
Substituting $X=(X'+v)/(1+vX')$ and $Y=Y'/[\gamma(1+vX')]$ into the chord equation gives
\[
\gamma(\cos\theta-vp)X'+\sin\theta\,Y'=A_v,
\qquad A_v=\gamma(p-v\cos\theta).
\]
The squared length of the coefficient vector is $\gamma^2(\cos\theta-vp)^2+\sin^2\theta=A_v^2+1-p^2$.
Its positive square root $H_v=\sqrt{A_v^2+s^2}$ normalizes both the oriented unit normal $n_{\theta'}=(\gamma(\cos\theta-vp),\sin\theta)/H_v$ and the signed offset $p'=A_v/H_v$.
It follows that the image half-opening $0<\alpha'<\pi$ satisfies $\cos\alpha'=p'$ and $s'=\sin\alpha'=s/H_v$.
To apply the interval differential operator after the boost, we need the rate at which its image half-opening changes when either endpoint moves.
Let $f_v$ be the increasing boundary map induced by $T_v$, with lifted endpoints $u=\theta-\alpha$, $w=\theta+\alpha$.
Then, $\alpha'=[f_v(w)-f_v(u)]/2$.
Varying $\theta$ moves the two endpoints in the same direction, whereas varying $\alpha$ moves them in opposite directions.
Using $u=\theta-\alpha$ and $w=\theta+\alpha$ gives $\partial_\theta\alpha'=[f_v'(w)-f_v'(u)]/2$ and $\partial_\alpha\alpha'=[f_v'(w)+f_v'(u)]/2$.
The difference of the squares is $-f_v'(u)f_v'(w)$, while the two second derivatives of $\alpha'$ agree because differentiating either endpoint twice removes the sign difference.
To evaluate the product, represent a boundary angle $x$ by the null vector $\ell(x)=(1,\cos x,\sin x)$ in the embedding of Eq.~\eqref{eq:klein_metric}.
The boost sends this vector to $b_v(x)\ell(f_v(x))$, where the time component is $b_v(x)=\gamma(1-v\cos x)>0$.
In the derivative of $b_v(x)\ell(f_v(x))$, all terms containing the null vector $\ell$ drop out of the squared norm because $\ell\cdot\ell=\ell\cdot\ell'=0$.
The boost preserves $\ell'\cdot\ell'=1$, leaving $[b_v(x)f_v'(x)]^2=1$ and the positive sign since the endpoint order is preserved.
The same boost preserves $\ell(u)\cdot\ell(w)=-2\sin^2\alpha$, so $b_v(u)b_v(w)\sin^2\alpha'=\sin^2\alpha$.
These two norm identities give
\begin{equation}
f_v'(x)=\frac{1}{b_v(x)},\qquad
\sin^2\alpha'=\sin^2\alpha\, f_v'(u)f_v'(w).
\label{eq:boost_boundary_scale}
\end{equation}
Combining Eq.~\eqref{eq:boost_boundary_scale} with the endpoint derivatives gives
\begin{equation}
(\partial_\theta^2-\partial_\alpha^2)\alpha'=0,
\qquad
s^2[(\partial_\theta\alpha')^2-(\partial_\alpha\alpha')^2]=-s'^2.
\label{eq:boost_chain_rule}
\end{equation}
In endpoint coordinates $(u,w)$, the boost multiplies the area element by $f_v'(u)f_v'(w)$.
Changing back to midpoint and half-opening introduces the same constant Jacobian on both sides, leaving $(s'/s)^2$.
Converting both half-openings to offsets, $|dp|=s\,d\alpha$ and $|dp'|=s'\,d\alpha'$, gives
\[
|dp'|\,d\theta'=(s'/s)^3\,|dp|\,d\theta,
\qquad
\frac{|dp'|\,d\theta'}{(1-p'^2)^{3/2}}=
\frac{|dp|\,d\theta}{(1-p^2)^{3/2}} .
\]
The last expression is an invariant measure on chords. It permits angular data taken in the two frames to be compared with the same weighting.
Each unoriented chord occurs twice over $0\le\theta<2\pi$, since $(p,\theta)$ and $(-p,\theta+\pi)$ label the same line. \\

\noindent
The chain rule for $s_2(\alpha')$ has a term proportional to $\partial_{\alpha'}s_2$ and a term proportional to $\partial_{\alpha'}^2s_2$.
The first identity in Eq.~\eqref{eq:boost_chain_rule} cancels the former, and the second fixes the coefficient of the latter.
Thus, the interval operator $-\DD=2+\sin^2\!\alpha(\partial_\theta^2-\partial_\alpha^2)$ acting on $s_2(\alpha')$ gives $2s_2(\alpha')-s'^2\partial_{\alpha'}^2s_2(\alpha')$.
Eq.~\eqref{eq:raycurvature} therefore equates the unnormalized chord integral in the boosted state to that of its image chord in the original state.
Dividing by $s$ and using $I_0=s'G_0(p')$ gives $G_v=(s'/s)G_0(p')$.
Substituting $p'=A_v/H_v$ and $s'=s/H_v$ into Eq.~\eqref{eq:mode_abel} gives the measured $G_v$ in the first expression below, while the second gives its rescaled density $F_v=e_v/d^2$,
\begin{equation}
G_v(p,\theta)=\frac{3B(1-p^2)[8A_v^2-(1-p^2)]}{32[1-p^2+A_v^2]^{5/2}},
\qquad F_v=\frac{3Bd}{4\pi D_v^3}\left(2-\frac{3d^2}{D_v^2}\right).
\label{eq:boosted_cft_ray}
\end{equation}
The subscript $0$ denotes the unboosted profile of Eq.~\eqref{eq:modedensity}, and the chord function is extended by zero for $|p|\ge1$.
The isometry preserves the slice and its future unit normal, so its pullback transforms the normal energy as $e_v(y)=e_0(T_v y)$.
Writing $y'=T_v y$ and $d'=\sqrt{1-|y'|^2}$ gives $d'=d/D_v$ and hence $F_v(y)=D_v^{-2}F_0(y')$, with $F_0=e_0/d'^2$ evaluated at $y'$.
Inserting Eq.~\eqref{eq:mode_abel} for $F_0(y')$ gives the second expression in Eq.~\eqref{eq:boosted_cft_ray}. \\

\noindent
For $B>0$ and $|p|<1$, the denominator of $G_v$ is strictly positive, so Eq.~\eqref{eq:boosted_cft_ray} gives a negative reading precisely when $8\gamma^2(p-v\cos\theta)^2<1-p^2$.
Using $p'^2=A_v^2/(A_v^2+s^2)$ reduces this condition to $|p'|<1/3$, the same central band of image chords as in Eq.~\eqref{eq:mode_abel}.
Every chord through the displaced center has $p=v\cos\theta$, so its image is a diameter with $A_v=0$ and $s'=1$.
All these chords have the same hyperbolic integral $I_v=sG_v=-3B/32$, while the Euclidean normalization in $G_v=I_v/s$ varies with angle.
An observer holding the width fixed sees a different pattern.
For example, of two perpendicular diameters $p=0$ in the observer's disk, the one with $\theta=\pi/2$ passes through the scalar center and gives a negative reading.
The other, with $\theta=0$, gives a positive reading for $8\gamma^2v^2>1$, or $|v|>1/3$. \\

\noindent
To see directly why this is the density reconstructed from the angular data, take Euclidean arclength coordinates $t,t'$ along the original and image chords.
Their common hyperbolic line element is $s\,|dt|/d^2=s'\,|dt'|/d'^2$, which implies
\[
|dt'|=\frac{s}{s'}D_v^{-2}|dt|,
\qquad F_v(y)\,|dt|=\frac{s'}s F_0(y')\,|dt'|.
\]
Integration along the original chord gives $\RR_E F_v(p,\theta)=(s'/s)G_0(p')=G_v(p,\theta)$, so the proposed displaced profile predicts each measured integral.
For fixed $|v|<1$, $D_v$ stays finite and positive with a positive lower bound, ensuring $F_v$ is integrable.
By the Radon inversion in Appendix~\ref{app:radoninversion}, two such profiles with equal integrals along every chord must coincide, including in regions of negative density. \\

\noindent
The same angular dependence follows by integrating the bulk density directly.
The Lorentz boost of the embedding in Eq.~\eqref{eq:klein_metric} preserves $q$ and geodesic arclength, so $I_v(\Gamma)=I_0(T_v\Gamma)$ for the original geodesic $\Gamma=\gamma_{p,\theta}$.
Let $t'$ denote Euclidean arclength along the image chord, measured from its closest point to the origin.
The squared radius there is $p'^2+t'^2$, so direct integration gives
\begin{equation}
\frac{I_0(T_v\Gamma)}{s'}
=\frac{3B}{4\pi}\int_{-s'}^{s'}[3(p'^2+t'^2)-1]\sqrt{s'^2-t'^2}\,dt'
=\frac{3B}{32}s'^2(9p'^2-1).
\label{eq:boosted_bulk_ray}
\end{equation}
The two elementary moments are $\int_{-s'}^{s'}\sqrt{s'^2-t'^2}dt'=\pi s'^2/2$ and $\int_{-s'}^{s'}t'^2\sqrt{s'^2-t'^2}dt'=\pi s'^4/8$.
Using $G_v=I_v/s$ and $G_0=I_0/s'$, this gives $(s'/s)G_0(p')$, exactly the CFT expression in Eq.~\eqref{eq:boosted_cft_ray} for every $\theta$, $|p|<1$, and $|v|<1$. \\

\noindent
If the scalar center and all polygon vertices move together by an isometry, their relative positions and the enclosed proper energy are preserved.
If the endpoints stay fixed while the scalar center moves, the region collects a different part of the profile.
For $y'=T_vy$, direct differentiation gives $d(y')=d(y)/D_v$ and $|\det(\partial y'/\partial y)|=D_v^{-3}$, hence $d\mu(y')=d\mu(y)$.
For a polygon $Q$ with fixed, cyclically ordered boundary angles $x_1,\ldots,x_{N_v}$, the pullback thus gives $\int_Q e_v\,d\mu=\int_{T_v Q}e_0\,d\mu$.
Applying Eq.~\eqref{eq:modepolygon} to the image polygon expresses the response at those fixed physical endpoints as
\[
\frac{cV_{v,Q}''}{L^2B}=\frac94\sum_{j=1}^{N_v}
\sin\!\left(2[f_v(x_{j+1})-f_v(x_j)]\right).
\]
Here, $V_{v,Q}''$ denotes the state Hessian at fixed $v$ and $Q$, and the increasing lift satisfies $x_{N_v+1}=x_1+2\pi$ and $f_v(x+2\pi)=f_v(x)+2\pi$.
If instead the endpoints are chosen as those of $Q=T_v^{-1}Q_{\rm trap}$ for each boost, the image gaps are precisely the original trapezoid gaps and the change of variables gives
\begin{equation}
\frac{cV_Q''}{L^2B}
=\frac{24\pi}{B}\int_Qe_v\,d\mu
=\frac{24\pi}{B}\int_{Q_{\rm trap}}e_0\,d\mu
=\frac{3024}{625}.
\label{eq:boosted_exact_volume}
\end{equation}

\subsection{Sign Crossing at Fixed Polygon Volume Response}
\label{app:quartet-sign-derivation}

\noindent
For the centered square in the dimension-$3/2$ field phase, the four equal gaps $\pi/2$ give $V_Q''=0$ in Eq.~\eqref{eq:modepolygon}.
We can see what the cancellation hides by computing three integrals of the same density $e$.
The signed proper energy $T_Q=\int_Q e\,d\mu$ controls the area, whereas $E_Q^-=\int_Q\max(-e,0)\,d\mu$ counts all negative contributions by magnitude.
The entropy measurements supply the weighted integral $u=\int_Q(W_Q/M_Q)e\,d\mu$, where $M_Q=\max_Q W_Q$.
For the square, $T_Q=0$ yet $-u/B=(108-75\sqrt2)/32>0$.
Indeed, symmetry gives $R''=2d_\square$ for the quartet contrast and $u=R''/(2\pi)$ at cross ratio $\eta=1/2$.
The negative region is a disk of radius $1/\sqrt3$, fully contained in the square.
Eq.~\eqref{eq:modedensity} therefore gives $E_Q^-=(3B/2)\int_0^{1/\sqrt3}(1-3r^2)r\,\mathrm dr=B/8$.
Keep this preparation fixed and move two endpoints to $-g$ and $g$, retaining the other two at $0$ and $\pi$, with $0<g\leq\pi/2$.
The cyclic order $(-g,0,g,\pi)$ gives gaps $g,g,\pi-g,\pi-g$.
In Eq.~\eqref{eq:modepolygon}, the terms $2\sin(2g)$ and $2\sin(2\pi-2g)$ cancel for every $g$, so both the volume Hessian and $T_Q$ vanish throughout the family.
Reflection about the diameter through $0$ and $\pi$ exchanges endpoints $1$ and $3$ while leaving the other two fixed.
The two opposite-side products are therefore equal, so each supplies half the vacuum diagonal product,
\[
\eta=\frac{\sin(g/2)\sin[(\pi-g)/2]}{\sin g\,\sin(\pi/2)}
=\frac12.
\]
In the mode-centered Klein disk, diagonal $13$ lies on $X=\cos g$ and diagonal $24$ lies on $Y=0$, so the crossing has radius $r_\times=\cos g$.
For this scalar tangent, all interval entropy slopes vanish, so the squared slope term in Eq.~\eqref{eq:ptolemy_hessian} is zero.
At $\eta=1/2$, Eq.~\eqref{eq:quartet_linear_main} counts both diagonal Hessians with coefficient $+1$ and each side Hessian with coefficient $-1/2$.
The diagonals have half-openings $g$ and $\pi/2$, and the four sides have half-openings $g/2,g/2,(\pi-g)/2,(\pi-g)/2$, using purity to choose the shorter arcs.
Consequently, $R''/B=s_2(g)+s_2(\pi/2)-s_2(g/2)-s_2((\pi-g)/2)$.
Set $h(s)=s^{-1}-s/2+s^3/8$, $s_g=\sin g$, $s_h=\sin(g/2)$ and $c_h=\cos(g/2)$.
Using $2\cot g=\cot(g/2)-\tan(g/2)$ in Eq.~\eqref{eq:entropy_calibration} cancels the terms proportional to $g$ and leaves
\[
\frac{R''}{3\pi B}=h(s_h)+h(c_h)-h(s_g)-h(1)-\frac1{s_g}.
\]
The sign change along this entire family follows by expressing the right-hand side in terms of $z=s_h+c_h-1$, with $0<z\leq\sqrt2-1$.
The identities $s_h c_h=s_g/2$ and $s_g=z(z+2)$ reduce the inverse powers to $s_h^{-1}+c_h^{-1}-2/s_g=2/(z+2)$.
The cubic terms reduce using $s_h^3+c_h^3=(1+z)^3-3s_g(1+z)/2$.
Combining them with the constant and linear terms gives
\[
\frac{R''}{3\pi B}=-\frac{z^2}{16(z+2)}\,\mathcal P(z),
\qquad \mathcal P(z)=2z^5+16z^4+48z^3+65z^2+29z-18 .
\]
Every nonconstant coefficient of $\mathcal P$ is positive, so it increases strictly for $z>0$.
Its endpoint values $\mathcal P(0)=-18$ and $\mathcal P(\sqrt2-1)=2+5\sqrt2$ place exactly one root $z_\times$ in the interval.
Since $z$ increases with $g$ and the prefactor of $\mathcal P$ is negative, the contrast is positive below $g_\times=\arcsin[z_\times(z_\times+2)]\simeq0.85096$ radians and negative above it.
At $g=g_\times$, both the weighted and unweighted moments vanish even though the density in Eq.~\eqref{eq:modedensity} is nonzero on open subsets inside the polygon.
A zero for this quartet therefore leaves open how much positive and negative energy it encloses.
The explicit choice $g_0=\arcsin(9/16)$ has $r_0=\cos g_0=5\sqrt7/16$, $\cos(g_0/2)=(5+\sqrt7)/8$, and $\sin(g_0/2)=(5-\sqrt7)/8$.
These give $z_0=\sin(g_0/2)+\cos(g_0/2)-1=1/4$ and, by direct substitution, $R''/B=3007\pi/98304>0$ and $u/B=3007/196608$.
A second member of the same family illustrates the difference between the local density at the diagonal crossing and the energy averaged by the quartet weight.
For $z=1/3$, one has $\sin g=7/9$ and $3r_\times^2-1=5/27>0$, so Eq.~\eqref{eq:modedensity} gives positive density where the weight is largest.
Nevertheless, $\mathcal P(1/3)=212/243$ gives $R''/B=-53\pi/6804<0$ because the measured moment also includes the negative density elsewhere inside this polygon. \\

\noindent
For the quartet with $g=g_0$, the positive reading $u$ alone gives only $E_Q^-\ge[-u]_+=0$, where $[x]_+=\max(x,0)$.
The complementary weight $1-W_Q/M_Q$ lies in $[0,1]$ and gives $T_Q-u\ge-E_Q^-$.
Combining the positive reading with the independently determined $T_Q=0$ gives the stronger bound $E_Q^-\ge3007B/196608$.
The fifth-endpoint measurement in Fig.~\ref{fig:squarecertificate}(d) gives the opposite sign for this same state.

\section{Realizing Smooth Profiles with Fixed Interval Data}
\label{app:finite}

\noindent
The constructions behind Sec.~\ref{sec:sharp} answer two versions of the measurement question.
For data at quadratic order, we prescribe the quartet reading and signed polygon energy independently using a time-even field with vanishing normal momentum.
A boundary transformation then sets the individual entropy Hessians to be consistent with the prescribed quartet reading.
For data at nonzero amplitude, we start with a time-odd field whose normal momentum gives positive density in an open disk.
We change the metric inside that disk and adjust the local scalar momentum to satisfy the constraint, keeping the sampled lengths fixed while increasing the enclosed area.
The final two subsections derive the concentration and measurement-error estimates for the negative-energy bound in Sec.~\ref{sec:signed-energy-bound} and the scalar charge bounds used in Sec.~\ref{sec:energy}.

\subsection{Time-Even Packets and Sharp Signed-Energy Bound}
\label{sec:energy-packets}
\noindent
The input is a pair of numbers: the signed energy $T_Q$ of a polygon and its normalized quartet reading $u$.
To prove sharpness, we approximate any negative-energy magnitude $\mathcal N>N_*$ while realizing both inputs exactly.
The positive energy is then fixed by $\mathcal P=\mathcal N+T_Q$.
The four terms in Eq.~\eqref{eq:joint} give $\mathcal N>0$, $\mathcal N>-T_Q$, $\mathcal N>-u$ and $\mathcal N>u-T_Q$.
The first two inequalities make both packet strengths positive, and the last two give $-\mathcal N<u<\mathcal P$.
This is just the range of $\mathcal P s_+-\mathcal N s_-$ as the two normalized weights $s_+,s_-$ vary independently from zero to one, excluding the endpoints.
Except for $u=T_Q=0$, the solution set $\mathcal P s_+-\mathcal N s_-=u$ is a line segment containing a point with $s_+\ne s_-$.
Continuity of $k$ along a ray from the crossing to any side then gives two interior points $y_\pm$ with $k(y_\pm)=s_\pm$.
When both moments are zero, vacuum already attains the infimum zero.
Point packets of opposite signed energies at these locations would give exactly the desired readings.
We approximate them inside $Q$ by smooth scalar profiles and then adjust their amplitudes to keep both readings exact. \\

\noindent
For the negative packet, write $L\rho$ for hyperbolic distance to $y_-$, choose a nondecreasing $C^\infty$ cutoff $b(t)$ equal to zero for $t\le0$ and one for $t\ge1$, and set
\begin{equation}
b_\delta(t)=b(2t/\delta-1),\quad
\rho_{\mathrm c}=2|\ln\delta|/\mu_*,
\quad \varphi(\rho)=b(\rho_{\mathrm c}+1-\rho)
\exp\!\left(-\mu_*\int_0^\rho b_\delta(t)dt\right)
\end{equation}
for $0<\delta<1$, taken small.
Because the radial derivative vanishes near $\rho=0$, this defines a smooth field on the disk.
Away from the outer taper, $\varphi'=-\mu_* b_\delta\varphi$.
For time-even data, the energy there is $2e/L=\mu_*^2(b_\delta^2-1)\varphi^2$.
In the core $\rho\le\delta/2$, the field is constant, and only its negative mass term contributes, giving $2e/L=-\mu_*^2$.
The density stays nonpositive until $\rho=\delta$.
Between $\delta$ and $\rho_{\mathrm c}$, the derivative has magnitude $\mu_*\varphi$, so its positive gradient energy cancels the negative mass term point by point.
In the outer taper, its value is nonnegative and proportional to the decaying exponential squared times $b'^2+2\mu_* b b'$, with $b,b'$ evaluated at $\rho_{\mathrm c}+1-\rho$.
On the taper, $\int_0^\rho b_\delta(t)\,dt\geq\rho-\delta$, so the squared exponential is at most $\exp[-2\mu_*(\rho_{\mathrm c}-\delta)]=O(\delta^4)$.
Although the taper moves to large radius, the area of $Q$ is finite, so the total positive taper energy inside $Q$ remains $O(\delta^4)$.
For small enough $\delta$, the core disk lies entirely in $Q$ and has area $2\pi[\cosh(\delta/2)-1]$ in $d\mu$.
Its negative energy has magnitude of order $\delta^2$ and therefore outweighs the taper.
Rescaling the field by a constant then gives total signed energy $-1$ in $Q$. \\

\noindent
Although the energy vanishes between the core and the outer taper, the field itself still occupies this region.
A second field placed there would produce cross terms, so we cut a small hole about $y_+$ before inserting the positive packet.
With $Lr$ the distance to $y_+$ and $0<\epsilon<1$, set
\[
\chi_\epsilon(r)=b\!\left(\frac{\ln(r/\epsilon^2)}{|\ln\epsilon|}\right),
\qquad \chi_\epsilon(0)=0 .
\]
It is zero for $r\le\epsilon^2$ and one for $r\ge\epsilon$.
On the intervening annulus, $|D\chi_\epsilon|\le C/(r|\ln\epsilon|)$.
Near $y_+$, the hyperbolic area element approaches $r\,dr\,d\theta$.
The radial integral of the squared cutoff gradient is therefore bounded by a constant times $|\ln\epsilon|^{-2}\int_{\epsilon^2}^{\epsilon}dr/r=1/|\ln\epsilon|$, which tends to zero.
Expanding $|D(\chi_\epsilon\varphi)|^2$ and using Cauchy--Schwarz on the cross term shows that $\int_Q|e(\chi_\epsilon\varphi)-e(\varphi)|\,d\mu$ tends to zero at fixed $\delta$.
Choose $\epsilon$ small enough to make this error at most $\delta^3$. \\

\noindent
Inside $r<\epsilon^2/2$, take a smooth, nonconstant compact bump of fixed height and radial scale $\epsilon^2$.
In two spatial dimensions, shrinking the bump on this scale reduces its area by a factor of order $\epsilon^4$. It increases its squared gradient by the inverse factor.
Its integrated gradient energy therefore tends to a strictly positive constant, while the mass term contributes only $O(\epsilon^4)$.
Let $e_+$ and $e_-$ be the densities of the positive bump and the punctured negative packet after multiplying each field by a positive constant to impose $\int_Qe_\pm d\mu=\pm1$.
For each $\delta$, choose $\epsilon=\epsilon(\delta)$ small enough to satisfy the hole estimate and $\epsilon(\delta)\leq\delta$.
Normalization amplifies the small tails of the negative packet because its unnormalized signed energy is itself small.
To bound the resulting error in the readings, let $D_\pm(r)$ denote the disk of dimensionless hyperbolic radius $r$ about $y_\pm$.
Write $\omega_\pm(r)=\sup_{y\in Q\cap D_\pm(r)}|k(y)-s_\pm|$, which tends to zero as $r\downarrow0$ by continuity.
For the positive bump, the unnormalized gradient integral has a positive limit, and its integrated negative part is at most a constant times $\epsilon^4$.
Normalization to signed mass one therefore gives $E_Q^-(e_+)=O(\epsilon^4)$ and $\int_Q |e_+|\,d\mu=1+2E_Q^-(e_+)$.
For the other packet, the signed mass before normalization is negative of order $\delta^2$.
The taper and the hole contribute at most $O(\delta^4+\delta^3)$ to its unnormalized absolute mass outside $D_-(\delta)$.
Within $D_-(\delta)$, the density is nonpositive.
Consequently, $\int_{Q\setminus D_-(\delta)} |e_-|\,d\mu=O(\delta)$ after normalization, and $\int_Qe_-\,d\mu=-1$ gives $E_Q^-(e_-)=1+O(\delta)$.
For $m_\pm=\int_Q k e_\pm\,d\mu$, subtract $s_\pm\int_Qe_\pm\,d\mu$ and use $|k-s_\pm|\leq1$ outside the respective packet disks to obtain \null
\[
\begin{aligned}
|m_+-s_+|&\leq\omega_+(\epsilon^2/2)\,[1+O(\epsilon^4)],\\
|m_-+s_-|&\leq\omega_-(\delta)\,[1+O(\delta)]+O(\delta).
\end{aligned}
\]
Thus, the localized weights converge to the desired values even though the negative packet has a distant taper. \\

\noindent
The fields have disjoint supports.
Multiplying them by $\sqrt{A_+}$ and $\sqrt{A_-}$ therefore adds their stress tensors with coefficients $A_+$ and $A_-$.
Requiring the exact moments for each sufficiently localized profile gives
\begin{align}
A_+-A_-&=T_Q,&m_+A_++m_-A_-&=u,\\
A_-&=\frac{u-m_+T_Q}{m_++m_-},&A_+&=T_Q+A_-.
\end{align}
The denominator tends to $s_+-s_-\ne0$.
Subtracting the target equations $\mathcal P-\mathcal N=T_Q$ and $s_+\mathcal P-s_-\mathcal N=u$ from the two measured moment equations gives the exact adjustment \null
\[
A_+-\mathcal P=A_--\mathcal N
=-\frac{\mathcal P(m_+-s_+)+\mathcal N(m_-+s_-)}{m_++m_-}.
\]
At fixed $\mathcal N>N_*$ and $s_+\ne s_-$, the denominator stays away from zero, and the numerator tends to zero.
The corrected squared amplitudes therefore approach $\mathcal N$ and $\mathcal P$, both positive.
Disjoint supports give $E_Q^-(A_+e_++A_-e_-)=A_+E_Q^-(e_+)+A_-E_Q^-(e_-)$ exactly, so the total approaches $\mathcal N$ while the two target moments remain exact.
To reach the infimum, take $\mathcal N_j=N_*+1/j$ for integers $j\geq1$.
For each $j$, choose widths small enough for the negative-energy error to be less than $1/j$ before increasing $j$.
This order allows the target packet masses and their separation in normalized weight to shrink along the sequence.
Each finite-width packet has finite global charge, even when its proper energy inside $Q$ is negative.
The distant taper carries the positive global energy required for this to happen.
Write $a_\delta>0$ for the field multiplier that normalizes its signed energy in $Q$ to $-1$ after the hole has been cut.
The unnormalized negative core has integrated energy of order $\delta^2$, bounded above and below in magnitude by positive multiples of $\delta^2$.
The errors from the taper and the hole are $O(\delta^4)$ and $O(\delta^3)$, respectively, so $a_\delta^2\asymp\delta^{-2}$ as $\delta\downarrow0$.
Let $\chi_-$ be the unit-radius hyperbolic distance from the global origin to $y_-$, and use angular coordinate $0\leq\vartheta<2\pi$ about $y_-$.
The hyperbolic cosine law gives $\int_0^{2\pi}N\,d\vartheta=2\pi\cosh\chi_-\cosh\rho$, whereas the polar area element is $\sinh\rho\,d\rho\,d\vartheta$.
On the outer shell $\rho_{\mathrm c}<\rho<\rho_{\mathrm c}+1$, its contribution to the normalized global charge is therefore
\[
\begin{aligned}
\mathcal E_{\rm sh}
&=\pi L a_\delta^2\cosh\chi_-
\int_{\rho_{\mathrm c}}^{\rho_{\mathrm c}+1}
\!\sinh\rho\cosh\rho\,
\exp\!\left[-2\mu_*\int_0^\rho b_\delta(t)\,dt\right]
(b'^2+2\mu_* b b')\,d\rho,\\
\mathcal E_{\rm sh}
&\asymp \delta^{-2}\exp[2(1-\mu_*)\rho_{\mathrm c}]
=\delta^{\,2-4/\mu_*}.
\end{aligned}
\]
Here, $b$ and $b'$ in the first line are evaluated at $\rho_{\mathrm c}+1-\rho$.
The shell has unit width, $\int_0^\rho b_\delta(t)\,dt=\rho+O(\delta)$ throughout it, and its nonnegative taper factor has a strictly positive integral independent of $\delta$.
For fixed $y_\pm$ and $0<\mu_*<1$, the positive charge in the remote shell diverges as the negative core shrinks.
Only a small part of the shell lies in $Q$, where its normalized proper energy is $O(\delta^2)$.
The lapse stays bounded near the core, so its weighted negative energy stays bounded, and the normalized error from cutting the hole near $y_+$ is $O(\delta)$.
The sharp local bound thus permits a growing global charge along this sequence.
When the charge is fixed independently, the global bound imposes an additional restriction. \\

\noindent
Each of these smooth, compact profiles also gives constraint-satisfying initial data at sufficiently small state amplitude.
Prescribe the even initial field $\lambda\Phi(\varrho,\theta)$, vanishing normal momentum, and reflection metric $h=L^2(d\varrho^2+U(\varrho,\theta)^2d\theta^2)$, where $\varrho\geq0$, $\theta\in[0,2\pi)$ is periodic, and the polar origin lies outside the field support.
In this gauge, $K_{\mathrm G}=-U_{\varrho\varrho}/(L^2 U)$ and
\begin{equation}
\frac{U_{\varrho\varrho}}{U}
=1-4\pi G_N\lambda^2
\left[(\partial_\varrho\Phi)^2
+\frac{(\partial_\theta\Phi)^2}{U^2}
-\mu_*^2\Phi^2\right].
\end{equation}
Near the origin, the matter field vanishes, and the regular vacuum solution is $U=\sinh\varrho$.
The equation contains no angular derivatives of $U$, so the given scalar profile supplies an ordinary differential equation for the metric along each radial ray.
On the compact radial support away from the origin, smooth dependence on the data keeps $U$ and $\partial_\varrho U$ positive for sufficiently small $|\lambda|$.
Outside the support, $U_{\varrho\varrho}=U$, so $U\sim A(\theta)\exp(\varrho)$ with $A(\theta)>0$.
Choosing $\exp(-\varrho)/A(\theta)$ as the boundary defining function fixes the boundary angular metric.
The amplitude range can shrink as a packet narrows, so we first take state derivatives for each fixed smooth profile.

\subsection{Matching Individual Entropy Hessians}
\label{sec:matching-hessians}
\noindent
Prescribe the six individual entropy Hessians $d_{ij}$ for four ordered endpoints, one for each pair $1\leq i<j\leq4$.
Because the scalar packets have zero linear stress and metric response, the quartet measures the combination $u=C_{\eta_Q}d/(4\pi M)$.
Once the packets match this combination, five independent adjustments remain.
The boundary reparametrization $f_\lambda(x)=x+\lambda^2v(x)/2$ supplies them while preserving the first-order scalar profile.
Its effect on the entropy Hessian is obtained from the vacuum entropy alone, since shifting the endpoints in an existing $O(\lambda^2)$ correction by $O(\lambda^2)$ only affects higher orders.
For endpoints $x_i,x_j$, the state-independent chord term is $(c/3)\ln\sin[(f_\lambda(x_j)-f_\lambda(x_i))/2]$ up to constants.
The cutoff term is $-(c/6)\ln[f_\lambda'(x_i)f_\lambda'(x_j)]$.
Using $f_0'=1$ and $\partial_\lambda f_\lambda|_{\lambda=0}=0$ gives the Hessian shift
\begin{equation}
H_{ij}=\frac c6\big[(v_j-v_i)k_{ij}+q_i+q_j\big],\qquad
k_{ij}=\cot\frac{x_j-x_i}{2},\quad v_i=v(x_i),\quad q_i=-v'(x_i).
\label{eq:endpointjets}
\end{equation}
The transformed vacuum satisfies the Ptolemy relation for each $\lambda$, so $C_{\eta_Q}H=0$.
For four endpoints, this one condition on six entropy Hessians leaves five linearly independent shifts compatible with the same quartet reading.
To establish that a boundary map can realize every such shift, we first solve for its values and first derivatives at an arbitrary finite number $N\ge4$ of endpoints.
Keep the ordered lifts $x_1<\cdots<x_N<x_1+2\pi$ and write $h_{ij}=6(d_{ij}-S''_{ij})/c$ for $1\le i<j\le N$.
Compare data with the same first entropy derivatives at these endpoints, as in the zero-slope scalar construction above.
The matched first derivatives fix the part of each quartet Hessian that is quadratic in the entropy slopes.
We show that the boundary map can produce the desired shifts in the individual second derivatives exactly when those shifts leave all the quartet Hessians unchanged.
Necessity follows because the order-$\lambda^2$ shift is that of a transformed vacuum, which has vanishing Ptolemy defect for every quartet.
The terms quadratic in the common first derivatives cancel when the two sets of quartet Hessians are subtracted. \\

\noindent
Conversely, suppose that all quartet contrasts already agree.
We solve explicitly for a smooth boundary map producing the desired individual shifts, beginning with $v_1=v_2=v_3=0$.
These three choices use the freedom to add a vacuum symmetry without changing any of the interval readings.
The corresponding generators are $1$, $\cos x$, and $\sin x$.
Each of these functions obeys $(v_j-v_i)k_{ij}=v'(x_i)+v'(x_j)$ and therefore gives zero Hessian shift in Eq.~\eqref{eq:endpointjets}.
A nonzero linear combination of the three functions has at most two distinct zeros on the circle, so their coefficients can be chosen to set three specified values to zero.
At these three endpoints, the term proportional to $v_j-v_i$ vanishes, leaving the measured shifts $h_{12}=q_1+q_2$, $h_{13}=q_1+q_3$, and $h_{23}=q_2+q_3$.
Adding the first two equations and subtracting the third determines $2q_1$, which gives
\[
q_1=\tfrac12(h_{12}+h_{13}-h_{23}),\qquad
q_2=h_{12}-q_1,\qquad q_3=h_{13}-q_1 .
\]
For each new endpoint $j=4,\ldots,N$, both the value $v_j$ and derivative $-q_j$ are unknown.
Subtracting the equations for pairs $1j$ and $2j$ eliminates $q_j$, and inserting the resulting $v_j$ into the first equation gives
\[
v_j=\frac{h_{1j}-q_1-h_{2j}+q_2}{k_{1j}-k_{2j}},\qquad
q_j=h_{1j}-q_1-v_j k_{1j}.
\]
The two cotangents have distinct arguments in $(0,\pi)$, so the denominator never vanishes for distinct ordered endpoints. \\

\noindent
This construction matches the $2N-3$ pairs incident on endpoint $1$ or $2$.
For each $3\le i<j\le N$, the remaining pair $ij$ appears in the quartet $12ij$ together with five already matched pairs.
With the lower-order entropy data fixed, a residual normalized shift $h_{ij}^{\rm res}$ changes $D_{ij}''(0)$ by $D_{ij}^0 h_{ij}^{\rm res}/2$, where $D_{ij}^0$ is the vacuum chord factor.
It follows from $T_{12ij}=D_{1i}D_{2j}-D_{12}D_{ij}-D_{1j}D_{2i}$ in Sec.~\ref{sec:quartet-rank} that the residual changes $T_{12ij}''(0)$ by $-D_{12}^0D_{ij}^0 h_{ij}^{\rm res}/2$.
The quartet contrasts agree by hypothesis, and neither vacuum chord factor vanishes, so $h_{ij}^{\rm res}=0$.
There are $(N-2)(N-3)/2$ independent conditions, one for each remaining pair, in agreement with the quartet rank in Sec.~\ref{sec:quartet-rank}.
For $N=4$, the last condition fixes $h_{34}$, proving that every target vector in the five-dimensional kernel is realized. \\

\noindent
To realize the endpoint values and derivatives as a single smooth generator, choose smooth cutoffs $\psi_i$ with disjoint supports that equal one near $x_i$ for $1\le i\le N$.
In each support, use the local angular coordinate $x-x_i$ and extend the resulting product by zero away from that support.
Then $v(x)=\sum_{i=1}^{N}\psi_i(x)[v_i-q_i(x-x_i)]$ is periodic and satisfies $v(x_i)=v_i$ and $v'(x_i)=-q_i$.
Because $f_\lambda'=1+\lambda^2v'/2$, the explicit bound $\lambda^2\|v'\|_\infty<2$ makes $f_\lambda$ an increasing circle reparametrization.
The reparametrization starts at second order and preserves the original scalar tangent and its quadratic global energy, as in Sec.~\ref{sec:energy}.

\subsection{Positive Scalar Background for Exact Finite-Amplitude Data}
\label{app:detailF1}

\noindent
To construct two geometries with equal sampled entropies at nonzero amplitude, first build a background that can accommodate a small local change of either sign in its density.
Fix the finite list of proper intervals and the target vacuum polygon $P_0$ of positive area.
Choose a nonnegative smooth compactly supported function $b$ that is positive on an open disk inside $P_0$.
Its dimensionless squared profile $e=b^2$ gives a physical energy coefficient $e/L^3$ that stays positive on that disk.
We use $C_c^\infty(U)$ for smooth functions of compact support in an open set $U$. \\

\noindent
Our initial data prescribe the exact scalar field, its derivative along the future unit normal $n$, and the extrinsic curvature as
\begin{equation}
\phi_\lambda=0,\qquad
n^\mu\partial_\mu\phi_\lambda=\lambda L^{-3/2}\sqrt{2e},
\qquad K_{ab}=0\quad(a,b=1,2),
\end{equation}
where the normal derivative means $\sum_{\mu=0}^2 n^\mu\partial_\mu$.
The spatial gradient and the potential energy vanish at this instant.
The physical energy density is therefore exactly $\rho_\lambda=\lambda^2\rho_2$ with $\rho_2=e/L^3$, while the momentum density vanishes.
The momentum constraint is satisfied, and the combined symmetry $t\mapsto-t$, $\phi\mapsto-\phi$ preserves the gravitational and scalar initial data.
Consequently, the metric is reflection symmetric on any development with symmetry-preserving boundary conditions and unique evolution. \\

\noindent
We solve the remaining Hamiltonian constraint in the conformal form $h_\lambda=\exp(2\omega)h_0$, where $h_0$ is the hyperbolic metric of radius $L$ and $\omega\to0$ at spatial infinity.
With $\Delta_{h_0}=\operatorname{div}_{h_0}\nabla_{h_0}$, the transformed Gaussian curvature is $\exp(-2\omega)(-L^{-2}-\Delta_{h_0}\omega)$.
The Hamiltonian constraint identifies this curvature with $-L^{-2}+8\pi G_N\lambda^2\rho_2$ and gives the exact elliptic equation
\begin{equation}
-\Delta_{h_0}\omega+L^{-2}(e^{2\omega}-1)
=8\pi G_N\lambda^2\rho_2 e^{2\omega}.
\label{eq:app_no_go_exact}
\end{equation}
The linearized operator is $\mathcal A_{\mathrm{ell}}=-\Delta_{h_0}+2/L^2$.
On continuous functions vanishing at infinity, the inverse of this shifted Laplacian has the heat-kernel representation
\[
\mathcal A_{\mathrm{ell}}^{-1}g=\int_0^\infty \exp(-2t/L^2)\exp(t\Delta_{h_0})g\,dt.
\]
Here, $\exp(t\Delta_{h_0})g$ is the profile obtained by diffusing $g$ for time $t\geq0$, that is, the solution of $\partial_t u=\Delta_{h_0}u$ with $u(0)=g$.
Acting with $\mathcal A_{\mathrm{ell}}$ on the integral produces the integrand $-\partial_t[\exp(-2t/L^2)\exp(t\Delta_{h_0})g]$.
The diffusion preserves decay at spatial infinity, and the maximum principle bounds the magnitude of the diffused profile by $\|g\|_\infty$ at every time.
The exponential factor therefore makes the term at $t=\infty$ vanish, leaving $g$ at $t=0$.
Integrating the same bound in time gives $\|\mathcal A_{\mathrm{ell}}^{-1}g\|_\infty\leq(L^2/2)\|g\|_\infty$.
Eq.~\eqref{eq:app_no_go_exact} is equivalent to
\begin{equation}
\omega=\mathcal A_{\mathrm{ell}}^{-1}\!\left[
8\pi G_N\lambda^2\rho_2e^{2\omega}
-L^{-2}(e^{2\omega}-1-2\omega)\right].
\end{equation}
We work on the closed ball $\|\omega\|_\infty\leq C\lambda^2$ of continuous functions vanishing at infinity, with $C>4\pi G_NL^2\|\rho_2\|_\infty$.
On this ball, the source term is $O(\lambda^2)$ and the nonlinear remainder $\exp(2\omega)-1-2\omega$ is $O(\lambda^4)$.
The inverse-operator bound and the choice of $C$ therefore keep the image inside the ball for sufficiently small $|\lambda|$.
Differentiation with respect to $\omega$ brings down either a factor of $\lambda^2$ from the source or a factor of $\exp(2\omega)-1=O(\lambda^2)$ from the remainder.
The map thus reduces the sup-norm distance between two iterates by a factor $O(\lambda^2)<1$, giving a unique solution in the ball.
Elliptic regularity gives spatial smoothness, and invertibility of the linearized operator gives smooth dependence on $\lambda^2$ by the implicit function theorem.
In particular, $\omega=\lambda^2\omega_2+O(\lambda^4)$, where
\begin{equation}
\left(-\Delta_{h_0}+\frac2{L^2}\right)\omega_2=8\pi G_N\rho_2 .
\label{eq:app_no_go_constraint}
\end{equation}
The Green function gives the metric response to a unit point source normalized using the physical area measure $dA_0$.
Rotational symmetry about its source $y'$ makes the response at $y$ a function of their dimensionless separation $\eta=d_{h_0}(y,y')/L$ alone.
Set $u=\cosh\eta$.
Away from $y'$, the radial Laplacian is $L^{-2}(\partial_\eta^2+\coth\eta\,\partial_\eta)$, so the homogeneous equation for a radial function $F(u)$ reduces to
\[
(u^2-1)F''(u)+2uF'(u)-2F(u)=0,\qquad u>1 .
\]
One solution is $F(u)=u$.
For a second solution, substitute $F(u)=uH(u)$ into the homogeneous equation.
The terms proportional to $H$ cancel because $u$ is already a solution, and the remaining equation is $\partial_u[u^2(u^2-1)H'(u)]=0$.
Taking the decaying solution proportional to $u\int_u^\infty [t^2(t^2-1)]^{-1}dt$ and splitting the integrand as $(t^2-1)^{-1}-t^{-2}$ gives
\[
G_2(\eta)=\frac{1}{2\pi}\left[\frac{u}{2}\ln\frac{u+1}{u-1}-1\right],
\qquad u=\cosh\eta .
\]
To fix the point-source normalization, take a small circle around the source.
As $\eta\downarrow0$, $G_2=-(2\pi)^{-1}\ln\eta+O(1)$ and $-2\pi\sinh\eta\,\partial_\eta G_2\to1$.
The latter expression is the outward flux for $-\Delta_{h_0}$ through a circle of physical radius $L\eta$. \\

\noindent
The integral representation makes $G_2$ positive for every $\eta>0$.
As $u\to\infty$, expansion of the logarithm gives $G_2=(6\pi u^2)^{-1}+O(u^{-4})$.
Thus, for the compact scalar source used here,
\[
\omega_2(y)=8\pi G_N\int_{\mathbb H^2}
G_2\!\left(d_{h_0}(y,y')/L\right)\rho_2(y')\,dA_0(y').
\]
For a nonzero nonnegative source, the integral is positive even outside the matter.
At a distance $L\eta_0$ from a fixed origin, it decays as $O(\exp(-2\eta_0))$ as $\eta_0\to+\infty$ at fixed $L$.
Thus, placing extra positive matter in an unsampled disk generally changes the metric along sampled chords as well.
In Appendix~\ref{app:finiteexact}, we instead prescribe a metric change supported inside the disk and solve for the accompanying redistribution of matter.
The nonlinear background satisfies a related bound.
Nonnegativity of $\rho_2$ and Eq.~\eqref{eq:app_no_go_exact} exclude a negative minimum of $\omega$.
Outside the source, $\mathcal A_{\mathrm{ell}}\omega=-L^{-2}(\exp(2\omega)-1-2\omega)\leq0$.
If $\eta$ denotes radial geodesic distance divided by $L$, then $\mathcal A_{\mathrm{ell}}(\cosh^{-2}\eta)=6L^{-2}\cosh^{-4}\eta$.
Comparison on the exterior with a sufficiently large multiple of $\cosh^{-2}\eta$ proves $\omega=O(\exp(-2\eta))$, the normalizable metric falloff.
The baseline thus has zero first metric and boundary-stress variations.
The EWP interpretation requires an additional state-preparation and evolution hypothesis.
We assume coherent states realize both sets of initial data, and their regular Einstein--scalar developments exist on a common finite comparison slab with fixed boundary sources.
Uniform continuity at the vacuum gives the field bound of Appendix~\ref{app:potentialbound} for sufficiently small $|\lambda|$.
On the regular fixed RT phase, Appendix~\ref{app:globalmaximum} then identifies each polygon area with its EWP volume.

\subsection{Compact Metric Deformation and Sampled Entropies}
\label{app:finiteexact}

\noindent
Start from the background solution $h_{b,\lambda}=\exp(2\omega_b(\lambda))h_0$ of Eq.~\eqref{eq:app_no_go_exact}, whose scalar density is $\rho_{b,\lambda}=\lambda^2b^2/L^3$.
The task is to preserve the minimum lengths for the chosen endpoints while increasing the area inside the target polygon.
A finite union of curves has empty interior here because the listed curves are smooth geodesics.
We can therefore place an open disk $U$ with closure in $\{b>0\}\cap\operatorname{int}P_0$ away from every queried vacuum geodesic and every polygon edge.
For a queried union of intervals, include every component in its chosen minimizing pairing.
Choose nonnegative $g\in C_c^\infty(U)$ positive somewhere, so its support is a positive distance from all listed curves.
On a fixed smooth RT phase, the curves depend continuously on amplitude and endpoints.
For all sufficiently small amplitudes and throughout a neighborhood of each queried endpoint pair, the curves still miss the support and $U$ remains inside the polygon after shrinking it if necessary. \\

\noindent
Set $h_{g,\lambda}=\exp(2t)h_{b,\lambda}$ with $t=\varepsilon\lambda^2 g$ and $\varepsilon>0$.
At points where $g>0$ and $\lambda\ne0$, the metric stretches each infinitesimal length by $\exp(t)>1$ and each infinitesimal area by its square.
Outside the support of $g$, both factors equal one.
For $\Delta_b=\operatorname{div}_{h_{b,\lambda}}\nabla_{h_{b,\lambda}}$, the transformation of Gaussian curvature and the Hamiltonian constraint give the new exact density
\begin{equation}
8\pi G_N\rho_{g,\lambda}
=e^{-2t}\left[8\pi G_N\rho_{b,\lambda}
-\Delta_b t+L^{-2}(e^{2t}-1)\right].
\label{eq:nonlinearhiddenmatter}
\end{equation}
A local increase in the conformal factor can lower the density through the curvature term $-\Delta_b t$.
Placing the change where the baseline scalar already carries positive energy leaves room for this decrease.
Write $b_{\min}=\min_{\operatorname{supp}g}b>0$ and $C_g=\sup_{\operatorname{supp}g}[\Delta_{h_0}g]_+$ to bound it uniformly.
In two spatial dimensions, $\Delta_b g=\exp(-2\omega_b)\Delta_{h_0}g$, and the baseline solution satisfies $\omega_b\geq0$.
Thus, $\Delta_b g\leq C_g$ throughout the support.
Because $t=\varepsilon\lambda^2g\geq0$, the term $L^{-2}(\exp(2t)-1)$ is nonnegative, and Eq.~\eqref{eq:nonlinearhiddenmatter} yields
\[
\frac{\rho_{g,\lambda}}{\lambda^2}
\geq \exp(-2t)
\left(\frac{b_{\min}^2}{L^3}
-\frac{\varepsilon C_g}{8\pi G_N}\right)
\qquad (\lambda\ne0,\ x\in\operatorname{supp}g).
\]
The single choice $\varepsilon C_g<8\pi G_N b_{\min}^2/L^3$ keeps the rescaled density strictly positive there for all sufficiently small $|\lambda|$.
Outside $\operatorname{supp}g$, the two ratios agree exactly.
The quotient at $\lambda=0$ is its smooth extension in $\lambda^2$ obtained from Eq.~\eqref{eq:nonlinearhiddenmatter}.
Consequently, $\phi_{g,\lambda}=0$ and $n^\mu\partial_\mu\phi_{g,\lambda}=\lambda\sqrt{2\rho_{g,\lambda}/\lambda^2}$ are smooth real initial data.
The square root is smooth on a neighborhood where the ratio is strictly positive and equals $\sqrt2 b/L^{3/2}$ outside the support.
The momentum constraint still holds with $K_{ab}=0$, and the combined time and scalar sign reversal makes the metric reflection symmetric.
The same state-preparation and regular-development assumptions as above identify these classical initial data with states and their polygon areas with EWP volumes.
For classical scaling, one can hold $8\pi G_N b^2/L$ fixed as $c$ increases. \\

\noindent
Fix two queried endpoints and impose the same boundary cutoff in the two metrics.
Let $\gamma_b$ be a baseline minimizing curve between these regulated endpoints in the prescribed homology class, and let $\ell_b$ and $\ell_g$ denote lengths in the two metrics.
For any admissible competitor $\Gamma$, the line elements obey $ds_g=\exp(t)\,ds_b$, while $t=0$ along $\gamma_b$.
Subtracting the common length of $\gamma_b$ separates the two contributions to the excess length of any competitor, giving
\begin{equation}
\ell_g(\Gamma)-\ell_g(\gamma_b)
=\bigl[\ell_b(\Gamma)-\ell_b(\gamma_b)\bigr]
+\int_\Gamma(e^t-1)\,ds_b .
\label{eq:compact-bump-length-gap}
\end{equation}
The first term is nonnegative because $\gamma_b$ minimized the baseline length.
The second is nonnegative because $t\geq0$, and it vanishes on the original minimizer.
At $\lambda\ne0$, a competitor ties only if it was already a minimizer and spends zero arclength where $g>0$.
In particular, a unique minimizer stays unique.
For a union, summing Eq.~\eqref{eq:compact-bump-length-gap} over the components compares all candidate collections, including alternative endpoint pairings.
The chosen pairing still attains the minimum because every component misses the deformation.
This avoidance persists for sufficiently distant cutoffs, and the boundary collars give identical endpoint counterterms.
Dividing the common renormalized minimum by $4G_N$ thus gives identical classical RT entropies throughout the chosen endpoint neighborhoods.
The same length comparison protects the polygon edges, which enclose the same coordinate region $P_{b,\lambda}$ in both geometries.
Denote its reflection areas by $A_g(\lambda)$ and $A_b(\lambda)$ and, under the comparison hypotheses, its maximal volumes by $V_g(\lambda)$ and $V_b(\lambda)$.
The difference comes entirely from the area element inside $U$, where each of the two independent tangent lengths gains a factor $\exp(t)$.
Integrating the resulting factor $\exp(2t)$ over the common region gives
\begin{equation}
A_g(\lambda)-A_b(\lambda)
=\int_U(e^{2\varepsilon\lambda^2 g}-1)\,dA_{b,\lambda}>0
\qquad (0<|\lambda|\text{ sufficiently small}),
\label{eq:nonlinearhiddenvolume}
\end{equation}
where $dA_{b,\lambda}$ is the baseline area element.
Uniformly on the support, $\exp(2\varepsilon\lambda^2g)-1=2\varepsilon\lambda^2g+O(\lambda^4)$ and $dA_{b,\lambda}=dA_0+O(\lambda^2)$.
The first term in the area difference is therefore $2\varepsilon\lambda^2\int_Ug\,dA_0$, giving
\begin{equation*}
A_g''(0)-A_b''(0)=4\varepsilon\int_U g\,dA_0>0 .
\end{equation*}
Under the comparison hypotheses, the reflection areas equal the maximal volumes at these amplitudes.
Any estimator $\widehat A_2$ assigns one number to the common data.
The triangle inequality puts at least one of its two errors at or above half the separation of the true Hessians, namely $2\varepsilon\int_U g\,dA_0$.
Choosing the midpoint attains this error for both states.
At a fixed nonzero amplitude, the same argument gives half the area difference in Eq.~\eqref{eq:nonlinearhiddenvolume}, with a value that depends on the query list and profiles.
Here both states have nonnegative density, so the construction distinguishes the magnitudes of their area Hessians.
The opposite-sign examples in Sec.~\ref{sec:energy} use signed densities and match the six entropies through quadratic order. \\

\noindent
The same ambiguity can be read off from the spatial weights at quadratic order, without comparing finite-amplitude lengths.
In a disk $U$ with closure inside $P_0$ and disjoint from all queried geodesics, every signed lapse $w_{ij}$ has a fixed sign.
If the disk lies on the positive side of a queried geodesic, the interval weight $[w_{ij}]_+$ equals the smooth lapse $w_{ij}$ there.
As a linear coordinate on the unit vacuum hyperboloid, this lapse satisfies $\bar\Delta w_{ij}=2w_{ij}$, where $\bar\Delta$ is its spatial Laplacian.
On the other side of the geodesic, $[w_{ij}]_+$ vanishes identically, so $(2-\bar\Delta)[w_{ij}]_+=0$ on the disk in either case.
Consequently, every sampled $K_Q$, a linear combination of those positive parts, obeys the same homogeneous equation on $U$. \\

\noindent
For the conformal perturbation $t=\varepsilon\lambda^2g$ above, expansion of Eq.~\eqref{eq:nonlinearhiddenmatter} gives the dimensionless quadratic-density change
\[
\delta e=e_g-e_b=\frac{\varepsilon L}{8\pi G_N}(2-\bar\Delta)g .
\]
A measurement integrates the local density change against its own weight $W$.
For any $W$ smooth on $U$, integration by parts moves the Laplacian onto $W$ without a boundary term because $g$ is compactly supported there, giving
\[
\int_{\mathbb H^2}W\,\delta e\,d\mu
=\frac{\varepsilon L}{8\pi G_N}\int_U g(2-\bar\Delta)W\,d\mu .
\]
Putting $W=[w_{ij}]_+$ gives zero for every sampled wedge moment, and the same result holds for every linear combination of those weights.
For $W=\mathbf 1_{P_0}$, the weight equals one throughout $U$, giving $(2-\bar\Delta)W=2$ there.
The target energy therefore changes by $\int_{P_0}\delta e\,d\mu=\varepsilon L(4\pi G_N)^{-1}\int_Ug\,d\mu>0$.
With $c=3L/(2G_N)$ and $dA_0=L^2d\mu$, multiplication by $24\pi/c$ gives $\delta A''/L^2=4\varepsilon\int_U g\,d\mu$, as follows independently from Eq.~\eqref{eq:nonlinearhiddenvolume}.
The global vacuum time-translation lapse $\cosh\eta$ also satisfies $(2-\bar\Delta)\cosh\eta=0$, so $\int_{\mathbb H^2}\cosh\eta\,\delta e\,d\mu=0$.
The lapse $\cosh\eta$ is positive everywhere, so this cancellation requires the nonzero difference $\delta e$ to take both signs.
Both states can still have nonnegative density everywhere, since the sign here indicates which state has more energy at a given point.
This cancellation explains how the deformation changes the unweighted polygon energy while preserving the energy measured at the boundary. \\

\noindent
The intrinsic area change can also be evaluated at fixed $\lambda$.
The area elements satisfy $dA_{g,\lambda}=\exp(2t)dA_{b,\lambda}$.
The difference of the two curvature densities in Eq.~\eqref{eq:nonlinearhiddenmatter} then gives
\begin{equation}
8\pi G_N\bigl(\rho_{g,\lambda}dA_{g,\lambda}
-\rho_{b,\lambda}dA_{b,\lambda}\bigr)
=\left[-\Delta_b t+L^{-2}(e^{2t}-1)\right]dA_{b,\lambda}.
\label{eq:local-curvature-density-difference}
\end{equation}
For unit weight on the common polygon, the integral of $\Delta_b t$ vanishes by compact support, and the remaining term gives $(A_g-A_b)/L^2$ in Eq.~\eqref{eq:GBexact}.
For the global vacuum lapse $N=\cosh\eta$, the cancellation of the quadratic energy moment has a different finite-amplitude consequence.
Let $\mathcal I_N(\lambda)=\int_{\mathbb H^2} N(\rho_{g,\lambda}\,dA_{g,\lambda}-\rho_{b,\lambda}\,dA_{b,\lambda})$ denote the difference of the two matter integrals with this fixed weight.
Integration by parts transfers $\Delta_b$ from $t$ to $N$ without a boundary term, and $\Delta_b N=e^{-2\omega_b}\Delta_{h_0}N=2L^{-2}e^{-2\omega_b}N$.
Consequently, Eq.~\eqref{eq:local-curvature-density-difference} gives
\begin{equation}
8\pi G_N L^2\mathcal I_N(\lambda)
=\int_U N\left[(e^{2t}-1-2t)+2t(1-e^{-2\omega_b})\right]dA_{b,\lambda} .
\label{eq:compact-bump-vacuum-lapse}
\end{equation}
Both terms in brackets are nonnegative for $t,\omega_b\geq0$, and the first is strictly positive wherever $t>0$.
Since $t$ and $\omega_b$ are $O(\lambda^2)$ on $U$, the first term starts at order $t^2$ and the second at order $t\omega_b$, both $O(\lambda^4)$.
Thus, Eq.~\eqref{eq:compact-bump-vacuum-lapse} gives zero difference in the quadratic matter charges of Sec.~\ref{sec:energy} and a positive difference in the vacuum-lapse matter integrals at nonzero amplitude.
Beyond quadratic order, the exact conserved charge also depends on the geometry's response.
It is fixed by the boundary stress and is equal for these two states because the geometries agree in a collar of the boundary on the measurement slice.
The area difference obtained by integrating Eq.~\eqref{eq:local-curvature-density-difference} with unit weight concerns the reflection slice of each constraint solution.
The corresponding boundary interval entropies and maximal polygon volumes require the state-realization and comparison assumptions stated above. \\

\noindent
The matter and metric Cauchy data, as well as the boundary sources, coincide in a boundary collar.
Local uniqueness gives the same near-boundary evolution around time zero, whose expansion fixes the renormalized boundary stress in the chosen coordinates \cite{deHaro:2000vlm}.
For a queried single interval, integrating this common stress against the known vacuum modular weight gives the same modular expectation.
Together with the common ordinary entropy, this yields the same leading relative entropy to vacuum throughout the endpoint neighborhood.
For a queried union, the length comparison establishes equality of its classical RT entropy at nonzero amplitude. \\

\noindent
A finite protocol can choose its next interval from its previous readings.
For a deterministic choice rule, record its queries, answers, and stopping decision on the baseline at a fixed small nonzero amplitude, including time-zero stress if used.
Choose $U$ inside $\{b>0\}\cap\operatorname{int}P_{b,\lambda}$ away from the recorded geodesics and polygon edges, then choose $\varepsilon$ small enough for that disk.
The first reading agrees in the two geometries, so the rule selects the same second query in each.
Iteration gives the same sequence of readings and the same stopping decision.
For queries of state derivatives at the vacuum, use the corresponding vacuum chords to choose the disk.
The positive clearance gives agreement throughout an open neighborhood of each endpoint pair and hence agreement of all endpoint derivatives there.
Such local information is called a germ.
A real-analytic function on a connected branch is fixed by its germ. In contrast, a smooth function can agree on an open set and differ elsewhere. \\

\noindent
A fixed list of intervals admits an indistinguishable pair of states, but the pair depends on the list.
For any such pair, the nonzero compact density difference has a chord with $\XX\delta e\ne0$ by injectivity of the line transform in Appendix~\ref{app:tomography}.
Continuity gives an open set of distinguishing chords, so a dense set of exact chord measurements separates this pair.
A different finite list can leave a different disk unmeasured.
For noisy samples, converting their spacing into an error bound also requires regularity assumptions on the profiles (Appendix~\ref{app:tomography}).

\subsection{Concentration and Measurement Errors for Quartet Readings}
\label{app:finite_measurement_errors}

\noindent
For the normalized quartet weight $k=W/M$, the readings $u$ and $T_Q$ measure the weighted and total energy in $Q$, respectively.
Eq.~\eqref{eq:joint} gives the sharp lower bound $N_*$ on the negative part $E_Q^-$, and its derivation expresses the gap $E_Q^--N_*$ as the sum of the cancellations in the weights $k$ and $1-k$. \\

\noindent
When $u(T_Q-u)<0$, the amount by which the bound is missed restricts where the two signs of the energy can lie.
Choose $w=k$ if $u>0$ and $w=1-k$ if $T_Q-u>0$, so that $P_w-N_w>0$ and $P_{1-w}-N_{1-w}<0$.
The signs give $\min(P_w,N_w)=N_w$ and $\min(P_{1-w},N_{1-w})=P_{1-w}$, so the exact difference reduces to
\[
E_Q^- - N_* =
\int_Q \left(w[-e]_+ +(1-w)[e]_+\right)d\mu .
\]
For any $0<\xi<1$, one has $w\geq\xi$ on the first region below and $1-w\geq\xi$ on the second, so
\[
\int_{Q\cap\{w\geq\xi\}}[-e]_+\,d\mu
+\int_{Q\cap\{w\leq 1-\xi\}}[e]_+\,d\mu
\leq \frac{E_Q^- - N_*}{\xi}.
\]
Thus, along any sequence with the two moments fixed and $E_Q^-\to N_*$, the negative energy in $w\geq\xi$ and the positive energy in $w\leq1-\xi$ both tend to zero for every fixed $\xi$.
For a positive quartet reading, the positive energy must therefore concentrate where $k$ approaches one and the negative energy where $k$ approaches zero.
A negative quartet reading interchanges these locations. \\

\noindent
With vanishing first metric variation and stress variation, $R''=C_{\eta_Q} S''$, where $C_{\eta_Q}$ acts on the six endpoint pairs $1\le i<j\le4$.
Two diagonal coefficients are one, two side coefficients are $-\eta_Q$, and two are $-(1-\eta_Q)$, so their absolute values sum to four.
If each measured $S_{ij}''$ differs from its true value by at most $\sigma$, then
\begin{equation}
\widetilde u=\frac{C_{\eta_Q}\widetilde S''}{4\pi M},\qquad
|\widetilde u-u|\le d_u=\frac{\sigma}{\pi M}.
\end{equation}
Suppose the other information available is an independent upper bound $T_Q\le T_+$, and put $u_-=\widetilde u-d_u$ and $u_+=\widetilde u+d_u$.
The right-hand side of Eq.~\eqref{eq:joint} is nonincreasing in $T_Q$, so its smallest value compatible with this upper bound uses $T_Q=T_+$.
For any real $T$, the expression $[-u]_++[u-T]_+$ equals $[-T]_+$ throughout the interval between $0$ and $T$ and increases by the distance to that interval outside it.
Minimizing it over $u_-\leq u\leq u_+$ therefore gives
\[
E_Q^-\ \geq\ \max\{0,-T_+,-u_+,u_--T_+\}.
\]
For example, if the smallest possible weighted reading exceeds the largest possible total, their difference forces negative energy into the complementary weight $1-k$.
This is the strongest lower bound from these two interval constraints alone, since the fixed-moment infimum in Eq.~\eqref{eq:joint} can be approached for the minimizing pair.
To connect these error bounds to measurements of the entropy itself, suppose that $y_j$ is the measured entropy at $\lambda=jh$, $j=-1,0,1$, with absolute error at most $s_E$.
At step size $h>0$, the centered differences $(y_1-y_{-1})/(2h)$ and $(y_1-2y_0+y_{-1})/h^2$ estimate the slope and the Hessian, respectively.
Their respective absolute errors as estimates of $S'(0)$ and $S''(0)$ are bounded by
\[
\frac{s_E}{h}+\frac{K_3h^2}{6},\qquad
\frac{4s_E}{h^2}+\frac{K_4h^2}{12},
\]
when $|\partial_\lambda^m S|\le K_m$ on $[-h,h]$ for $m=3,4$.
The error in the second derivative can grow as the sample spacing shrinks because the measurement errors at the three sample points enter with absolute coefficients $1,2,1$ before division by $h^2$.
The $K_3$ and $K_4$ terms bound the truncation errors and grow with $h$.
For example, if $s_E,K_4>0$ and their bounds hold over the needed interval, the quoted Hessian error bound is smallest at $h^4=48s_E/K_4$, where the two error terms are equal.
For the truncation errors, the first centered difference of the exact entropy averages $S'(u)$ over $-h\leq u\leq h$ with density $1/(2h)$.
Integrating the fundamental theorem of calculus twice shows that the second difference averages $S''(u)$ over the same interval with the triangular density $(h-|u|)/h^2$.
Both densities are even, so the terms linear in $u$ in either Taylor expansion integrate to zero.
Their respective second moments are $h^2/3$ and $h^2/6$.
Multiplying these moments by $K_3/2$ and $K_4/2$ gives the two stated truncation bounds. \\

\noindent
If a matched first-order boundary graviton is present, the measured slopes supply the additional square in Eq.~\eqref{eq:ptolemy_hessian}.
Write $d=I_1'(0)-I_2'(0)=S'_{12}+S'_{34}-S'_{14}-S'_{23}$, where all four entropy slopes are evaluated at the vacuum.
Suppose every entropy slope and every entropy Hessian have absolute errors bounded by $\sigma_1$ and $\sigma_2$, respectively, and the endpoints are known.
The corresponding measured slope contrast obeys $|\widetilde d-d|\leq\omega$ with $\omega=4\sigma_1$.
Using $M^2=\eta_Q(1-\eta_Q)$ and $t_c=3/c$, form the corrected estimator
\[
\widehat u=\frac{C_{\eta_Q}\widetilde S''-t_c M^2\widetilde d^{\,2}}{4\pi M},
\qquad
|\widehat u-u|\leq
\frac{4\sigma_2+t_c M^2\omega(2|\widetilde d|+\omega)}{4\pi M}.
\]
The second term follows from $|\widetilde d^2-d^2|\leq|\widetilde d-d|(2|\widetilde d|+|\widetilde d-d|)$.
Consequently, the same negative-energy certificate applies with $\widehat u$ and this error bound, including when the linear response differs between the four sides.

\subsection{Bounds from Conserved Scalar Energy}
\label{app:finite_charge_estimates}

\noindent
For a scalar with $\mu_*^2=-M_\phi^2 L^2\in(0,1)$, Sec.~\ref{sec:energy} identifies the measured quadratic charge with $\mathcal E=\int_{\mathbb H^2}Ne\,d\mu$ when the boundary sources are fixed, the first metric variation vanishes, and the field has standard source-free falloff.
Here $N=\cosh\chi$ is the global lapse, $D$ is the covariant derivative on the unit hyperbolic disk, $\phi_1$ is the field tangent, and $\pi_\phi$ is its dimensionless normal momentum.
To find the measured charge bounds, separate the squared field from its gradient and momentum contributions by defining
\begin{equation}
P_\phi=\int_{\mathbb H^2}N\phi_1^2d\mu,\qquad
A_\phi=\int_{\mathbb H^2}N(|D\phi_1|^2+\pi_\phi^2)d\mu
\end{equation}
so that $2\mathcal E/L=A_\phi-\mu_*^2P_\phi$.
The negative mass term is controlled if the gradient and momentum pay at least the cost $P_\phi$.
To prove that inequality, introduce the radial vector field $Y=\tanh\chi\,\partial_\chi$.
The radial divergence is $(\sinh\chi)^{-1}\partial_\chi(\sinh\chi\,N\tanh\chi)=2N$.
Integrating $\operatorname{div}(NY)\phi_1^2$ by parts gives $P_\phi=-\int_{\mathbb H^2}N\phi_1Y\cdot D\phi_1\,d\mu$.
Let $Y^\flat$ denote the covector obtained from $Y$ with the unit hyperbolic metric.
Expanding the square below, using this integration by parts for its cross term and $1-|Y|^2=1/\cosh^2\chi$, gives
\[
A_\phi-P_\phi=\int_{\mathbb H^2}N
\left(\left|D\phi_1+\phi_1Y^\flat\right|^2+\pi_\phi^2
+\frac{\phi_1^2}{\cosh^2\chi}\right)d\mu .
\]
If either smooth initial datum $\phi_1$ or $\pi_\phi$ is nonzero, one of the last two terms is positive on an open set, so $A_\phi>P_\phi$.
Writing $2\mathcal E/L=(1-\mu_*^2)A_\phi+\mu_*^2(A_\phi-P_\phi)$ therefore gives $(1-\mu_*^2)A_\phi\leq2\mathcal E/L$ and proves strict positivity of the charge for every nonzero tangent in this scalar sector.
Pointwise, the negative part of $e$ is bounded by $L\mu_*^2\phi_1^2/2$, and its positive part by $L(|D\phi_1|^2+\pi_\phi^2)/2$.
Since $N\ge1$, extending either integral from $Q$ to the entire disk gives
\[
E_Q^-\le\frac{L\mu_*^2P_\phi}{2}
\le\frac{\mu_*^2\mathcal E}{1-\mu_*^2},\qquad
\int_Q[e]_+\,d\mu\le\frac{LA_\phi}{2}
\le\frac{\mathcal E}{1-\mu_*^2}.
\]
A weight between zero and one can collect at most the full positive part of the energy or the full magnitude of its negative part.
The bounds just obtained therefore apply to both the total $T_Q=\int_Qe\,d\mu$ and the complementary reading $T_Q-u=\int_Q(1-k)e\,d\mu$.
Writing $E_{\rm lo}=-\mu_*^2\mathcal E/(1-\mu_*^2)$ and $E_{\rm hi}=\mathcal E/(1-\mu_*^2)$, one has $T_Q\in[E_{\rm lo},E_{\rm hi}]$ and $T_Q\in[u+E_{\rm lo},u+E_{\rm hi}]$.
Their intersection gives Eq.~\eqref{eq:cappedvolume}. \\

\noindent
The preceding lower bound used $N\geq1$ in estimating the unweighted negative energy.
We can keep more information about the radius by retaining the $\phi_1^2/\cosh^2\chi$ term in the positive-square identity.
Set $b_*=1-\mu_*^2>0$ and write $2\mathcal E/L=(A_\phi-P_\phi)+b_*P_\phi$.
The two terms proportional to $\phi_1^2$ then combine to give
\[
\begin{aligned}
\frac{2\mathcal E}{L}
={}&\int_{\mathbb H^2} N\bigl( |D\phi_1+\phi_1Y^\flat|^2
+\pi_\phi^2\bigr)\,d\mu\\
&+\int_{\mathbb H^2}\bigl(b_* N+N^{-1}\bigr)\phi_1^2\,d\mu .
\end{aligned}
\]
Since $b_*N+N^{-1}\geq2\sqrt{b_*}$, the unweighted integral of $\phi_1^2$ is at most $\mathcal E/(L\sqrt{b_*})$.
Combining this estimate with $[-e]_+\leq L\mu_*^2\phi_1^2/2$ bounds $E_Q^-$ by $\mu_*^2\mathcal E/(2\sqrt{b_*})$.
Both $T_Q$ and $T_Q-u$ are at least $-E_Q^-$ because their weights lie between zero and one.
Taking the larger of the two lower endpoints yields Eq.~\eqref{eq:global-negative-improved}.

\section{Finite-Central-Charge Pairings and Limits of Reconstruction}
\label{app:vacuum-support}

\noindent
The vacuum measurement in Sec.~\ref{sec:vacuum-pairing} identifies an endpoint pairing by shifting the interval ends.
We first bound the error in each endpoint test and determine how far the endpoints can move while a common pairing still minimizes.
We propagate the resulting uncertainties to the full pairing matrix, illustrate the area count for three intervals, and follow the change of pairing through a two-interval transition.
A free Dirac sector supplies an example where the convergence rate affects a smoothed estimate of area.
The final parts examine the additional inputs needed to infer a matter-induced area change or a quantum recovery fidelity.

\subsection{Conditions for Identifying Pairings}
\label{app:detailG2}

\noindent
At finite central charge, even exact entropy measurements need not give the zeros and ones predicted by the geodesic formula.
The largest displacement of an exact ratio from its ideal pairing entry is $r_c=\max_{1\leq i,j\leq m}|\Delta_{ij}e_c|/D_{ij}^{(c)}$.
For the total area, the errors in several entries can act together, so we also use $\delta_c=\|R^{(c)}-P\|_2$, where the operator norm $\|M\|_2$ is the greatest length $\|Mx\|_2$ attained for a Euclidean unit vector $x$.
Applying the matrix to the unit vector along column $j$ bounds each entry in that column by $\delta_c$.
Conversely, the squared operator norm is at most the sum of all $m^2$ squared entries of the error matrix, each at most $r_c^2$.
Consequently,
\begin{equation}
r_c\leq\delta_c\leq m r_c .
\end{equation}
At fixed $m$ and fixed widths, $D_{ij}^{(c)}/c\to d_{ij}>0$ as $c\to+\infty$.
Eq.~\eqref{eq:top_exact} then shows that the ratios converge to $P$ precisely when $\Delta_{ij}e_c=o(c)$ for every pair.
A correction to an individual entropy can be of order $c$ and still cancel if it depends on just one of the two shifts. \\

\noindent
One sufficient input is the leading RT expansion on this finite set of configurations.
With all three limits taken as $c\to+\infty$, it reads
\begin{equation}
\frac{S_c(W)}c\longrightarrow\sum_{k=1}^{m}s_0(u_k^W,v_{\pi(k)}^W),
\qquad
\frac{s_c(u,v)}c\longrightarrow s_0(u,v),
\qquad
\frac{D_{ij}^{(c)}}c\longrightarrow d_{ij}>0 .
\label{eq:top_leadingRT}
\end{equation}
Here, $s_0$ is the limiting single-interval entropy per unit central charge.
Subtracting the sum of these single-interval limits from the union limit gives $e_c(W)/c\to0$.
There are finitely many sampled unions at fixed $m$, so even their largest remainder $E_c=\max_W|e_c(W)|$ obeys $E_c/c\to0$.
Each four-entropy difference has error at most $4E_c$, and division by the smallest denominator $D_*^{(c)}=\min_{1\leq i,j\leq m}D_{ij}^{(c)}$ in Eq.~\eqref{eq:top_exact} gives
\begin{equation}
\delta_c\leq\frac{4m E_c}{D_*^{(c)}}\longrightarrow0 .
\end{equation}
For fixed widths, $E_c=O(1)$ would imply an $O(c^{-1})$ error, but the leading entropy formula alone gives no convergence rate.
With exactly known entropies, $r_c<1/2$ is sufficient for rounding every ratio to its correct pairing entry.
With uncertain entropies, the observer can instead ask whether twice the measured numerator exceeds the measured denominator, avoiding division by a small signal.
Let $\widetilde N_{ij}$ and $\widetilde D_{ij}$ denote the measured differences, and let $\eta_U,\eta_I$ bound the error in each constituent union or single-interval entropy, respectively.
An independent estimate $|\Delta_{ij}e_c|\le b_{ij}$ for each $1\le i,j\le m$ accounts for the difference between the exact entropy and the geodesic prediction, even with perfect measurements.
For $Z_{ij}=2\widetilde N_{ij}-\widetilde D_{ij}$, Eq.~\eqref{eq:top_exact} and the triangle inequality give
\[
\big|Z_{ij}-(2P_{ij}-1)D_{ij}^{(c)}\big|
\leq 2b_{ij}+8\eta_U+4\eta_I .
\]
The factor eight counts four union readings used twice, and the factor four counts four single-interval readings used once.
A shared edge predicts $Z_{ij}=+D_{ij}^{(c)}$, and separate edges predict $Z_{ij}=-D_{ij}^{(c)}$, before either source of error.
If the reference signal $D_{ij}^{(c)}$ is known to exceed the error bound, no allowed error can reverse the sign.
One can also decide directly from the data whenever the interval obtained by adding and subtracting the error bound from the measured $Z_{ij}$ lies wholly above or below zero.
Reducing measurement noise tightens this interval only down to the residual uncertainty from $b_{ij}$.
The large-$c$ formula in Eq.~\eqref{eq:top_leadingRT} requires a common minimizing pairing and, in its derivation by the CFT replica method, assumptions about dominance and continuation \cite{Hartman:2013mia}.
For sequences with increasing $m$, shrinking widths, or a nearby transition, the errors must be bounded uniformly along the sequence. \\

\noindent
The required shift must also preserve the shortest configuration, which can be checked using the known vacuum endpoints.
For each admissible noncrossing pairing $\rho$, let $F_\rho=(1/3)\sum_{k=1}^{m}\ln(|v_{\rho(k)}-u_k|/\epsilon)$ be its predicted entropy divided by $c$, up to a common constant.
If $\pi$ uniquely minimizes this quantity, its advantage over the nearest competitor is $g=\min_{\rho\ne\pi}(F_\rho-F_\pi)>0$, where $\rho$ ranges over admissible pairings.
Here, $g$ is an entropy separation, and the lengths entering the width test are $d_0=\min_{1\le i,j\le m}|v_j-u_i|$, the shortest initial left-to-right distance, and $h=\max_{1\le i\le m}\{a_i,b_i\}$, the largest shift.
Keep the strips disjoint and take $2h<d_0$.
At most two endpoints move, each by at most $h$, so the absolute changes in all boundary separations of a pairing sum to at most $2h$.
Each separation stays at least $d_0-2h$, so the slope of its logarithm is bounded by $1/(d_0-2h)$.
Applying this estimate to the candidate and its competitor gives
\[
|F_\rho(W)-F_\rho|\le\frac{2h}{3(d_0-2h)},\qquad
F_\rho(W)-F_\pi(W)\ge g-\frac{4h}{3(d_0-2h)}
\quad(\rho\ne\pi),
\]
where both length sums can move as the endpoints shift.
Keeping the lower bound on their difference positive gives the sufficient width condition $h<3g d_0/(4+6g)$.
Close to a tie, this condition favors narrower strips, but the signal itself shrinks with both widths.
For fixed nonzero separation $d=v_j-u_i$, expanding Eq.~\eqref{eq:pair-strip-denominator} as $(a_i+b_j)/|d|\to0$ gives
\[
D_{ij}^{(c)}
=\frac{c\,a_i b_j}{3d^2}
\left[1+O\!\left(\frac{a_i+b_j}{|d|}\right)\right].
\]
Halving both widths therefore reduces the leading signal to one quarter, even with the same precision on each entropy.
A useful choice of widths must keep the pairing fixed while leaving enough signal to resolve it. \\

\noindent
A branch change can mimic a shared edge even with exact measurements.
For example, take $I_1=[0,18]$ and $I_2=[27,45]$ in a common length unit, with exterior widths $a_1=b_2=4$ at the outer ends and $a_2=b_1=1$ at the inner ends.
Compare the product of endpoint separations for the disconnected pairing with that for the connected pairing, in that order.
Before shifting, the products are $18^2=324$ and $45\cdot9=405$, so the two outer ends belong to different geodesics.
Shifting either outer end alone gives $22\cdot18=396$ and $49\cdot9=441$, preserving that assignment.
Shifting both gives $22^2=484$ and $53\cdot9=477$, so the shortest configuration now joins the two outer ends.
Following the original two geodesics throughout would give zero mixed response, since neither sees both shifts.
The actual entropy in the jointly shifted configuration is smaller by $(c/3)\ln(484/477)$ at leading order, which adds this amount to $N_{12}^{(c)}$ because the joint entropy enters with a minus sign.
Using $v_2-u_1=45$ in Eq.~\eqref{eq:pair-strip-denominator} then gives
\[
\lim_{c\to+\infty}R_{12}^{(c)}
=\frac{\ln(484/477)}{\ln[49^2/(45\cdot53)]}>1 .
\]
This ratio exceeds one because it compares the gain from changing the minimizing configuration with the signal from shifting a single edge.
Interpreting it as a partner assignment would join the outer ends in the original union, where its shorter configuration had them on separate edges.

\subsection{Pairing Separation and Propagation of Measurement Errors}
\label{app:pairing-error-proofs}

\noindent
To locate the individual polygons, compare the full matrix of endpoint labels with each admissible pairing.
For two distinct candidates $P$ and $P'$, orthogonality gives $\|P-P'\|_2=\|\Id-P^{\mathsf T}P'\|_2$.
The product $P^{\mathsf T}P'$ compares the two assignments and has at least one nontrivial cycle.
On a cycle of length $k\ge2$, the largest singular value is $2$ for even $k$ and $2\cos[\pi/(2k)]$ for odd $k$, by taking the integer $a$ nearest to $k/2$ in $2|\sin(\pi a/k)|$.
Every odd $k\ge3$ therefore gives a value at least $\sqrt3$, and for $m=2$ the only nontrivial cycle has length two.
It follows that
\begin{equation}
\|P-P'\|_2\ge 2r_m,\qquad
r_m=\begin{cases}
1,&m=2,\\
\sqrt3/2,&m\ge3 .
\end{cases}
\label{eq:pair-permutation-separation}
\end{equation}
Suppose that a certified error $\varepsilon<r_m$ bounds $\|R-P\|_2$ for the true pairing.
For every competitor $P'$, Eq.~\eqref{eq:pair-permutation-separation} and the triangle inequality give $\|R-P'\|_2\ge2r_m-\varepsilon>\varepsilon$.
Minimizing $\|R-P'\|_2$ over the finite set of admissible noncrossing pairings thus selects the unique true pairing.
Because $\sin(\pi/m)\le r_m$ for $m\ge2$, the error bound for Eq.~\eqref{eq:pair-threshold} also suffices to recover the partners from the full matrix.
For example, with six intervals and a certified error of $3/5$, the singular-value count lacks its guarantee since $3/5>\sin(\pi/6)=1/2$.
The matrix is still within $3/5<\sqrt3/2$ of the true pairing, so comparing the individual candidates identifies the partners and hence the area. \\

\noindent
For measured entropies, the geometric correction bound and the instrument errors together determine whether we can certify the area count.
Suppose each measured denominator $\widetilde D_{ij}$ is positive and form $\widetilde R_{ij}=\widetilde N_{ij}/\widetilde D_{ij}$.
Write $n_{ij}=\widetilde N_{ij}-N_{ij}^{(c)}$ and $t_{ij}=\widetilde D_{ij}-D_{ij}^{(c)}$ for the two four-term errors.
Their absolute values are bounded by $4\eta_U$ and $4\eta_I$, respectively, regardless of correlations between the repeated measurements.
Subtracting $P_{ij}$ before division gives the exact relation
\[
\widetilde R_{ij}-P_{ij}
=\frac{\Delta_{ij}e_c+n_{ij}-P_{ij}t_{ij}}{\widetilde D_{ij}}.
\]
Since $P_{ij}$ is either zero or one, a sufficient bound on the operator norm follows by summing the squared entrywise bounds,
\[
\|\widetilde R-P\|_2
\leq
\left[\sum_{i=1}^{m}\sum_{j=1}^{m}
\left(\frac{b_{ij}+4\eta_U+4\eta_I}{\widetilde D_{ij}}\right)^2
\right]^{1/2}.
\]
If this upper bound is below $\sin(\pi/m)$, no allowed error can move a singular value across the counting threshold, and $H_m(\widetilde R)$ gives the exact integer $m-C(P)$.
For known vacuum endpoints and central charge, Eq.~\eqref{eq:pair-strip-denominator} gives the denominators without a measurement so that one can set $\eta_I=0$.
The union entropies still require a bound on their finite-$c$ remainder, either directly on $|\Delta_{ij}e_c|$ or through $b_{ij}=4E_c$. \\

\noindent
A smooth approximation to $H_m$ assigns a weight to each squared singular value.
The weight should vanish at zero and equal one at every positive value allowed for a cycle of length at most $m$.
The finite set of such values determines the roots of a polynomial,
\begin{align}
\Lambda_m&=\{4\sin^2(\pi a/k):2\leq k\leq m,\ 1\leq a<k,\ a,k\in\mathbb Z\},\nonumber\\
p_m(x)&=\prod_{\zeta\in\Lambda_m}(1-x/\zeta),
\qquad A(R)=(\Id-R)(\Id-R)^{\mathsf T}.
\end{align}
Repeated values are included once in the product.
For each real $m\times m$ matrix $R$, define
\begin{equation}
Q_m(R)=\operatorname{Tr}\!\left[A(R)\big(A(R)+p_m(A(R))^2\big)^{-1}\right].
\end{equation}
The identity $v^{\mathsf T}A(R)v=\|(\Id-R)^{\mathsf T}v\|_2^2$ for $v\in\mathbb R^m$ makes all eigenvalues $x$ of $A(R)$ nonnegative.
In an orthonormal eigenbasis, the matrix trace is a sum of the scalar responses $x/[x+p_m(x)^2]$.
This response vanishes at $x=0$, since $p_m(0)=1$, and equals one at every allowed positive value $x\in\Lambda_m$, since $p_m(x)=0$ there.
The denominator stays positive at all $x\geq0$, so the inverse exists for any real $R$ and the trace is real analytic with values in $[0,m]$.
A singular value that starts at zero can grow by $O(\delta_c)$, so its squared value and its weight are both $O(\delta_c^2)$.
For a positive squared value $x=\zeta+O(\delta_c)$, the missed weight is $p_m(x)^2/[x+p_m(x)^2]$.
Since $p_m(\zeta)=0$ and $x$ stays away from zero, this weight loss is also $O(\delta_c^2)$.
Adding the $m$ weights gives $Q_m(R^{(c)})=\nu+O_m(\delta_c^2)$.
The subscript $m$ allows the error constant to depend on the number of intervals.

\subsection{Examples with Several Competing Pairings}
\label{app:pairing-branch-examples}

\noindent
For example, take $I_1=[0,4]$, $I_2=[5,9]$, and $I_3=[10,14]$ in a common length unit.
The pairing $u_1\leftrightarrow v_3$, $u_2\leftrightarrow v_1$, $u_3\leftrightarrow v_2$ joins the two outer ends and the two gaps.
Its product of boundary separations is $14\cdot1\cdot1=14$, smaller than the other four admissible products $36$, $36$, $64$, and $336$.
Its nearest competitor lies higher by $g=\ln(36/14)/3$ in leading entropy per unit central charge.
For all six strip widths equal to $h=1/10$, the shortest separation is $d_0=1$ and $4h/[3(d_0-2h)]=1/6<g$.
The measurements therefore stay on the same minimizing branch.
For the inner endpoints $u_2=5$ and $v_1=4$, the signed separation is $d=-1$, and the reference signal from a single edge is $(c/3)\ln(81/80)$.
Since these endpoints share an edge, the union measurement has this same leading signal.
Pairing $u_2$ with either $v_2$ or $v_3$ instead tests two different edges, so both mixed signals vanish at leading order.
The limiting second row is therefore $(1,0,0)$, and the other two rows give the partner columns $3$ and $2$ for $u_1$ and $u_3$, respectively.
Reading these rows as successive steps gives the walk $u_1\to v_3\to u_3\to v_2\to u_2\to v_1\to u_1$.
It closes after visiting six distinct endpoints, so there is one ideal hexagon with area $(6-2)\pi L^2=4\pi L^2$.
To recover the same number from the ratios, let $\boldsymbol 1=(1,1,1)^{\mathsf T}$ and write $\Id_3$ for the identity of size three.
For this pairing, $P+P^{\mathsf T}$ has a one at each off-diagonal entry and zeros on the diagonal, so
\[
(\Id_3-P)^{\mathsf T}(\Id_3-P)
=2\Id_3-P-P^{\mathsf T}
=3\Id_3-\boldsymbol 1\boldsymbol 1^{\mathsf T}.
\]
The matrix annihilates $\boldsymbol 1$ and multiplies each vector perpendicular to it by three, so the singular values of $\Id_3-P$ are $0,\sqrt3,\sqrt3$.
For this hexagon, two singular values exceed the threshold $\sin(\pi/3)$, giving two area units.
For example, if all nine normalized entries have absolute errors at most $1/5$, the operator norm of the error is at most $3/5<\sin(\pi/3)$.
The same count, and hence the same area $4\pi L^2$, then follows from those imprecise entries.
This result uses the disk homology class and pure vacuum geometry.
Additional winding, horizon, or brane boundaries change the area calculation. \\

\noindent
For two intervals, the cross ratio is
\begin{equation}
x=\frac{(v_1-u_1)(v_2-u_2)}{(u_2-u_1)(v_2-v_1)},\qquad 0<x<1 .
\end{equation}
Writing the two products as $(v_1-u_1)(v_2-u_2)$ for the disconnected pairing and $(v_2-u_1)(u_2-v_1)$ for the connected pairing gives their ratio $x/(1-x)$.
For $x<1/2$, the disconnected product is smaller; its two edges are the individual-interval edges, and the uncovered area vanishes.
For $x>1/2$, the connected product is smaller, and its two-cycle encloses a quadrilateral of area $2\pi L^2$ \cite{Headrick:2010zt}.
At $x=1/2$, the two pairings have the same leading entropy on $U$.
The outward shifts used here select a definite side of the crossing.
Indeed, let $l_1=v_1-u_1$, $l_2=v_2-u_2$, and $s=u_2-v_1>0$.
Then
\[
x=\frac{l_1}{l_1+s}\frac{l_2}{l_2+s}.
\]
An outer-end shift increases one length, $l_1$ or $l_2$, while keeping the gap $s$ fixed.
An inner-end shift increases that length and decreases $s$.
Both factors lie between zero and one, and every nonzero allowed shift strictly increases their product.
Every shifted region therefore selects the connected pairing, which is also a minimizer on the original union and so qualifies for Eq.~\eqref{eq:top_exact}.
The unshifted entropy alone leaves the uncovered area ambiguous between zero and $2\pi L^2$.
The finite shifts specify a side of the transition and measure the quadrilateral on that side.

\subsection{Convergence Rates and Dirac-Sector Example}
\label{app:detailG4}

\noindent
A small singular value in finite-$c$ data can be the displaced image of an exact zero.
Consider counting each singular value $s$ with weight $s^2/(s^2+c^{-1})$, which is one half at $s=c^{-1/2}$.
If this scale falls faster than the displacement size, the prescription counts a vacuum zero as a positive signal.
To quantify this effect, evaluate the resulting trace $Q_c^{\rm soft}(R)=\operatorname{Tr}[A(R)(A(R)+c^{-1}\Id)^{-1}]$ on the two-interval matrices
\begin{equation}
P=\Id_2,\qquad R^{(c)}=\Id_2+c^{\alpha-1}(J_2-\Id_2),\qquad 0<\alpha<1 ,
\end{equation}
where $\Id_2$ is the two-dimensional identity and every entry of $J_2$ is one.
The two singular values of $J_2-\Id_2$ are one, so both eigenvalues of $A(R^{(c)})$ equal $c^{2\alpha-2}$.
As $c\to+\infty$, this gives
\begin{equation}
Q_c^{\rm soft}(R^{(c)})
=\frac{2 c^{2\alpha-1}}{1+c^{2\alpha-1}}
\longrightarrow
\begin{cases}
0,&0<\alpha<1/2,\\
1,&\alpha=1/2,\\
2,&1/2<\alpha<1 .
\end{cases}
\end{equation}
In all three cases, $R^{(c)}$ approaches the identity and both singular values of $\Id_2-R^{(c)}$ vanish in the limit. Hence, neither direction contributes a vacuum area unit.
At $\alpha=1/2$, each signal equals the moving cutoff $c^{-1/2}$, so the smooth rule counts two half units as one spurious area unit.
For $\alpha>1/2$, the cutoff shrinks faster than either signal, and both contribute a full unit in the limit.
For fixed $m$, suppose $R^{(c)}\to P$ in operator norm as $c\to+\infty$, and order the singular values $s_a(c)=\sigma_a(\Id-R^{(c)})$ from largest to smallest for $1\leq a\leq m$.
Recall that $\nu=m-C(P)$.
The nonzero limiting singular values have indices $1\leq a\leq\nu$.
For sufficiently large $c$, each such $s_a(c)$ is at least $\gamma_m/2$, so its contribution $1-[1+c s_a(c)^2]^{-1}$ differs from one by at most $4/(c\gamma_m^2)$.
Subtracting the limiting rank from the trace gives
\[
Q_c^{\rm soft}(R^{(c)})-\nu
=\sum_{a=\nu+1}^{m}\frac{c s_a(c)^2}{1+c s_a(c)^2}
-\sum_{a=1}^{\nu}\frac{1}{1+c s_a(c)^2},
\]
where a sum over no indices equals zero.
The second sum tends to zero because its singular values stay bounded away from zero.
The first has a fixed number $C(P)\geq1$ of nonnegative terms, so it tends to zero precisely when its largest term, the one from $s_{\nu+1}(c)$, does.
Consequently, $Q_c^{\rm soft}(R^{(c)})\to\nu$ if and only if $\sqrt c\,s_{\nu+1}(c)\to0$.
If only the full matrix error is known, $s_{\nu+1}(c)\leq\delta_c$ gives the sufficient condition $c\delta_c^2\to0$. \\

\noindent
An independent set of free boundary fields can produce just such a slowly decaying signal in a physical entropy measurement.
For independent massless Dirac fields with total central charge $c_{\rm f}$, the vacuum multi-interval entropy on the line takes the form \cite{Casini:2008wt}
\begin{equation}
S_{\rm f}(U)=\frac{c_{\rm f}}3\left[
\sum_{i,j=1}^{m}\ln|v_j-u_i|
-\sum_{1\leq i<j\leq m}\ln|u_j-u_i|
-\sum_{1\leq i<j\leq m}\ln|v_j-v_i|-m\ln\epsilon\right].
\end{equation}
All lengths are expressed in a common unit.
Every term involving only left endpoints or only right endpoints drops out of $\Delta_{ij}S_{\rm f}$.
Of the mixed logarithms, only $\ln|v_j-u_i|$ depends on both shifted endpoints.
It gives exactly the single-interval denominator, so the Dirac sector produces a one for every pair, including pairs that share no edge in the leading gravitational geometry.
Let an independent sector of central charge $c_{\rm h}$ have ratio matrix $R_{\rm h}$ on the same configurations.
The numerators and denominators add separately in the product state, and Eq.~\eqref{eq:pair-strip-denominator} weights the two denominators in proportion to their central charges.
Thus, with $c=c_{\rm h}+c_{\rm f}$ and $J_m$ the all-ones matrix,
\begin{equation}
R^{(c)}=(1-\eta_c)R_{\rm h}+\eta_c J_m,\qquad
\eta_c=c_{\rm f}/c .
\end{equation}
In a two-interval disconnected phase, suppose $R_{\rm h}=\Id_2+O(1/c_{\rm h})$ and take $c_{\rm f}\sim c^\alpha$ with $1/2<\alpha<1$.
Then, $\Id_2-R^{(c)}=-\eta_c(J_2-\Id_2)+o(\eta_c)$, and both singular values equal $\eta_c+o(\eta_c)$.
The leading pairing is still the identity, but $Q_c^{\rm soft}\to2$ since $c\eta_c^2\to\infty$.
The fixed threshold in Eq.~\eqref{eq:pair-threshold} still gives zero in this example for sufficiently large $c$. \\

\noindent
For a general pairing with fixed $m\ge2$, suppose $\|R_{\rm h}-P\|_2=O(c_{\rm h}^{-1})$ and set $R_0=(1-\eta_c)P+\eta_c J_m$.
A vector fixed by $P$ is constant on each cycle.
One such vector is constant on all intervals, and the remaining $C(P)-1$ independent directions can be chosen so their entries sum to zero.
The all-ones matrix $J_m$ multiplies the constant vector by $m$ and annihilates each zero-sum vector.
On these directions, the singular values of $\Id-R_0$ are therefore $(m-1)\eta_c$ once and $\eta_c$ with multiplicity $C(P)-1$.
The remaining directions are the nonconstant Fourier modes of the cycles, on which $J_m$ vanishes.
A mode with eigenvalue $\exp(\mathrm{i}\varphi)$ there has singular value $|1-(1-\eta_c)\exp(\mathrm{i}\varphi)|$ for $\Id-R_0$, which stays positive as $\eta_c\to0$.
These subspaces are orthogonal and preserved by the matrix and its transpose, so they account for all $m$ singular values. \\

\noindent
Suppose $c_{\rm f}/\sqrt c\to\ell$ for a finite $\ell\ge0$, with $c_{\rm h}\to\infty$ and $m$ fixed.
Then $\eta_c\to0$ and $\sqrt c\,\eta_c\to\ell$.
The exact product ratio differs from $R_0$ by $(1-\eta_c)(R_{\rm h}-P)$, whose norm is $O(c^{-1})$ because $1-\eta_c=c_{\rm h}/c$.
Multiplying the resulting singular-value errors by $\sqrt c$ still gives a vanishing error.
Substitution in the trace yields
\[
\lim_{c\to+\infty}Q_c^{\rm soft}(R^{(c)})
=\nu+\frac{(m-1)^2\ell^2}{1+(m-1)^2\ell^2}
+(C(P)-1)\frac{\ell^2}{1+\ell^2}.
\]
If instead $c_{\rm f}/\sqrt c\to\infty$ while $c_{\rm f}/c\to0$, each of the $C(P)$ small singular values contributes 1 to the limiting trace, giving $m$.
In both scalings, the limiting pairing is $P$, and the dimensionless vacuum area inferred from the fixed threshold is still $2\pi\nu$ under the disk hypotheses.
The rate assumption on the independent sector is additional input to this example.

\subsection{Relation to Energy Reconstruction}
\label{app:scope}
\label{app:detailH1}

\noindent
Let $\Omega_0$ be the union of the nondegenerate polygons traced out by the cycles of the vacuum pairing $P$.
For a cycle of length $k$, Eq.~\eqref{eq:GBexact} gives the vacuum area $(2k-2)\pi L^2$, and summing over cycles gives Eq.~\eqref{eq:pairing-vacuum-area} \cite{Fujiki:2026ucr,Abt:2017pmf,Abt:2018ywl}.
Equal cycle counts can therefore give equal vacuum areas at different positions in the disk.
When matter is present, the endpoint labels tell us which part of the energy distribution lies inside $\Omega_0$. \\

\noindent
Now follow the same RT branch in a smooth coherent scalar family and let $A_\Omega(\lambda)$ be the sum of the reflection-slice areas of these components.
The topology term in Eq.~\eqref{eq:GBexact} is constant along this branch.
Since $\rho_\lambda=\lambda^2 e/L^3+o(\lambda^2)$, its integral starts at order $\lambda^2$ and the moving edges can be replaced by their vacuum positions at this order under the falloff assumptions of Appendix~\ref{app:detailA6}.
Substituting $dA_0=L^2d\mu$ and $8\pi G_N/L=12\pi/c$ gives
\begin{equation}
\frac{A_\Omega(\lambda)}{L^2}
=2\pi[m-C(P)]
+\frac{12\pi}{c}\lambda^2\int_{\Omega_0} e\,d\mu
+o(\lambda^2).
\label{eq:cycle-matter-expansion}
\end{equation}
When the reflection polygons maximize area in the specified Cauchy classes throughout the family, the same expression gives their total maximal volume $V_\Omega(\lambda)$.
Its Hessian is therefore $V_\Omega''(0)/L^2=(24\pi/c)\int_{\Omega_0}e\,d\mu$.
In this formula, the vacuum endpoints and $P$ specify the integration region, and the continuously varied single-interval entropies determine its energy integral through Eq.~\eqref{eq:boundarywindow}.
For the centered, rotationally invariant $\Delta=3/2$ family of definite parity, the two-width data of Sec.~\ref{sec:square-calibration} suffice for this last step.
A coherent excitation also changes the whole-slice volume \cite{Bernamonti:2020bcf}.
For a polygon, the relative ordering of the field and momentum responses depends on how much of each phase's energy distribution lies inside its edges. \\

\noindent
Even complete chord data leave an ambiguity about the scalar that produced them.
Consider two nonzero smooth real momentum packets $f_1$ and $f_2$ supported in disjoint compact regions of the disk.
Set the initial field $\phi_1=0$ in both cases and choose the rescaled momentum to be either $\pi_\phi^{(1)}=f_1+f_2$ or $\pi_\phi^{(2)}=f_1-f_2$.
Both choices give odd scalar reflection parity and zero spatial momentum flux on the slice.
Reversing the sign on the second packet would change the energy by $(L/2)[(f_1+f_2)^2-(f_1-f_2)^2]=2L f_1f_2$ in Eq.~\eqref{eq:energy-density}.
This product vanishes at every point because the packets do not overlap.
An overall sign change cannot reverse only the second packet, so two distinct initial conditions for scalar evolution have the same chord integrals and polygon-area Hessians. \\

\noindent
An interval has two endpoints, so an entropy measured for all intervals is a function on a two-dimensional space.
Equations on this space connect endpoint variations to a source in the bulk \cite{Nozaki:2013vta,Czech:2016tqr,deBoer:2015kda}.
Earlier reconstructions use small boundary regions to measure energy \cite{Lin:2014hva}, while quantum Fisher information, which measures the distinguishability of nearby states to quadratic order, is related to canonical energy \cite{Lashkari:2015hha}.
Mixed endpoint derivatives also define the Crofton measure, which weights geodesics when expressing a curve length as an integral over its intersections with them \cite{Czech:2015qta}.
Related inverse problems recover homogeneous stationary metrics and metric perturbations near AdS$_3$ from interval data \cite{Chae:2026wnp,Cao:2020uvb}.
Here, a first-order change of boundary stress can accompany the scalar excitation, so we use the second-order stress kernel and vacuum descendant relative entropy to subtract its contribution \cite{Beach:2016ocq,Hollands:2019czd,Panebianco:2019plp}.
For a volume regulated below a single RT surface, a connection with Fisher information has been proposed \cite{Banerjee:2017qti}.
To predict a polygon response for a general density, the interval centers and widths must vary so that the geodesics pass through different parts of the polygon.
The local entropy source permits this calculation using entropy measurements alone.
For the two scalar phases, the direct area calculation follows the moving edges in the bulk. At the same time, the CFT two-point function gives the entropy responses at the boundary.

\subsection{Normalized Conditional Mutual Information and Recovery}
\label{app:detailH2}

\noindent
On a fixed vacuum branch, a normalized response resolved on the zero side of the threshold identifies endpoints on separate geodesics.
To interpret the same ratio as a statement about recovering quantum information, one must specify what the receiver knows and which correlations the recovered state should retain.
Consider two independent strings of fair bits, $Z$ of length $n$ and $W$ of length $k$, with $n,k$ positive integers.
Keep a reference copy $X=(Z,W)$ inaccessible and ask a receiver with $B=Z$ to recreate a second copy $Y=(Z,W)$.
To realize the two entropy measurements for this example, put $X$ and $Y$ in the two strips $X_1$ and $Y_1$, a register with a fixed value in $I_1$, and $Z$ in $I_2$.
Additional unobserved registers can purify the state.
In the numerator of Eq.~\eqref{eq:top_def}, conditioning on $I_1$ and $I_2$ reveals $Z$ and leaves $k$ bits shared by the strips.
In the denominator, conditioning on $I_1$ alone reveals neither string and leaves $n+k$ shared bits.
Expressing one bit as $\ln 2$ in natural entropy units gives
\begin{equation}
N_{11}=k\ln2,\qquad D_{11}=(n+k)\ln2,\qquad
R_{11}=\frac{k}{n+k}.
\end{equation}
Thus, $R_{11}$ tells us the fraction of the shared bits that are missing.
A fraction of one half, for example, could represent one missing bit in two or fifty missing bits in a hundred. \\

\noindent
Copying $z$ and drawing a fresh uniform $w$ gives $Y$ exactly the right distribution when examined alone.
The obstruction is that $B$ holds no record of the original $w$ so that the new string can agree with the one in the inaccessible $X$ only by chance.
For a quantum channel $\mathcal R:B\to BY$, denote its output given $z\in\{0,1\}^n$ by $\sigma_z=\mathcal R(|z\rangle\langle z|)$.
The output depends only on $z$, while the target $|t_{z,w}\rangle=|z\rangle_B|z,w\rangle_Y$ depends on $w\in\{0,1\}^k$ as well.
For target $\tau$ and output $\sigma$, the root fidelity is $F(\tau,\sigma)=\operatorname{Tr}\sqrt{\sqrt\tau\,\sigma\sqrt\tau}$, equal to one for identical states and zero for orthogonal supports.
The untouched copy in $X$ labels orthogonal blocks by $(z,w)$, each with probability $2^{-(n+k)}$.
Within each block, the target is a definite string, so its root fidelity with $\sigma_z$ is the square root of the probability the output assigns to that string.
Averaging gives
\[
F\!\left(\rho_{XBY},(\mathrm{id}_X\otimes\mathcal R)(\rho_{XB})\right)
=2^{-(n+k)}\sum_{z\in\{0,1\}^n}\sum_{w\in\{0,1\}^k}
\sqrt{\langle t_{z,w}|\sigma_z|t_{z,w}\rangle},
\]
where $\mathrm{id}_X$ is the identity channel on $X$.
For each $z$, the $2^k$ distinct targets are orthogonal, so their probabilities in $\sigma_z$ sum to at most one.
Cauchy--Schwarz bounds the sum of square roots by $2^{k/2}$, so averaging over the $2^k$ strings gives $F\leq2^{-k/2}$ for every recovery channel.
A receiver who keeps $z$ and draws $w$ uniformly assigns probability $2^{-k}$ to every target and attains the bound.
Thus, the optimal root fidelity is $2^{-k/2}=\exp[-I(X:Y\mid B)/2]$, which saturates the recovery inequality in this example \cite{Fawzi:2015eym}.
Increasing the number $n$ of known bits drives $R_{11}$ to zero both when $k=1$ and when $k=\lfloor\sqrt n\rfloor$.
The optimal fidelity stays at $1/\sqrt2$ in the first case. It tends to zero in the second case because, in the second family, the number of missing bits grows. \\

\noindent
A receiver who can communicate with a sender faces a different problem.
In entanglement-assisted state redistribution of a finite-dimensional pure state on $XYBA$, the sender holds $YA$ and must transfer $Y$ to the receiver holding $B$, preserving the correlations with the inaccessible $X$.
The optimal asymptotic rate is $I(X:Y\mid B)/(2\ln2)$ communicated qubits per copy \cite{Devetak:2008oqm}.
If one groups $r$ independent copies into a single task, for any positive integer $r$, the ratio stays at $k/(n+k)$ and the number of missing bits rises to $rk$.
The same block calculation gives optimal root fidelity $2^{-rk/2}$ even for a joint recovery channel. In contrast, additivity gives rate $rk/2$ qubits per grouped copy.
Knowing only the fraction thus loses the scale needed for either prediction.
For a measured endpoint pair with $D_{ij}^{(c)}>0$, keeping the denominator restores that scale, since the conditional mutual information in natural units is $N_{ij}^{(c)}=D_{ij}^{(c)}R_{ij}^{(c)}$.
With a finite-dimensional regulator, recovery of $Y_j$ from $U$ while $X_i$ remains inaccessible has optimal root fidelity at least $\exp[-D_{ij}^{(c)}R_{ij}^{(c)}/2]$ \cite{Fawzi:2015eym}.
For the same state, the redistribution protocol with purification and entanglement assistance has the exact rate $D_{ij}^{(c)}R_{ij}^{(c)}/(2\ln 2)$ communicated qubits per copy. \\

\noindent
Even after that scale is restored, the conditional mutual information need not determine the optimal recovery fidelity.
Keep $B=Z$ and $X=Y=(Z,W)$, but take finite alphabets $\mathcal Z$ and $\mathcal W$ with a general distribution $p_z p_{w|z}$, where $\sum_{z\in\mathcal Z}p_z=1$ and $\sum_{w\in\mathcal W}p_{w|z}=1$ for each $z$ of positive probability.
For a probability vector $p$ on a finite alphabet $\mathcal A$, write $H(p)=-\sum_{a\in\mathcal A}p_a\ln p_a$ with $0\ln0=0$.
Because $X$ and $Y$ are identical copies, conditioning on $Z$ leaves $N_{11}=\sum_{z:p_z>0}p_zH(p_{\cdot|z})$, while the deterministic denominator register gives $D_{11}=H(p_Z)+N_{11}$.
For each observed $z$, define $q_{w|z}=\langle t_{z,w}|\sigma_z|t_{z,w}\rangle$.
Orthogonality of the targets gives $\sum_{w\in\mathcal W}q_{w|z}\leq1$.
The block fidelity maximized by the channel at that $z$ obeys
\[
\sum_{w\in\mathcal W}p_{w|z}\sqrt{q_{w|z}}
\leq\left(\sum_{w\in\mathcal W}p_{w|z}^2\right)^{1/2}
\left(\sum_{w\in\mathcal W}q_{w|z}\right)^{1/2}.
\]
To attain equality at fixed $z$, choose $\sqrt{q_{w|z}}$ proportional to $p_{w|z}$ and normalize the probabilities to one.
This gives $q_{w|z}=p_{w|z}^2/\sum_{w'\in\mathcal W}p_{w'|z}^2$.
The receiver realizes it by keeping the observed value $z$ and sampling a fresh $w$ with these probabilities, giving greater weight to the more likely strings.
Substitution gives the largest attainable root fidelity,
\begin{equation}
F_{\rm opt}=\sum_{z:p_z>0}p_z
\left(\sum_{w\in\mathcal W}p_{w|z}^2\right)^{1/2}.
\label{eq:copied-register-optimal-fidelity}
\end{equation}
Here, $z$ ranges over the positive-probability values in $\mathcal Z$.
The optimal fidelity depends on the sum of squared conditional probabilities in each branch. In contrast, the conditional mutual information uses the average of their Shannon entropies. \\

\noindent
For an explicit comparison, take $Z$ to be a fair bit and $\mathcal W=\{0,1,2,3\}$.
In the first state, $W$ is uniform on $\{0,1\}$ for either value of $Z$.
In the second state, $W=0$ when $Z=0$ and $W$ is uniform on all four values when $Z=1$.
Thus, the missing information is always one bit in the first state and is zero or two bits with equal probability in the second.
These averages give the same $N_{11}=\ln2$, $D_{11}=2\ln2$, ratio $1/2$, and redistribution rate of one half qubit per copy.
For the first state, Eq.~\eqref{eq:copied-register-optimal-fidelity} gives root fidelity $1/\sqrt2$ on both branches.
For the second, the receiver knows the answer on one branch and guesses among four equally likely answers on the other, giving average root fidelity $(1+1/2)/2=3/4$.
For both states, the receiver lacks one bit on average, giving the same assisted transfer rate.
In the second state, however, half the observed values of $Z$ identify the missing value of $W$ with certainty, which raises its average recovery fidelity.
A polygon cycle count combines several such entropy ratios and specifies neither a recovery map nor which registers a single receiver could use.

\subsection{Domain of Classical Results}
\label{app:detailH3}

\noindent
To interpret a reflection-slice area as a maximal volume, we specify which spacelike fillings compete, require a regular causal development, and impose the timelike Ricci convergence condition used in the comparison theorem \cite{Headrick:2014cta,Burnett:1995rf}.
On a slice without time-reflection symmetry, its extrinsic curvature also contributes to the constraint.
A nonzero matter background, a noncoherent quantum state, different boundary falloff, or a higher-curvature action can change the relation between the entropy variation and canonical energy.
For sampled data, reconstructing a continuum density also requires quantitative information about its smoothness and the measurement error.
At finite $c$, bulk quantum entropy contributes to the boundary entropy, and more general mixed states require a prescription for the corresponding bulk region \cite{Fujiki:2026ucr}.

\end{document}